\PassOptionsToPackage{unicode,bookmarksnumbered,pdfpagelabels,pagebackref,hypertexnames=false,linktocpage=true}{hyperref}
\PassOptionsToPackage{hyphens}{url}
\PassOptionsToPackage{dvipsnames,svgnames,x11names}{xcolor}
\documentclass[
  english,
  12pt,
  letterpaper,
  phd]{ucbthesis}
\usepackage{xcolor}
\usepackage{amsmath,amssymb}
\usepackage{iftex}
\ifPDFTeX
  \usepackage[T1]{fontenc}
  \usepackage[utf8]{inputenc}
  \usepackage{textcomp} % provide euro and other symbols
\else % if luatex or xetex
  \usepackage{unicode-math} % this also loads fontspec
  \defaultfontfeatures{Scale=MatchLowercase}
  \defaultfontfeatures[\rmfamily]{Ligatures=TeX,Scale=1}
\fi
\usepackage{lmodern}
\ifPDFTeX\else
\fi
\IfFileExists{upquote.sty}{\usepackage{upquote}}{}
\IfFileExists{microtype.sty}{% use microtype if available
  \usepackage[]{microtype}
  \UseMicrotypeSet[protrusion]{basicmath} % disable protrusion for tt fonts
}{}
\makeatletter
\@ifundefined{KOMAClassName}{% if non-KOMA class
  \IfFileExists{parskip.sty}{%
    \usepackage{parskip}
  }{% else
    \setlength{\parindent}{0pt}
    \setlength{\parskip}{6pt plus 2pt minus 1pt}}
}{% if KOMA class
  \KOMAoptions{parskip=half}}
\makeatother
\usepackage{color}
\usepackage{fancyvrb}

\DefineVerbatimEnvironment{Highlighting}{Verbatim}{commandchars=\\\{\}}
\newenvironment{Shaded}{}{}

\newcommand{\AttributeTok}[1]{\textcolor[rgb]{0.49,0.56,0.16}{#1}}

\newcommand{\BuiltInTok}[1]{\textcolor[rgb]{0.00,0.50,0.00}{#1}}
\newcommand{\CharTok}[1]{\textcolor[rgb]{0.25,0.44,0.63}{#1}}
\newcommand{\CommentTok}[1]{\textcolor[rgb]{0.38,0.63,0.69}{\textit{#1}}}

\newcommand{\ControlFlowTok}[1]{\textcolor[rgb]{0.00,0.44,0.13}{\textbf{#1}}}

\newcommand{\DecValTok}[1]{\textcolor[rgb]{0.25,0.63,0.44}{#1}}

\newcommand{\ExtensionTok}[1]{#1}
\newcommand{\FloatTok}[1]{\textcolor[rgb]{0.25,0.63,0.44}{#1}}
\newcommand{\FunctionTok}[1]{\textcolor[rgb]{0.02,0.16,0.49}{#1}}

\newcommand{\KeywordTok}[1]{\textcolor[rgb]{0.00,0.44,0.13}{\textbf{#1}}}
\newcommand{\NormalTok}[1]{#1}
\newcommand{\OperatorTok}[1]{\textcolor[rgb]{0.40,0.40,0.40}{#1}}
\newcommand{\OtherTok}[1]{\textcolor[rgb]{0.00,0.44,0.13}{#1}}

\newcommand{\SpecialCharTok}[1]{\textcolor[rgb]{0.25,0.44,0.63}{#1}}

\newcommand{\StringTok}[1]{\textcolor[rgb]{0.25,0.44,0.63}{#1}}
\newcommand{\VariableTok}[1]{\textcolor[rgb]{0.10,0.09,0.49}{#1}}

\usepackage{longtable,booktabs,array}
\usepackage{caption}
\usepackage{calc} % for calculating minipage widths
\usepackage{etoolbox}
\makeatletter
\patchcmd\longtable{\par}{\if@noskipsec\mbox{}\fi\par}{}{}
\makeatother
\usepackage{graphicx}
\makeatletter
\newsavebox\pandoc@box
\newcommand*\pandocbounded[1]{% scales image to fit in text height/width
  \sbox\pandoc@box{#1}%
  \Gscale@div\@tempa{\textheight}{\dimexpr\ht\pandoc@box+\dp\pandoc@box\relax}%
  \Gscale@div\@tempb{\linewidth}{\wd\pandoc@box}%
  \ifdim\@tempb\p@<\@tempa\p@\let\@tempa\@tempb\fi% select the smaller of both
  \ifdim\@tempa\p@<\p@\scalebox{\@tempa}{\usebox\pandoc@box}%
  \else\usebox{\pandoc@box}%
  \fi%
}
\def\fps@figure{htbp}
\makeatother
\usepackage[inkscape=false]{svg}
\NewDocumentCommand\citeproctext{}{}
\NewDocumentCommand\citeproc{mm}{%
  \begingroup\def\citeproctext{#2}\cite{#1}\endgroup}
\makeatletter
 \let\@cite@ofmt\@firstofone
 \def\@biblabel#1{}
 \def\@cite#1#2{{#1\if@tempswa , #2\fi}}
\makeatother
\newlength{\cslhangindent}
\newlength{\csllabelwidth}
\newenvironment{CSLReferences}[2] % #1 hanging-indent, #2 entry-spacing
 {\begin{list}{}{%
  \setlength{\itemindent}{0pt}
  \setlength{\leftmargin}{0pt}
  \setlength{\parsep}{0pt}
  \ifodd #1
   \setlength{\leftmargin}{\cslhangindent}
   \setlength{\itemindent}{-1\cslhangindent}
  \fi
  \setlength{\itemsep}{#2\baselineskip}}}
 {\end{list}}
\usepackage{calc}
\newcommand{\CSLBlock}[1]{\hfill\break\parbox[t]{\linewidth}{\strut\ignorespaces#1\strut}}

\ifLuaTeX
\usepackage[bidi=basic,shorthands=off]{babel}
\else
\usepackage[bidi=default,shorthands=off]{babel}
\fi
\ifPDFTeX
\else
\babelfont{rm}[Extension=.otf,UprightFont=*-regular,BoldFont=*-bold,ItalicFont=*-italic,BoldItalicFont=*-bolditalic]{texgyreschola}
\fi
\ifLuaTeX
  \usepackage{selnolig} % disable illegal ligatures
\fi
\providecommand{\tightlist}{%
  \setlength{\itemsep}{0pt}\setlength{\parskip}{0pt}}
\usepackage{amsthm}
\usepackage{amsthm}
\usepackage{mathtools}

\usepackage{physics}
\usepackage{siunitx}

\usepackage{pgfplots}
\usepgfplotslibrary{groupplots,dateplot}
\usetikzlibrary{patterns,shapes.arrows}
\pgfplotsset{compat=newest}
\newcommand{\hideFromPandoc}[1]{#1}
\hideFromPandoc{
    \let\Begin\begin
    \let\End\end
}
\cftsetindents{part}{0em}{2.5em}
\cftsetindents{chapter}{0em}{2.5em}
\setpnumwidth{1.8em}
\setrmarg{2.8em}
\usepackage{fvextra}
\usepackage{newunicodechar}
\newunicodechar{^^^^2009}{\,}
\begingroup
    \catcode`θ=\active \gdef θ{{\rmfamily\char"03B8}}
    \catcode`φ=\active \gdef φ{{\rmfamily\char"03C6}}
    \catcode`π=\active \gdef π{{\rmfamily\char"03C0}}
\endgroup
\newcommand{\greekincode}{\catcode`θ=\active \catcode`φ=\active \catcode`π=\active}
\DefineVerbatimEnvironment{Highlighting}{Verbatim}{breaklines,breakafter=/,breaknonspaceingroup,commandchars=\\\{\},codes=\greekincode}
\theoremstyle{plain}
\newtheorem{Theorem}{Theorem}

\newtheorem{Corollary}[Theorem]{Corollary}

\theoremstyle{definition}
\newtheorem{Definition}{Definition}
\newtheorem{Assumption}{Assumption}
\theoremstyle{remark}

\makeatletter
\@ifpackageloaded{subcaption}{}{\usepackage{subcaption}}
\@ifpackageloaded{caption}{}{\usepackage{caption}}
\AtBeginDocument{%

}
\AtBeginDocument{%

}
\newcounter{pandoccrossref@subfigures@footnote@counter}
\newenvironment{pandoccrossrefsubfigures}{%
\setcounter{pandoccrossref@subfigures@footnote@counter}{0}
\begin{figure}\centering%
\gdef\global@pandoccrossref@subfigures@footnotes{}%
\DeclareRobustCommand{\footnote}[1]{\footnotemark%
\stepcounter{pandoccrossref@subfigures@footnote@counter}%
\ifx\global@pandoccrossref@subfigures@footnotes\empty%
\gdef\global@pandoccrossref@subfigures@footnotes{{##1}}%
\else%
\g@addto@macro\global@pandoccrossref@subfigures@footnotes{, {##1}}%
\fi}}%
{\end{figure}%
\addtocounter{footnote}{-\value{pandoccrossref@subfigures@footnote@counter}}
\@for\f:=\global@pandoccrossref@subfigures@footnotes\do{\stepcounter{footnote}\footnotetext{\f}}%
\gdef\global@pandoccrossref@subfigures@footnotes{}}
\@ifpackageloaded{float}{}{\usepackage{float}}
\floatstyle{ruled}
\@ifundefined{c@chapter}{\newfloat{codelisting}{h}{lop}}{\newfloat{codelisting}{h}{lop}[chapter]}
\floatname{codelisting}{Listing}

\AtEndPreamble{%
\@ifpackageloaded{cleveref}{}{\usepackage{cleveref}}
\crefname{figure}{fig.}{figs.}
\Crefname{figure}{Fig.}{Figs.}
\crefname{table}{tbl.}{tbls.}
\Crefname{table}{Tbl.}{Tbls.}
\crefname{equation}{eq.}{eqns.}
\Crefname{equation}{Eq.}{Eqns.}
\crefname{listing}{lst.}{lsts.}
\Crefname{listing}{Lst.}{Lsts.}
\crefname{section}{sec.}{secs.}
\Crefname{section}{Sec.}{Secs.}
\crefname{codelisting}{\cref@listing@name}{\cref@listing@name@plural}
\Crefname{codelisting}{\Cref@listing@name}{\Cref@listing@name@plural}
}
\makeatother
\usepackage{bookmark}
\IfFileExists{xurl.sty}{\usepackage{xurl}}{} % add URL line breaks if available
\makeatletter
\@ifundefined{xmpquote}{\newcommand{\xmpquote}[1]{#1}}{}
\makeatother
\hypersetup{
  pdftitle={CMB Data Analysis in the example of the POLARBEAR experiment and the LiteBIRD experiment},
  pdfauthor={Kolen Cheung},
  pdflang={en},
  pdfkeywords={\xmpquote{dissertation}, \xmpquote{PhD}, \xmpquote{DE-CDSE}, \xmpquote{CMB}, \xmpquote{B-mode}, \xmpquote{POLARBEAR}, \xmpquote{LiteBIRD}},
  colorlinks=true,
  linkcolor={blue},
  filecolor={blue},
  citecolor={blue},
  urlcolor={blue},
  pdfcreator={LaTeX via pandoc}}

\ifXeTeX

\fi
\ifPDFTeX
\pdftrailerid{}
\pdftrailer{/ID [ <7e049db3d997ec98b644a35638fe3d9f> <7e049db3d997ec98b644a35638fe3d9f> ]}
\fi
\ifLuaTeX
\pdfvariable trailerid {[ <7e049db3d997ec98b644a35638fe3d9f> <7e049db3d997ec98b644a35638fe3d9f> ]}
\fi

\title{CMB Data Analysis in the example of the POLARBEAR experiment and the LiteBIRD experiment}
\author{Kolen Cheung}
\degreesemester{Fall}
\degreeyear{2022}
\degree{Doctor of Philosophy}
\cochairs{Professor Adrian Lee}{Doctor Julian Borrill}{}
\othermembers{Professor Katherine Yelick \\ Professor Matt Pyle \\ }
\numberofmembers{4}
\field{Physics}
\emphasis{Computational \& Data Science \& Engineering}
\campus{Berkeley}

\begin{document}
\maketitle
\copyrightpage
\begin{abstract}
In this dissertation, I briefly go through
my subject areas---from Astrophysics, Cosmology, to Cosmic Microwave Background (CMB) radiation;
and Data Science \& High-Performance Computing.
I go through the beginning of the Universe briefly in \cref{sec:beginning-universe}
and introduce the Physics of the CMB in \cref{sec:CMB-radiation}.
Then I move on to how it can be observed experimentally, mathematically, and computationally in \cref{sec:CMB-observations}.
In \cref{sec:POLARBEAR,sec:polarbear-research,sec:LiteBIRD-research}, I write about 2 specific CMB experiments---POLARBEAR \& LiteBIRD, together with my research involved in them.

In \cref{sec:polarbear-research}, original research on a measurement of the CMB \(B\)-mode
at sub-degree scales is presented using the 3rd--5th seasons of POLARBEAR observations.
This demonstrates the capability of a single medium aperture telescope with a Continuously Rotating Half-Wave Plate (CRHWP) to target both the low-\(\ell\) and high-\(\ell\) \(B\)-mode science at \(50 \leq \ell \leq 3000\).

In \cref{sec:LiteBIRD-research}, the effect of detector crosstalk systematics on the CMB power-spectra is studied in detail in the example of the LiteBIRD experiment.
An analytical, map-domain solution is developed,
a computational simulation method is implemented,
and the two predictions are compared
using the LiteBIRD experiment design specification.
The general agreement of the two shows that the analytical model
well describes the LiteBIRD experiment,
and can be used to calculate the systematic effects without performing
time-consuming time-domain simulations with a high degree of accuracy.
This can be used to forecast the amount of crosstalk systematics
and to develop mitigation strategies.
A mitigation strategy is presented,
which is a hardware-based mitigation method
that minimizes the crosstalk systematic effects in the 2pt power-spectra.

While the author claims that this dissertation is original, unpublished, and is an independent work,
only those in \cref{sec:polarbear-research,sec:LiteBIRD-research} are original research.
Other materials in \cref{sec:beginning-universe,sec:CMB-radiation,sec:CMB-observations,sec:POLARBEAR}
are introductory materials summarizing well-known knowledge in the field.
In particular, many plots, graphs and visualizations from these parts are from existing publications
which should fall under ``fair use'' and are properly cited.
There are a few exceptions that images are obtained through internal circulations
including presentations from others in the collaborations
where original source is not known.
This is documented in \cref{sec:references}.
\end{abstract}

\begin{frontmatter}

\begin{dedication}
\null
\vfil
% \begin{center}

\Begin{center}

\textsc{\textbf{To Her}}
\End{center}

\vspace{12pt}

\emph{To my grandma Pick Yee Lau,}\\
who chose to care for her small grandson over a surgery for full recovery of her leg.

\emph{To my mother Joyce Tsui,}\\
who combats depression alone when her only son pursues a PhD study in a foreign country.

\emph{To my wife Ivy Cheung,}\\
who sacrifices her career and follows me till the end of the world.

\emph{To my beloved Belle the cat,}\\
who has accompanied me throughout my study, and is our only family in the US.

\vspace{12pt}

I miss you, grandma and Belle,\\
I wish you would be with me in the graduation,\\
and I look forward to meeting you again in the new heaven and new earth.

% \end{center}
\vfil
\null
\end{dedication}

{
\hypersetup{linkcolor=blue}
\setcounter{tocdepth}{6}
\tableofcontents
}
\listoffigures
\listoftables
\begin{acknowledgements}
I thank my advisor Adrian Lee.
Thank you for giving me the opportunity when no one else did.
Thank you for keeping my position funded.
Thank you for your patience.
Thank you for teaching me to work \emph{academically}.
Thank you for your feedback on my dissertation on such a tight schedule.
You have the ability to appreciate things when no one can and your smile is contagious.

I thank my co-advisor Julian Borrill,
who suggested the structure of the presentation of my qualifying exam
after a disastrous preparation.
It continues to form the backbone of this dissertation.
Thank you for the opportunity at the computational cosmology center (C3).
Thank you for being straight to the point and be an example of how to get things done.
Thank you for the wisdom you shared in casual conversations.

I thank Kathy Yelick who serves as the representative of my designated emphasis.
Despite being so busy as the Associate Laboratory Director for Computing Sciences at LBNL,
she did not hesitate to be on my qualifying exam and dissertation committee,
and she offered to meet when I faced difficulties.
Thank you for supporting my job hunt.

I thank my graduate advisor Holger Müller.
Without your dedicated work on my case,
I would not have even started the research in this dissertation, let alone finish it.
Thank you for your kindness, listening, and understanding.
We need more people like you, who listen.
You are my role model in how I treat others.

I thank Matt Pyle for his positivity who serves on my qualifying exam and dissertation committee.
Thank you for your interesting questions
which refine my dissertation.
I thank Uros Seljak who serves on my qualifying exam committee.
Thank you for having to participate twice in my presentations.

I thank the many people that I worked with in POLARBEAR\@.
Thank you Akito Kusaka who prepared me by foretelling
how long it would take me to graduate when I make the jump to this field.
Thank you for your insights---it is eyeopening to watch you ask questions in any kinds of meetings.
Your process of asking questions while thinking out loud
is almost like a master class showing exactly how we should think when new research is presented to us.
Thank you Christian~Reichardt for all the insights in CMB analysis and your advice to spend time wisely in managing my research.
Thank you Yuji Chinone for your mentorship.
Despite our differences,
I know you care about our research.
Thank you Neil Goeckner-Wald,
thanks for all the wonderful ideas and insights you shared.
Thank you for all the hard work.
You are smart and yet humble.

I also thank the people from the C3.
Thank you Ted Kisner and Reijo Keskitalo who shows me how CMB data analysis should be carried out properly.
Thank you Reijo for all the little things you shared, from personal to research experiences.
Thank you Giuseppe Puglisi for your work on LiteBIRD where my research has benefited from.
I cannot believe the day I told you ``I'm going to see you tomorrow'' before the COVID lockdown is the last day we met.
I hope to meet you again soon.

I would also like to thank all the people who have inspired me and shaped me to become who I am.
Thank you Wong Wai Kay, a senior schoolmate in SYSS\@.
Thank you for introducing me to all those wonderful books in mathematics,
and taught me how to play international bridge.
All these shaped the way I think to this day.
Thank you, my math teacher Mr.~Wong Chi Shing from SYSS\@.
Your exposition on mathematics as the universal language shaped how I appreciate mathematics till this day.
Thank you, my English teacher Mr.~Keith Chan from CHECSS\@.
You are a uniquely caring teacher who taught us life lessons through teaching English.
I hope I lived up to your expectation to no longer be a self-centered person.
I thank Prof.~Yang Zhiyu and Prof.~Ng Tai Kai from HKUST\@.
I thank for all the opportunities related to the physics olympiads
without which I would not be doing this today.
Thank you Prof.~Shengwang Du from HKUST
for sharing your experience in finding research project in graduate school
and your worldview.
Both are contributing factors into how I became a CMB physicist.
Thank you James Chu, a lovely elder who inspires young people, and a fellow physicist.
I should have listened to your advice earlier,
and not clung onto being a theorist.
Thank you Ted Peng for sharing an inspiration from a CMB physicist in 2010.
And little did I know, one day I would become a CMB physicist myself!

I thank a lifelong schoolmate, Lokman Tsui.
We shared the same primary school,
secondary school, undergraduate school, and graduate school.
We became companions since the day we met in physics olympiads.
At a time we were to conquer String Theory together.
I wish we could have cracked it.
I miss you in the academia.

Lastly,
I thank all my loved ones---my grandma, my mother, my wife, my soon-to-be born baby, and our cat Belle.
This dissertation is dedicated to you.
I thank my father,
who gives me all the freedom I need
to make any decision I like.
I also thank my aunt Tsui Sau Hing
who should receive sainthood
for being the last person
who still cares for my mother.
She is the only reason I can stop worrying about my mother
and focus on my study abroad.
I thank my uncle General Cheung,
he treated me like his son when I was young.
I still kept the notebook he bought me
when I first studied abroad at UCB\@.
I have to thank my wife Ivy Cheung again,
who directly contributed to the successful completion of this dissertation.
How can I live without you?
It broke my heart when you chopped your finger accidentally on the day the COVID lockdown started.
Thanks for taking care of me all these years.
Thanks for enduring this ridiculous country where one holding an F2 spouse visa is not allowed to work.
Thanks for suffering in the broken US medical system where you cannot even get to see a doctor between my graduation and our departure of the US\@.
I am thankful we are getting out of here, as you always said we will one day do.
Thanks for letting me learn how to love.
Thanks for being poor with me.
Thanks for burning the midnight oil with me,
but seriously, you cannot do this anymore.
Thanks for cooking so much good food for me,
making me a proper US-sized person.
Thanks for pushing me to relax more and enjoy life once
in a while.
Thanks for keeping me down-to-earth
when I always dream big.
Thanks for listening to me
explaining physics to you.
I know you enjoy how soothing it is to you too,
helping you fall asleep faster than any medicine.
\end{acknowledgements}

\end{frontmatter}

\mainmatter
\pagestyle{headings}
\part{The beginning of our Universe}\label{sec:beginning-universe}

\chapter{The Big Bang}\label{the-big-bang}

\begin{figure}
\centering
\pandocbounded{\includegraphics[keepaspectratio,alt={Timeline of the Universe}]{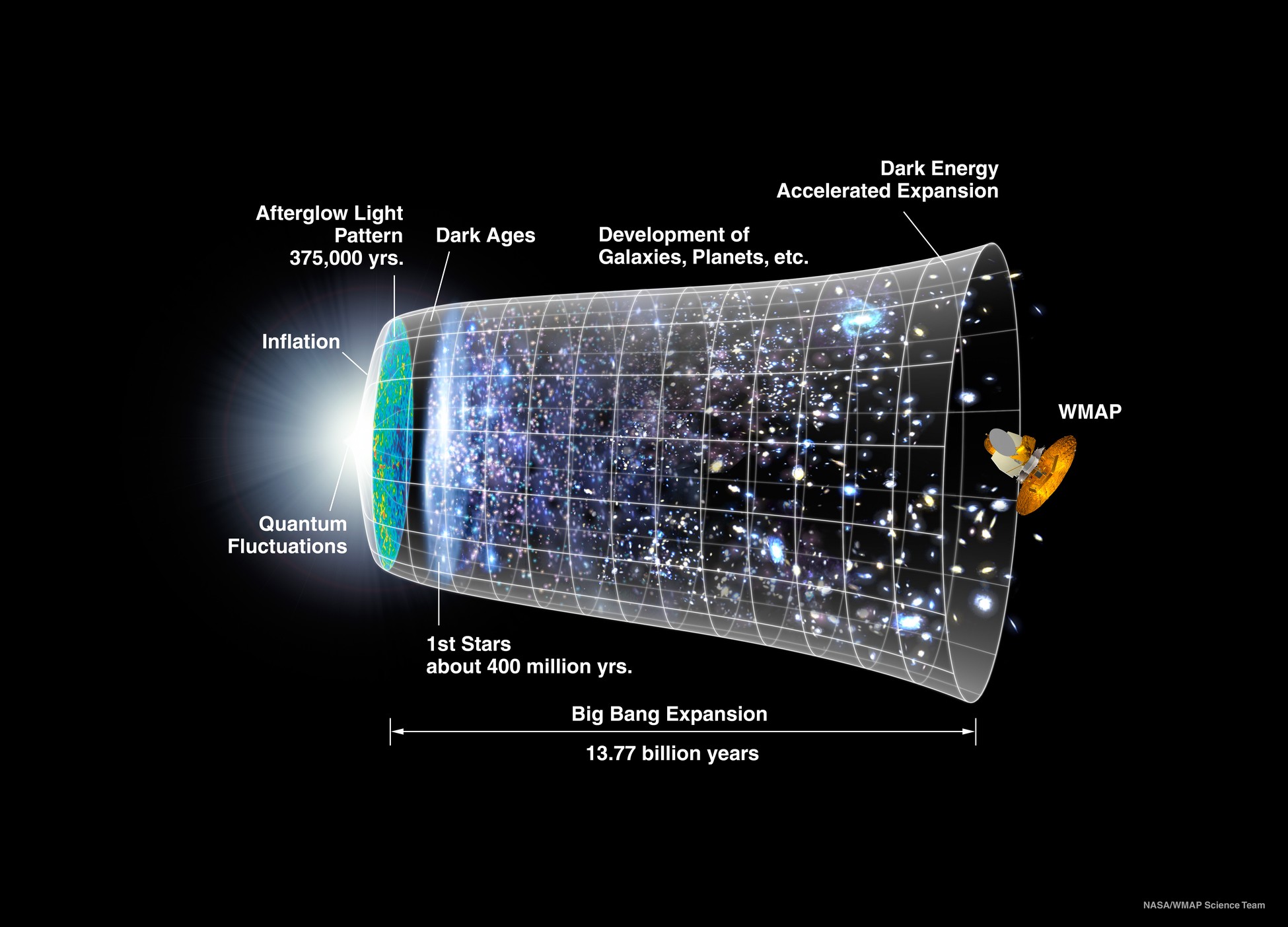}}
\caption[Timeline of the Universe]{Timeline of the Universe\footnotemark{}}
\end{figure}
\footnotetext{\citeproc{ref-nasa__wmap_science_team_timeline_nodate}{NASA / WMAP Science Team, \emph{Timeline of the Universe}}.}

The Universe starts with a Big Bang.

To a good approximation, it started with a uniform density soup of particles
and as they are so hot, they are in thermal equilibrium.
As the Universe expands they are diluted and cooled down,
and eventually they fell out of equilibrium and decoupled with others one by one.

One particularly important moment is when free electrons and protons start to form neutral Hydrogen
and it becomes transparent for photons to start free-streaming to us.
This is called recombination.
The time was around year 375,000.
The majority of these photons have not been re-scattered when they reached us.
Since they are now peaked at microwave range,
this is known as the Cosmic Microwave Background (CMB) Radiation.

However, the Universe was not completely uniform.
It started with tiny primordial fluctuations
which grew under the influence of gravity.
Before any substantial structure was formed,
there was the (cosmic) Dark Ages.
Then the first stars were born as the fluctuations grew non-linear.
And the Universe has grown more complex ever since.

In recent history of the Universe, the expansion is accelerating.
This is one of the key signatures of our Universe and is caused by Dark Energy.

That briefly summarized the Big Bang theory,
in particular, the ΛCDM model that is well-understood
via general relativity, thermodynamics, and particle physics,
and is confirmed observationally to high precision.
Basically, all known branches of fundamental physics lead to this successful standard model of cosmology.

\chapter{Fine tuning problems}\label{fine-tuning-problems}

However, the Big Bang theory left us unsatisfied for multiple fundamental problems.

For example, the best-fit ΛCDM model tells us
that the Universe is consistent with zero curvature today
whereas there are no fundamental reasons why it should be exactly zero.
And because as the Universe grows, curvature also grows,
the initial curvature might be incredibly close to zero---this is one example of a class of problems known as fine-tuning problems.

Another fine-tuning problem is the horizon problem.
Given our understanding in the ΛCDM model,
we can ask what was the size of region
that was in causal contact at the time the CMB formed.
It turns out it corresponds to about \(\ang{2}\) on our sky.
Yet, we found that the Universe (and the CMB) is pretty uniform across the whole sky.

\chapter{Cosmic inflation}\label{sec:inflation}

A solution to the fine tuning problems from last chapter is inflationary models.
The basic feature of inflation is
that the Universe initially goes through a period of rapidly accelerating expansion,
expanded by a factor as much as \(\sim e^{60}\).
The ``initial condition'' of ΛCDM model is the aftermath of this,
which explains why the curvature is so small ``initially''.
Also, the seemingly causally disconnected region
was once in causal contact before the inflation,
which explains the horizon problem.

Inflationary models also made some general predictions,
and some of these have been observationally verified,
such as the scalar spectral index \(n_s < 1\).

Although inflation provided explanations to these problems
and there are hints showing that inflation happened,
it is still unproven,
and the details of inflation are largely unknown.
Unlike the situation with the Big Bang model,
which has a concrete, parameterized ΛCDM model
that is derived from known fundamental physics,
there are plenty of theoretical models
that are beyond the Standard Models
and fit the general picture of our known Universe
without much experimental or observational evidence
as which one is the true fundamental theory of physics.

The relevant energy scales for inflationary physics cannot be reached experimentally
unless there is a fundamental paradigm shift in how we accelerate particles.
Therefore, observational probes are our best hope in studying inflationary physics.

In the next part, we are going to look at one such probe,
the Cosmic Microwave Background (CMB) Radiation.
It is the earliest direct signal from the Universe we can possibly observe.
Not only does it currently provide some of the most precise constraints in the known ΛCDM model,
but it also offers hope on probing inflationary physics
by observing the imprint of primordial gravitational waves on the CMB.

\part{The Cosmic Microwave Background (CMB) radiation}\label{sec:CMB-radiation}

\chapter{The CMB spectrum}\label{the-cmb-spectrum}

\begin{figure}
\centering
\pandocbounded{\includegraphics[keepaspectratio,alt={CMB blackbody spectrum}]{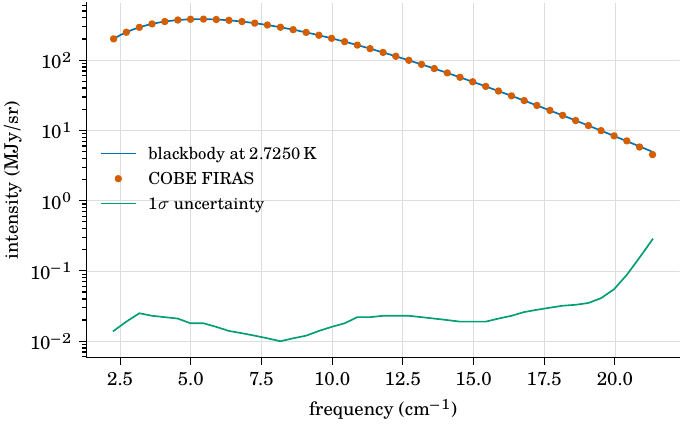}}
\caption{CMB blackbody spectrum}\label{fig:firas_monopole_spec_v1}
\end{figure}

As mentioned in \cref{sec:beginning-universe},
when the photons were last scattered during recombination,
they were in thermal equilibrium.
\Cref{fig:firas_monopole_spec_v1} shows the measurement of the CMB spectrum
by COBE's FIRAS instrument\footnote{\citeproc{ref-fixsen_cosmic_1996}{Fixsen et al., {``The Cosmic Microwave Background Spectrum from the Full \emph{COBE} FIRAS Data Set''}}; \citeproc{ref-nasa_cmb_nodate}{NASA, \emph{CMB Monopole Spectrum}}.}
which is in close agreement with a blackbody prediction
with a temperature of \(\SI{2.7255 \pm 0.0006}{\kelvin}\).
The last scattering happened at around \(z \approx 1100\),
corresponding to a temperature of \(\SI{\sim 3000}{\kelvin}\).

It also shows that the CMB is incredibly smooth.
No matter which part of the sky we are looking at,
it is the same temperature within this error-bar.

\chapter{The CMB anisotropy}\label{the-cmb-anisotropy}

\begin{figure}
\centering
\pandocbounded{\includegraphics[keepaspectratio,alt={Intensity}]{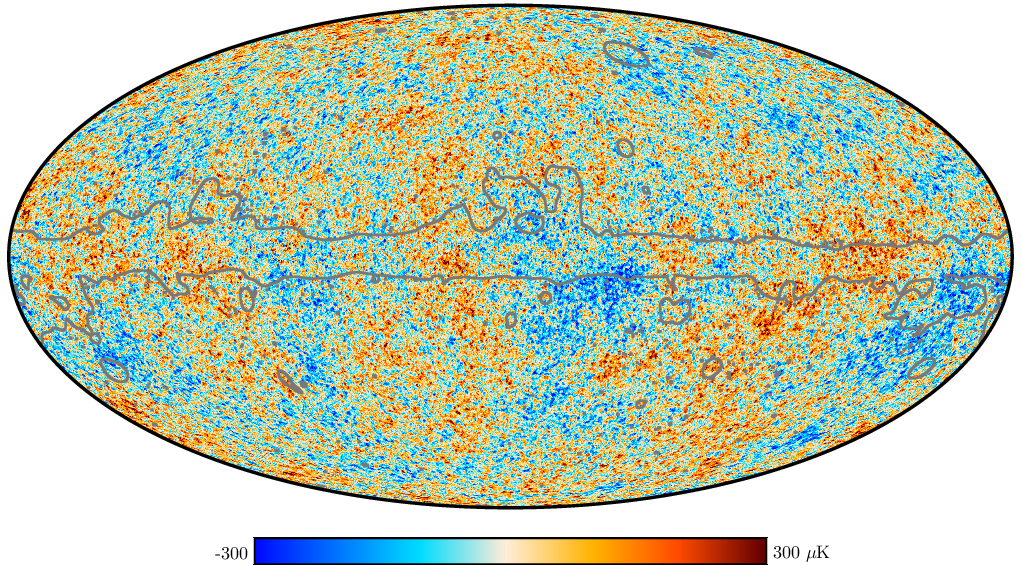}}
\caption[Intensity]{Intensity\footnotemark{}}\label{fig:Planck_2018_T_CMB}
\end{figure}
\footnotetext{\citeproc{ref-planck_collaboration_planck_2020-2}{Planck Collaboration, Aghanim, Akrami, Arroja, et al., {``\emph{Planck} 2018 Results''}}.}

In fact, as seen in the intensity map of CMB in \cref{fig:Planck_2018_T_CMB},
the fluctuation of the intensity is about 1 part in \(\num{e5}\).

Moreover, it is known that light waves can be polarized.
The primary mechanism to generate polarization is through
Thomson scattering of the anisotropic intensity.
As Thomson scattering generates only linear polarization,
circular polarization is therefore not expected to be present\footnote{This assumption has been experimentally constrained, such as in Nagy et al., \citeproc{ref-nagy_new_2017}{{``A New Limit on CMB Circular Polarization from SPIDER''}}.
  There exist known mechanisms to generate non-zero but tiny contributions of circular polarization in the CMB,
  such as
  Zarei et al., \citeproc{ref-zarei_generation_2010}{{``Generation of Circular Polarization of the CMB''}};
  Mohammadi, \citeproc{ref-mohammadi_evidence_2014}{{``Evidence for Cosmic Neutrino Background from CMB Circular Polarization''}};
  King and Lubin, \citeproc{ref-king_circular_2016}{{``Circular Polarization of the CMB''}};
  Sadegh, Mohammadi, and Motie, \citeproc{ref-sadegh_generation_2018}{{``Generation of Circular Polarization in CMB Radiation via Nonlinear Photon-Photon Interaction''}};
  Montero-Camacho and Hirata, \citeproc{ref-montero-camacho_exploring_2018}{{``Exploring Circular Polarization in the CMB Due to Conventional Sources of Cosmic Birefringence''}};
  Inomata and Kamionkowski, \citeproc{ref-inomata_circular_2019}{{``Circular Polarization of the Cosmic Microwave Background from Vector and Tensor Perturbations''}}.}.
Therefore, we can focus our attention to the linear polarization of the CMB.

\begin{figure}
\centering
\pandocbounded{\includegraphics[keepaspectratio,alt={Polarization}]{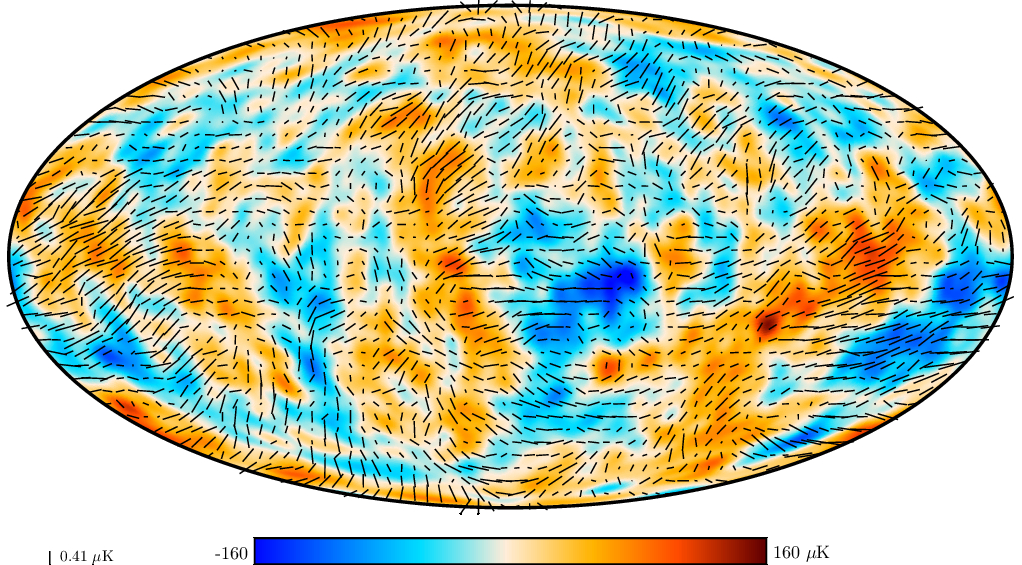}}
\caption[Polarization]{Polarization\footnotemark{}}\label{fig:Planck_2018_Pol_CMB}
\end{figure}
\footnotetext{\citeproc{ref-planck_collaboration_planck_2020-2}{Planck Collaboration, Aghanim, Akrami, Arroja, et al., {``\emph{Planck} 2018 Results''}}.}

\begin{figure}
\centering
\pandocbounded{\includegraphics[keepaspectratio,alt={B-mode template}]{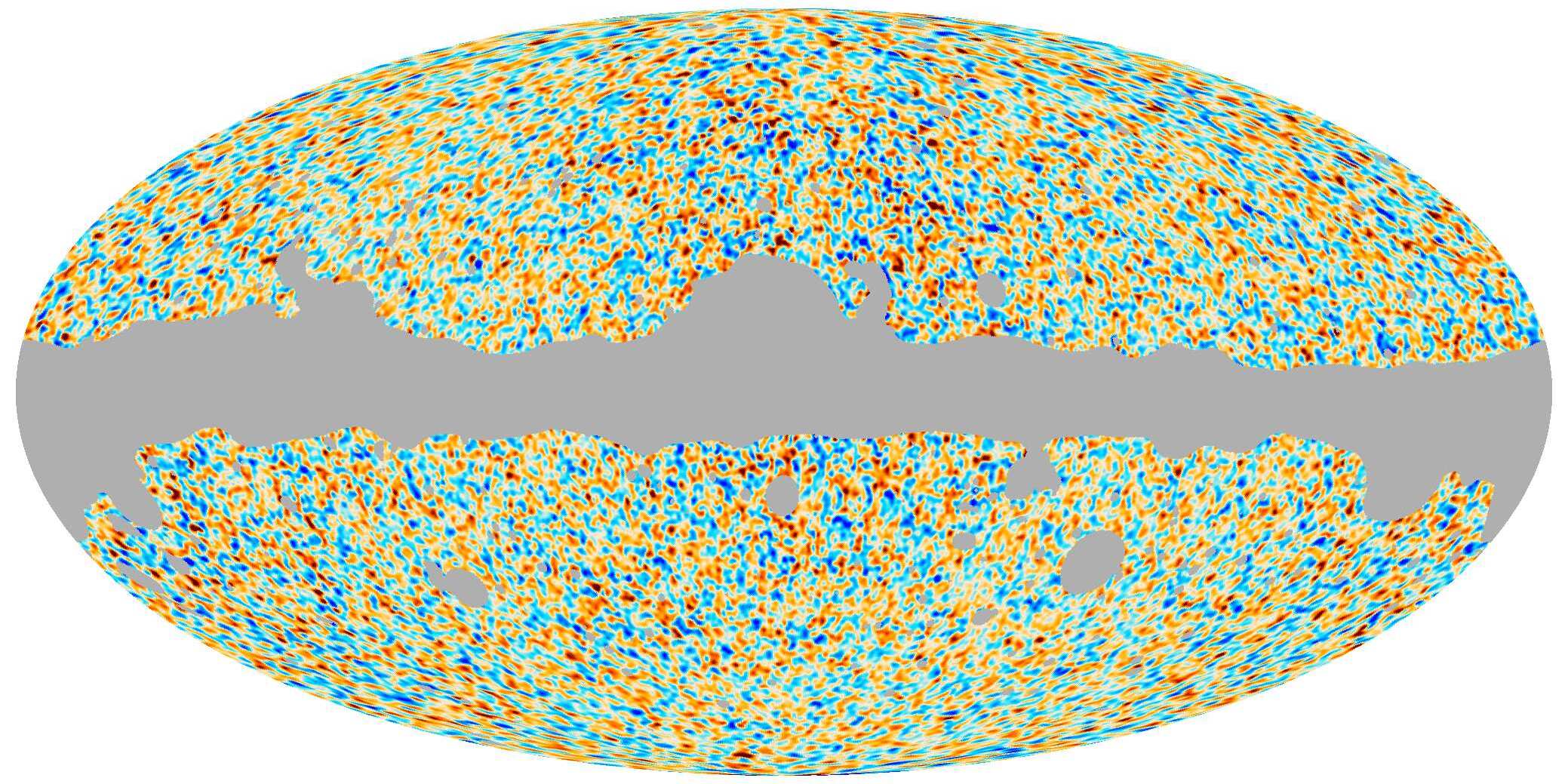}}
\caption[B-mode template]{B-mode template\footnotemark{}}\label{fig:BB-template-60-arcmin}
\end{figure}

\Cref{fig:Planck_2018_Pol_CMB} shows the smoothed intensity
overlaid with the polarization arrows indicating the polarization direction.
First, we can see that the arrows and color are quite correlated.
And the arrows mainly have divergent patterns,
but not so much of a curl.
In fact, it is a good idea to decompose the \(2\) degree of freedom from the linear polarization
into this divergence and curl pattern, known as the E-mode and B-mode of the CMB,
as this is independent of the choice of coordinate system.
In \cref{fig:BB-template-60-arcmin}, we can visualize B-mode
using a template constructed by Planck.

\footnotetext{\citeproc{ref-planck_collaboration_planck_2016-1}{Planck Collaboration, Ade, et al., {``\emph{Planck} Intermediate Results''}}.}

Putting all together,
\(T\), \(E\) \& \(B\) uniquely characterizes the CMB.

\chapter{The CMB power-spectra}\label{the-cmb-power-spectra}

As these maps are on a \(2\)-sphere,
we can perform spherical harmonics transformation on these,
and calculate the \(2\)-pt correlation function between them.

We can form \(TT\), \(EE\), \(BB\), and \(TE\), \(EB\), \(TB\) combinations out of them.
And by underlying parity symmetry, we expect \(EB\) and \(TB\) to be exactly zero\footnote{Similar to the absence of circular polarization in the CMB,
  and in fact most if not all null-hypothesis in physics,
  it can and should be experimentally tested, and discovery can lead to new physics.
  This phenomenon of non-zero \(TB, EB\) is known as cosmic birefringence,
  and upper limits have been placed by experiments such as Planck Collaboration, Aghanim, et al., \citeproc{ref-planck_collaboration_planck_2016-2}{{``\emph{Planck} Intermediate Results''}}.
  Potential new physics that can produce cosmic birefringence includes
  parity violation, Lorentz violation, or axion-photon mixing (\citeproc{ref-particle_data_group_review_2022}{Particle Data Group et al., {``Review of Particle Physics''}}).}.
Hence, we are left with \(TT\), \(TE\), \(EE\) \& \(BB\).

\begin{figure}
\centering
\pandocbounded{\includegraphics[keepaspectratio,alt={CMB Power-Spectra}]{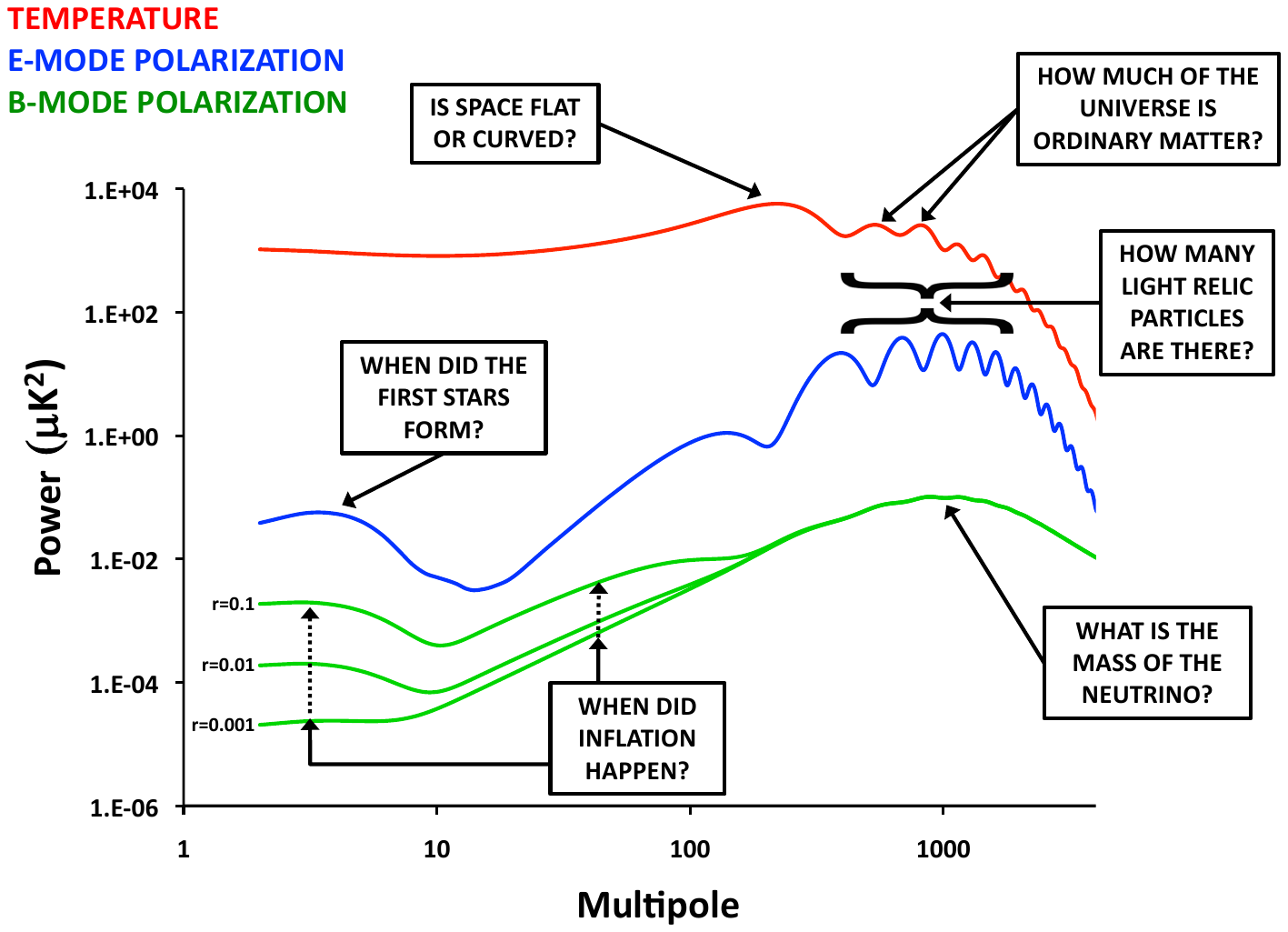}}
\caption[CMB Power-Spectra]{CMB Power-Spectra\footnotemark{}}\label{fig:annotated_spectra}
\end{figure}
\footnotetext{\citeproc{ref-borrill_cmb_nodate}{Borrill, \emph{CMB Power-Spectra}, n.d.}}

\cref{fig:annotated_spectra} shows the \(2\)-pt correlations
(also known as power-spectra) we theoretically expect.
This only depends on the initial parameters unlike the maps.

In this plot we start to see how CMB can be used to extract information about our Universe.
Here the horizontal axis is the multipole, roughly corresponding to inverse angular scale.
For example, on the top left the low-\(\ell\) \(TT\) has a Sachs-Wolfe Plateau, corresponding to the horizon scale.
On the right, \(TT\), \(EE\) has these oscillatory peaks, a signature of the Baryon Acoustic Oscillations (BAO) between gravity and pressure of all matters. Hence it is sensitive to spatial curvature and matter compositions including matter, dark matter, and dark energy.
On the bottom, we see that \(BB\) is the weakest, as it can only be contributed by gravitational effects.
For example, on the right we have the high-\(\ell\) \(BB\) dominated by the gravitational lensing of \(EE\), and is sensitive to the mass of neutrino.

\chapter{The primordial B-mode}\label{sec:primordial-BB}

Everything that has been mentioned so far are measured to various degree of accuracy.
But, on the lower-left, there is something that has yet to be measured.
This is the extreme low-\(\ell\) \(BB\), large scale region
where no other known form of physics can generate a significant signal of,
except for primordial gravitational waves.

This is a tensor perturbation expected from inflation.
In other words, while the inflation may have many subtle consequences
in different parts of the CMB, this is the region where we expect
the highest signal-to-noise ratio to detect the signature of inflation.
It gives a rare chance for us to glimpse at a direct evidence of inflation.
Moreover, it may offer us the capability to start falsifying different inflationary models.

\part{CMB Observations}\label{sec:CMB-observations}

\chapter{Foreground of the CMB}\label{foreground-of-the-cmb}

Now that we know what knowledge we might gain from the CMB,
we turn our attention to how we might observe and analyze it.

\section{Intensity}\label{intensity}

We already peeked ahead at \cref{fig:Planck_2018_T_CMB},
this time we want to focus on some peculiar grey lines on the maps.
These represent masking regions within which we actually did not observe the CMB.

The reason is simply that when we look for the Cosmic Microwave Background,
there are foreground contaminations from many sources.
For example, there is a big horizontal region, corresponding to our own galaxy.

\begin{figure}
\centering
\pandocbounded{\includegraphics[keepaspectratio,alt={Foreground Intensity maps}]{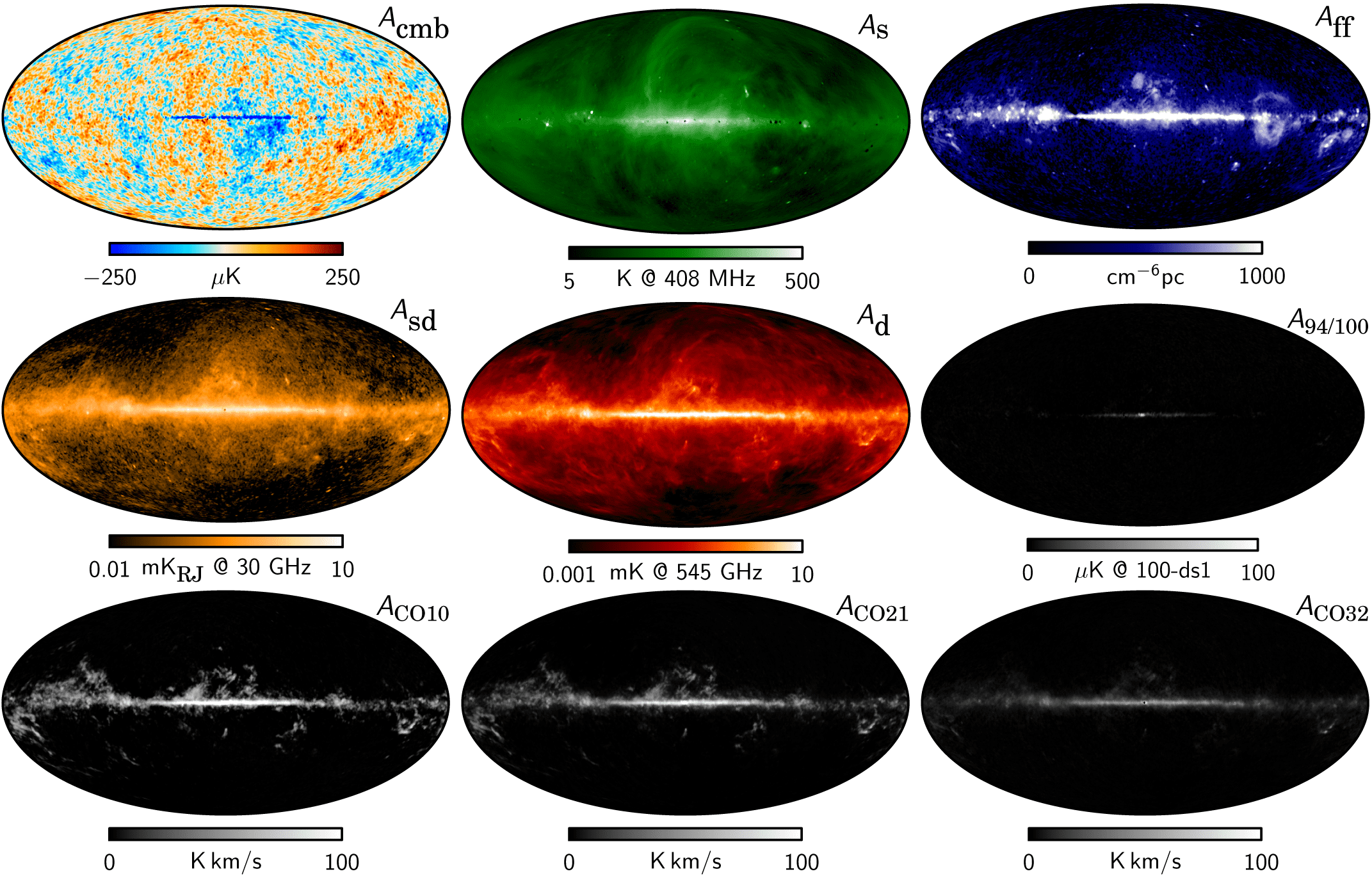}}
\caption[Foreground Intensity maps]{Foreground Intensity maps\footnotemark{}}\label{fig:2015_FGAmpl}
\end{figure}
\footnotetext{\citeproc{ref-planck_collaboration_planck_2016}{Planck Collaboration, Adam, et al., {``Planck 2015 Results''}}.}

Here in \cref{fig:2015_FGAmpl} we can see maps of different components.
We are not going into the details here, but it suffices to say
that around \(\SI{20}{\percent}\) of the sky is too bright corresponding to
the masked out region in \cref{fig:Planck_2018_T_CMB}.
And also different components may be stronger at different frequencies.

\begin{figure}
\centering
\pandocbounded{\includegraphics[keepaspectratio,alt={Foreground Intensity spectra}]{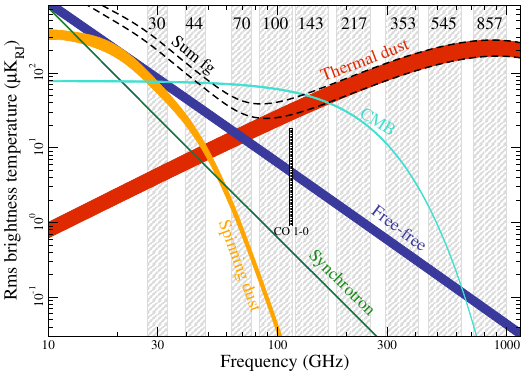}}
\caption[Foreground Intensity spectra]{Foreground Intensity spectra\footnotemark{}}\label{fig:2015_FGSpectra}
\end{figure}
\footnotetext{\citeproc{ref-planck_collaboration_planck_2016}{Planck Collaboration, Adam, et al., {``Planck 2015 Results''}}.}

It is more obvious in power-spectra domain,
here in \cref{fig:2015_FGSpectra} we can see the foreground components
have different frequency dependence from the CMB,
and importantly for the Intensity, at some frequencies the CMB
is brighter than all foregrounds combined.

\section{Polarization}\label{sec:polarization-foreground}

\begin{figure}
\centering
\pandocbounded{\includegraphics[keepaspectratio,alt={Foreground Synchrotron Polarization}]{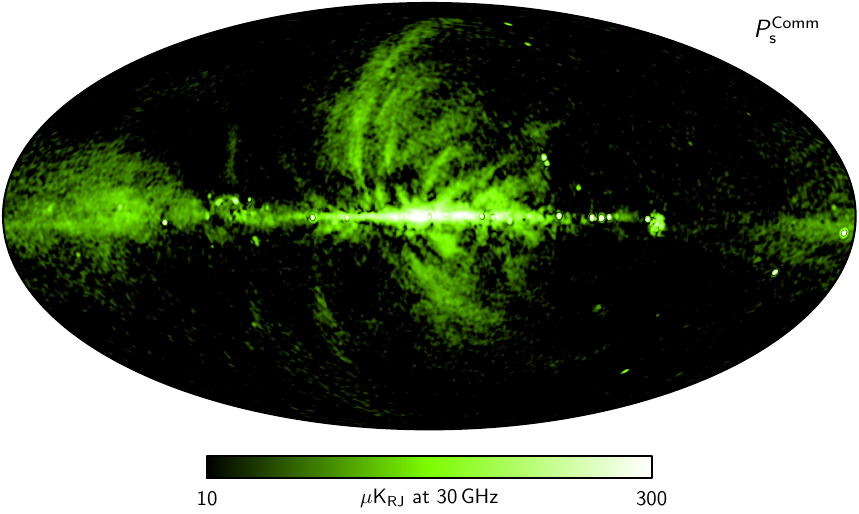}}
\caption[Foreground Synchrotron Polarization]{Foreground Synchrotron Polarization\footnotemark{}}\label{fig:Planck_2018_synch_P_rc6_40arc_n1024_v5}
\end{figure}
\footnotetext{\citeproc{ref-planck_collaboration_planck_2020-1}{Planck Collaboration, Akrami, Ashdown, et al., {``\emph{Planck} 2018 Results''}}.}

\begin{figure}
\centering
\pandocbounded{\includegraphics[keepaspectratio,alt={Foreground Dust Polarization}]{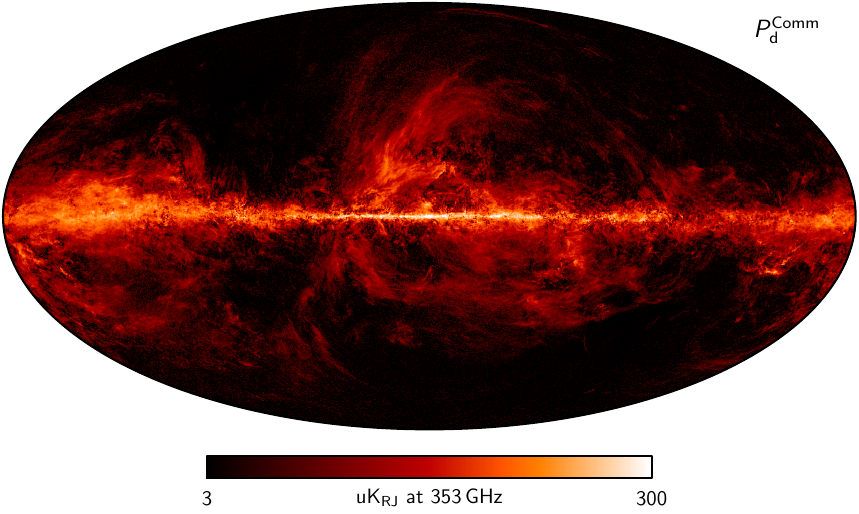}}
\caption[Foreground Dust Polarization]{Foreground Dust Polarization\footnotemark{}}\label{fig:Planck_2018_dust_P_rc6_5arc_n1024_v5}
\end{figure}
\footnotetext{\citeproc{ref-planck_collaboration_planck_2020-1}{Planck Collaboration, Akrami, Ashdown, et al., {``\emph{Planck} 2018 Results''}}.}

Here in \cref{fig:Planck_2018_synch_P_rc6_40arc_n1024_v5} \& \cref{fig:Planck_2018_dust_P_rc6_5arc_n1024_v5},
we see the situation is quite different for polarizations.
First, there are only 2 sources of significant foreground polarization---synchrotron and dust.

\begin{figure}
\centering
\pandocbounded{\includegraphics[keepaspectratio,alt={Foreground Polarization spectra}]{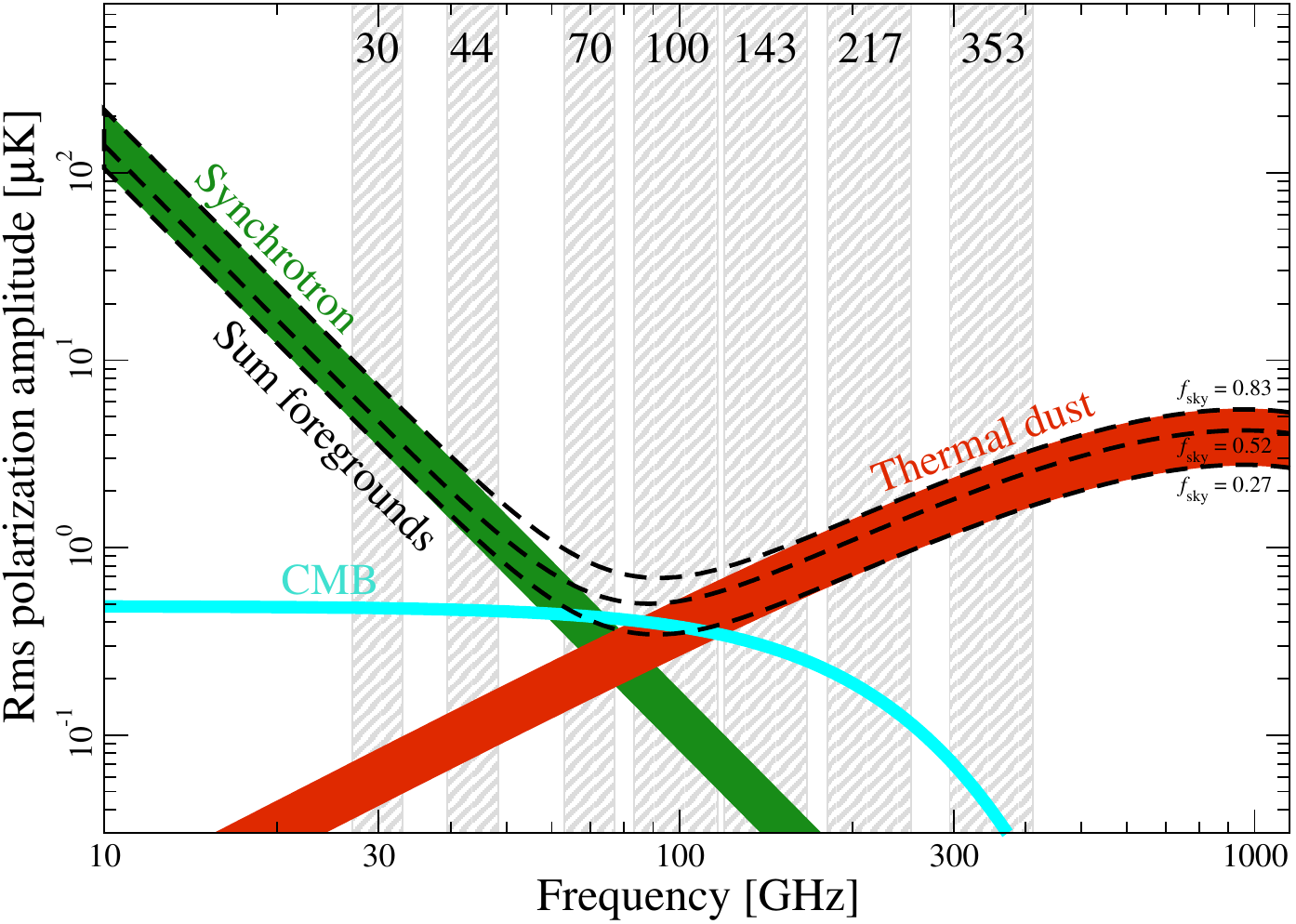}}
\caption[Foreground Polarization spectra]{Foreground Polarization spectra\footnotemark{}}\label{fig:Planck_2018_P_FGs}
\end{figure}
\footnotetext{\citeproc{ref-planck_collaboration_planck_2020-1}{Planck Collaboration, Akrami, Ashdown, et al., {``\emph{Planck} 2018 Results''}}.}

Second, as it becomes obvious in power-spectra domain in \cref{fig:Planck_2018_P_FGs},
nowhere does the CMB dominate over foreground contaminants.
Therefore, in measuring the polarization of the CMB,
it is particularly important to measure in multiple frequencies.

\chapter{Noise in CMB observations}\label{sec:noise-modulation}

Now let us switch gears to talk about the noise properties.
This has important implications on
how the CMB is observed,
how CMB experiments are designed,
and even how CMB data is analyzed.

\section{Atmospheric noise}\label{sec:atmosphere}

When we talk about observation frequencies,
ground based CMB experiments have extra constraints here.

\begin{figure}
\centering
\pandocbounded{\includegraphics[keepaspectratio,alt={Atmospherical windows}]{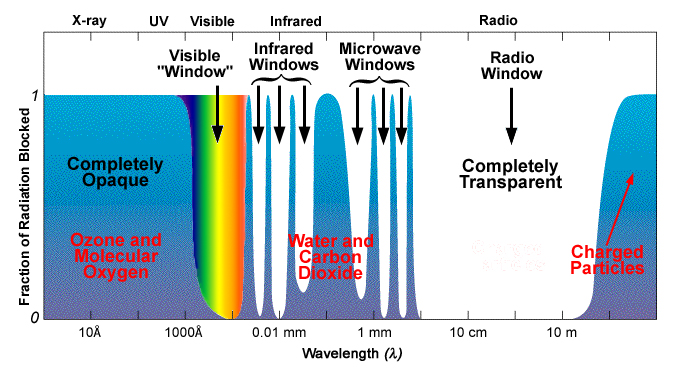}}
\caption{Atmospherical windows}\label{fig:atmosphericalWindows}
\end{figure}

Take a look at \cref{fig:atmosphericalWindows}, we see that our choice
of observation frequencies is limited by the atmospheric windows.

\begin{figure}
\centering
\pandocbounded{\includegraphics[keepaspectratio,alt={PWV throughout the World}]{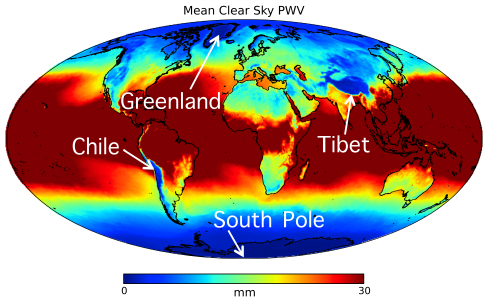}}
\caption[PWV throughout the World]{PWV throughout the World\footnotemark{}}\label{fig:PWV}
\end{figure}
\footnotetext{\citeproc{ref-barron_optimization_2018}{Barron et al., {``Optimization Study for the Experimental Configuration of CMB-S4''}}.}

Not only that, from the precipitable water vapor (PWV) level across the world shown in \cref{fig:PWV}, we see that only some regions in the world favor CMB observations.
It is because higher PWV,
corresponding to red color in the figure,
results in more absorption
and therefore is less ideal for CMB experiments.

In fact, this is so important that all current ground based CMB experiments are in the low PWV regions shown as dark blue here\footnote{This includes 4 regions---Chile, Antarctica, Greenland, and Tibet.
  The South Pole in Antarctica and the Atacama Desert in Chile are well-established
  sites for CMB observations for reasons outlined in Barron et al., \citeproc{ref-barron_optimization_2018}{{``Optimization Study for the Experimental Configuration of CMB-S4''}}.
  3 more sites---Ali in Tibet, Dome A in Antarctica, and Summit Camp in Greenland---are explored in Kuo, \citeproc{ref-kuo_assessments_2017}{{``Assessments of Ali, Dome A, and Summit Camp for Mm-Wave Observations Using MERRA-2 Reanalysis''}}.
  It concludes that, except for Greenland where a high level of liquid water clouds is detected,
  all remaining 4 sites are promising for CMB observations,
  where Ali in Tibet is the only one on the list which is ``in the Northern Hemisphere that will complement well-established Southern sites''.
  Lastly, it should be noted that existing infrastructure in the South Pole and the Atacama Desert
  makes it favorable for future deployments.}.

\section{\texorpdfstring{\(1/f\)-noise}{1/f-noise}}\label{f-noise}

Having mentioned the external factors involved in CMB experiments,
we will briefly go through the actual instrument CMB experimentalists can design.

There is a lot of interesting physics involved in the design of these instruments.
For example, it involves optical design of the telescope.
And in detectors, there is Transition Edge Sensor (TES) bolometer
which utilizes superconductivity to achieve photon-noise limited sensitivity.
In detector readout, Superconducting Quantum Interference Devices (SQUIDs) are used
to couple TESes together to achieve multiplexing,
which enables us to pack a lot of detectors in a focal plane
to beat down the noise.

But few of the characteristics of an instrument are as ubiquitous and have as profound implications as
the following one we are going to mention: \(1/f\) noise.

\begin{pandoccrossrefsubfigures}

\begin{subfigure}{0.5\linewidth}\centering\pandocbounded{\includegraphics[keepaspectratio]{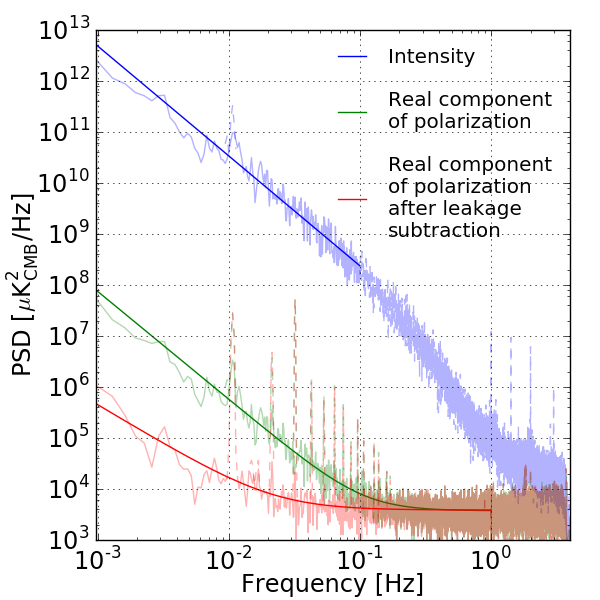}}\caption{\label{fig:018-004}}\end{subfigure}%
\begin{subfigure}{0.5\linewidth}\centering\pandocbounded{\includegraphics[keepaspectratio]{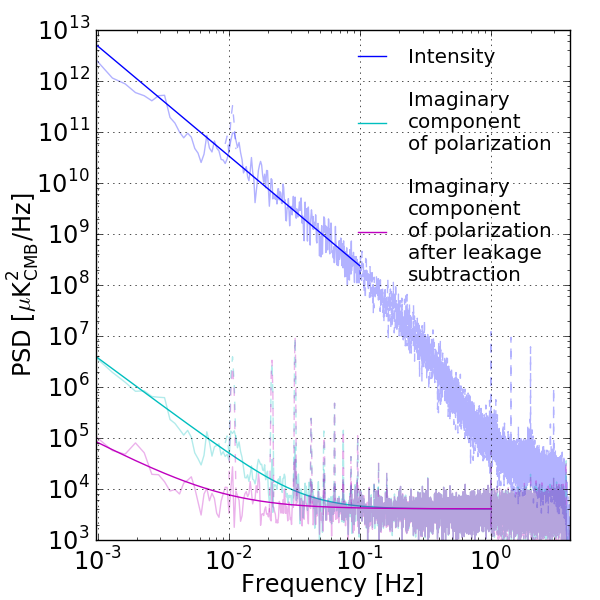}}\caption{\label{fig:018-006}}\end{subfigure}%

\caption{``PSDs of coadded timestreams for all focal plane detectors for the real part (left panel) and for the imaginary part (right panel). The blue line shows the intensity fluctuation and the green (cyan) and red (magenta) lines show the real (imaginary) part of the polarization signal before and after leakage subtraction, respectively. The spikes at the harmonics of \(\SI{0.01}{\hertz}\) are the scan synchronous signals.''\footnote{\citeproc{ref-takakura_performance_2017}{Takakura et al., {``Performance of a Continuously Rotating Half-Wave Plate on the POLARBEAR Telescope''}}.}}\label{fig:HWP-PSD}

\end{pandoccrossrefsubfigures}

From \cref{fig:HWP-PSD}, we can see an example spectrum of a detector TOD from POLARBEAR.

\begin{equation}\protect\phantomsection\label{eq:noise}{N \left[ 1 + \left( \frac{f_\text{knee}}{f} \right)^\alpha \right]}\end{equation}

\Cref{eq:noise} describes the empirical form of noise like this.
The first term is a white noise floor,
and the second term is a straight line with slope \(\alpha\) on the log-log scale.
This is the infamous \(1/f\) noise.

There are multiple sources that can lead to this empirical form,
including atmosphere, electronics, temperature drifts, etc.

In this specific example, the atmosphere is the dominant contribution.
But this is a fairly generic feature that appears in all our CMB measurements,
and this characteristic affects how we design our experiments.

\section{Scanning Strategy}\label{sec:scanningStrategy}

The main implication here is that we cannot fixate on a point in space and measure the CMB,
because in this static limit, where the frequency approaches \(0\),
the noise level virtually diverges leading to a large uncertainty.

Instead, the rule of the game here is to modulate the signal,
meaning that we make it change.
That is, we change the signal such that it is no longer at zero frequency
and away from the singularity.

That is why all CMB experiments have to design a scanning strategy
to scan across the sky.
This process is also called a differential measurement
as effectively we are measuring the difference in the signal,
the anisotropy of the CMB.

Because the CMB is not a scalar,
having multiple components,
cross-linking is also important.
Cross-linking basically means that you revisit the same pixel
with a different signal composition
so that you can eliminate the degeneracy
and solve for different signal components.

We are not going into the details of designing scanning strategy here,
but we now focus on an interesting twist on scanning strategy below.

\section{Modulation of the CMB signal via HWP}\label{modulation-of-the-cmb-signal-via-hwp}

In measuring the CMB,
\(1/f\) noise is \emph{less} of a problem in measuring the Intensity,
simply because the signal is big at low-\(\ell\),
corresponding to low-frequency in the time-domain,
which means we can afford larger noise-level there.
But this is not the case for CMB polarization.

CMB polarization can be measured by pair-differencing---it takes 2 detectors at \(\ang{90}\)
and takes the difference between them.
But any non-ideality between the 2 is going to
create Intensity to Polarization (I-to-P) leakage.
For example, the 2 detectors might not be exactly at \(\ang{90}\),
or might have different beam shape or band pass.

These make CMB polarization measurements more challenging.

One idea to solve these problems
is to modulate the signal again,
but to much higher frequencies
utilizing polarization itself.

\begin{figure}
\centering
\pandocbounded{\includegraphics[keepaspectratio,alt={HWP}]{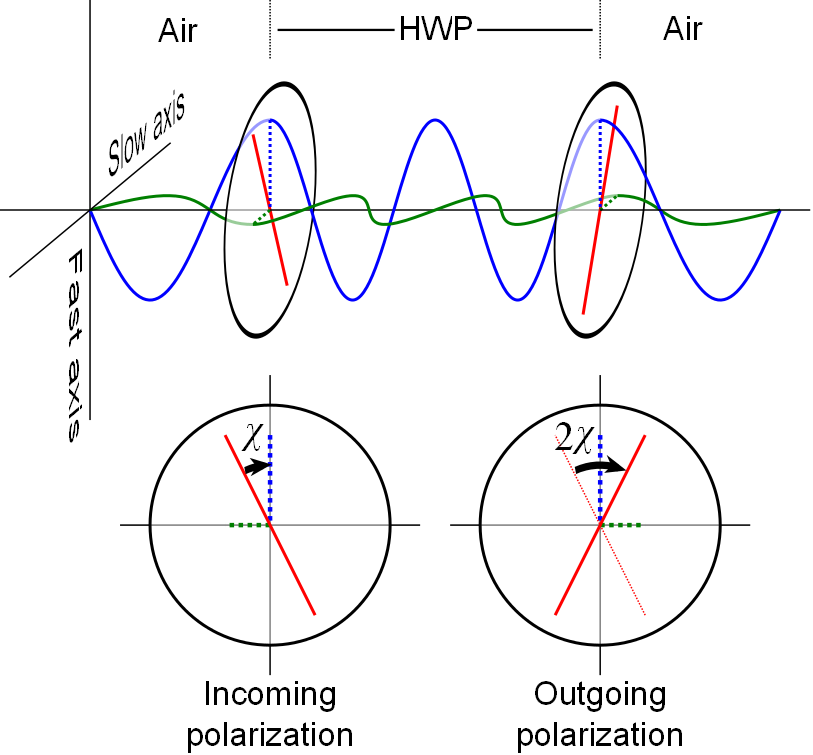}}
\caption{HWP}\label{fig:003-003}
\end{figure}

\Cref{fig:003-003} is a Half-Wave Plate (HWP)\@.
What it does is that when light is transmitted through a HWP,
the polarization angle will be reflected.
We can then put the HWP in Continuous Rotation (CRHWP),
and it will modulate any polarized components on the sky-side of the HWP
by \(4\) times the frequency we rotate it.
Note that this means the unpolarized components including atmospheric noise
are left unmodulated.
This separates the \(1/f\) infrared divergence region
from the polarized signal at \(4f\).

In \cref{fig:HWP-PSD},
we see the effect of HWP demodulation,
one of the techniques to reconstruct the original polarized components
from the modulated signals which is detailed in \cref{sec:CRHWPDemodulation}.
We can see from the middle curve that \(1/f\) noise is hugely suppressed.

This kills \(2\) birds with one stone---it modulates the signals away from the \(1/f\) infrared divergence,
and also enables a single detector to see both polarization modes,
hence eliminating the need of pair-differencing
and the associated systematics.

But there is still residual I-to-P leakage due to HWP systematics.
It can include detector non-linearity, off-axis mirror, etc.
It can be further suppressed in analysis using leakage subtraction
as seen in the lowest curve.

In addition to these systematics,
mitigation of \(1/f\) noise using HWP modulation is also limited
by the presence of polarized noise components.
HWP modulation can modulate such noise too,
so that the signal is no longer separated from the infrared divergence.
Takakura et al.\footnote{\citeproc{ref-takakura_measurements_2019}{{``Measurements of Tropospheric Ice Clouds with a Ground-Based CMB Polarization Experiment, POLARBEAR''}}.} provides one such example,
where ice crystals in tropospheric ice clouds
can create such polarized \(1/f\) noise components.

\chapter{CMB data}\label{cmb-data}

With the help of experimentalists,
let us say that we have observed the CMB
and have our data.
What would it look like
and how we should process it?

\begin{equation}\protect\phantomsection\label{eq:scanning}{d_t = P_{tp} m_p + n_t}\end{equation}

\Cref{eq:scanning} represents scanning the sky
and observing a signal, where

\begin{description}
\tightlist
\item[\(m_p\)]
sky signal where \(p\) is our pixelization scheme.
\item[\(P_{tp}\)]
pointing matrix which represents
where we are pointing at the sky,
and also which signal components we are picking up
if it is sensitive to polarization.
\item[\(n_t\)]
noise in time-domain.
\item[\(d_t\)]
the observed timestream of data (TOD).
\end{description}

Now our task is to invert this process and obtain the map.

\chapter{CMB mapmaking}\label{sec:CMBMapmaking}

\section{The ``naïve physicist''}\label{the-nauxefve-physicist}

The ``naïve physicist'' looks at this and says,
``it is simple to reconstruct the CMB maps.
Wherever the telescope is pointing to,
we can simply average down the signal to the corresponding pixel.''

Except that it would not work, take a look at \cref{fig:007-040}:

\begin{figure}
\centering
\pandocbounded{\includegraphics[keepaspectratio,alt={Binned map}]{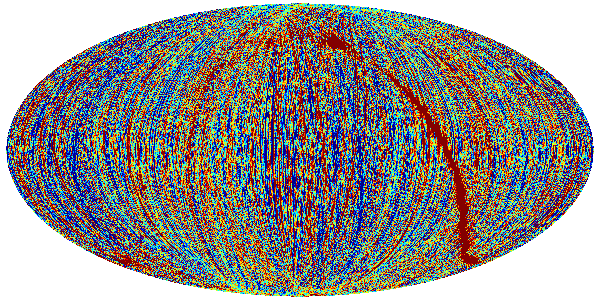}}
\caption[Binned map]{Binned map\footnotemark{}}\label{fig:007-040}
\end{figure}
\footnotetext{\citeproc{ref-keihanen_making_2010}{Keihänen et al., {``Making Cosmic Microwave Background Temperature and Polarization Maps with MADAM''}}.}

We can see visible vertical stripes here.
These stripes will appear
in the scanning directions
because of the infamous \(1/f\) noise.
The high amplitude, long time-scale noise
is manifest in the next scanning line
which now has a vastly different amplitude.

No wonder this is literally called naïve mapmaker.

\section{The ``naïve mathematician''}\label{the-nauxefve-mathematician}

The ``naïve mathematician'' looks at \cref{eq:scanning} and says,
``wait a minute, I recognize this,
this is just a standard linear regression problem,
and we can write down the solution immediately
by Generalized Least Squares (GLS).''

Indeed, by GLS, we know the solution to \cref{eq:scanning} is\footnote{How may \(N_{tt'}\) be estimated?
  In the framework of GLS, it is part of the model.
  That is, it is assumed to be a known quantity.
  In practice, one needs to build a model to describe
  the noise properties and then estimate the parameters
  of the noise model using the data.
  Note that in CMB observations,
  the signal is so faint that the TOD is dominated by the noise.
  In the special case that the noise model is uncorrelated white noise,
  one can simply estimate the noise variance using FFT on the TOD utilizing this fact.
  Incidentally, this technique is used in POLARBEAR\@.
  See Keihänen et al., \citeproc{ref-keihanen_making_2010}{{``Making Cosmic Microwave Background Temperature and Polarization Maps with MADAM''}} for another assumption and technique to model
  and estimate the noise covariance.}

\begin{align}
N_{tt'}             &= \left\langle n_t n_{t'} \right\rangle        \nonumber               \\
W_{pp'}             &= P_{pt}^T N^{-1}_{tt'} P_{t'p'}               \nonumber               \\
\hat{m}_p           &= W^{-1}_{pp'} P_{p't}^T N^{-1}_{tt'} d_{t'}   \label{eq:map_estimator}   \\
\text{Cov}(\hat{m}) &= W^{-1}                                       \label{eq:map_covariance}
\end{align}

Moreover, it can be shown that this solution is unbiased,
and lossless in the sense that
the Fisher information matrix
with respect to the input parameters is preserved.

This is called maximal likelihood mapmaker,
which is one example of optimal mapmakers that are lossless.

Indeed, this has been the mapmaker used in CMB experiments
such as COBE (1990s) in the beginning.
But as pointed out by Borrill,\footnote{\citeproc{ref-borrill_challenge_1999}{{``The Challenge of Data Analysis for Future CMB Observations,''} 1999}.} this has been computationally prohibitive for about 2 decades,
because of the \(O(N_p^3)\) behavior of the algorithm involved in
eq.~\ref{eq:map_estimator} \&~\ref{eq:map_covariance}.\footnote{Actual time complexity is \(O(N_p^3)\) with a larger prefactor
  involving multiple Cholesky decompositions in later steps. See Borrill, \citeproc{ref-borrill_challenge_1999}{} for details.}

Today, a minority of CMB experiments still use this to solve for the map,
but not the covariance.
Hence their mapmaking is unbiased,
but not lossless.

\section{Naïve meeting naïve becomes not so naïve}\label{nauxefve-meeting-nauxefve-becomes-not-so-nauxefve}

One day the naïve physicist met the naïve mathematician,
and they realized when the noise is uncorrelated white noise,
then the naïve mapmaker is the optimal mapmaker.

But we have seen that the noise is far from white.
We could, however, cheat and make it white.

The trick is to high-pass-filter our data
to remove excessive low frequency noise.
Then, as we have seen earlier, CRHWP modulation
already makes the noise quite white to begin with.

But now we have a new problem,
since we filtered our data,
the map is now biased.
And in practice, we typically have thrown away
on the order of dozens of percent of information
which is the cost to pay for this naïveness.

\begin{figure}
\centering
\pandocbounded{\includegraphics[keepaspectratio,alt={Biased mapmaking}]{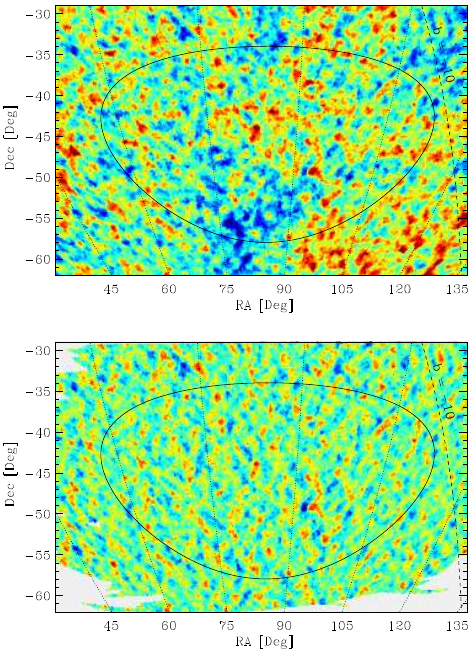}}
\caption[Biased mapmaking]{Biased mapmaking\footnotemark{}}\label{fig:biased-map}
\end{figure}
\footnotetext{\citeproc{ref-hivon_master_2002}{Hivon et al., {``MASTER of the Cosmic Microwave Background Anisotropy Power Spectrum''}}.}

\Cref{fig:biased-map} shows us an example of high-pass filtering,
we can see that the biased map below has much less large scale structure
than the original input above.
While the map is biased,
its effect can be corrected in the power-spectrum domain
via a method called MASTER devised in the same paper.

The tradeoff for the lossy mapmaking is
now we have superior computational requirement.
The naïve mapmaker, as we are simply binning the data,
is a streaming algorithm,
it has \(O(N_t)\) time complexity,
and is embarrassingly parallel.

For example, this is the mapmaker of choice in the POLARBEAR experiment.

\section{Unbiased mapmaking}\label{unbiased-mapmaking}

But unbiased mapmaking is desirable
if it is achievable.
At the very least it is a cost tradeoff
between data and computational time.
To illustrate, it might be cheaper to build more detectors for ground-based experiments,
but for satellite-based experiments,
we definitely want to squeeze the last bit of information
from the data.
Not to mention unbiased maps can provide
useful information within and beyond the CMB communities.

There are many unbiased mapmakers
with different iterative solutions,
examples are one used in a subset of POLARBEAR analysis,\footnote{\citeproc{ref-poletti_making_2017}{Poletti et al., {``Making Maps of Cosmic Microwave Background Polarization for B-Mode Studies''}}.} and the Madam mapmaker\footnote{\citeproc{ref-keihanen_making_2010}{Keihänen et al., {``Making Cosmic Microwave Background Temperature and Polarization Maps with MADAM''}}.} used in Planck analysis
and will be used in many future CMB experiments including LiteBIRD.

They are often solving the mapmaking equation iteratively
to achieve a lower time complexity,
therefore they are unbiased
and often asymptotically optimal.

\begin{figure}
\centering
\pandocbounded{\includegraphics[keepaspectratio,alt={Noiseless map}]{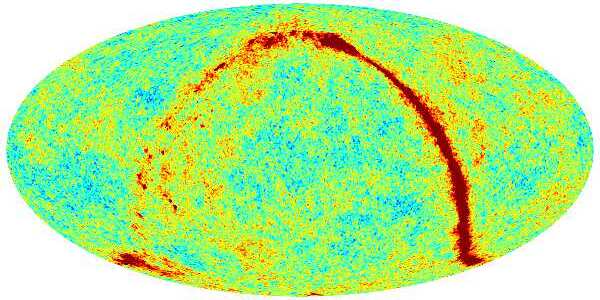}}
\caption[Noiseless map]{Noiseless map\footnotemark{}}\label{fig:madam-before}
\end{figure}
\footnotetext{\citeproc{ref-keihanen_making_2010}{Keihänen et al., {``Making Cosmic Microwave Background Temperature and Polarization Maps with MADAM''}}.}

\begin{figure}
\centering
\pandocbounded{\includegraphics[keepaspectratio,alt={Destriped map}]{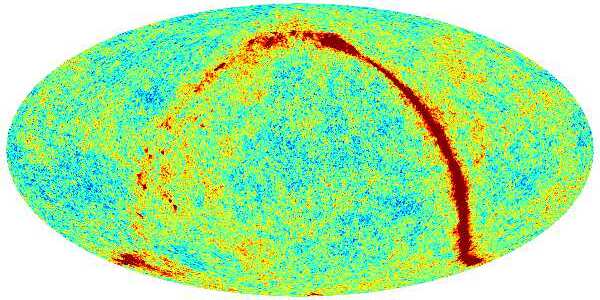}}
\caption[Destriped map]{Destriped map\footnotemark{}}\label{fig:madam-after}
\end{figure}
\footnotetext{\citeproc{ref-keihanen_making_2010}{Keihänen et al.}}

From the example of Madam mapmaker in \cref{fig:madam-before} \& \cref{fig:madam-after},
we can see that the two are very similar,
meaning that the destriped map successfully
recovers the noiseless map,
comparing to the visible stripes in \cref{fig:007-040}.

\chapter{Systematics characterization and mitigation strategies}\label{systematics-characterization-and-mitigation-strategies}

\chaptermark{Systematics characterization and mitigation}

\section{Timestream processing}\label{timestream-processing}

Now that we know how we generally obtain a map,
and we understand its statistical properties,
we turn our attention to systematics.

\begin{figure}
\centering
\pandocbounded{\includegraphics[keepaspectratio,alt={Cosmic ray seen by Planck}]{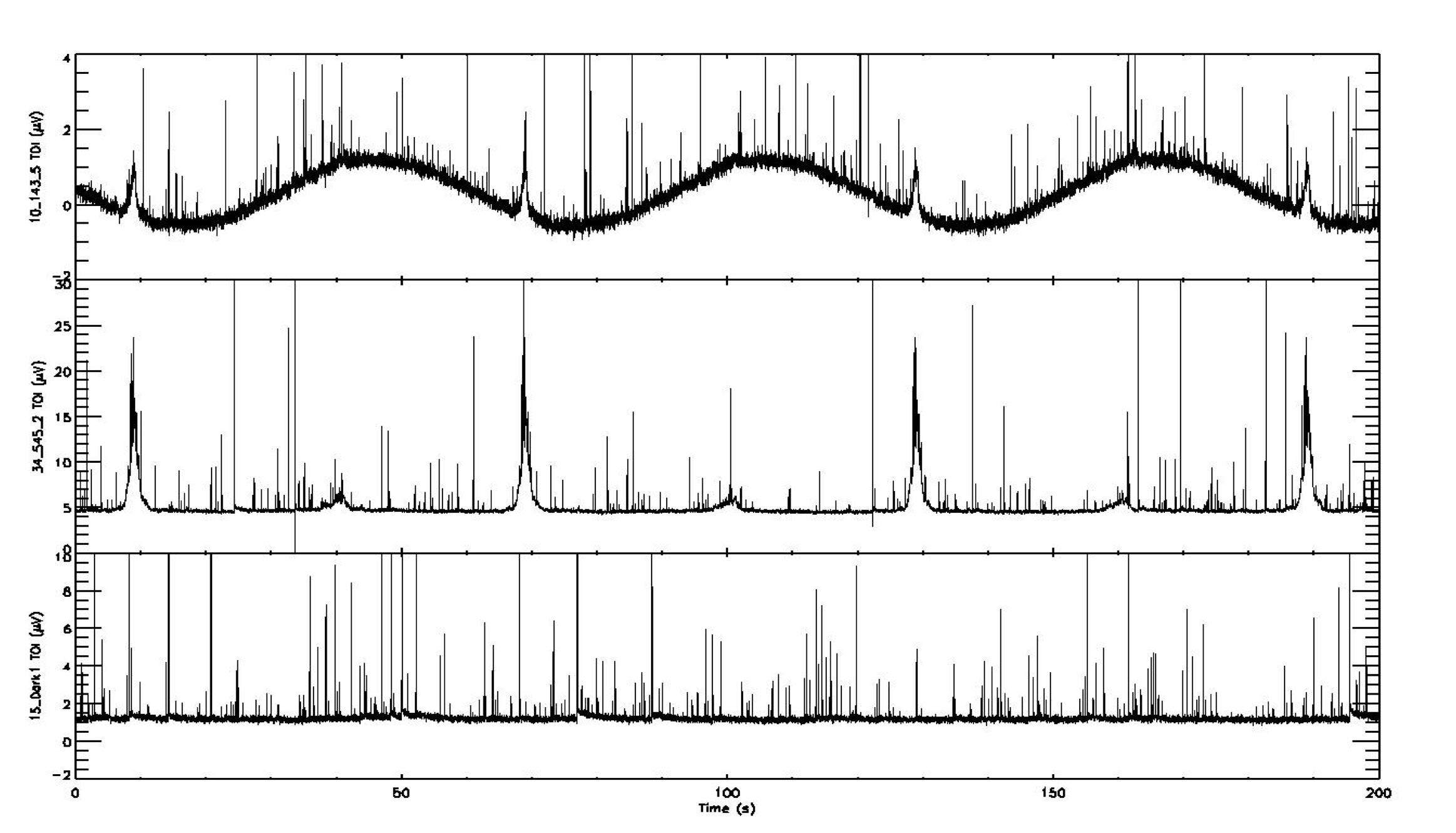}}
\caption{Cosmic ray seen by Planck}\label{fig:cosmic-ray}
\end{figure}

In fact, before we pass the TOD to the mapmaker,
we need to do some cleanup first.
For example, detectors have glitches
from sources such as cosmic rays as in \cref{fig:cosmic-ray},
which are not part of our sky signal
and thus should be removed.

There is also scan-synchronous signal,
such as ground pickup.
As the telescope is scanning the sky,
the signal from the surroundings of the telescope
can contaminate the sky signal it is observing.
One mitigation strategy is to construct a ground template
and subtract that from the TOD.

Other examples that we have already mentioned are
I-to-P leakage subtraction,
high pass filtering,
foreground masking when building templates.

\section{Map-domain processing}\label{map-domain-processing}

Having the TOD made into maps,
not only did we perform data compression,
but also it boosts the signal-to-noise ratio.
Hence, we can perform further systematic mitigation.

It can be as simple as masking out the foreground
such as point sources and our galaxy as in \cref{fig:Planck_2018_T_CMB}.
And if we have multi-frequency observations,
we can extract foreground components as in
\cref{fig:2015_FGAmpl},
\cref{fig:Planck_2018_synch_P_rc6_40arc_n1024_v5},
and \cref{fig:Planck_2018_dust_P_rc6_5arc_n1024_v5}.

\chapter{CMB power-spectra}\label{cmb-power-spectra}

We have previously mentioned that
we can obtain the power-spectra
by performing spherical harmonics transformations
and form \(TT\), \(TE\), \(EE\), \(BB\) spectra.
But we have not mentioned why.

One reason is that maps are not unique.
Our Universe is a random realization of
fundamental physics and its parameters,
and we as observers in the Universe
are a random point-of-view.
Changing either of these will change the map
but not the power-spectra,
as it is probing on the expectation values
of these random fields.

\begin{figure}
\centering
\pandocbounded{\includegraphics[keepaspectratio,alt={TT spectrum}]{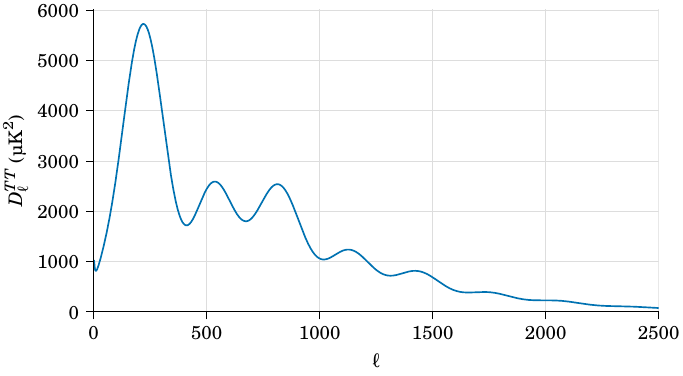}}
\caption{\(TT\) spectrum}\label{fig:planck_2018_TT}
\end{figure}

\begin{figure}
\centering
\pandocbounded{\includegraphics[keepaspectratio,alt={Intensity map (real): the CMB temperature measured by Planck (2018 SMICA map, inpainted)}]{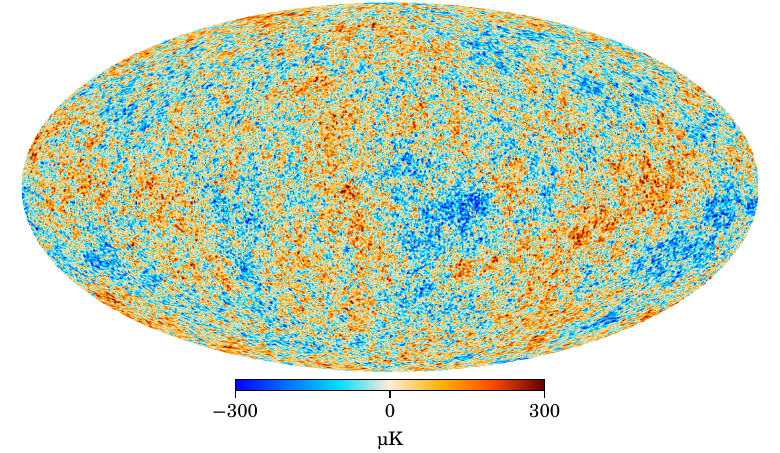}}
\caption[Intensity map (real): the CMB temperature measured by \emph{Planck} (2018 SMICA map, inpainted)]{Intensity map (real): the CMB temperature measured by \emph{Planck} (2018 SMICA map, inpainted)\footnotemark{}}\label{fig:planck_2018_measured}
\end{figure}
\footnotetext{\citeproc{ref-planck_collaboration_planck_2020-2}{Planck Collaboration, Aghanim, Akrami, Arroja, et al., {``\emph{Planck} 2018 Results''}}.}

\begin{figure}
\centering
\pandocbounded{\includegraphics[keepaspectratio,alt={Intensity map (random): a random realization of the spectrum in , at the resolution of }]{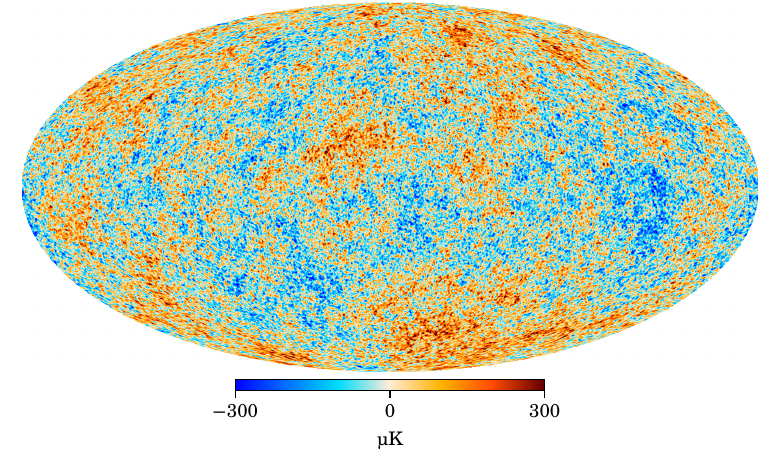}}
\caption{Intensity map (random): a random realization of the spectrum in \cref{fig:planck_2018_TT}, at the resolution of \cref{fig:planck_2018_measured}}\label{fig:planck_2018_random}
\end{figure}

Here is a simple example that from the same power-spectrum in \cref{fig:planck_2018_TT},
the one we measured here in the Universe is \cref{fig:planck_2018_measured},
but \cref{fig:planck_2018_random} is another random realization that yields the same spectrum!

Furthermore, theoretically we expect
the primordial random fields to be nearly Gaussian,
which is also not violated observationally.
This means that the power-spectra as
2nd order statistics are sufficient statistics.
That is, this is a lossless data compression,
which further helps us to boost the signal-to-noise ratio.

This means that we can perform further systematic mitigation,
such as unresolved point sources,
gravitational lensing of the CMB by the foreground,
or we can construct templates for systematic effects
that are not mitigated in earlier steps.

\section{Beyond 2-pt}\label{beyond-2-pt}

Although we said that 2-pt statistics are sufficient statistics
for the primordial CMB, the CMB we observed has
non-linear, hence non-Gaussian distortion,
for example by the gravitational lensing from the foreground.
While this distorts all power-spectra,
this affects B-mode the most as we have seen earlier.

In fact, the lensing potential can be constructed to describe the gravitational lensing effects,
which can then be used together with the \(E\)-mode
to produce a lensing \(B\)-mode template,
which is what we have seen in \cref{fig:BB-template-60-arcmin}.

\chapter{Simulations of CMB experiments}\label{sec:cmbSimulations}

\begin{figure}
\centering
\pandocbounded{\includegraphics[keepaspectratio,alt={Analysis pipelines}]{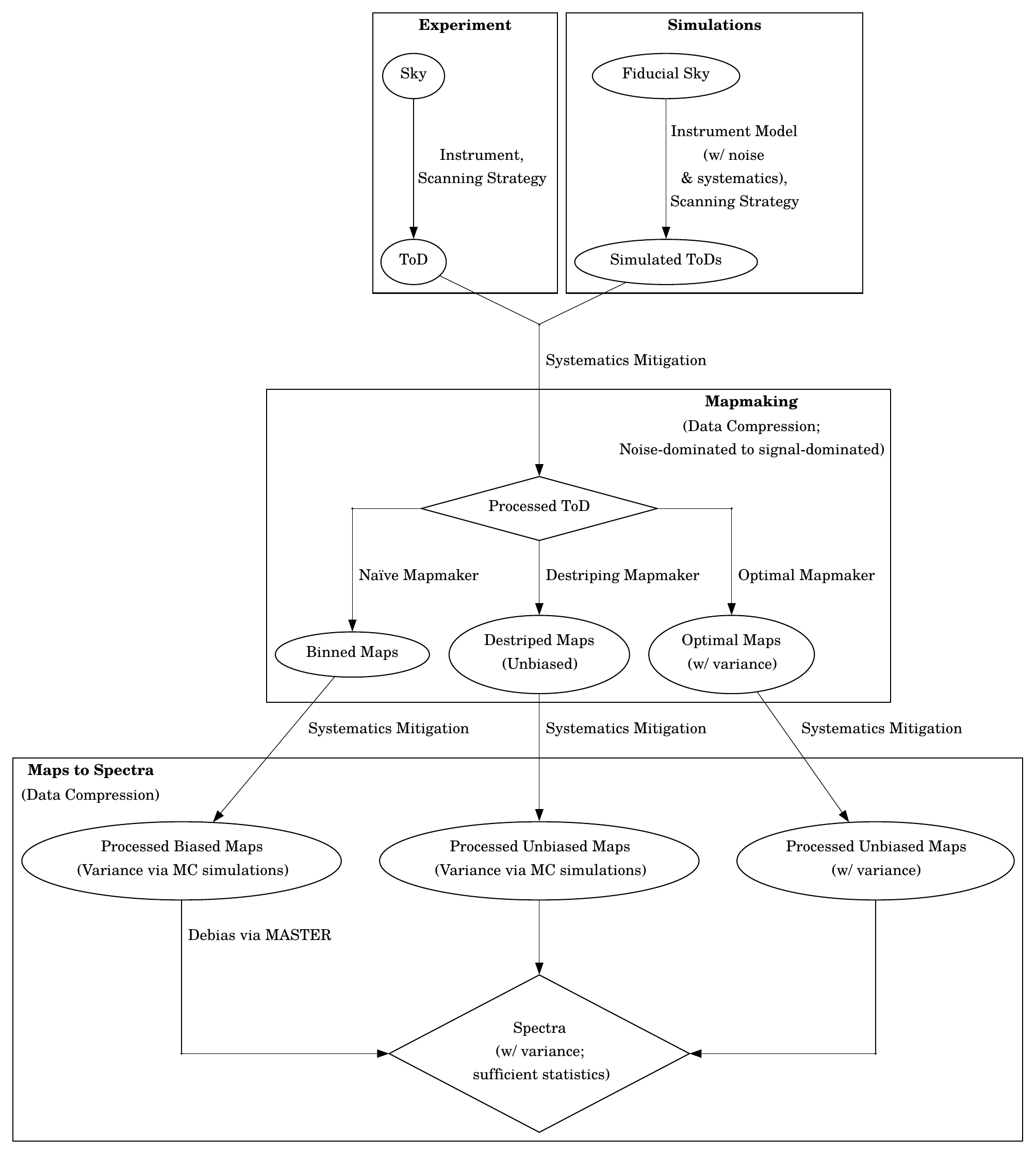}}
\caption{Analysis pipelines}\label{fig:analysis}
\end{figure}

Now that we have gone through all the major components needed
for CMB data analysis,
we can now look at the big picture of a general analysis pipeline
in \Cref{fig:analysis}.

The only missing piece so far
is the ability to simulate everything we said
with an instrument model that captures
the important characteristics such as
scanning strategy,
noise,
systematic effects,
etc.

There are many reasons we would need this.
But the fundamental reason in our formulation here
that we need simulation is that
our mapmakers do not produce the covariance information
due to computational constraints.

Therefore, we need to perform Monte Carlo (MC) simulations.
We inject random noise with physical characteristics
and use Monte Carlo method to estimate the statistical error
in our final power-spectra.
The typical number would be \(O(100-1000)\),
which means that computationally we are trading off
a higher time complexity to a lower one with a larger pre-factor.

Once we have this facility in place,
we can use this to perform pipeline validation,
often early in the process of designing an experiment,
to demonstrate the correctness of the pipeline,
and the capability of handling the observed dataset.

Also, we can use this to simulate systematic effects
by injecting them into the data
and measuring the outcome
such as any biases and its associated error,
and then devise systematic mitigation strategies
and quantify their effectiveness.

Being able to simulate the statistical and systematic error,
we can use this to perform forecasts
which can inform experimental design.

\chapter{Posterior estimation from the data and its likelihood}\label{posterior-estimation-from-the-data-and-its-likelihood}

\chaptermark{Posterior estimation}

Now that is where the fun begins.
From here, we can start to infer about our universe.

\begin{figure}
\centering
\pandocbounded{\includegraphics[keepaspectratio,alt={Likelihood \& Inference}]{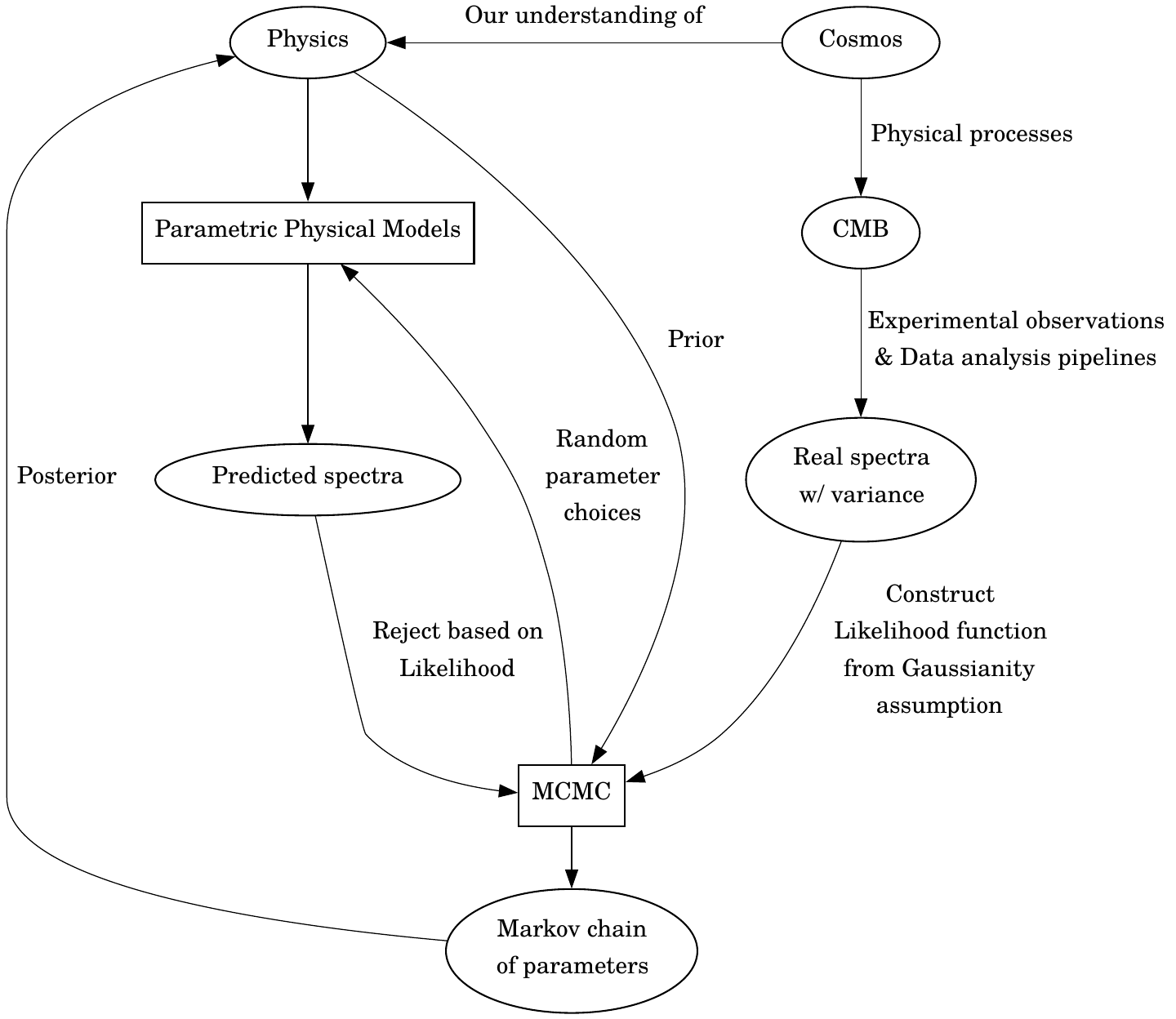}}
\caption{Likelihood \& Inference}\label{fig:likelihood}
\end{figure}

In \Cref{fig:likelihood},
the pipeline enables us to obtain
the CMB power-spectra with covariance information.

Our theoretical understanding of the Universe
enables us to build parametric Physical models,
which include cosmological parameters such as
those in the ΛCDM model including
matter compositions \& spatial curvature,
new physics,
or parameterized systematics
such as foreground components
\& their spectral indices.

Gathering all the pieces,
we can perform Markov chain Monte Carlo (MCMC) method
to ``invert'' from observed power-spectra to parameters.
It works like this:

\begin{enumerate}
\def\labelenumi{\arabic{enumi}.}
\tightlist
\item
  construct a likelihood function
  using the measured power-spectra
  assuming Gaussianity,
\item
  write down the prior information from what we already know in physics,
  such as the Universe is flat from other measurements,
\item
  make random parameter choices and use our model to predict the output power-spectra,
\item
  accept or reject the choice based on the likelihood and prior,
\item
  obtain a Markov chain of parameters,
\end{enumerate}

which gives us an approximate
maximum a posteriori (MAP) probability estimation.

This leads to
refinements of our understanding of the Universe
or even discovery of new physics
depending on the significance of its parameters.

\chapter{Current measurements}\label{current-measurements}

Now let us take a look at the current measurements of the CMB.

\begin{figure}
\centering
\pandocbounded{\includegraphics[keepaspectratio,alt={CMB Power-spectra summary}]{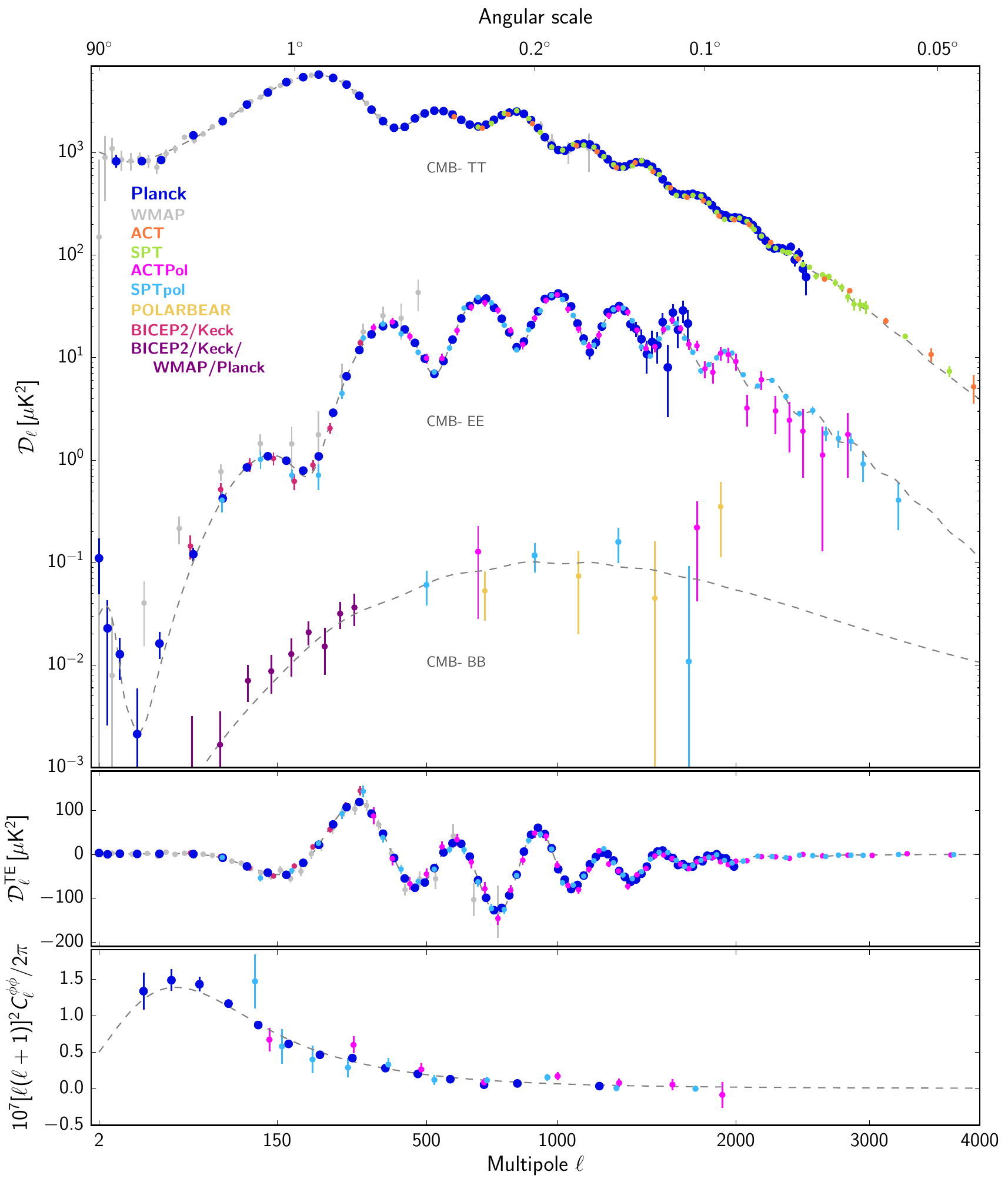}}
\caption[CMB Power-spectra summary]{CMB Power-spectra summary\footnotemark{}}\label{fig:2018_compilation_v5}
\end{figure}
\footnotetext{\citeproc{ref-planck_collaboration_planck_2020-2}{Planck Collaboration, Aghanim, Akrami, Arroja, et al., {``\emph{Planck} 2018 Results''}}.}

In \cref{fig:2018_compilation_v5},
we can see the spectra.
We can see most of these are measured very accurately
over a wide range of scales.
In particular, the \(TT\) spectrum
at \(\ell \lesssim 1600\)
is measured approaching
the cosmic variance limit
meaning that observationally we cannot further
improve the error-bar beyond a factor of a few fundamentally.

As we explore each of the curves,
we can see boundaries that can be pushed in future CMB experiments.
For example, the tails and the lensing constraints can be improved,
resulting in better understanding of Neutrino physics
such as the effective number of species,
sum of Neutrino masses,
and its mass hierarchy.

Furthermore, we can see that the B-mode polarization,
where evidence is first reported by POLARBEAR,
is the least constrained.
Not shown on the graph,
the primordial B-mode at \(\ell \lesssim 100\) is not measured yet.
This is the science goal of many upcoming CMB experiments including LiteBIRD\@.
This primordial B-mode is often parameterized by
the tensor-to-scalar-ratio \(r\).
Let us see what can be achieved by measuring this.

\begin{figure}
\centering
\pandocbounded{\includegraphics[keepaspectratio,alt={}]{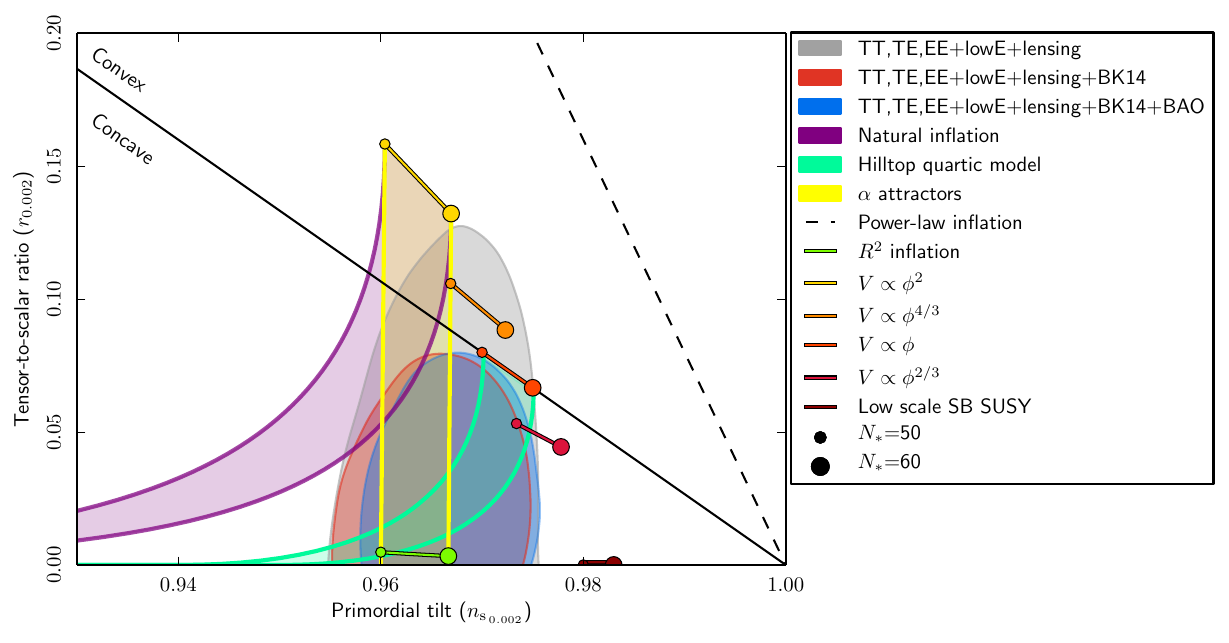}}
\caption[]{\footnotemark{}}\label{fig:Planck2018_TestTestV4BK14LegacyV4_120mm}
\end{figure}
\footnotetext{\citeproc{ref-planck_collaboration_planck_2020}{Planck Collaboration, Akrami, Arroja, et al., {``\emph{Planck} 2018 Results''}}.}

In \cref{fig:Planck2018_TestTestV4BK14LegacyV4_120mm},
the 2 axes are \(r, n_s\).
As we mentioned in \cref{sec:inflation},
we know that \(n_s < 1\)
is a sign of inflation.
But it alone cannot break
the degeneracy in inflationary models
as seen in this plot.
This \(r\) offers us
a chance to glimpse into the details
or even to falsify
inflationary models.

\chapter{Next stage of CMB experiments}\label{sec:next-stage}

In order to achieve these goals,
there are a lot of upcoming challenges.

Regardless of the science goal
of a particular CMB experiment,
a common enemy is noise.
The remaining battlegrounds
as we have seen are weaker signals.
And as we are probing weaker and weaker signals,
we need to achieve lower noise levels.

\begin{figure}
\centering
\pandocbounded{\includegraphics[keepaspectratio,alt={Exponential CMB data growth. The dotted line represents the projected growth in the post-Moore's law era.}]{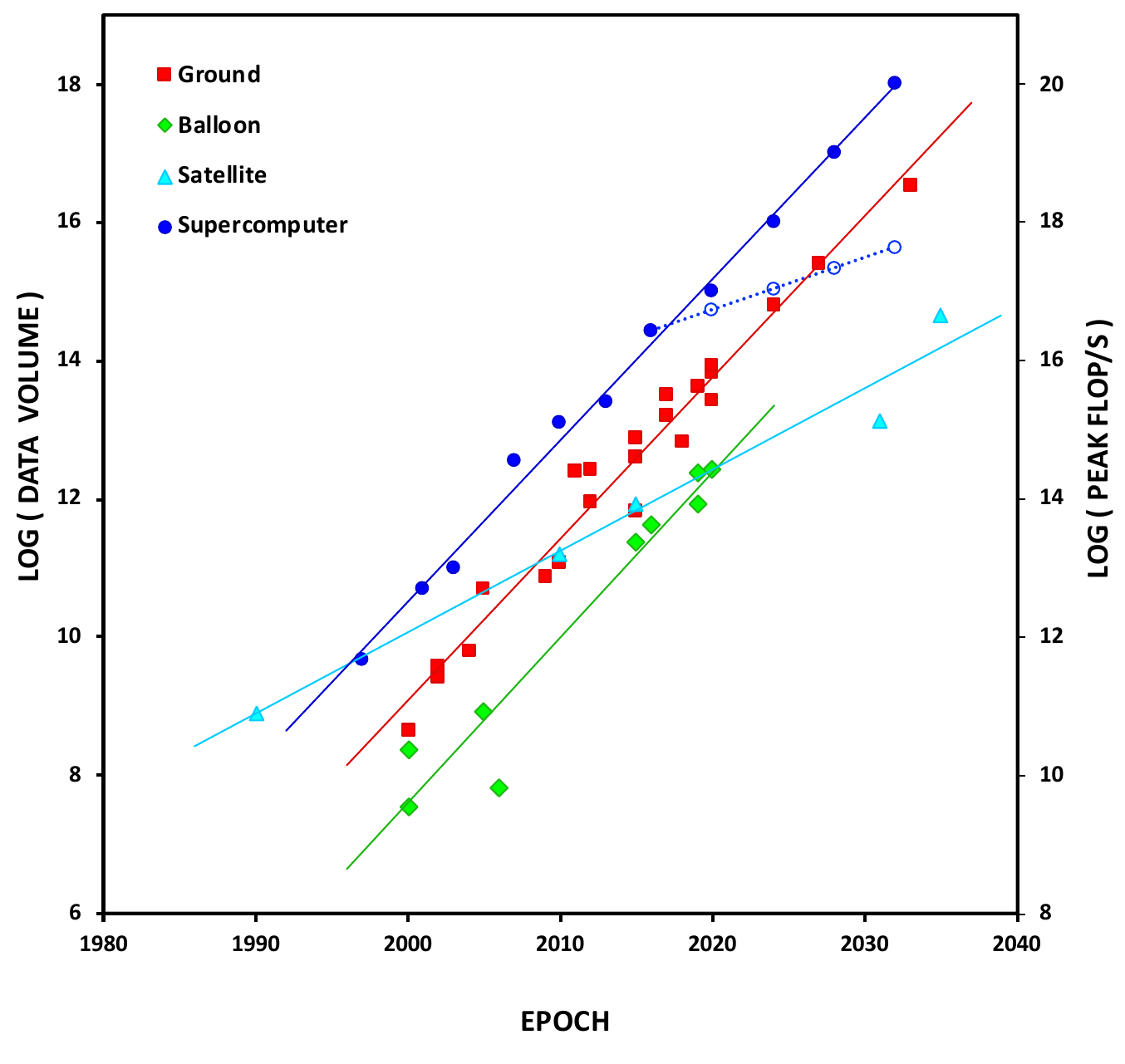}}
\caption[Exponential CMB data growth. The dotted line represents the projected growth in the post-Moore's law era.]{Exponential CMB data growth.\footnotemark{} The dotted line represents the projected growth in the post-Moore's law era.}\label{fig:exponential-cmb-data-growth}
\end{figure}

As our detectors are already photon-noise limited,
it means that we need to increase the number of detectors.
This means inevitably the CMB data volumes will grow,
and the growth is exponential as we can see in \Cref{fig:exponential-cmb-data-growth}.

\footnotetext{\citeproc{ref-borrill_exponential_nodate}{Borrill, \emph{Exponential CMB Data Growth}}.}

In general, all the datasets are growing exponentially
across different sectors.
On the other hand, HPC growth is slowing down
as we are approaching the end of Moore's law,
basically because we are hitting a Physical limit there.

In fact, if we look at the ground-based observations,
the CMB data growth will soon surpass
the HPC growth.

This means we need to stay on the cutting edge
in hardware and software
to deliver the computational requirements
of CMB analysis.

Moreover, systematics do not average down.
As we are probing weaker signals,
systematic effects that were once negligible
might matter now.
This imposes challenges to both
experimental design
and analysis software design
to minimize, quantify, and mitigate these systematic effects.

With the big picture in our mind,
let us focus on my research
in 2 specific CMB experiments---the POLARBEAR experiment and the LiteBIRD experiment.
In \cref{sec:POLARBEAR},
we will briefly introduce the POLARBEAR experiment,
a ground based CMB experiment in the past decade.
In \cref{sec:polarbear-research},
we will focus on my research topic on the large-patch high-\(\ell\)
\(B\)-mode measurement.
Lastly, in \cref{sec:LiteBIRD-research},
we will study the crosstalk systematic effects
in a future satellite-based CMB experiment LiteBIRD.

\part{The POLARBEAR experiment}\label{sec:POLARBEAR}

\chapter{The story behind the name POLARBEAR}\label{sec:whats-in-a-name}

To introduce the POLARBEAR experiment,
let us start with its name,
which has a story too good
to be skipped.

For a long while I did not know its meaning.
From some casual conversation,
I thought it was because POLARBEAR
started at the University of California, Berkeley (UCB)
and Cal Bears\footnote{The athletic teams representing the UCB is the California Golden Bears, often called Cal Bears in short. See \url{https://www.calbears.com/}.} is its identity,
and we measured the polarization in the CMB,
so obviously POLARBEAR it is.

That was until I was preparing for this
and tried searching for its actual etymology.
A quick search on ``acronym POLARBEAR cosmology''
will land you on
\href{https://www.cfa.harvard.edu/~gpetitpas/Links/Astroacro.html}{Dumb Or Overly Forced Astronomical Acronyms Site (DOOFAAS)},
a site created by Dr.~Glen R. Petitpas from Harvard
dedicated to dumb astronomical acronyms.
POLARBEAR stands for
\textbf{POLAR}ization of the \textbf{B}ackground millim\textbf{E}ter b\textbf{A}ckground \textbf{R}adiation.

Except that it is wrong.
Probably the misunderstanding originated from a 2003 article\footnote{\citeproc{ref-delabrouille_polarization_2003}{Delabrouille et al., {``Polarization Experiments''}}.}
that somehow thought it was true
without any citations on how they knew that.

It turns out it stands for
\textbf{POLAR}ization of the \textbf{B}ackground \textbf{R}adiation,
which is defined rarely in some of the POLARBEAR papers.\footnote{\citeproc{ref-ade_polarbear_2015}{Ade et al., {``POLARBEAR Constraints on Cosmic Birefringence and Primordial Magnetic Fields''}}.}

However, that is not the end of the story.
How should it be typeset?
I have seen the following forms used in various POLARBEAR publications:

\begin{enumerate}
\def\labelenumi{\arabic{enumi}.}
\tightlist
\item
  PolarBeaR in the early days to emphasize its construction;
\item
  POLARBEAR, \textsc{polarbear} or sometimes \textsc{Polarbear}
  which uses \textsc{small caps} as a stylistic typographical choice; and
\item
  Polarbear very occasionally.
\end{enumerate}

Polarbear, and its stylistic derivative \textsc{Polarbear},
is clearly wrong as it represents a capitalization of a non-existing word ``polarbear''.
PolarBeaR does not share wide-spread use.
POLARBEAR and \textsc{polarbear} are both correct,
while the former is used more often.

Hence, POLARBEAR is used consistently as the name of the experiment throughout.

\chapter{The POLARBEAR telescope}\label{sec:medium-aperture-telescope}

\begin{figure}
\centering
\pandocbounded{\includegraphics[keepaspectratio,alt={POLARBEAR Huan Tran Telescope at the James Ax Observatory located at Cerro Toco in the Atacama Desert in Northern Chile}]{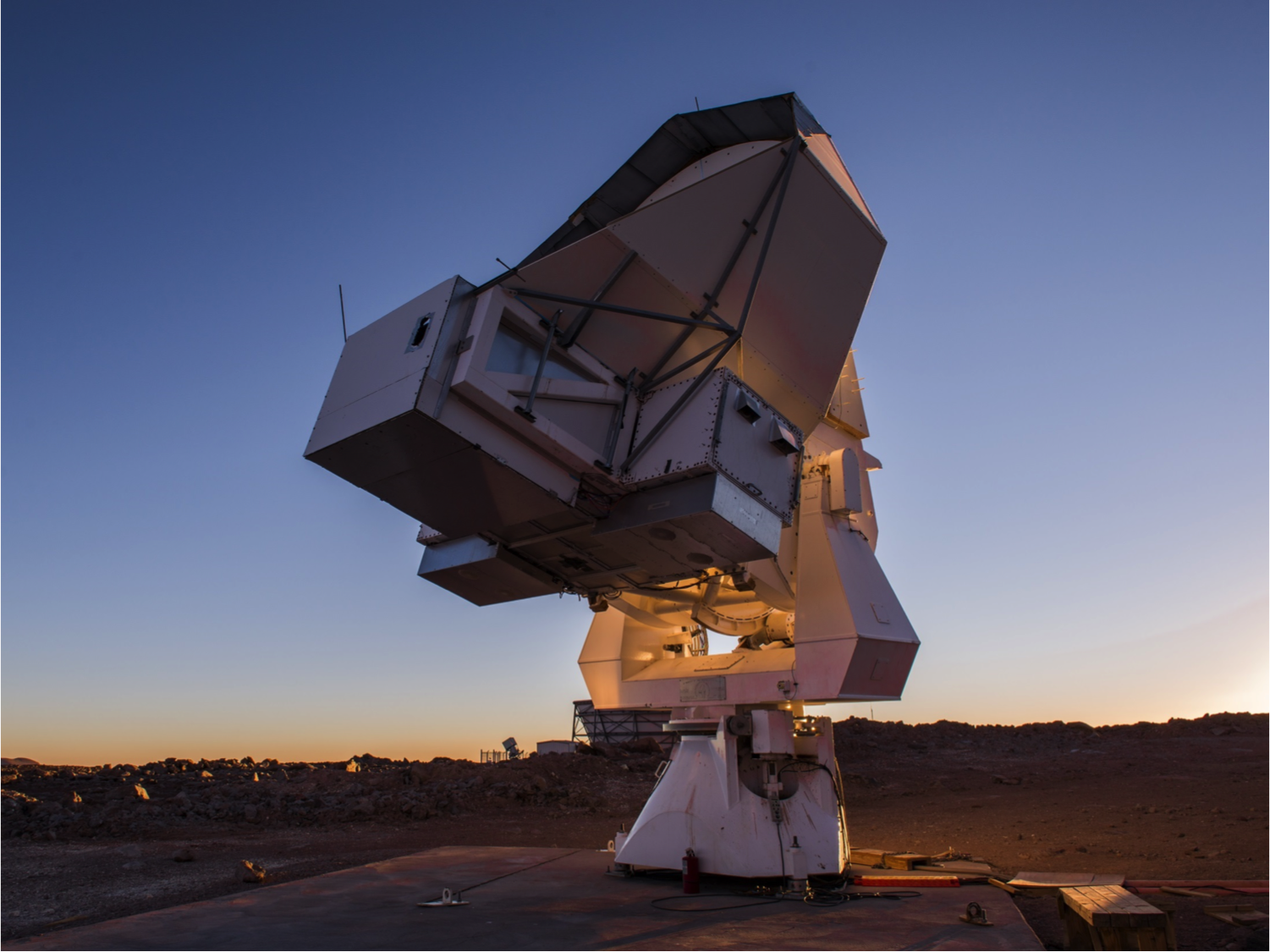}}
\caption[POLARBEAR Huan Tran Telescope at the James Ax Observatory located at Cerro Toco in the Atacama Desert in Northern Chile]{POLARBEAR Huan Tran Telescope at the James Ax Observatory located at Cerro Toco in the Atacama Desert in Northern Chile\footnotemark{}}\label{fig:image20}
\end{figure}
\footnotetext{\citeproc{ref-polarbear_collaboration_polarbear_nodate}{POLARBEAR collaboration, \emph{POLARBEAR Huan Tran Telescope at the James Ax Observatory Located at Cerro Toco in the Atacama Desert in Northern Chile}}.}

\begin{figure}
\centering
\pandocbounded{\includegraphics[keepaspectratio,alt={Cross section of POLARBEAR (left) and detail of the area around the prime focus (right).}]{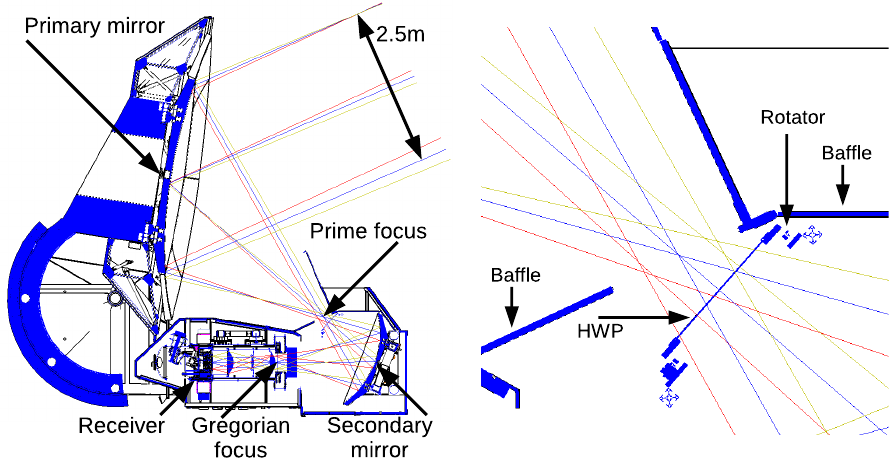}}
\caption[Cross section of POLARBEAR (left) and detail of the area around the prime focus (right).]{Cross section of POLARBEAR (left) and detail of the area around the prime focus (right).\footnotemark{}}\label{fig:PB17-HWP-fig-1}
\end{figure}
\footnotetext{\citeproc{ref-takakura_performance_2017}{Takakura et al., {``Performance of a Continuously Rotating Half-Wave Plate on the POLARBEAR Telescope''}}.}

POLARBEAR is a CMB experiment installed on the
\(\SI{2.5}{\metre}\) aperture Huan Tran Telescope as shown in \cref{fig:image20},
located at the James Ax Observatory
in the Atacama Desert in Chile\footnote{\citeproc{ref-the_polarbear_collaboration_measurement_2020}{The POLARBEAR Collaboration et al., {``A Measurement of the Degree-Scale CMB \emph{B}-Mode Angular Power Spectrum with POLARBEAR''}}.}
(see \cref{fig:PWV} and \cref{sec:atmosphere}),
with a \(\SI{3.5}{\arcminute}\) beam.
As shown in \cref{fig:PB17-HWP-fig-1},
it has an off-axis Gregorian design
with the advantage of
a clear aperture.\footnote{\citeproc{ref-kermish_polarbear_2012}{Kermish et al., {``The POLARBEAR Experiment''}}.} It packs 1274 transition edge sensor (TES) bolometers
observing at \(\SI{150}{\GHz}\).

Detectors are cooled to \(\SI{250}{\milli\kelvin}\),
whereas the lenses are at \(\SI{4}{\kelvin}\).
It achieves
\(\text{NET}_\text{array} = \SI{23}{\micro\kelvin \sqrt{\second}}\).

POLARBEAR has 5 seasons of observations.
In the 1st--2nd seasons (2012--2014),
it observed 3 different patches on the sky
with a total of \(\SI{25}{\degree^2}\) sky coverage.
It achieved a sensitivity of \(\SI{8.3}{\micro\kelvin\arcminute}\) in one of the patches named RA23.
A B-mode spectrum was constrained at \(500 \leq \ell \leq 2100\),
where the 1st season of data provided the first direct evidence of polarization lensing.\footnote{\citeproc{ref-ade_measurement_2014}{Ade et al., {``Measurement of the Cosmic Microwave Background Polarization Lensing Power Spectrum with the POLARBEAR Experiment''}}.}

After the 2nd season,
a Continuously Rotating Half-Wave Plate (CRHWP) was installed
to modulate the polarization signal
so that the \(1/f\) noise was reduced
to target large scale, low-\(\ell\) signal.
Optimally, it should be placed as far skyward as possible
to avoid I-to-P leakage caused by the asymmetry
in the optical system.
But the limitation of HWP technology
had put its maximum size to
about \(\SI{0.5}{\metre}\) in diameter.
Due to this constraint,
the optimal point to place it is
between the primary and secondary mirrors\footnote{\citeproc{ref-takakura_performance_2017}{Takakura et al., {``Performance of a Continuously Rotating Half-Wave Plate on the POLARBEAR Telescope''}}.}
as shown in \cref{fig:PB17-HWP-fig-1}.

\begin{figure}
\centering
\pandocbounded{\includegraphics[keepaspectratio,alt={POLARBEAR sky coverage}]{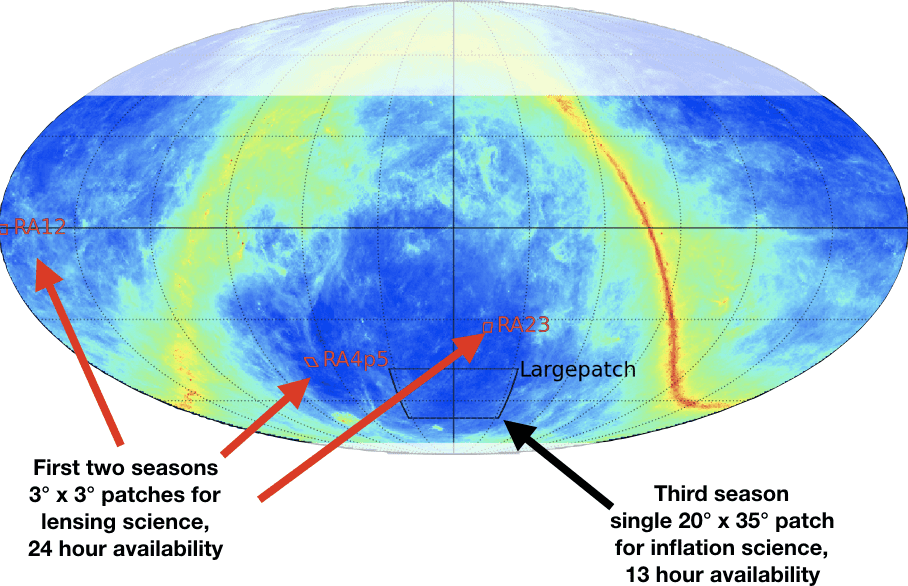}}
\caption{POLARBEAR sky coverage}\label{fig:pb-sky-coverage}
\end{figure}

In the remaining 3rd--5th seasons (2014--2017),
it observed a sky area of \(\SI{670}{\degree^2}\),
as shown in \cref{fig:pb-sky-coverage}.
The presence of the CRHWP and the larger sky area
enables us to target low-\(\ell\) science.
The resultant observed \(\ell\)-range is
\(50 \leq \ell \leq 3000\).
We called this informally the ``large-patch''.
This unique design
of a medium aperture telescope
with a CRHWP enables us
to target both low-\(\ell\) for primordial B-mode
and high-\(\ell\) for lensing B-mode.
It achieves a sensitivity of \(\SI{32}{\micro\kelvin\arcminute}\).

\chapter{The POLARBEAR null-test framework}\label{sec:POLARBEAR-null-test-framework}

Kendrick Smith first proposed a null-test framework for systematics detection
in CMB observations\footnote{\citeproc{ref-chinone_polarbear_2014}{Chinone, {``POLARBEAR Null-Test Framework''}}.}
in the QUIET experiment in 2011.\footnote{\citeproc{ref-quiet_collaboration_first_2011}{QUIET Collaboration et al., {``FIRST SEASON QUIET OBSERVATIONS''}}.} This concept of a null-test framework is used in POLARBEAR extensively.

It is based on null hypothesis testing.
If we can construct a null estimator
that should be identically zero in the
absence of a certain systematic effect,
then a ``discovery'' of this estimator being non-zero
implies the existence of a systematic effect
that is not well controlled.
Mathematically,

\[\hat{C}_{b}^{\text{null}} = \hat{C}_{b}^{A A} + \hat{C}_{b}^{B B} - 2 \hat{C}_{b}^{A B} ,\]

where \(A, B\) represent the two halves of a null-split\footnote{Here, each null-split refers to a choice of splitting the data into two halves---\(A\) \& \(B\).},
that is, \(A\) and \(B\) together contain the entirety of the observed dataset.
All the \(\hat{C}\) here are de-biased signal power-spectra,
defined by the procedure in \cref{sec:MASTER}.
By construction, we have \(\left\langle \hat{C}_{b}^{\text{null}} \right\rangle = 0\) in the absence of systematics\footnote{That is, the signal does not contain a systematic component that is different between \(A\) \& \(B\).},
hence a null estimator.

\Cref{fig:null-correlation} shows us
both the null-splits involved
and their correlations in our large-patch analysis.

Each null-split is chosen to be a physical variable
that can be used to split the data into two halves.
It is particularly useful as a diagnostic device
to reveal unknown systematic effects.

For example, \texttt{QU\_\allowbreak{}PIXEL} splits the data
into two halves based on the polarization angles
of the detectors fabricated on the wafer.
This can help us to check against
some unknown fabrication systematics.

In addition,
POLARBEAR is a blinded experiment
to avoid observer bias
and confirmation bias.
Only after
passing the whole null-test suite
and other unblinding criteria
such as Kolmogorov--Smirnov test,
we are allowed to look at
our science product---the measured power-spectra.

\section{Null split definitions}\label{sec:null-test-definitions}

The following are the null-splits used in the large-patch null-test suite,
first appeared in The POLARBEAR Collaboration et al.\footnote{\citeproc{ref-the_polarbear_collaboration_measurement_2020}{{``A Measurement of the Degree-Scale CMB \emph{B}-Mode Angular Power Spectrum with POLARBEAR''}}.}

\begin{description}
\tightlist
\item[FIRST\_SECOND]
``First half versus second half''

the dataset is split into two equal-weight halves chronologically to probe for time dependent miscalibration or changes in the instrument.
\item[RS\_MIDDLE, RM\_SET, SM\_RISE]
``middle versus rising and setting'',
``setting versus rising and middle'',
``Rising versus middle and setting''

the three different CES types are split in all possible combinations to detect elevation-dependent miscalibration or residual ground synchronous signal.
\item[LR\_SUBSCAN]
``Left-going versus right-going subscans''

the dataset is split in half according to the direction of motion of the telescope to test for microphonic or magnetic pickup in the data.
\item[GAIN\_BY\_CES]
``High gain versus low gain observations''

the dataset is split into observations with above and below average mean detector gain coefficients to search for problems with the gain calibration.
\item[PWV]
``High PWV versus low PWV''

the dataset is split by PWV as measured by the nearby APEX radiometer to check for loading or weather dependent effects.
\item[LEAK\_BY\_BOLO]
``Mean temperature to polarization leakage by channel''

split the dataset into detectors that see small and large temperature leakage coefficients to test the subtraction and search for residual contamination.
\item[LEAK\_BY\_CES]
``Mean temperature to polarization leakage by observation''

split the dataset into observations that see small and large temperature leakage coefficients to test the subtraction and search for residual contamination.
\item[AMP\_2F\_BY\_BOLO, AMP\_4F\_BY\_BOLO]
`` 2f amplitude by channel'',
`` 4f amplitude by channel''

split the data by HWP signal amplitude to check for problems removing the HWP structure or systematic contamination coupling into the data through these terms.
\item[AMP\_2F\_BY\_CES, AMP\_4F\_BY\_CES]
`` 2f amplitude by observation'',
`` 4f amplitude by observation''

split the data by HWP signal amplitude to check for problems removing the HWP structure or systematic contamination coupling into the data through these terms.
\item[QU\_PIXEL]
``Q versus U pixels''

each detector wafer is fabricated with two sets of polarization angles. We split the data into the two pixel types to check for problems in the device fabrication.
\item[SUN\_HORIZON, MOON\_HORIZON]
``Sun above or below the horizon'',
``Moon above or below the horizon''

we split observations based on whether or not the sun or moon is up to check for residual sidelobe contamination.
\item[SUN\_DIST, MOON\_DIST]
``sun boresight distance'',
``moon boresight distance''

we split observations based on the sun or moon boresight distance to check for residual sidelobe contamination.
\item[THS\_BHS, LHS\_RHS]
``Top half versus bottom half'', ``left half versus right half''

we split detectors by the boresight axis of the telescope to check for optical distortion and problems due to far sidelobes.
\item[TOP\_BOTTOM]
``Top versus bottom bolometers''

with a continuous HWP each bolometer TOD independently measures Q and U\@. We explicitly separate detector pairs to check for temperature aliasing or device mismatch.
\end{description}

\chapter{The POLARBEAR analysis pipelines}\label{sec:polarbear-mapmaking-requirements}

Because of this null-test setup
and the unblinding criteria,
it means that we need to be able to run
the pipeline for all these null-tests combinations
and to rerun it repeatedly
until we can pass the null-tests.
We need to figure out a way to fail fast,
speculate often on how to mitigate the systematics,
and repeat.

\section{Naïve Mapmaking}\label{nauxefve-mapmaking}

That is why the naïve mapmaker\footnote{see \cref{sec:CMBMapmaking}.}
is the mapmaker of choice.
Since it is just binning the data,
per TOD we can perform binning
all at once with different null-splits.
This makes the computational cost of
having more null-splits almost free.
(In fact, in one paper with the small-patch dataset,\footnote{\citeproc{ref-the_polarbear_collaboration_measurement_2017}{The POLARBEAR Collaboration et al., {``A Measurement of the Cosmic Microwave Background \emph{B}-Mode Polarization Power Spectrum at Subdegree Scales from Two Years of POLARBEAR Data''}}.} both the naïve mapmaker and an unbiased mapmaker are used
and they demonstrated they can only manage the full null-tests suite
on the naïve mapmaker pipeline because of computational constraints,
and the results from both pipelines are consistent.)

Per Constant Elevation Scan (CES),
which is about an hour of TOD observation\footnote{Constant Elevation Scan refers to the fact
  that those hour-long observations are scanning the sky at constant elevation.},
mapmaking can be performed independently.

Mathematically, we can make it abstract and write

\begin{equation}
m^{i} = \mathcal{M} \mathopen{} \left \{ d^{i} \right \} \mathclose{} , i = 1 , \dots , n_{\text{CES}} , \label{eq:mapmaking_illustration}
\end{equation}

where \(d\) represents the TOD, \(m\) represents the map,
\(i\) represents the \(i\)-th CES, and
\(n_\text{CES} = 3391\) represents the total number of CES\@.
\(\mathcal{M}\) represents the mapmaking procedure that turns TOD into maps.
While this procedure involves a lot of operations including
filtering and TOD processing,
the key feature here is that it is a mathematically well-defined operator.

Note that here many indices are suppressed,
explicitly, we have \(m^{i , X , n}_{p}\),
where \(X\) represents the different \(I, Q, U\) components,
\(n\) represents different null-splits, such as \(1A\) for the first half of the 1st null-split,
\(3B\) for the second half of the 3rd null-split, etc.
\(p\) represents the pixel index of the map, which is inherently a \(2\)-dimensional object.
As we take the flat-sky approximation here,
we can understand \(p\) as enumerating over
\((1, 1), (1, 2), \ldots, (2, 1), (2, 2), \ldots\).

Finally, note that \(m\) here is a random variable.
Mathematically,

\[m^{i , X , n}_{p} \sim \mathcal{N} \mathopen{} \left ( \mu^{i , X , n}_{p}, 1 / w^{i , X , n}_{p} \right ) \mathclose{} .\]

Uncorrelated Gaussian white noise in the TOD is assumed
in order for this to be strictly true.\footnote{In the limit that this assumption is exact,
  naïve mapmaker is the optimal mapmaker (See \cref{sec:CMBMapmaking}.)
  This implies that we need filtering to turn \(1/f\)-noise into approximately white noise,
  and deviation from this assumption results in less-than-optimal but usable maps,
  where usability is subject to other criteria including null-test passing and de-biasing.}
Here \(μ\) is what people usually called the map, where \(w\) is statistically called precision---the inverse of the variance,
and is usually called the weights among physicists.

\section{Coadd}\label{coadd}

After that, another pipeline called coadd
simply adds individual CES maps together,
weighted by inverse-variance.

Mathematically, first we should mention the variables that we propagate in the mapmaking and coadd pipelines to be \(M , w , \text{ where } M \equiv w \mu\).

And then we write down the inverse-variance weighting scheme,

\begin{align}
w^{I}	&	 = \sum_{i \in I} w^{i}	\label{eq:coaddByInvVar1}	\\
\mu^{I}	&	 = \frac{\sum_{i \in I} w^{i} \mu^{i}}{\sum_{i \in I} w^{i}}	\nonumber	\\
 \Rightarrow M^{I}	&	 = \sum_{i \in I} M^{i}	\label{eq:coaddByInvVar2}
\end{align}

Here the \(I\) index represents a map-bundle---a collection of CES maps
that together form a complete map of the patch we observed.
Each map bundle amounts to
about 10 days' worth of observation,
and we have 38 map-bundles in total.

This explains the verb ``coadd'' and
why we keep track of the combination of \(M, w\),
where \(μ = \frac{M}{w}\) is the quantity of interest.
It also explains why naïve mapmaking hugely simplifies the pipelines computationally.
To put it in programming terms,
the naïve mapmaker can be implemented using the MapReduce paradigm where
the mapmaking-pipeline \emph{maps} each TOD per CES into a map,
and the coadd-pipeline \emph{reduces} these maps into map bundles.

Note again that many indices are suppressed and can be expanded to \(m^{I , X , n}_{p}\).
Note that per map bundle \& null-split,
this is embarrassingly parallel.

\section{Cross-pseudo-power-spectra}\label{sec:pseudoSpectra}

The cross-pseudo-power-spectra,
often called pseudo-spectra for brevity,
between the null-split map bundles
will be computed by
the pseudo-spectra pipeline.\footnote{This is called pseudo
  because it is so far biased
  and will be corrected for in later steps.}

Schematically,

\[\tilde{C}^{X^{\prime} Y^{\prime} , n}_{b} = \mathcal{P} \mathopen{} \left \{ m^{I , X , n}_{p}, m^{J , Y , n}_{p} \right \} \mathclose{} ,\]

where \(\mathcal{P}\) is a mathematically well-defined operator\footnote{See The POLARBEAR Collaboration et al., \citeproc{ref-ade_measurement_2014-1}{{``A Measurement of the Cosmic Microwave Background \emph{B}-Mode Polarization Power Spectrum at Sub-Degree Scales with POLARBEAR''}}, eq.~16.}
that turns map-bundles into cross-pseudo-power-spectra.
Note that \(X' = T, E, B\) while \(X = I, Q, U\) where \(T \equiv I\).
For the purpose here, we would note only that \(\mathcal{P}\) involves Fourier transform,
a procedure to ``rotate'' \(Q, U\) components into \(E, B\), as well as a procedure to obtain a pure B-mode estimator
that suppresses E-to-B leakage.\footnote{\citeproc{ref-smith_pseudo-_2006}{Smith, {``Pseudo- C ℓ Estimators Which Do Not Mix E and B Modes''}}.} Computationally, we note that per null-split,
the spectra can be computed independently.

\section{Monte Carlo (MC) simulations}\label{monte-carlo-mc-simulations}

And the whole process will be repeated,
once with the real data,
and \(n_\text{sim}\) times each with a random simulation\footnote{See \cref{sec:cmbSimulations}.}.

Since all these are embarrassingly parallel,
it should be any HPC (High-Performance Computing) programmer's dream.

And because of the simplicity of the pipeline,
it is a policy in POLARBEAR to implement all these in pure Python.

\section{Splitting the pipeline into two}\label{sec:splitting-ell}

All these worked fabulously in small-patch.
But when processing the large-patch dataset,
it failed spectacularly.
Simply put,
the software had not been expecting
nor tested against the larger map size\footnote{Large-patch is about 2 orders of magnitude larger than small-patch.
  When maps are made at the same resolution of \(\SI{2}{\arcminute}\),
  small-patch would be hundreds of pixels wide, and large-patch would be thousands of pixels wide.}.

In the early days of large-patch analysis,
it was decided to make maps with much larger pixels
at \(\SI{10}{\arcminute}\) which is good enough for \(\ell \lesssim 600\).
This decision split the pipeline into two,
the low-\(\ell\) pipeline for \(50 \leq \ell \leq 600\)
targeting primordial B-mode,
and the high-\(\ell\) pipeline for \(600 \leq \ell \leq 3000\),
targeting lensing B-mode, E-mode, and the lensing potential.

In addition to ``solving'' the computational problem in the interim,
splitting such a large \(\ell\)-range into two
enables each pipeline to have their own optimization and mitigation strategies,
such as filtering, data-selection, etc.

\chapter{CRHWP demodulation}\label{sec:CRHWPDemodulation}

The simplicity and efficiency in POLARBEAR's mapmaking strategy was discussed earlier,
but a detail was glossed over about how different \(I, Q, U\) map components can be decomposed during mapmaking.

One commonly used scheme is called pair-differencing\footnote{See The POLARBEAR Collaboration et al., \citeproc{ref-ade_measurement_2014-1}{{``A Measurement of the Cosmic Microwave Background \emph{B}-Mode Polarization Power Spectrum at Sub-Degree Scales with POLARBEAR''}}, eq.~14, 15.}.
But as the large-patch is observed after the installation of CRHWP,
a huge simplification can be used instead.

We already know that CRHWP modulates the signal
to higher frequency such that
the polarization signal is away from the \(1/f\) noise.

\begin{figure}
\centering
\pandocbounded{\includegraphics[keepaspectratio,alt={Frequency-domain based demodulation}]{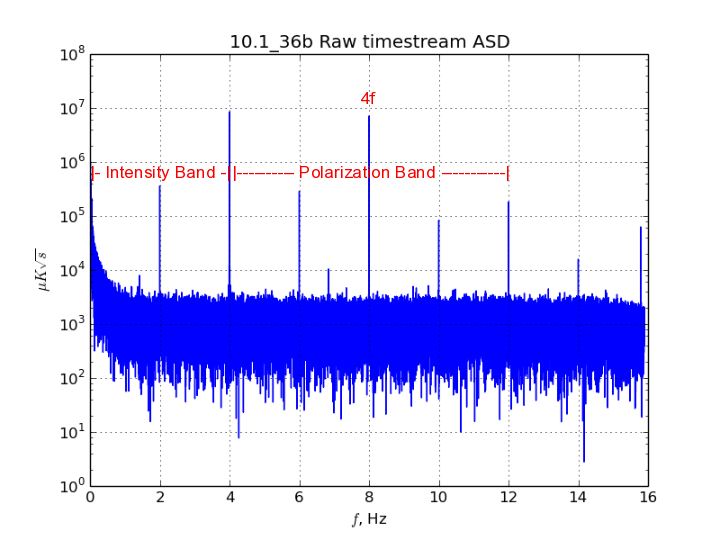}}
\caption{Frequency-domain based demodulation}\label{fig:sample_ASD-6462}
\end{figure}

\cref{fig:sample_ASD-6462} illustrates
a frequency-domain based demodulation scheme.
The signal is nicely separated in the frequency domain
provided that we choose the scanning speed
and rotational speed carefully.\footnote{\citeproc{ref-takakura_performance_2017}{Takakura et al., {``Performance of a Continuously Rotating Half-Wave Plate on the POLARBEAR Telescope''}}.} This allows us to obtain the intensity signal \(I\)
by low pass filtering,
and the polarization signal
first by high pass filtering,
then by a translation in the frequency domain
because the HWP is rotating at constant speed.
And if we do this in phasor then we can obtain
\(Q\) \& \(U\) as the real and imaginary parts respectively.

And all these are obtained in the time-domain.
So it enables a workflow that the TOD is pre-processed with lower level processing
including demodulation,
and the life cycle of null-test iterations normally starts from this processed TOD\@.
The mapmaking step is then very simple, as the components are already separated.

This concludes the brief overview of the POLARBEAR analysis pipelines,
for more details, see for example
The POLARBEAR Collaboration et al.\footnote{\citeproc{ref-ade_measurement_2014-1}{{``A Measurement of the Cosmic Microwave Background \emph{B}-Mode Polarization Power Spectrum at Sub-Degree Scales with POLARBEAR''}}.} Takakura et al.\footnote{\citeproc{ref-takakura_performance_2017}{{``Performance of a Continuously Rotating Half-Wave Plate on the POLARBEAR Telescope''}}.} The POLARBEAR Collaboration et al.\footnote{\citeproc{ref-the_polarbear_collaboration_measurement_2020}{{``A Measurement of the Degree-Scale CMB \emph{B}-Mode Angular Power Spectrum with POLARBEAR''}}.}

\part{\texorpdfstring{A Measurement of the CMB \(B\)-mode Angular Power Spectrum at Sub-degree Scales from \(670\) Square Degrees of POLARBEAR Data}{A Measurement of the CMB B-mode Angular Power Spectrum at Sub-degree Scales from 670 Square Degrees of POLARBEAR Data}}\label{sec:polarbear-research}

\chapter{Sensitivity projection}\label{sensitivity-projection}

As we have mentioned in \cref{sec:medium-aperture-telescope},
in the 3rd--5th seasons, POLARBEAR observed
a sky area of \(\SI{670}{\degree^2}\).
This dissertation focuses on the high-\(\ell\) \(B\)-mode science product from this observation.
Let us start with a back-of-the-envelope calculation to see what result we should be expecting.

Since we have a \(\SI{3.5}{\arcminute}\) beam,
We can estimate the maximum \(\ell\) that we can resolve by using the rule of thumb

\begin{equation}\protect\phantomsection\label{eq:degree-to-l}{Δθ \sim \frac{180^\circ}{\ell}.}\end{equation}

This is because for the spherical harmonics used in decomposing the CMB\footnote{See \cref{sec:sphericalHarmonics} for details.},

\[Y_{\ell }^{m}(\theta ,\varphi )={\sqrt {{\frac {(2\ell +1)}{4\pi }}{\frac {(\ell -m)!}{(\ell +m)!}}}}\,P_{\ell }^{m}(\cos {\theta })\,e^{im\varphi } .\]

Since \(|m| \leq \ell\), the smallest periodicity of the magnitude in the \(φ\)-direction is exactly \(\frac{180^\circ}{\ell}\).

So, from a resolution of a \(\SI{3.5}{\arcminute}\) beam,
we expect

\[\ell_\text{max} = \frac{180^\circ}{\SI{3.5}{\arcminute}} \sim 3000.\]

For the minimum \(\ell\) we can measure,
that is subject to different criteria.
First, our large-patch is roughly \(\ang{45}\) wide,
so using a similar estimate seems to suggest we can measure down to \(\ell \sim 4\).
But there are other factors that come into play.
The most important one is the \(\ell_\text{knee}\),
a characteristic \(\ell\) scale determined by the presence of low frequency noise. It turns out that \(\ell_\text{knee} = 90\), which is the topic of the low-\(\ell\) paper.\footnote{\citeproc{ref-the_polarbear_collaboration_measurement_2020}{The POLARBEAR Collaboration et al., {``A Measurement of the Degree-Scale CMB \emph{B}-Mode Angular Power Spectrum with POLARBEAR''}}.}

As mentioned in \cref{sec:splitting-ell},
the pipelines are split into two at \(\ell = 600\).
This work focuses on the high-\(\ell\) range of \(600\)--\(3000\).

In this \(\ell\)-range,
we are observing the lensing \(B\)-mode.
This lensing \(B\)-mode is predicted by,
as the name suggests,
the lensing effect of the \(E\)-mode power-spectrum that becomes a \(B\)-mode pattern.
It has first been observed by POLARBEAR in its first two seasons of data in the range of \(500 \leq \ell \leq 2100\).\footnote{\citeproc{ref-the_polarbear_collaboration_measurement_2017}{The POLARBEAR Collaboration et al., {``A Measurement of the Cosmic Microwave Background \emph{B}-Mode Polarization Power Spectrum at Subdegree Scales from Two Years of POLARBEAR Data''}}.} A natural question to ask is how the sensitivity of this project would compare to that.

To do this,
we need to know how the sensitivity should scale with sky-area and map-depth.

\section{Knox Formula}\label{sec:Knox-formula}

According to Hivon et al.\footnote{\citeproc{ref-hivon_master_2002}{{``MASTER of the Cosmic Microwave Background Anisotropy Power Spectrum''}}.} eq.~16-17,

\begin{equation}\protect\phantomsection\label{eq:knox_formula}{\Delta C_l \approx \sqrt{\frac{2}{\nu_l}} \left( C_l + \frac{N_l}{B_l^2} \right)}\end{equation}

\[\nu_b \approx (2 l + 1) \Delta l f_\text{sky, eff.}\]

\[f_\text{sky, eff.} \equiv f_\text{sky} \frac{w_2^2}{w_4}\]

where it is understood that the \(l\)-bins are half open intervals: \(b = [l - \frac{\Delta l}{2}, l + \frac{\Delta l}{2})\).

Here, \(C_\ell\) is the power-spectrum,
\(Δ C_\ell\) is its error-bar,
\(ν\) is the d.o.f.,
\(B_\ell\) is the beam,
\(N_\ell\) is the noise spectrum,
\(f_\text{sky}\) is the fraction of the sky observed.
\(f_\text{sky, eff.}\) is the effective \(f_\text{sky}\)
adjusted by the 2nd and 4th moment
(\(w_2\) and \(w_4\) respectively)
of the mask used in calculating the power-spectra from the maps.

Note that this formula is exact given \(\ell\) if the underlying random variables
(including the \(C_\ell\) term and the noise term)
are all Gaussian and independent.
\(ν = 2 \ell + 1\) per \(\ell\),
and the factor of \(f_\text{sky}\) adjusts for the fraction of the number of modes lost in not measuring the full sky.
Hence the validity of this formula depends on how well these assumptions hold in practice.

Understanding what it means,
we can estimate \(ν\) better,\footnote{\citeproc{ref-hamilton_sensitivity_2008}{Hamilton et al., {``Sensitivity of a Bolometric Interferometer to the Cosmic Microwave Backgroud Power Spectrum''}}, eq.~20.}

\[\nu_b \equiv \sum_{l - \frac{\Delta l}{2}}^{l + \frac{\Delta l}{2} - 1} (2l + 1) f_\text{sky, eff.} = 2 l \Delta l f_\text{sky, eff.}.\]

That is, the one Hivon et al.\footnote{\citeproc{ref-hivon_master_2002}{{``MASTER of the Cosmic Microwave Background Anisotropy Power Spectrum''}}.} used over-estimated the error-bar a little bit, which becomes relatively negligible when \(\ell\) is large.

Now, let us move on to see how we can project the sensitivity knowing only how the noise and \(f_\text{sky}\) scale. Assuming noise-limited,

\begin{equation}\protect\phantomsection\label{eq:noiseScaling}{\Delta C_b \sim N_b \sqrt{ \frac{ 2 }{ \nu_b } } \propto \frac{n^2}{\sqrt{f_\text{sky}}}.}\end{equation}

Here we introduce the map-depth \(n\) which is the square-root of the noise spectrum, and also usually put in a different unit.
Typically, the noise spectrum has a steradian implicitly in its unit\footnote{That is, the unit is simply \(\unit{\micro\kelvin^2}\) with the steradian not written explicitly.}.
But map-depth is usually put in the unit of
\(\unit{\micro\kelvin \arcminute}\)
so a conversion from \(\unit{\steradian}\) to
\(\unit{\arcminute^2}\) (square-arcminute)
is needed.

\subsection{Method 1: Scaling by the change of sky area only}\label{method-1-scaling-by-the-change-of-sky-area-only}

Consider only the change in sky coverage,
if we \emph{decrease} the sky area by a factor of \(r\),

\[f_\text{sky} \rightarrow \frac{ f_\text{sky} }{r},\]

\[n \rightarrow \frac{n}{ \sqrt{r} },\]

\[Δ C_\ell \rightarrow \frac{Δ C_\ell}{\sqrt{r}},\]

where the scaling of \(n\) is deduced from the scaling of the standard error of the mean, assuming the total amount of observation stays constant.

That is, the signal-to-noise ratio is increased by \(\sqrt{r}\).

In the first two seasons, POLARBEAR observed a sky area of \(\SI{25}{\degree^2} \rightarrow r = 25 / 670\).
The signal-to-noise ratio is expected to decrease by \(0.19\).

Since The POLARBEAR Collaboration et al.\footnote{\citeproc{ref-the_polarbear_collaboration_measurement_2017}{{``A Measurement of the Cosmic Microwave Background \emph{B}-Mode Polarization Power Spectrum at Subdegree Scales from Two Years of POLARBEAR Data''}}.} reports a statistical error of \(0.26\) on \(A_L\),
we expect it to be around \(1.3\) in this measurement.

But we can do slightly better.
To account for the difference in the number of seasons (2 vs.~3),
deducing from the standard error of the mean again,
the signal-to-noise ratio will decrease by
\(0.19 \times \sqrt{\frac{3}{2}} = 0.23\),
leading to an error of \(A_L\) to be around \(1.1\).

All these should have been known before the beginning of the 3rd season of observation\footnote{While the slight difference in the use of \(\ell\)-range is not accounted for so far, it turns out to be negligible.}.
And since \(A_L\) simply is the ratio of what we observed comparing to what we expect,
its fiducial value is \(1\),
That is, we should have expected the sensitivity to be less than \(1σ\).

Before we move on, let us improve this calculation slightly by using the actual number of observation hours.
The POLARBEAR Collaboration et al.\footnote{\citeproc{ref-the_polarbear_collaboration_measurement_2017}{{``A Measurement of the Cosmic Microwave Background \emph{B}-Mode Polarization Power Spectrum at Subdegree Scales from Two Years of POLARBEAR Data''}}.}
reports \(\SI{2800}{\hour}\) of observations,
while we have \(\SI{2985}{\hour}\).
That is, it is much less data volume than we would expect from the increase from \(2\) to \(3\) seasons. One of the reasons is the eruption of a nearby volcano on October 30th, 2015,\footnote{\citeproc{ref-the_polarbear_collaboration_measurement_2020}{The POLARBEAR Collaboration et al., {``A Measurement of the Degree-Scale CMB \emph{B}-Mode Angular Power Spectrum with POLARBEAR''}}.} among other mechanical problems.
This means the signal-to-noise ratio is decreased by a factor of \(0.19 \times \sqrt{\frac{2985}{2800}} = 0.20\) instead, resulting in an error of \(A_L\) to be around \(1.3\).

\subsection{Method 2: Estimated using the actual map depth in map domain}\label{method-2-estimated-using-the-actual-map-depth-in-map-domain}

The map depth \(n\) can be estimated in the map domain directly.

Naïvely, this can be estimated by the deepest part of the map.
Internally, it is reported to be
\(\SI{4.5}{\micro \kelvin \arcminute}\)\footnote{Hereafter for the RA23 patch out of 3 different patches only. See The POLARBEAR Collaboration et al., \citeproc{ref-the_polarbear_collaboration_measurement_2017}{{``A Measurement of the Cosmic Microwave Background \emph{B}-Mode Polarization Power Spectrum at Subdegree Scales from Two Years of POLARBEAR Data''}} for its definition.}
from the first two seasons of data,
and \(\SI{16}{\micro \kelvin \arcminute}\) for the later three seasons.

Using \cref{eq:noiseScaling},
we would expect the error-bar to be \(2.4\) times more,
resulting in \(Δ A_L = 0.63\).

But this number is highly deceptive for two reasons.
First, the absolute gain was not known at the time.
Taking into account the absolute gain in hindsight,
the map depth is \(\SI{18.4}{\micro \kelvin \arcminute}\),
resulting in \(Δ A_L = 0.84\).

\begin{figure}
\centering
\pandocbounded{\includegraphics[keepaspectratio,alt={Normalized map depth illustrating the scan pattern, The POLARBEAR Collaboration et al.}]{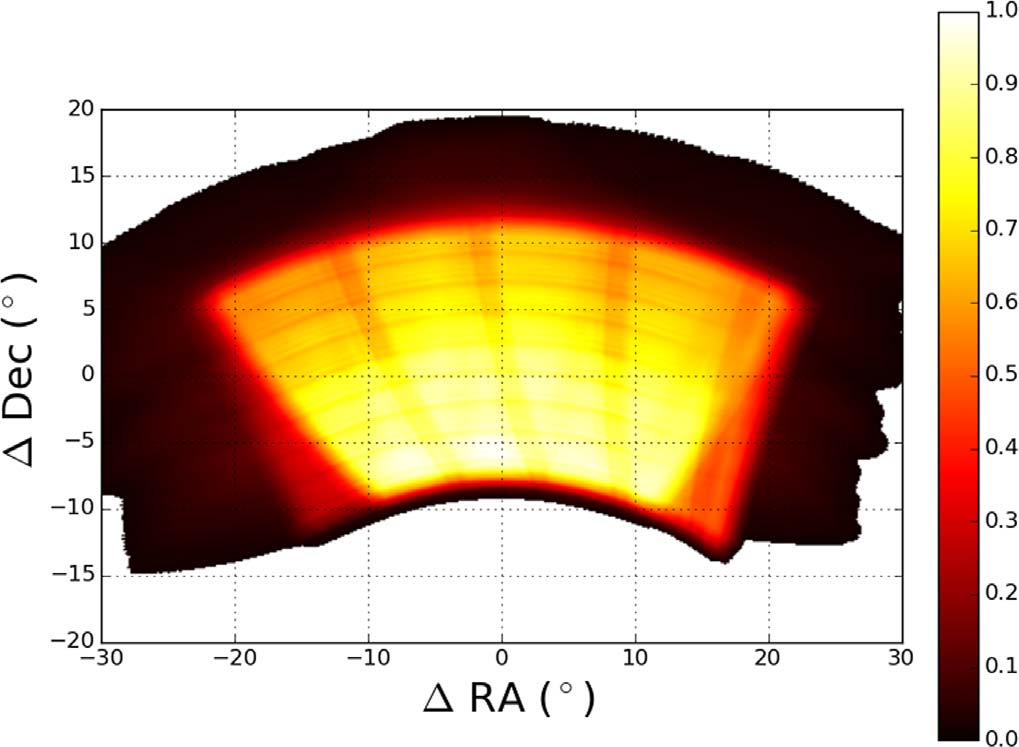}}
\caption[Normalized map depth illustrating the scan pattern, The POLARBEAR Collaboration et al.]{Normalized map depth illustrating the scan pattern, The POLARBEAR Collaboration et al.\footnotemark{}}\label{fig:the_polarbear_collaboration_measurement_2020-003-000}
\end{figure}

The more serious problem is that the map depth is highly irregular as seen in \cref{fig:the_polarbear_collaboration_measurement_2020-003-000},
while the first two seasons have very uniform maps.

\footnotetext{\citeproc{ref-the_polarbear_collaboration_measurement_2020}{{``A Measurement of the Degree-Scale CMB \emph{B}-Mode Angular Power Spectrum with POLARBEAR''}}.}

So a better metric would be to compare by their weighted average map depth.
This is known internally to be \(\SI{5.5}{\micro \kelvin \arcminute}\)
for the first two seasons
and \(\SI{23.8}{\micro \kelvin \arcminute}\) for last three.
This amounts to \(Δ A_L = 0.94\).

\subsection{Method 3: Estimated using the actual map depth in power-spectra domain}\label{method-3-estimated-using-the-actual-map-depth-in-power-spectra-domain}

\begin{figure}
\centering
\pandocbounded{\includegraphics[keepaspectratio,alt={EE noise spectrum (\textbackslash unit\{\textbackslash micro \textbackslash kelvin \textbackslash arcminute\})}]{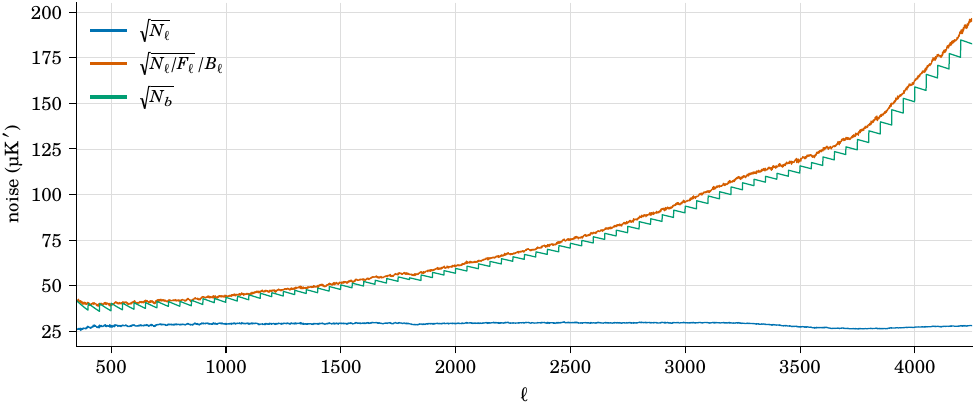}}
\caption{\(EE\) noise spectrum (\(\unit{\micro \kelvin \arcminute}\))}\label{fig:EE_noise_spectrum}
\end{figure}

\begin{figure}
\centering
\pandocbounded{\includegraphics[keepaspectratio,alt={BB noise spectrum (\textbackslash unit\{\textbackslash micro \textbackslash kelvin \textbackslash arcminute\})}]{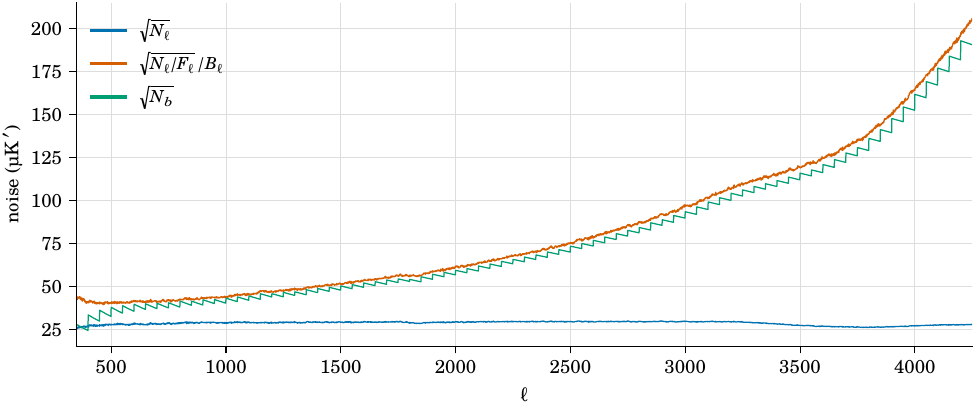}}
\caption{\(BB\) noise spectrum (\(\unit{\micro \kelvin \arcminute}\))}\label{fig:BB_noise_spectrum}
\end{figure}

In \cref{fig:EE_noise_spectrum} and \cref{fig:BB_noise_spectrum},
we plotted the noise spectrum.
\(\sqrt{N_\ell}\) is the noise-spectrum,
and the \(\sqrt{N_b}\) is the noise-spectrum after de-biasing using the MASTER algorithm,\footnote{\citeproc{ref-hivon_master_2002}{Hivon et al., {``MASTER of the Cosmic Microwave Background Anisotropy Power Spectrum''}}.} binned at a bin-width of 50.
Since our science product is the de-biased power-spectra,
\(\sqrt{N_b}\) should be the focus here.

Even if we are using the smallest value at \(\ell = 525\)
which equals \(\SI{36.2}{\micro \kelvin \arcminute}\),
comparing to an internally provided value from the first two seasons at
\(\SI{6.3}{\micro \kelvin \arcminute}\),
it would project to be \(Δ A_L = 1.7\).

\subsection{Lesson learnt}\label{lesson-learnt}

To summarize,
we should have projected the sensitivity to be
\(Δ A_L = 1.1\) before the observation even started,
and \(Δ A_L = 1.7\) in the beginning of the analysis.

But, all these were unknown to anyone in the collaboration
at the time, including the very author who spent a few years
in analysis. Hence, this is perhaps the most important
but simple physics one should understand and calculate.
Sometimes we got buried in details and got lost in the big picture.

However, all is not lost.
As we shall soon see,
this experiment shows that we can measure the polarization of the CMB
in a wide range of \(50 \leq \ell \leq 3000\)\footnote{For
  \(50 \leq \ell \leq 600\), \(E\)-mode is only used for calibration.}
using a single medium aperture telescope.

\chapter{Reproducible research}\label{sec:reproducible-research}

\section{Introduction}\label{introduction}

Distilled to the essence, one may define science as a discipline to predict.
In this aspect, reproducibility can be viewed as the cornerstone of science---without reproducibility,
one cannot reliably predict even the outcome of a repeated experiment, and
it is not science.

However, the term replication crisis
or reproducibility crisis has been coined in the recent history of science.\footnote{\citeproc{ref-ioannidis_why_2005}{Ioannidis, {``Why Most Published Research Findings Are False''}}; \citeproc{ref-schooler_metascience_2014}{Schooler, {``Metascience Could Rescue the `Replication Crisis'''}}.}

There are two layers of reproducibility.
The first is reproducibility in the general sense which
focuses on the reproducibility of the experiment in general,
and is a standard expectation in scientific publications.
The second is restricted to a subset of reproducibility in an experiment,
called computationally reproducible research or often simply reproducible research.\footnote{\citeproc{ref-alston_beginners_2021}{Alston and Rick, {``A Beginner's Guide to Conducting Reproducible Research''}}; \citeproc{ref-peng_reproducible_2011}{Peng, {``Reproducible Research in Computational Science''}}.}

Regarding the general reproducibility,
it is central to any CMB experiment.
For example, systematic effects are an important area that many CMB physicists deeply worry about.
Systematic effects can be viewed as one form of reproducibility requirement---that is,
when constructing an independent experiment which comes with a different set of systematics,
they better reproduce the results from each other by excluding any significant systematic effects.
In POLARBEAR, the null-test framework is used because it helps to avoid these systematic effects.
POLARBEAR is designed to be a ``Blind Analysis as a Correction for Confirmatory Bias'',\footnote{\citeproc{ref-lilienfeld_blind_2017}{MacCoun and Perlmutter, {``Blind Analysis as a Correction for Confirmatory Bias in Physics and in Psychology''}}.} which is one example of many forms of bias that result in non-reproducible science.
The general agreements between results from CMB experiments indicate that the general reproducibility crisis is not a major problem here.

However, in terms of computationally reproducible research,
CMB experiments fail in this regard to various extents,
especially in the first generations of ground-based CMB experiments including POLARBEAR.

What is (computationally) reproducible research? Peng\footnote{\citeproc{ref-peng_reproducibility_2015}{{``The Reproducibility Crisis in Science''}}.} said,

\begin{quote}
There are two major components to a reproducible study: that the raw data from the experiment are available; and that the statistical code and documentation to reproduce the analysis are also available.
\end{quote}

Hutson\footnote{\citeproc{ref-hutson_artificial_2018}{{``Artificial Intelligence Faces Reproducibility Crisis''}}.} mentioned in an article focusing on Artificial Intelligence (AI) with the following subtitle: ``Unpublished code and sensitivity to training conditions make many claims hard to verify'',
while Miyakawa\footnote{\citeproc{ref-miyakawa_no_2020}{{``No Raw Data, No Science''}}.} even claim in the title: ``No raw data, no science: another possible source of the reproducibility crisis''.

To summarize, (computationally) reproducible research focuses on what happens after an experiment gathers data.
Reproducing an experiment is expensive, making the entry barrier to reproduce very high.
Reproducible research requires everything but the building of the experiment itself to be completely reproducible
in the sense that the observed data and software written to come to the conclusion of the research
are open for anyone to examine and come to an independent conclusion.
A prerequisite of this would require all raw data collected to be released openly,
and all essential software that is used in a publication to be made at the very least source-available
if not open-source\footnote{Source-available simply means the source is available for audit,
  open-source has additional requirements concerning the rights of the users receiving the source codes as well.}.

In this sense, none of the CMB experiments are reproducible
to the best of the author's knowledge.
For example,
all WMAP data are made available in LAMBDA\footnote{\citeproc{ref-bennett_nine-year_2013}{Bennett et al., {``NINE-YEAR \emph{WILKINSON MICROWAVE ANISOTROPY PROBE} ( \emph{WMAP} ) OBSERVATIONS''}}.} but not all its associated
code and recipe for complete reproducibility\footnote{Incidentally, because of the availability of
  full WMAP data, a third party effort has been made to claim the first
  reproducible CMB map in Bobin, Sureau, and Starck, \citeproc{ref-bobin_cosmic_2016}{{``Cosmic Microwave Background Reconstruction from WMAP and \emph{Planck} PR2 Data''}}.}.
Planck releases many of their data products and software packages, but not all.\footnote{\citeproc{ref-noauthor_planck_nodate}{{``Planck Mission Products - Planck Legacy Archive Wiki''}}.} Simons Observatory releases most of its software in
a \href{https://github.com/simonsobs}{public GitHub organization},\footnote{\citeproc{ref-koopman_simons_2020}{Koopman et al., {``The Simons Observatory''}}; \citeproc{ref-li_simons_2021}{Li et al., {``The Simons Observatory''}}.} but no commitment has been made yet for complete (computational) reproducibility.\footnote{\citeproc{ref-ade_simons_2019}{Ade et al., {``The Simons Observatory''}}.} Not to mention POLARBEAR has released their data products sparingly,
has made no raw data available to the public,
nor most, if not all, of the software they used internally.

While this is not something the author can change,
the author decomposes the different extents of reproducibility in the following way,
and makes every effort to make it as reproducible as humanly possible given the constraints.

First, we already mentioned that (computationally) reproducible research
has components of data and software.
Second, we can define the degree of reproducibility for different levels of audiences---the general public, and internal collaborators.
This can be illustrated in \cref{tbl:reproducible}.

\needspace{7\baselineskip}

\begin{longtable}[]{@{}
  >{\raggedright\arraybackslash}p{(\linewidth - 4\tabcolsep) * \real{0.1250}}
  >{\raggedright\arraybackslash}p{(\linewidth - 4\tabcolsep) * \real{0.5750}}
  >{\raggedright\arraybackslash}p{(\linewidth - 4\tabcolsep) * \real{0.3000}}@{}}
\caption{Different extents of (computationally) reproducible research}\label{tbl:reproducible}\tabularnewline
\toprule\noalign{}
\begin{minipage}[b]{\linewidth}\raggedright
\end{minipage} & \begin{minipage}[b]{\linewidth}\raggedright
Collaborators
\end{minipage} & \begin{minipage}[b]{\linewidth}\raggedright
General public
\end{minipage} \\
\midrule\noalign{}
\endfirsthead
\toprule\noalign{}
\begin{minipage}[b]{\linewidth}\raggedright
\end{minipage} & \begin{minipage}[b]{\linewidth}\raggedright
Collaborators
\end{minipage} & \begin{minipage}[b]{\linewidth}\raggedright
General public
\end{minipage} \\
\midrule\noalign{}
\endhead
\bottomrule\noalign{}
\endlastfoot
Data & Is the data accessible to internal collaborators? Is the data format documented? & Is the data available to the public? \\
Software & Is the software available to internal collaborators? Is the recipe to reproduce a product available and documented, including the software environment? & Is the software used in research released as open-source software? \\
\end{longtable}

\section{Making this research more reproducible}\label{making-this-research-more-reproducible}

\subsection{Software}\label{software}

As mentioned above,
the raw data used in POLARBEAR is not available to the general public.
The software that the bulk of data analysis depends on resides in a software repository named AnalysisBackend, and is not made source-available.

While there is nothing the author can do to change this,
the author makes an effort to factor the code in a way that most newly written code
does not reside in AnalysisBackend but in other repositories such that
the author can freely open-source them without being subject to the agreements
of any existing authors in the original repositories.

Because of that, most of the descriptions here
focus on reproducibility to internal collaborators.

While POLARBEAR's data is in principle available to any internal collaborators,
they reside in a project directory in the \href{https://docs.nersc.gov/filesystems/community/}{Community File System (CFS)} at NERSC\@.
Due to the lack of maintainers in that space,
any collaborators can write to that space with a default permission that is not readable to anyone but the one who writes the data.
Hence, while all files in this space are supposed to be group-readable among all collaborators,
some files such as some instrumental metadata are not readable.
While this does not impact the results of analysis,
it does cause trouble in, for example, staging raw data to scratch space
for further analysis,
and is a sign of not designing with the goal of reproducibility towards even internal collaborators.
While I can volunteer to manage the project account,
since there is no maintainer,
there is not much I can do to change the situation.

To make my contribution reproducible to internal collaborators,
I set up a GitHub organization for internal collaborators.
Using that platform,
I provide a repository for all the scripts I submitted to NERSC---this resembles how a lab notebook is used in a laboratory setting,
whereas I recorded every script I ran at NERSC so that everything I did is traceable, a first step towards being reproducible.

Soon I found that the software environment can be part of the puzzle
that makes something not reproducible.
At first it was provided by another collaborator as \href{https://github.com/tskisner/hpcports}{HPCPorts}, but
as it is abandoned and no longer updated,
I volunteered to maintain a software environment
to run our software stack at NERSC\@.
To make this software environment reproducible,
I had to automate the building of the software environments
(which is basically what HPCPorts did) in the form of scripts.
Eventually, I rewrote it as a Python package called \href{https://github.com/ickc/python-pmpm}{pmpm} which stands for Python manual package manager.

Another pain point in reproducibility is the software we used---AnalysisBackend.
It has no maintainer,
because of that, every collaborator just creates their own branch
and merging their contribution is often not done.
One consequence of this is that on many occasions in our analysis meetings,
after discussing an analysis problem for a long while,
someone would often start to realize the root of the problem and ask,
``wait a minute, which branch are you using?''.
While this situation cannot be changed in AnalysisBackend without top-down management,
the main collaborators working on large-patch analysis
try to converge on a common branch.
While each of us still maintains our own branch,
we mostly keep the differences between these branches minimal.

Since I opt for writing in separate repositories as much as possible,
there I can opt for semantic versioning\footnote{\href{https://semver.org/}{Semantic Versioning}}
with changelogs and documentations.

Regarding reproducibility between collaborators,
oftentimes the pipeline scripts are passed
with a long chain of command line arguments
that is hard to trace the differences because there can be
hundreds of options involved.
A better way would be to put these configuration arguments
in a format that is more tractable.
In order to do this while still being able to run
arbitrary command line programs with arguments from any software
including AnalysisBackend,
\href{https://github.com/ickc/yaml2cli}{yaml2cli} is written to
``organize cli args by YAML''
which can generate bash scripts including SLURM batch scripts
from a configuration written in YAML\@.
YAML is chosen because of its readability and features
such as alias that facilitate ``code reuse'' within the configuration file.
See \cref{sec:POLARBEAR-configuration} for examples used in this analysis.

\subsection{Data}\label{data}

Data generated in my analysis also uses the semantic versioning scheme.
See \cref{sec:PB-null-changelog} for the versions
of null-tests performed in this analysis.

The data container itself should be chosen for performance
and stable archival purpose.
While many collaborators like to save data in the Python
pickle format due to its ease of use,
``you should be wary of using pickles for long-term storage where the underlying code is not highly stable''\footnote{\href{https://wiki.python.org/moin/UsingPickle}{UsingPickle - Python Wiki}}.
HDF5\footnote{\href{https://www.hdfgroup.org/solutions/hdf5/}{The HDF5® Library \& File Format - The HDF Group}}
is a standard data container format designed for this
purpose. It has wide adoption in the field of cosmology.

Not only is it important to use a stable
container for archival purpose,
how data is stored should be documented
carefully to ensure it is used and interpreted correctly.
In this analysis,
the data released internally are documented
and the documentation is made available in \cref{sec:POLARBEAR-data-release}.

\chapter{Computational challenges}\label{sec:computational-challenges}

\section{Orientation}\label{orientation}

In the beginning,
I was told to just run the whole pipeline
with a smaller pixel at \(\SI{2}{\arcminute}\)
and see what the null-test failures are.
It was null-test v0.1.0\footnote{I adopt \href{https://semver.org}{semantic versioning} in my null-test runs. See \cref{sec:PB-null-changelog}.}.

However,
I encountered multiple job failures
because the nodes generated out-of-memory errors and/or exceeded the requested allocation time.
It involved a lot of manual labor
including fixing and rerunning whatever
corrupted files when the jobs
got killed.

Long story short,
with 25 realizations\footnote{1 for the real data, 8 each for 3 kinds of simulations.},
it took me 1 month and the following amount of NERSC hours\footnote{NERSC hour
  is a unit of computer time at the
  National Energy Research Scientific Computing Center (NERSC)\@.
  It was calibrated to the computational power of 1 CPU core in an hour
  for an older NERSC system called Hopper.
  Exactly how much 1 NERSC hour is worth depends on calibration
  and it has been changed over time.
  During the time this analysis was done,
  on a Cori Haswell system where each node holds two Intel Xeon Processor E5-2698 v3 processors with \(\SI{128}{\giga\byte}\) memory (See \href{https://web.archive.org/web/20221130094825/https://docs.nersc.gov/systems/cori/\#haswell-compute-nodes}{Cori - NERSC Documentation}),
  80 NERSC hours were charged when using 1 such node for an hour.}:

\begin{enumerate}
\def\labelenumi{\arabic{enumi}.}
\tightlist
\item
  mapmaking: 33690 NERSC hours
\item
  coadd: 55339 NERSC hours
\item
  final-coadd: 86 NERSC hours
\item
  mode-coupling matrices: \textasciitilde70 NERSC hours
\item
  pseudo-spectra: \textasciitilde60,000 NERSC hours
\end{enumerate}

I actually got the last one for free using the big-memory node at NERSC\footnote{There
  are 2 such big-memory nodes at NERSC with \(\SI{750}{\giga\byte}\).
  During the time of analysis, it was free to use but very congested.}.
Had it been charged, it would have been a total of \textasciitilde150,000 NERSC hours.

And if you are laughing right now,
you understand that mapmaking
is supposed to be the only expensive part.
But here,
co-adding is more expensive
than mapmaking
(where it is only doing weighted-sum),
and pseudo-spectra pipeline
is more expensive than co-add.

To put that time cost into perspective,
I ran approximately 31 different
null-test versions\footnote{around 30--40 depending on how you count. See \cref{sec:PB-null-changelog}.}
that involve at least the post-mapmaking pipelines.

Not only that,
near the end of the analysis
where we are closer to passing the null-tests,
we increased our total number of realizations
from 25 to 97.
And we used \(769\) realizations in the final run.
Had I been using the original pipeline for this,
I would have used
\textasciitilde5.2 million NERSC hrs
and 31 months of turnaround time
for this alone.
So by the time of publishing this dissertation, I would have been nowhere near passing the null-tests,
and would have exhausted my NERSC allocation per year a dozen times already.

\section{Low lying fruit}\label{sec:low-lying-fruits}

Essentially embarrassing parallelism can only be \(\SI{100}{\percent}\) effective
if you can feed each processor a job and the load is perfectly balanced.
It turns out both of these are far from true in our pipelines
with our large-patch map size requirements.

For example,
in the mapmaking pipeline,
recall from \cref{eq:mapmaking_illustration},
\(m^{i} = \mathcal{M} \mathopen{} \left \{ d^{i} \right \} \mathclose{} , i = 1 , \dots , n_{\text{CES}}\),
the default configuration
is to run an MPI job with as many processes as the number of CESes where \(n_\text{CES} = 3391\).
So each job is waiting for the longest running job to finish.

In the end,
I wrapped the AnalysisBackend mapmaking command line interface with a Python function interface\footnote{It is a peculiarity in the design of AnalysisBackend
  that mapmaking can only be called from the command line using \texttt{argparse}.}
and call it from an MPI script.
With naïve random scheduling
and using fewer MPI processes,
the load balancing issue is solved
and the total NERSC hours needed is decreased by around a factor of \(5\).

Another example is the coadd-pipeline from \cref{eq:coaddByInvVar1} \& \cref{eq:coaddByInvVar2},
\(w^{I , X , n}_{p} = \sum_{i \in I} w^{i , X , n}_{p} , M^{I , X , n}_{p} = \sum_{i \in I} M^{i , X , n}_{p}\).
It is very inefficient, requiring \(\SI{40}{\gibi\byte}\) of memory per process,
and has a bad reading pattern\footnote{It loads all combinations of \(X, n, p\) and \(i \in I\) in memory first.}.
By rewriting it from scratch,
it has a more balanced read pattern,
memory use is reduced to \(\SI{\sim 135}{\mebi\byte}\),
and since the problem is IO bound,
multithreading is used,
resulting in negligible NERSC hours
needed comparing to mapmaking.
Notice also from the inherent parallelism involved\footnote{I decided not to dispatch over \(X\) due to how the datasets are stored and the fact that there is enough parallelism.},
\(I = 1, \ldots, 38\), \(n = 1A, 1B, \ldots, 21A, 21B\),
with a total of \(1535\)\footnote{It does not exactly equal \(38 \times 21 \times 2\) because some of the null-split maps would be empty.} embarrassingly parallel tasks.

The last remaining computational problems
are the mode-coupling matrices\footnote{See \cref{sec:modeCoupling}.} and maps-to-pseudo-spectra pipelines\footnote{See \cref{sec:pseudoSpectra}.},
which I call the ``post-mapmaking pipeline'' in short.
While the former does not seem to be expensive,
it is a blocking dependency of the next step,
making it also a target to shave the turn-around time.

I have a few iterations
in accelerating these pipelines.
Earlier attempts are more Pythonic
and trying to be backward compatible.
But these ``duct-tape'' solutions
do not work well
as they rely on the big-memory node
which is a bottleneck
both in terms of scaling out
the applications to more simulations
and its availability\footnote{I often
  need to wait from days to weeks to get into the queue.}.

\section{Accelerating the post-mapmaking pipeline}\label{accelerating-the-post-mapmaking-pipeline}

Post-mapmaking pipeline includes everything after obtaining the maps,
including the calculation of mode-coupling matrices
and calculating the pseudo-spectra from maps.

Both of these are examples
where it cannot be written efficiently
using pure Python code that heavily relies on vectorization\footnote{The de facto standard
  in writing pure Python code in scientific computing is to use Numpy and the ecosystem built around it including Scipy.
  Numpy defines an array interface with syntactic sugars in Python
  with operations built around the concept of vectorization,
  and most of the operations are written as C modules,
  resulting in mostly efficient code.
  Limitations on its effectiveness originate from the Python data model,
  the generality of this array interface which is compiled AOT,
  and how effective vectorized programming can be on a particular algorithm.}.

\begin{figure}
\centering
\pandocbounded{\includegraphics[keepaspectratio,alt={Illustration of 2D FFT to 1D power-spectra}]{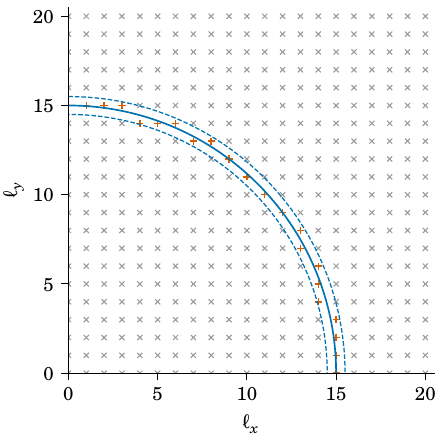}}
\caption{Illustration of 2D FFT to 1D power-spectra}\label{fig:annular_15_15}
\end{figure}

For example,
in \cref{fig:annular_15_15},
when calculating the pseudo-spectra from maps,
the maps are first converted to frequency domain using Fast Fourier Transform (FFT),
and since we are interested in \(\ell\),
corresponding to the magnitude of the frequencies in the 2D-FFT,
the values need to be binned down to individual \(\ell\).

\begin{figure}
\centering
\pandocbounded{\includegraphics[keepaspectratio,alt={Illustration of 2D FFT to 1D power-spectra with intermediate bin-width}]{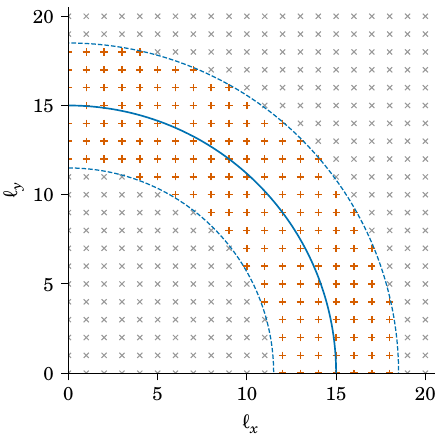}}
\caption{Illustration of 2D FFT to 1D power-spectra with intermediate bin-width}\label{fig:annular_12_18}
\end{figure}

In fact, the original code relies on vectorization
to first create a mask per \(\ell\),
and calculate the averages by indexing arrays by this mask.
Obviously it is very inefficient,
but it also means that in order to get the job done in a reasonable amount of time,
the averaging is done across an intermediate bin-width
as shown in \cref{fig:annular_12_18}.

\begin{figure}
\centering
\pandocbounded{\includegraphics[keepaspectratio,alt={Null test v0.6.2  null-spectra at finer bin-width}]{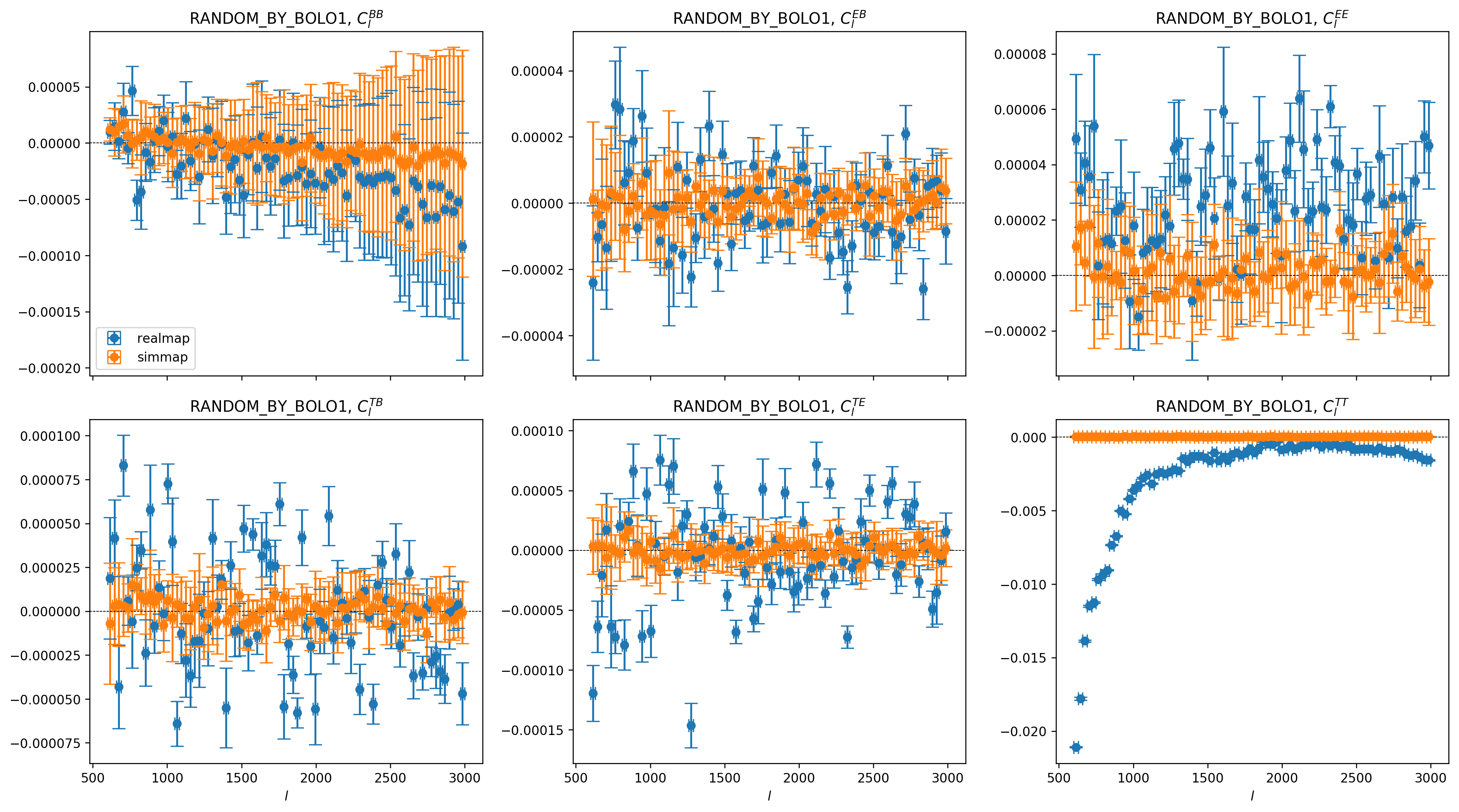}}
\caption{Null test v0.6.2 \texttt{RANDOM\_\allowbreak{}BY\_\allowbreak{}BOLO1} null-spectra at finer bin-width}\label{fig:0_6_2-RANDOM_BY_BOLO1-fine}
\end{figure}

Besides the computational problem,
I found that the resultant binned null-spectra
from this pipeline have different qualitative behavior
depending on the bin-width used,
for example, comparing \cref{fig:0_6_2-RANDOM_BY_BOLO1} \& \cref{fig:0_6_2-RANDOM_BY_BOLO1-fine}.
I suspect the use of this intermediate bin-width might be where the problem lies
in its implementation in AnalysisBackend.

The obvious first step is to proof-read
the whole codebase
to see where the problem lies and also
how I might speed it up.
Eventually,
I found that it was easier
to completely rewrite it.

In order for it to be performant,
and as we have a policy of using Python only,
I evaluated Cython and Numba,
and in the end I chose Numba
because it is easier to write Numba code
that is guaranteed to release the GIL (\texttt{nogil}),
is compiled to machine code (\texttt{nopython}),
and is parallelized (\texttt{parallel}).
Since Numba uses LLVM to just-in-time (JIT) compile the function,
distribution is also effortless which is a bonus\footnote{In traditional ahead-of-time (AOT) compilation, the pre-compiled binaries should be compiled to optimize per architecture for each system.
  Since JIT compiler compiles at run-time which sees the CPU,
  it auto-optimizes for the target architecture without extra work,
  such as the Cori II KNL system which uses Intel Xeon Phi processors.}.

This code is
\href{https://github.com/ickc/TAIL}{released as open source software on GitHub named TAIL},
which stands for This Ain't Interpreted Language.

\begin{codelisting}

\caption{Binning 2D spectra to 1D in Numba}\label{lst:bin_psd}

\begin{Shaded}
\begin{Highlighting}[]
\CommentTok{\# https://github.com/ickc/TAIL/blob/3b4e49d6d5a4/tail/pseudospectra.py}
\AttributeTok{@jit}\NormalTok{(nopython}\OperatorTok{=}\VariableTok{True}\NormalTok{, nogil}\OperatorTok{=}\VariableTok{True}\NormalTok{, parallel}\OperatorTok{=}\VariableTok{True}\NormalTok{)}
\KeywordTok{def}\NormalTok{ \_bin\_psd(pixel\_size, l\_max, psd):}
\NormalTok{    N }\OperatorTok{=}\NormalTok{ psd.shape[}\DecValTok{0}\NormalTok{]}
\NormalTok{    freq }\OperatorTok{=}\NormalTok{ fftfreq(N, pixel\_size)}
\NormalTok{    n }\OperatorTok{=}\NormalTok{ l\_max }\OperatorTok{+} \DecValTok{1}
\NormalTok{    psd\_1d }\OperatorTok{=}\NormalTok{ np.zeros(n)}
\NormalTok{    hit }\OperatorTok{=}\NormalTok{ np.zeros(n, dtype}\OperatorTok{=}\NormalTok{np.int64)}
\NormalTok{    pi\_2 }\OperatorTok{=}\NormalTok{ np.pi }\OperatorTok{*} \FloatTok{2.}
    \ControlFlowTok{for}\NormalTok{ i }\KeywordTok{in}\NormalTok{ prange(N):}
\NormalTok{        freq\_i }\OperatorTok{=}\NormalTok{ freq[i]}
        \ControlFlowTok{for}\NormalTok{ j }\KeywordTok{in} \BuiltInTok{range}\NormalTok{(N):}
\NormalTok{            freq\_j }\OperatorTok{=}\NormalTok{ freq[j]}
\NormalTok{            l }\OperatorTok{=} \BuiltInTok{int}\NormalTok{(}\BuiltInTok{round}\NormalTok{(}
\NormalTok{                pi\_2 }\OperatorTok{*}\NormalTok{ np.sqrt(}
\NormalTok{                    freq\_i }\OperatorTok{*}\NormalTok{ freq\_i }\OperatorTok{+}
\NormalTok{                    freq\_j }\OperatorTok{*}\NormalTok{ freq\_j}
\NormalTok{                )}
\NormalTok{            ))}
\NormalTok{            idx }\OperatorTok{=}\NormalTok{ l }\ControlFlowTok{if}\NormalTok{ l }\OperatorTok{\textless{}}\NormalTok{ l\_max }\ControlFlowTok{else}\NormalTok{ l\_max}
\NormalTok{            hit[idx] }\OperatorTok{+=} \DecValTok{1}
\NormalTok{            psd\_1d[idx] }\OperatorTok{+=}\NormalTok{ psd[i, j]}
\NormalTok{    psd\_1d }\OperatorTok{=}\NormalTok{ psd\_1d[:}\OperatorTok{{-}}\DecValTok{1}\NormalTok{]}
\NormalTok{    hit }\OperatorTok{=}\NormalTok{ hit[:}\OperatorTok{{-}}\DecValTok{1}\NormalTok{]}
\NormalTok{    ak\_to\_dl\_scale }\OperatorTok{=}\NormalTok{ N }\OperatorTok{*}\NormalTok{ pixel\_size}
\NormalTok{    ak\_to\_dl\_scale }\OperatorTok{*=} \FloatTok{0.5} \OperatorTok{*}\NormalTok{ ak\_to\_dl\_scale }\OperatorTok{/}\NormalTok{ np.pi}
    \ControlFlowTok{for}\NormalTok{ i }\KeywordTok{in} \BuiltInTok{range}\NormalTok{(l\_max):}
\NormalTok{        hit\_ }\OperatorTok{=}\NormalTok{ hit[i]}
\NormalTok{        psd\_1d[i] }\OperatorTok{=}\NormalTok{ (}
\NormalTok{            (psd\_1d[i] }\OperatorTok{*}\NormalTok{ ak\_to\_dl\_scale }\OperatorTok{/}\NormalTok{ hit\_) }\OperatorTok{*}
\NormalTok{            (i }\OperatorTok{*}\NormalTok{ (i }\OperatorTok{+} \DecValTok{1}\NormalTok{))}
\NormalTok{        ) }\ControlFlowTok{if}\NormalTok{ hit\_ }\OperatorTok{\textgreater{}} \DecValTok{0} \ControlFlowTok{else}\NormalTok{ np.nan}
\NormalTok{    fill\_nan(psd\_1d, fill\_boundary}\OperatorTok{=}\VariableTok{False}\NormalTok{)}
    \ControlFlowTok{return}\NormalTok{ psd\_1d}
\end{Highlighting}
\end{Shaded}

\end{codelisting}

For example,
one function from the pseudo-spectra pipeline is shown in \cref{lst:bin_psd}.
Its style is quite C-like,
and the \texttt{prange} there is similar to \texttt{omp\ parallel\ for} to explicitly parallelize the loop.

\begin{codelisting}

\caption{Python as Metaprogramming language of Numba}\label{lst:get_M_gen}

\begin{Shaded}
\begin{Highlighting}[]
\CommentTok{\# https://github.com/ickc/TAIL/blob/3b4e49d6d5a4/tail/modecoupling.py\#L128{-}L203}
\KeywordTok{def}\NormalTok{ \_get\_M\_gen(Mtype, pure}\OperatorTok{=}\StringTok{\textquotesingle{}hybrid\textquotesingle{}}\NormalTok{):}
    \ControlFlowTok{if}\NormalTok{ Mtype }\OperatorTok{==} \StringTok{\textquotesingle{}TTTT\textquotesingle{}}\NormalTok{:}
\NormalTok{        \_J }\OperatorTok{=}\NormalTok{ \_J\_t}
    \ControlFlowTok{else}\NormalTok{:}
\NormalTok{        \_J\_p }\OperatorTok{=}\NormalTok{ \_get\_J\_p(Mtype, pure}\OperatorTok{=}\StringTok{\textquotesingle{}hybrid\textquotesingle{}}\NormalTok{)}

        \AttributeTok{@jit}\NormalTok{(nopython}\OperatorTok{=}\VariableTok{True}\NormalTok{, nogil}\OperatorTok{=}\VariableTok{True}\NormalTok{)}
        \KeywordTok{def}\NormalTok{ \_J(k1, k2, k3):}
            \ControlFlowTok{return}\NormalTok{ \_J\_t(k1, k2, k3) }\OperatorTok{*} \OperatorTok{\textbackslash{}}
\NormalTok{                \_J\_p(k1, k2, k3)}

    \AttributeTok{@jit}\NormalTok{(nopython}\OperatorTok{=}\VariableTok{True}\NormalTok{, nogil}\OperatorTok{=}\VariableTok{True}\NormalTok{)}
    \KeywordTok{def}\NormalTok{ simps(W, k1, k2):}
\NormalTok{        k3\_min }\OperatorTok{=}\NormalTok{ np.}\BuiltInTok{abs}\NormalTok{(k1 }\OperatorTok{{-}}\NormalTok{ k2)}
\NormalTok{        k3\_max }\OperatorTok{=}\NormalTok{ k1 }\OperatorTok{+}\NormalTok{ k2}
\NormalTok{        result }\OperatorTok{=} \FloatTok{0.}
        \ControlFlowTok{for}\NormalTok{ i, k3 }\KeywordTok{in} \BuiltInTok{enumerate}\NormalTok{(}
            \BuiltInTok{range}\NormalTok{(k3\_min }\OperatorTok{+} \DecValTok{1}\NormalTok{, k3\_max)}
\NormalTok{        ):}
\NormalTok{            result }\OperatorTok{+=}\NormalTok{ (}\DecValTok{2} \OperatorTok{{-}}\NormalTok{ i }\OperatorTok{\%} \DecValTok{2}\NormalTok{) }\OperatorTok{*} \OperatorTok{\textbackslash{}}
\NormalTok{                \_J(k1, k2, k3) }\OperatorTok{*}\NormalTok{ W[k3] }\OperatorTok{*}\NormalTok{ k3}
        \ControlFlowTok{return}\NormalTok{ result}

    \AttributeTok{@jit}\NormalTok{(nopython}\OperatorTok{=}\VariableTok{True}\NormalTok{, nogil}\OperatorTok{=}\VariableTok{True}\NormalTok{, parallel}\OperatorTok{=}\VariableTok{True}\NormalTok{)}
    \KeywordTok{def}\NormalTok{ \_get\_M(W, l\_max, dl):}
\NormalTok{        bin\_width }\OperatorTok{=}\NormalTok{ dl }\OperatorTok{//} \DecValTok{2}
\NormalTok{        n }\OperatorTok{=}\NormalTok{ l\_max }\OperatorTok{//}\NormalTok{ dl}
\NormalTok{        M }\OperatorTok{=}\NormalTok{ np.empty((n, n))}
        \ControlFlowTok{for}\NormalTok{ i }\KeywordTok{in}\NormalTok{ prange(n):}
\NormalTok{            k1 }\OperatorTok{=}\NormalTok{ bin\_width }\OperatorTok{+}\NormalTok{ dl }\OperatorTok{*}\NormalTok{ i}
            \ControlFlowTok{for}\NormalTok{ j }\KeywordTok{in} \BuiltInTok{range}\NormalTok{(n):}
\NormalTok{                k2 }\OperatorTok{=}\NormalTok{ bin\_width }\OperatorTok{+}\NormalTok{ dl }\OperatorTok{*}\NormalTok{ j}
\NormalTok{                M[i, j] }\OperatorTok{=}\NormalTok{ k2 }\OperatorTok{*}\NormalTok{ simps(W, k1, k2)}
        \ControlFlowTok{return}\NormalTok{ M}

    \ControlFlowTok{return}\NormalTok{ \_get\_M}
\end{Highlighting}
\end{Shaded}

\end{codelisting}

An innovation in this code is the use
of the Python language as a meta-programming language of the Numba language, for example as shown in \cref{lst:get_M_gen}.
The reason is that only a subset of the Python language
can be compiled to machine code using Numba.
So by writing higher order functions using the Python language
which returns a Numba-jit-compiled function,
we can achieve generality while being performant.

\section{Results}\label{results}

In terms of turnaround time,
note that each step of the pipeline
is independent
and each has a fundamentally
different number of parallel jobs to do.
There are
\(n_\text{sim} \times n_\text{CES}\) mapmaking tasks,
\(n_\text{sim} \times n_\text{map bundle} \times n_\text{null-split}\) coadd tasks, and
\(n_\text{sim} \times n_\text{null-split}\) pseudo-spectra tasks.
A key is to find the right MPI job size
such that multiple pipelines can be put together
into one single job script with one \texttt{srun} after another\footnote{It is a limitation
  of the design in AnalysisBackend,
  which takes advantage of the embarrassingly parallel nature of our algorithms,
  to have each task read the inputs from disk and dump the outputs to disk.
  While this is an inefficient way of handling intermediate data
  by having IO on disk instead of holding it in memory,
  it does simplify the way to chain these independent pipelines together
  by using one \texttt{srun} after another when each pipeline has very different topology from each other.}.

By designing this carefully,
the full null-test pipelines
can be finished with 2 submissions to the NERSC queue via SLURM workload manager
to minimize the queue time
and hence the total turnaround time.

In the earlier phase
when these are put together
and we are still in the phase of solving null-test failures,
the turnaround time is shortened from 1 month to typically a day,
running exactly the same number of simulations.

As a testimony of the effectiveness of this short turnaround time,
shortly after this is put in place,
in 2 weeks I finished 6 iterations of null-test suites,
leading to eventually the discovery and solution to our null-test failures.

In the final run with
\(769\) realizations,
instead of around \(5.2\) million NERSC hrs
and 31 months of turnaround time,
it only took me about \(440,000\) NERSC hrs
and a few days to finish the first few hundred simulations
(where I use the job array here to march through the queue more quickly.)
This is a reduction of a factor of \(12\) in compute time,
and a factor of a few hundreds in turnaround time.

Now that we have the computational challenge under control,
we can turn our attention to
what sort of problems we are facing
when we are working towards unblinding.

\chapter{Timestream of Data (TOD)}\label{sec:TODmisc}

As we have some general ideas about the pipelines from previous sections,
let us go back and revisit some of the details involved.
We should note that the low-\(\ell\) B-mode large-patch analysis is published already.
All large-patch analyses share similar pipelines.
The earlier a task is in the pipeline, the more it is shared among them.
So here we are not repeating all the details but will refer to
The POLARBEAR Collaboration et al.\footnote{\citeproc{ref-the_polarbear_collaboration_measurement_2020}{{``A Measurement of the Degree-Scale CMB \emph{B}-Mode Angular Power Spectrum with POLARBEAR''}}.} whenever possible.

We will start from the Timestream of Data (TOD)\@.
As we mentioned in \cref{sec:scanningStrategy},
when observing the CMB, which is inherently a 2D object,
we instead scan the sky and obtain a TOD first\footnote{Also see \cref{sec:CRHWPDemodulation}.}.

To expand upon \cref{eq:scanning} to show all complications that arise in real-world observations,

\begin{align}
m^{\prime}_{p}	&	 = \mathcal{B} \mathopen{} \left \{ m_{p} \right \} \mathclose{}	\label{eq:beamConvolution}	\\
d \mathopen{} \left ( t \right ) \mathclose{}	&	 = P_{t p} \left( m^{\prime I}_{p} + \eta \left( m^{\prime Q}_{p} \cos \mathopen{} \left ( 2 \psi_{t} \right ) \mathclose{} + m^{\prime U}_{p} \sin \mathopen{} \left ( 2 \psi_{t} \right ) \mathclose{} \right) \right) + N_{t}	\label{eq:TODObservation}	\\
d^{\prime} \mathopen{} \left ( t \right ) \mathclose{}	&	 = \mathcal{D} \mathopen{} \left \{ d \mathopen{} \left ( t \right ) \mathclose{} \right \} \mathclose{} \approx \left( 1 + g d \mathopen{} \left ( t \right ) \mathclose{} \right) d \mathopen{} \left ( t - \tau d \mathopen{} \left ( t \right ) \mathclose{} \right ) \mathclose{}	\label{eq:TODconvolution}
\end{align}

Here, \(m_p\) represents the physical CMB map on the sky.
And eventually \(d'(t)\) is the TOD recorded as raw data.

\Cref{eq:beamConvolution} represents the map that the telescope sees.
This normally means it is a beam-convolved map,
assuming the beam is symmetric for simplicity.

\Cref{eq:TODObservation} represents how the telescope scans the sky
and obtains a TOD\@.
Note that the HWP angle \(\psi\) and the pointing \(P\) are involved,
where both together constitute the scanning strategy
which determines the observed signal.
\(η\) represents the polarization efficiency of the detectors.
And \(N\) is the noise of the detectors.

Lastly, \cref{eq:TODconvolution}
represents a non-linear effect from the detectors
which makes the observed TOD a convolved version of the actual TOD\@.
The approximation is characterized in Takakura et al.\footnote{\citeproc{ref-takakura_performance_2017}{{``Performance of a Continuously Rotating Half-Wave Plate on the POLARBEAR Telescope''}}.} eq.~2.9.

In fact, these describe the TOD observed per detector,
so all parameters involved can depend on the individual detector.

While a scanning strategy determines \(P\) and \(\psi\),
the observing pointing \(P\) and HWP angle \(\psi\) are recorded.
For the pointing, as outlined in The POLARBEAR Collaboration et al.\footnote{\citeproc{ref-ade_measurement_2014-1}{{``A Measurement of the Cosmic Microwave Background \emph{B}-Mode Polarization Power Spectrum at Sub-Degree Scales with POLARBEAR''}}.} a \(5\)-parameter pointing model is used to reconstruct the pointing
to reduce error from the actual pointing.
In large-patch analysis,
we followed a similar procedure
to construct the pointing model using point sources.\footnote{\citeproc{ref-the_polarbear_collaboration_measurement_2020}{The POLARBEAR Collaboration et al., {``A Measurement of the Degree-Scale CMB \emph{B}-Mode Angular Power Spectrum with POLARBEAR''}}.}

\Cref{eq:beamConvolution} means that we need to characterize the beam as well.
The beam window function\footnote{This inherently assumes the beam is symmetric.}
is measured using dedicated observations of Jupiter,\footnote{\citeproc{ref-the_polarbear_collaboration_measurement_2020}{The POLARBEAR Collaboration et al., {``A Measurement of the Degree-Scale CMB \emph{B}-Mode Angular Power Spectrum with POLARBEAR''}}.} and an effective beam is used across all detectors although they differ slightly across wafers.
This is good enough as the differences between the beams are much less than the statistical error-bar at this noise level.

Then, we need to characterize the detectors including
the polarization efficiency and the effect of detector non-linearity.
The methodology of the latter is studied in Takakura et al.\footnote{\citeproc{ref-takakura_performance_2017}{{``Performance of a Continuously Rotating Half-Wave Plate on the POLARBEAR Telescope''}}.} Detector time constants are measured by using a thermal source on the telescope,
and polarization efficiency together with their polarization angles are calibrated
using Tau A observation,
as indicated in The POLARBEAR Collaboration et al.\footnote{\citeproc{ref-the_polarbear_collaboration_measurement_2020}{{``A Measurement of the Degree-Scale CMB \emph{B}-Mode Angular Power Spectrum with POLARBEAR''}}.}

After all these calibrations,
the TOD is deconvolved\footnote{This only applies
  to the real data but not the simulations as we do not simulate the detector non-linearity effects.}
as an inversion of \cref{eq:TODconvolution},
the mapmaker is effectively inverting \cref{eq:TODObservation},
and at a later stage the power-spectra are de-biased by the beam profile to invert the effect of \cref{eq:beamConvolution}.

\section{Time domain filtering}\label{time-domain-filtering}

One key difference between the low-\(\ell\) and high-\(\ell\) pipelines is illustrated below.

As the naïve mapmaker is used in our analysis,
filtering the TOD is one of the techniques
we can count on
as we are not performing un-biased mapmaking.
One of the benefits of splitting the large-patch analysis
into low-\(\ell\) and high-\(\ell\) pipelines
is that we can use different amounts of filtering to suppress
different levels of unwanted systematics.
In high-\(\ell\), we can afford heavier filtering from the low-frequency end,
as we do not need the low-\(\ell\) signals.

TOD filtering is an optimization problem.
If too little,
systematics are too large
and null-tests will fail.
If too much,
we are eating into the signal
and hurting the sensitivity.

One such example is the common-mode filter.
Common mode means the mode common to all detectors.
And it is estimated as the sample mean of those detectors at time \(t\).
It can be understood in terms of Principal Component Analysis (PCA)
that the common mode approximately corresponds to the principal component among the detectors.
The subtraction of the common mode helps reduce the amount of \(1/f\) noise.

\begin{figure}
\centering
\pandocbounded{\includegraphics[keepaspectratio,alt={A photograph of the fully assembled focal plane, support structure and milliKelvin wiring.}]{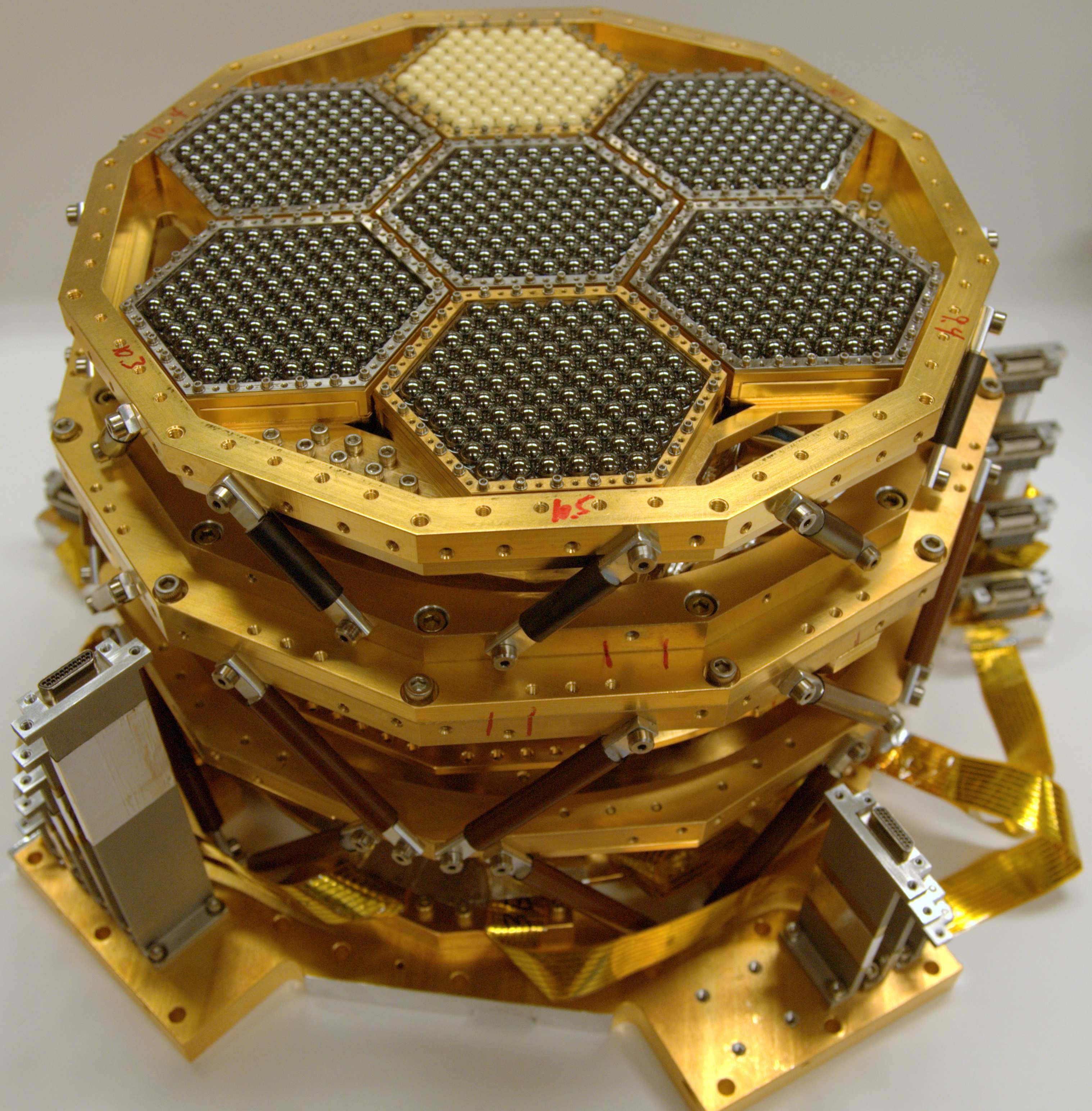}}
\caption[A photograph of the fully assembled focal plane, support structure and milliKelvin wiring.]{A photograph of the fully assembled focal plane, support structure and milliKelvin wiring.\footnotemark{}}\label{fig:POLARBEAR-wafer}
\end{figure}
\footnotetext{\citeproc{ref-kermish_polarbear_2012}{Kermish et al., {``The POLARBEAR Experiment''}}.}

Note that common mode is constructed among a set of detectors.
In low-\(\ell\), a single common mode is constructed over all detectors on the focal plane.
As seen in \cref{fig:POLARBEAR-wafer},
a single wafer corresponds to a smaller angular scale on the sky, compared to the full focal plane\footnote{A total of \(7\) wafers are on the focal plane, where a smaller region on the focal plane corresponds to a smaller field of view on the sky.}.
Hence, the high-\(\ell\) pipeline can afford more aggressive filtering by constructing a common mode per wafer.

Similarly,
we have polynomial filtering per sub-scan\footnote{A sub-scan corresponds to
  the duration where the telescope is constantly scanning in one direction.
  That is, each turn-around point marks the next sub-scan.} of the telescope.
An \(n\)-th order polynomial is fitted in the TOD within each sub-scan
and is subtracted from the TOD\@.
This attempts to remove the long time scale fluctuation from the \(1/f\) noise,
but also inevitably removes some of our large angular scale signal.
Again, the high-\(\ell\) pipeline can afford
a larger degree polynomial without eating too much into our science range here.
In this specific research,
a \(9\)-degree polynomial is fitted and subtracted from the TOD
for all 3 components \(I, Q, U\).

\chapter{Analytical methods}\label{analytical-methods}

We will go through the main analytical methods involved here.
The general framework is the same with previous B-mode papers from POLARBEAR.\footnote{\citeproc{ref-ade_measurement_2014-1}{The POLARBEAR Collaboration et al., {``A Measurement of the Cosmic Microwave Background \emph{B}-Mode Polarization Power Spectrum at Sub-Degree Scales with POLARBEAR''}}; \citeproc{ref-the_polarbear_collaboration_measurement_2017}{The POLARBEAR Collaboration et al., {``A Measurement of the Cosmic Microwave Background \emph{B}-Mode Polarization Power Spectrum at Subdegree Scales from Two Years of POLARBEAR Data''}}; \citeproc{ref-the_polarbear_collaboration_measurement_2020}{The POLARBEAR Collaboration et al., {``A Measurement of the Degree-Scale CMB \emph{B}-Mode Angular Power Spectrum with POLARBEAR''}}.} The basic approach will be outlined here, with more details provided whenever they differ.

\section{Cross-pseudo-power-spectra calculation}\label{sec:pseudoSpectraCalculation}

We briefly outlined the cross-pseudo-power-spectra in \cref{sec:pseudoSpectra}.
Equation-wise, it is well-documented in previous POLARBEAR publications on B-mode.
It is based on Hinshaw et al.\footnote{\citeproc{ref-hinshaw_firstyear_2003}{{``First‐Year \emph{Wilkinson Microwave Anisotropy Probe} ( \emph{WMAP} ) Observations''}}.}

We mentioned earlier 38 map-bundles were formed.
Each of them amounts to about 10 days' worth of observations,
and each forms a complete map of the observed patch of sky.

This cross-map approach assumes that each map-bundle
is an independent observation of the sky.
So the cross-pseudo-power-spectra would be noise-free and bias-free.
The validity of the assumption is from empirical knowledge
that the correlations in the TOD have a timescale of around \(\SI{20}{\minute}\),
and from empirical studies including the null-test suite.
Hence, this estimator provides a signal-only estimation.

The noise pseudo-power-spectra can be estimated
by first calculating the pseudo-power-spectra of the full map
(that is coadded from all 38 map-bundles),
often called the auto-power-spectra,
and subtracting the signal-only estimate from previous step.

\section{Smoothing and tapering of the maps}\label{smoothing-and-tapering-of-the-maps}

As the estimate of \(E\)-mode and \(B\)-mode involves 2nd order derivatives\footnote{See \cref{sec:CMBFlatSkyMath}.},
we need to make sure the input maps are at least twice differentiable.
Also, at the edge of the map, derivatives can be problematic.
Hence, a tapering is usually done to ensure gradual transition to zero.
Finally, in the calculation of pseudo-power-spectra,
we would want the region with a lower level of noise to be weighted more heavily to reduce the resultant error-bar.
Naturally, as we already have the weight map\footnote{To recall,
  the weight map is precision---the inverse of the variance at each pixel on the map.},
it is used for this purpose.

Putting all together, the weight map will be smoothed, tapered---the product of which is often called the weight-mask or simply mask---and will be multiplied to the map before performing
pseudo-power-spectra estimation in the previous section.

In previous POLARBEAR B-mode analyses,
a Hamming window with a scale of \(\SI{0.53}{\degree}\) is used to smooth the map,
and a \(C^2\)-tapering of \(\SI{0.5}{\degree}\) wide is performed at the edge.
These weight maps are calculated per map within the map-bundles.

In this analysis,
a Gaussian blur of \(\SI{3.3013338110314874}{\arcminute}\)\footnote{The number
  is chosen so that this Gaussian window is minimally different from the Hamming window above.}
is used to smooth the map,
and a \(C^\infty\)-tapering with a scale of \(\SI{1}{\degree}\) is applied at the edge.

\section{Monte Carlo Apodized Spherical Transform Estimator (MASTER)}\label{sec:MASTER}

MASTER is ``a fast and accurate method for estimation of the cosmic microwave background (CMB) anisotropy angular power spectrum''\footnote{Hivon et al., \citeproc{ref-hivon_master_2002}{{``MASTER of the Cosmic Microwave Background Anisotropy Power Spectrum''}}.}.

This is the central concept that enables the naïve mapmaker that
biases (filtering) is allowed but will be de-biased
in the power-spectra domain.
And the effect of the weight mask from the previous section
will also be accounted for. Mathematically,

\[\tilde{C}_\ell = \sum_{\ell'} M_{\ell \ell'} F_{\ell'} B^2_{\ell'} C_{\ell'}\]

Here, \(C\) is the original, un-biased (cross-)power-spectrum.
\(B\) is the beam profile,
\(F\) is the filter transfer function,
and \(M\) is the mode-coupling matrices.
\(\tilde{C}\) is the pseudo-(cross-)power-spectra
that we can measure in \cref{sec:pseudoSpectraCalculation}.

The beam is measured in previous steps\footnote{See \cref{sec:TODmisc}.}
and is known at this stage.
Mode-coupling matrices account for the fact that
multiplying a weight-mask would change the resultant power-spectra
and is the topic later in \cref{sec:modeCoupling}
and can be calculated analytically given the weight-mask.

Had we known the filter transfer function \(F\),
we could have used that to invert the formula
and would have obtained an unbiased estimate of \(C\).

The key idea in using this equation
is that we can perform Monte Carlo (MC) simulations to estimate \(F\).
Specifically, we can run MC simulations
that replicate the actual experiment
including any filtering effects, resulting in

\begin{equation}\protect\phantomsection\label{eq:MASTER_MC}{\tilde{C}^\text{MC}_\ell = \sum_{\ell'} M_{\ell \ell'} F_{\ell'} B^2_{\ell'} C^\text{MC}_{\ell'}}\end{equation}

Here, \(C^\text{MC}\) is a known input, \(\tilde{C}^\text{MC}\)
can be calculated as outlined in \cref{sec:pseudoSpectraCalculation}.
Hence we can solve this equation and obtain an estimate of \(F\).

In fact, since there are \(E \rightarrow B\) leakage and vice versa,
we can perform different MC simulations with one set of maps
having only \(T, E\) signals, and another set having only \(T, B\).
This allows us to estimate \(F^{EE \rightarrow BB}, F^{BB \rightarrow EE}\)
in addition to \(F^{TT \rightarrow TT}, F^{EE \rightarrow EE}, F^{BB \rightarrow BB}\).

Now, the devil is in the details.
How do we invert \cref{eq:MASTER_MC}?
Note that the mode-coupling matrices have a very high condition number
unless we have close to full-sky coverage\footnote{This
  can be understood by observing in the limit of full-sky
  coverage, \(M \rightarrow I\) the identity matrix.}.
Inverting it numerically is technically possible
but the resultant \(F\) would have huge numerical error making it not usable.

Hivon et al.\footnote{\citeproc{ref-hivon_master_2002}{{``MASTER of the Cosmic Microwave Background Anisotropy Power Spectrum''}}.} has mentioned an iterative approach
to estimate \(F\).
This approach inherently depends on
the number-of-iterations parameter \(n_\text{iter}\).
Note that when \(n_\text{iter} \rightarrow \infty\)
it should recover the \(F\) with high numerical error,
as the iterative approach is solving the inverse matrix problem iteratively.
\(n_\text{iter}\) has to be chosen carefully.
Hivon et al.\footnote{\citeproc{ref-hivon_master_2002}{{``MASTER of the Cosmic Microwave Background Anisotropy Power Spectrum''}}.} chose \(n_\text{iter} = 1\) for BOOMERANG,
and The POLARBEAR Collaboration et al.\footnote{\citeproc{ref-ade_measurement_2014-1}{{``A Measurement of the Cosmic Microwave Background \emph{B}-Mode Polarization Power Spectrum at Sub-Degree Scales with POLARBEAR''}}.} chose \(n_\text{iter} = 10\) for POLARBEAR's small-patch experiment.

Below we will describe another method
that has no such arbitrary parameter for the numerical solution.

\subsection{Modified MASTER}\label{modified-master}

The key idea is to move the term \(F\) to the left,

\begin{equation}\protect\phantomsection\label{eq:modifiedMASTER}{\tilde{C}'_\ell = \sum_{\ell'} F'_\ell M_{\ell \ell'} B^2_{\ell'} C_{\ell'}.}\end{equation}

The prescription here is clearly wrong.
The filter transfer function captures the effect
of the filtering performed on the TOD\@.
The mode-coupling effect
happens at a later stage when calculating the
pseudo-(cross-)power-spectra.
So only the order prescribed in MASTER is correct.

Analytically, if \(M\) is diagonal or \(F\) is a constant,
the modified MASTER and the original MASTER are the same.
We know that \(M\) is close to diagonal
(in the sense that main diagonal is dominant in magnitude)
and \(F\) is slowly varying across \(\ell\).

Empirically, it should be mentioned that
the estimate of \(F\) is never exact here,
because it is estimated using MC simulations,
and because it is numerically unstable to construct \(F\) exactly.

So phenomenologically, as long as the error made is small,
\cref{eq:modifiedMASTER} should be ok.

Note that if we define

\begin{equation}\protect\phantomsection\label{eq:modifiedFinMASTER}{\tilde{C}''_\ell = \sum_{\ell'} M_{\ell \ell'} F'_{\ell'} B^2_{\ell'} C_{\ell'}}\end{equation}

Then

\[\tilde{C}'_\ell - \tilde{C}''_\ell
= \sum_{\ell'} \left(
    F'_{\ell} M_{\ell \ell'} -
    M_{\ell \ell'} F'_{\ell'}
\right)
B^2_{\ell'} C_{\ell'}\]

So the difference is roughly speaking a commutator between \(M, F\).

To find out the error made when using the modified MASTER,
we can use MC simulations \cref{eq:MASTER_MC}. Note that in this equation,
both \(\tilde{C}\) and \(C\) are known.
We can then use \cref{eq:modifiedMASTER} to estimate \(F'\),
plug that in \cref{eq:modifiedFinMASTER},
and calculate the quantity

\[\frac{\tilde{C}''_\ell - \tilde{C}_\ell}{\tilde{C}_\ell}.\]

This represents the relative error made
in terms of the resultant pseudo-power-spectra
had we known the true \(F\) but used the modified MASTER to estimate \(F'\) instead.

\begin{figure}
\centering
\pandocbounded{\includegraphics[keepaspectratio,alt={Relative error made in TT pseudo-power-spectra using modified MASTER}]{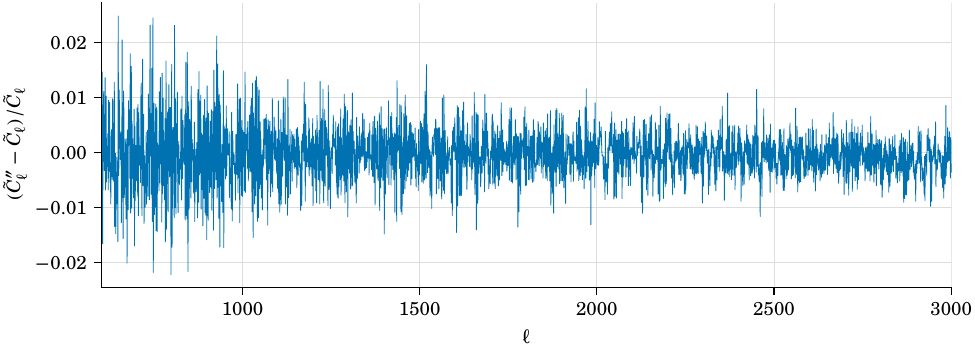}}
\caption{Relative error made in \(TT\) pseudo-power-spectra using modified MASTER}\label{fig:relErrTT}
\end{figure}

\begin{figure}
\centering
\pandocbounded{\includegraphics[keepaspectratio,alt={Relative error made in EE pseudo-power-spectra using modified MASTER}]{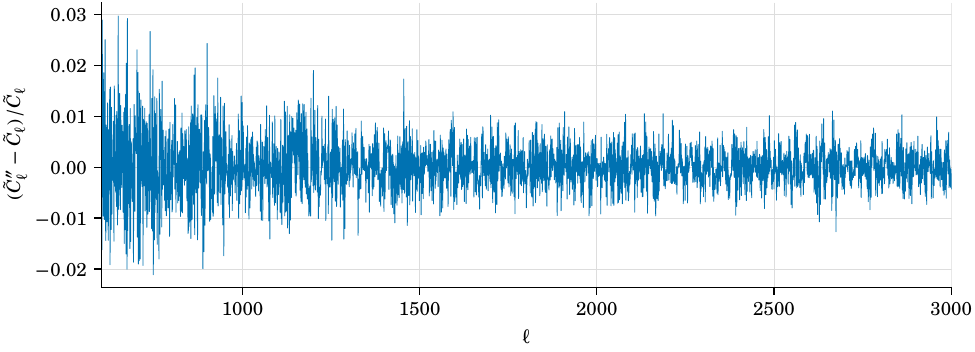}}
\caption{Relative error made in \(EE\) pseudo-power-spectra using modified MASTER}\label{fig:relErrEE}
\end{figure}

\begin{figure}
\centering
\pandocbounded{\includegraphics[keepaspectratio,alt={Relative error made in BB pseudo-power-spectra using modified MASTER}]{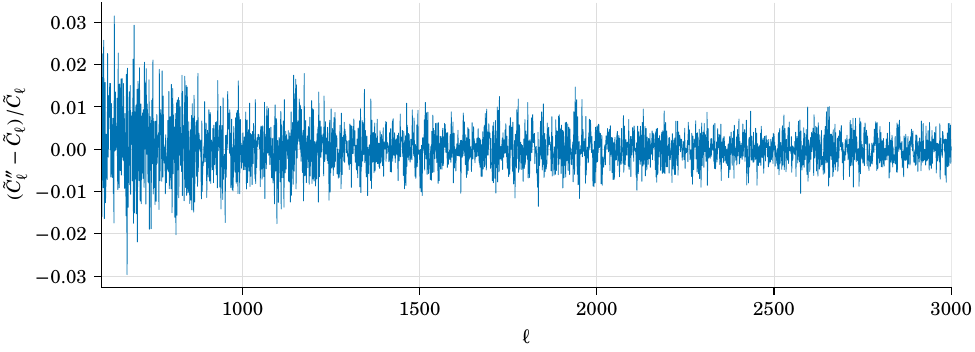}}
\caption{Relative error made in \(BB\) pseudo-power-spectra using modified MASTER}\label{fig:relErrBB}
\end{figure}

Estimating this relative error in the pseudo-power-spectra cases of \(TT, EE, BB\),
which is shown in \cref{fig:relErrTT}, \cref{fig:relErrEE}, \cref{fig:relErrBB} using 128 MC simulations,
we see that the relative error is small (RMS at \(\sim \SI{0.5}{\percent}\) across \(600 \leq \ell \leq 3000\)) and unbiased.
As the bin-width used in a later step is 100, the error would be much smaller.

Not only that, empirically,
I find that this relative error is roughly inversely proportional to \(F'\).
Since \(0 < F \leq 1\),
empirically as long as \(F'\) is not close to zero\footnote{This is a region
  that the de-biased power-spectra would have close to infinite error
  which is not usable anyway.},
the approximation performed by the modified MASTER should be small.

Note that the relative error mentioned here is comparing
between modified MASTER and the original MASTER,
in my pipeline I consistently always use the modified MASTER throughout the pipelines so the actual error made
should be even smaller.

To conclude,
we find that the modified MASTER is parameter-free requiring no tuning,
the error made is negligible,
and is good enough phenomenologically in this particular analysis.

Finally, it should be noted that
this modified MASTER may be used elsewhere.
It may have been mentioned in casual conversations with other CMB scientists.
The justification here may be new in the sense that
it is shown here explicitly why we can do this.

\section{Mode-coupling calculations}\label{sec:modeCoupling}

We will briefly go through how the mode-coupling matrices are estimated.

First, the mode-coupling matrix is often written as
\(M^{XYXY}\), or \(M^{XY \rightarrow XY}\),
representing the mode-coupling matrix from the \(XY\) power-spectra
to the \(XY\) pseudo-power-spectra.

\(XY\) can be any cross-spectra \(TT, EE, BB, TE, TB, EB\).

For the \(TT\) case,
Hivon et al.\footnote{\citeproc{ref-hivon_master_2002}{{``MASTER of the Cosmic Microwave Background Anisotropy Power Spectrum''}}.} has derived it in the appendix.
For the rest which includes polarizations,
Miller\footnote{\citeproc{ref-miller_mode_2013}{{``Mode Mixing Calculation on the Flat Sky''}}.} documented them in an internally circulated document,
and implemented it in AnalysisBackend.
In these equations,
the B-mode is assumed to be constructed by the pure B-mode estimator.\footnote{\citeproc{ref-smith_pseudo-_2006}{Smith, {``Pseudo- C ℓ Estimators Which Do Not Mix E and B Modes''}}.}

Later, I found some problems in AnalysisBackend regarding
the mode-coupling matrix calculations.

First, there is a bug that in the case of \(EB\),
the pure \(B\)-mode estimator is not used.
Second, there are various numerical problems in the calculation
that the matrices deviate from actual values when it is far away
from the main diagonal.

The mode-coupling matrix code is completely rewritten
to fix this, in addition to making it run much faster
as explained in \cref{sec:computational-challenges}.
\href{https://github.com/ickc/TAIL/blob/3b4e49d6d5a4/tail/modecoupling.py}{This module is released in TAIL}.

There is another subtlety related to the calculation of mode-coupling matrices.
In the previous pipelines of POLARBEAR \(B\)-mode analysis,
each map-bundle uses their own weight-mask,
which implies that the mode-coupling matrix calculated
here is not correct
(as the analytical formula assumes all share the same weight-mask.)
Phenomenologically, using MC simulations,
previous pipelines can recover unbiased power-spectra empirically.
Alternatively, we can use a single weight-mask (constructed using the fully coadded map.)
The disadvantage of this approach is that the weight-mask applied is less than optimal,
and therefore potentially reducing sensitivity in the end.
Both approaches have been compared in this study,
and no major differences between them are found.
In the end, the latter approach is taken as it is better safe (less than optimal)
than sorry (wrong mode-coupling matrices applied.)

\section{Zero padding of the map for the calculation of the power-spectra}\label{zero-padding-of-the-map-for-the-calculation-of-the-power-spectra}

In the calculation of the pseudo-(cross-)power-spectra in \cref{sec:pseudoSpectraCalculation},
because an FFT is needed to first transform the map into frequency domain,
zero-padding is required as FFT assumes periodicity.

However, how much zero-padding is needed?
If we simply estimate from \cref{eq:degree-to-l},
as the lowest \(\ell\) here is \(600\),
it corresponds to around \(\SI{18}{\arcminute}\)
on both ends.
This amounts to only \(9\) pixels wide
in \(\SI{2}{\arcminute}\) resolution.

However, as Kuo et al.\footnote{\citeproc{ref-kuo_highresolution_2004}{{``High‐Resolution Observations of the Cosmic Microwave Background Power Spectrum with ACBAR''}}.} said,

\begin{quote}
For a strictly stationary process this is not a problem since the correlation usually vanishes after a sufficiently long time lag. For piecewise stationary processes, a straightforward application of FFT produces false correlation for lag \(> n/2\), where \(n\) is the number of elements. This can be prevented by zero padding.
\end{quote}

So a \(2\times\) zero-padding is recommended
when calculating the pseudo-power-spectra.

The effects of zero-padding between
\(1.3\times\) and \(2\times\) are compared
in this study
and negligible differences are found
from simulations.
However, as a caution,
we still follow the standard practice of
having \(2\times\) zero-padding in the final analysis.

\section{Noise}\label{sec:noise}

Noise estimate is used in 2 places,
one in the coadd pipeline where
the noise is used to determine the weight,
and another when noise is injected in the simulations
to mimic the real observations
where the statistical fluctuations in
the resultant power-spectra are used
to estimate the statistical error-bar
of the power-spectra.

There are 2 noise pipelines.
In the first one,
we assume an uncorrelated white noise model,
and estimate them in the TOD\@.
This is always the noise model used
to construct the weight-maps.
Optionally, this white noise model
is also used in the simulation
to simulate noisy observations.

Another noise pipeline is called
sign-flipped noise.
Assuming the maps containing
signal and noise terms only,
if we randomly assign a sign
in the coadd process to the maps,
then the coadded map should on average
contain no signal and only have noise.

These 2 noise pipelines are compared,
and they are found to be in general agreement
in \(N^{EE}, N^{BB}\),
as the assumption of uncorrelated white noise
is very good there because of the CRHWP modulation
away from the \(1/f\) noise\footnote{The POLARBEAR Collaboration et al., \citeproc{ref-ade_measurement_2014-1}{{``A Measurement of the Cosmic Microwave Background \emph{B}-Mode Polarization Power Spectrum at Sub-Degree Scales with POLARBEAR''}}
  further compared between correlated
  and uncorrelated white noise models
  and found them to be consistent.}.
This is not true for the temperature maps
and its noise power-spectra \(N^{TT}\)
simply because of the \(1/f\) noise.

As explained before,
less than optimal construction of the weight-maps
only results in less than optimal (bigger error-bar)
but usable resultant maps.

To obtain a realistic error-bar
in simulations including the intensity map,
we opt for the sign-flip noise.
With this realistic estimate of statistical errors,
we shall soon see we can pass the null-test suite
even if we include the temperature-related power-spectra (including \(TT, TE, TB\))
for the first time in POLARBEAR\@.
These power-spectra are used for absolute gain calibration later.

\chapter{Statistical methods}\label{sec:statistical-methods}

\section{Absolute Gain Calibration}\label{absolute-gain-calibration}

Keating, Shimon, and Yadav\footnote{\citeproc{ref-keating_self-calibration_2013}{{``SELF-CALIBRATION OF COSMIC MICROWAVE BACKGROUND POLARIZATION EXPERIMENTS''}}.} shows that we can use the CMB itself to
perform absolute gain and angle calibration.
Absolute gain refers to an overall factor multiplying the power-spectra,
absolute angle refers to an overall angle rotating the linear polarization of the CMB\@.
In a later section,
we will show how they are estimated and corrected for in this analysis.

\subsection{Cosmic variance limit}\label{cosmic-variance-limit}

Let us first start by investigating a toy model where there is no noise.
This is often coined as the cosmic variance limit,
as it is an intrinsic uncertainty from the universe even if we can observe perfectly.

We have already seen the Knox formula \cref{eq:knox_formula},
in the limit of no noise,

\[\Delta C_\ell \approx \sqrt{\frac{2}{\nu_\ell}} C_\ell.\]

Defining absolute gain \(\alpha\)\footnote{Note that if we define absolute gain \(\beta\) in the TOD and hence the map, \(\alpha \equiv \beta^2\).} as

\[C_b^\text{obs.} = \alpha C_b^\text{Th.}\]

\[\alpha_b = \frac{C_b^\text{obs.}}{C_b^\text{Th.}}\]

\[\Delta \alpha_b \approx \sqrt{\frac{2}{\nu_b}}\]

Now, we assume a naïve OLS procedure assuming they are uncorrelated\footnote{It will not be exactly uncorrelated in our case since we have far from a full sky map.} random variables as a function of \(b\),
the corresponding standard error will then be

\begin{align*}
\Delta  &= \left( \sum_b \frac{1}{(\Delta \alpha_b)^2} \right)^{-\frac{1}{2}}   \\
    &= \left( \sum_b \frac{\nu_b}{2} \right)^{-\frac{1}{2}} \\
    &= \left( \sum_b \frac{\sum_{l \in b} (2l + 1) f_\text{sky, eff.}}{2} \right)^{-\frac{1}{2}}    \\
    &= \left( \sum_{l = l_\text{min}}^{l_\text{max}} (2l + 1) \frac{f_\text{sky, eff.}}{2} \right)^{-\frac{1}{2}}   \\
    &= \left( \frac{(l_\text{max} + 1 - l_\text{min}) (l_\text{max} + 1 + l_\text{min})}{2} f_\text{sky, eff.} \right)^{-\frac{1}{2}}   \\
\end{align*}

Note that this is independent of the actual binning \(b\) used.

Using our actual data,
\(f_\text{sky, eff.} \approx 0.0159\).

With \(l_\text{min} = 600, l_\text{max} + 1 = 3000, \Delta \approx 0.00381\).

\subsubsection{Method 1}\label{method-1}

Using \(C_l^{TT}\), the error of the absolute gain calibration \(\alpha\) is \(\Delta\).

\subsubsection{Method 2}\label{method-2}

Using \(C_l^{EE}\), the \(\alpha\) estimated above is the product of the actual absolute gain times the polarization efficiency squared. That is,

\[\hat{\alpha} = \frac{\alpha^E}{\eta^2}\]

\[\left( \frac{\Delta \alpha}{\alpha} \right)^2 = \left( \frac{\Delta}{\alpha^E} \right)^2 + \left( 2 \frac{\Delta \eta}{\eta} \right)^2\]

Assuming \(\alpha \sim 1, \frac{\Delta}{\alpha^E} \sim \Delta \approx 0.00381\).

From internally circulated data, \(\frac{\Delta \eta}{\eta} \approx 0.0151\).

\subsubsection{Method 3}\label{method-3}

Inverse variance weighted between method 1 \& 2.

Since \(2 \frac{\Delta \eta}{\eta} \gg \frac{\Delta}{\alpha}\), the error bar obtained here is virtually identical to method 1.

\subsubsection{Conclusion}\label{conclusion}

In the cosmic variance limit, using the temperature data to estimate the absolute gain is much more accurate as the uncertainty in the polarization efficiency dominates the uncertainty in the case of polarization\footnote{Note that this does not reduce statistical uncertainty
  in determining the absolute scale of the CMB polarizations \emph{per se}.
  This is because there are other contributions to this absolute scale,
  such as the ``absolute polarization efficiency''.
  If it can be shown that other kinds of systematics in determining
  this absolute scale are well controlled,
  then the benefits of getting a well-characterized absolute gain
  can result in meaningful reduction of statistical error in the CMB polarization measurements.}.

This toy model illustrates
why we would want to include the temperature power-spectrum
for absolute gain calibration.
With it, another bonus is that we can cross-check
if the polarization efficiency is correct
in the sense that the ``absolute polarization efficiency''
should be consistent with \(1\).
Later we will see how this compares with the absolute gain inferred using the actual data.

\section{POLARBEAR null-test framework in equations}\label{sec:null-framework-equations}

In \cref{sec:POLARBEAR-null-test-framework},
we already have an overview of the POLARBEAR null-test framework.

We see null-spectra are defined there.
During the early phase of running null-tests,
we can intuitively see
what null-tests are failing
where there are huge outliers.
But we can do better to decide this objectively.

It involves constructing null-estimators from null-power-spectra.
Exactly how these statistics are constructed is somewhat arbitrary.
The POLARBEAR Collaboration et al.\footnote{\citeproc{ref-ade_measurement_2014-1}{{``A Measurement of the Cosmic Microwave Background \emph{B}-Mode Polarization Power Spectrum at Sub-Degree Scales with POLARBEAR''}}.} described the 5 different statistics used
in the small-patch analysis over 3 different sky regions resulting
in a total of 15 statistics.
The definitions of the 5 statistics are somewhat vague,
and the way the 15 statistics are combined is not specified.

There is also an internally circulated document Chinone\footnote{\citeproc{ref-chinone_polarbear_2014}{{``POLARBEAR Null-Test Framework''}}.}
documenting this null-test framework in more details.

Below we will mathematically define these quantities
in this analysis.

\begin{Definition}
Null-power-spectra.

Define random variable \(\chi(X, n, b) \equiv \frac{\hat{C}^{Xn, \text{null}}_b}{\hat{σ}^{Xn}_b}\) as a function of

\(X\): power-spectrum. This includes \(TT, TE, TB, EE, EB, BB\).

\(n\): null-split, such as \texttt{AMP\_\allowbreak{}4F\_\allowbreak{}BY\_\allowbreak{}BOLO}, \ldots{}

\(b\): an \(\ell\)-bin, such as \(600 \leq \ell < 700\).

Note that by construction,
\(\chi\) is a standard score,
assuming the theoretical expected value of \(0\),
and estimated using the sample standard deviation
via MC simulations.
Asymptotically in \(n_\text{sim}\) and \(ν\), \(\chi \xrightarrow{d} N(0, 1)\).
\end{Definition}

\begin{Definition}\label{def-null-statistics}
Null-statistics.

Per spectrum, we collapse the null-power-spectra into the following 5 different ways,

\[Y_1(X) = \lvert \langle \chi(X, n, b) \rangle_{n, b} \rvert\]

\[Y_2(X) = \max_{n, b} \chi^2(X, n, b)\]

\[Y_3(X) = \max_{n} \sum_b \chi^2(X, n, b)\]

\[Y_4(X) = \max_{b} \sum_{n} \chi^2(X, n, b)\]

\[Y_5(X) = \sum_{n, b} \chi^2(X, n, b)\]

Consider sample space \(\Omega\).

\[Y_i: \Omega \rightarrow \mathbb{R}\]

\[Y_i: s \mapsto y_i\]

\[G_{Y_i}(y_i) \equiv P(Y_i > y_i)\]

Since \(G \equiv 1 - F\), \(G_{Y_i} \rightarrow U(0, 1)\)

Note that \(F\) is the cumulative distribution function (CDF)
in statistics, while \(G\) is one minus that, often called
probability-to-exceed (PTE).
\end{Definition}

In this analysis, the 5 statistics \(G_{Y_i}(y_i)\) are considered, by estimating the distribution of \(Y_i\) empirically through Monte Carlo simulation, and \(y_i\) from measured data. Additionally, 3 spectra are considered in the unblinding criteria: \(BB, EB, EE\). That is, there are a total of \(15\) p-values.

\begin{Theorem}\label{thm-null-statistics-iid}
Direct computation of null-statistics assuming \(\chi(X, n, b) \xrightarrow[i.i.d.]{d} N(0, 1)\).

\begin{align*}
G_{Y_{1}} \mathopen{} \left ( y_{1} \right ) \mathclose{}	&	 = \operatorname{erfc} \mathopen{} \left ( \sqrt{\frac{n_{\text{split}} n_{\text{b}}}{2}} \left| y_{1} \right| \right ) \mathclose{}	\\
G_{Y_{2}} \mathopen{} \left ( y_{2} \right ) \mathclose{}	&	 = \frac{n_{\text{split}} n_{\text{b}}}{2^{\frac{1}{2}} \Gamma \mathopen{} \left ( \frac{1}{2} \right ) \mathclose{}} \int_{y_{2}}^{\infty} \left( P \mathopen{} \left ( \frac{1}{2}, \frac{u}{2} \right ) \mathclose{} \right)^{n_{\text{split}} n_{\text{b}} - 1} \mathrm{e}^{- \frac{u}{2}} u^{\frac{1}{2} - 1} \,\mathrm{d} u	\\
G_{Y_{3}} \mathopen{} \left ( y_{3} \right ) \mathclose{}	&	 = \frac{n_{\text{split}}}{2^{\frac{n_{\text{b}}}{2}} \Gamma \mathopen{} \left ( \frac{n_{\text{b}}}{2} \right ) \mathclose{}} \int_{y_{3}}^{\infty} \left( P \mathopen{} \left ( \frac{n_{\text{b}}}{2}, \frac{u}{2} \right ) \mathclose{} \right)^{n_{\text{split}} - 1} \mathrm{e}^{- \frac{u}{2}} u^{\frac{n_{\text{b}}}{2} - 1} \,\mathrm{d} u	\\
G_{Y_{4}} \mathopen{} \left ( y_{4} \right ) \mathclose{}	&	 = \frac{n_{\text{b}}}{2^{\frac{n_{\text{split}}}{2}} \Gamma \mathopen{} \left ( \frac{n_{\text{split}}}{2} \right ) \mathclose{}} \int_{y_{4}}^{\infty} \left( P \mathopen{} \left ( \frac{n_{\text{split}}}{2}, \frac{u}{2} \right ) \mathclose{} \right)^{n_{\text{b}} - 1} \mathrm{e}^{- \frac{u}{2}} u^{\frac{n_{\text{split}}}{2} - 1} \,\mathrm{d} u	\\
G_{Y_{5}} \mathopen{} \left ( y_{5} \right ) \mathclose{}	&	 = Q \mathopen{} \left ( \frac{n_{\text{split}} n_{\text{b}}}{2}, \frac{y_{5}}{2} \right ) \mathclose{}
\end{align*}

\(P\) and \(Q\) are
the regularized lower incomplete gamma function
and the regularized upper incomplete gamma function
respectively.\footnote{See \citeproc{ref-technology_us_nist_2010}{Technology (U.S.), \emph{NIST Handbook of Mathematical Functions Hardback and CD-ROM}}, eq.~8.2.4.} \(n_\text{split}\) is the total number of null-splits,
\(n_\text{b}\) is the total number of \(b\)-bins.

It is trivial to prove these formulae
when we realize \(\max\) is a special case of ordered statistics where the general formula is known.
Proof is left as an exercise to the readers.
\end{Theorem}

With these formulae,
we do not need to estimate the PTE via MC simulations.
Hence, it is much faster to evaluate
and can act as an independent cross-check
by comparing both approaches.
We expect them to be similar,
but note that the latter approach
makes a qualitative assumption.
Hence, the PTEs from MC simulations
are the ones used in the unblinding criteria.

Note that all special functions are available
in Scipy\footnote{\href{https://web.archive.org/web/20230104064725/https://docs.scipy.org/doc/scipy/tutorial/general.html}{``SciPy is a collection of mathematical algorithms and convenience functions built on the NumPy extension of Python.''}},
and the integral can be evaluated numerically.
A Python implementation of these functions is available
\href{https://github.com/ickc/TAIL/blob/3b4e49d6d5a4/tail/analysis/stats.py}{as \texttt{p\_\allowbreak{}1}, \ldots, \texttt{p\_\allowbreak{}5} in the \texttt{stats} module in TAIL}.

\needspace{11\baselineskip}

\begin{longtable}[]{@{}llllll@{}}
\caption{PTE for the null-statistics defined in definition~\ref{def-null-statistics} estimated using MC simulations.}\label{tbl:p-values}\tabularnewline
\toprule\noalign{}
power-spectra & \(G_{Y_1}\) & \(G_{Y_2}\) & \(G_{Y_3}\) & \(G_{Y_4}\) & \(G_{Y_5}\) \\
\midrule\noalign{}
\endfirsthead
\toprule\noalign{}
power-spectra & \(G_{Y_1}\) & \(G_{Y_2}\) & \(G_{Y_3}\) & \(G_{Y_4}\) & \(G_{Y_5}\) \\
\midrule\noalign{}
\endhead
\bottomrule\noalign{}
\endlastfoot
BB & 82.8 \% & 40.2 \% & 52.1 \% & 46.9 \% & 14.5 \% \\
EB & 79.7 \% & 31.2 \% & 35.9 \% & 55.9 \% & 7.6 \% \\
EE & 36.5 \% & 97.7 \% & 90.4 \% & 99.4 \% & 83.4 \% \\
TB & 8.6 \% & 59.6 \% & 98.8 \% & 96.1 \% & 89.1 \% \\
TE & 17.0 \% & 84.2 \% & 59.2 \% & 86.1 \% & 85.9 \% \\
TT & 17.2 \% & 1.6 \% & 9.0 \% & 3.5 \% & 6.6 \% \\
\end{longtable}

\needspace{11\baselineskip}

\begin{longtable}[]{@{}llllll@{}}
\caption{PTE for the null-statistics defined in definition~\ref{def-null-statistics} estimated using theorem~\ref{thm-null-statistics-iid} assuming \(\chi(X, n, b) \xrightarrow[i.i.d.]{d} N(0, 1)\).}\label{tbl:p-values-iid}\tabularnewline
\toprule\noalign{}
power-spectra & \(G_{Y_1}\) & \(G_{Y_2}\) & \(G_{Y_3}\) & \(G_{Y_4}\) & \(G_{Y_5}\) \\
\midrule\noalign{}
\endfirsthead
\toprule\noalign{}
power-spectra & \(G_{Y_1}\) & \(G_{Y_2}\) & \(G_{Y_3}\) & \(G_{Y_4}\) & \(G_{Y_5}\) \\
\midrule\noalign{}
\endhead
\bottomrule\noalign{}
\endlastfoot
BB & 79.5 \% & 43.9 \% & 53.1 \% & 40.4 \% & 13.1 \% \\
EB & 76.2 \% & 32.8 \% & 36.7 \% & 53.1 \% & 5.8 \% \\
EE & 26.9 \% & 97.4 \% & 90.2 \% & 99.0 \% & 85.2 \% \\
TB & 4.8 \% & 61.6 \% & 98.2 \% & 96.3 \% & 90.6 \% \\
TE & 11.2 \% & 86.1 \% & 58.8 \% & 83.8 \% & 85.6 \% \\
TT & 2.3 \% & 0.0 \% & 0.0 \% & 0.1 \% & 1.2 \% \\
\end{longtable}

As an example,
here we show the comparison between the two estimates
in \cref{tbl:p-values} and \cref{tbl:p-values-iid}
using null-test v1.0.0.

The general agreement between the two is a cross-check
that they are behaving as we expected.
The disagreements from the null-statistics generated using the \(TT\) power-spectrum
is evidence that the assumption is not valid there.
\Cref{fig:bin-correlation} further confirms the i.i.d.\ assumption is invalid.
As explained in \cref{sec:null-framework-equations},
\cref{tbl:p-values} is used in the unblinding criteria.

\subsection{Kolmogorov--Smirnov test}\label{sec:ks-test}

From the real map, we obtained \(\chi^\text{real}(X, n, b)\). From n simulated maps, we obtained \(\chi^{\text{sim}, i}(X, n, b)\). A Kolmogorov--Smirnov test between \(\chi^\text{real}(X, n, b)\) and \(\chi^{\text{sim}, i}(X, n, b)\) is performed.
This is often to ensure the simulation is statistically
similar to the real observation
such that the passing of the null-test would not be
too optimistic\footnote{In other words, the PTE is a \(U[0, 1]\) random variable.}.

\subsection{Correlations between null-splits}\label{sec:null-correlation}

\begin{figure}
\centering
\includesvg[width=1\linewidth,height=\textheight,keepaspectratio]{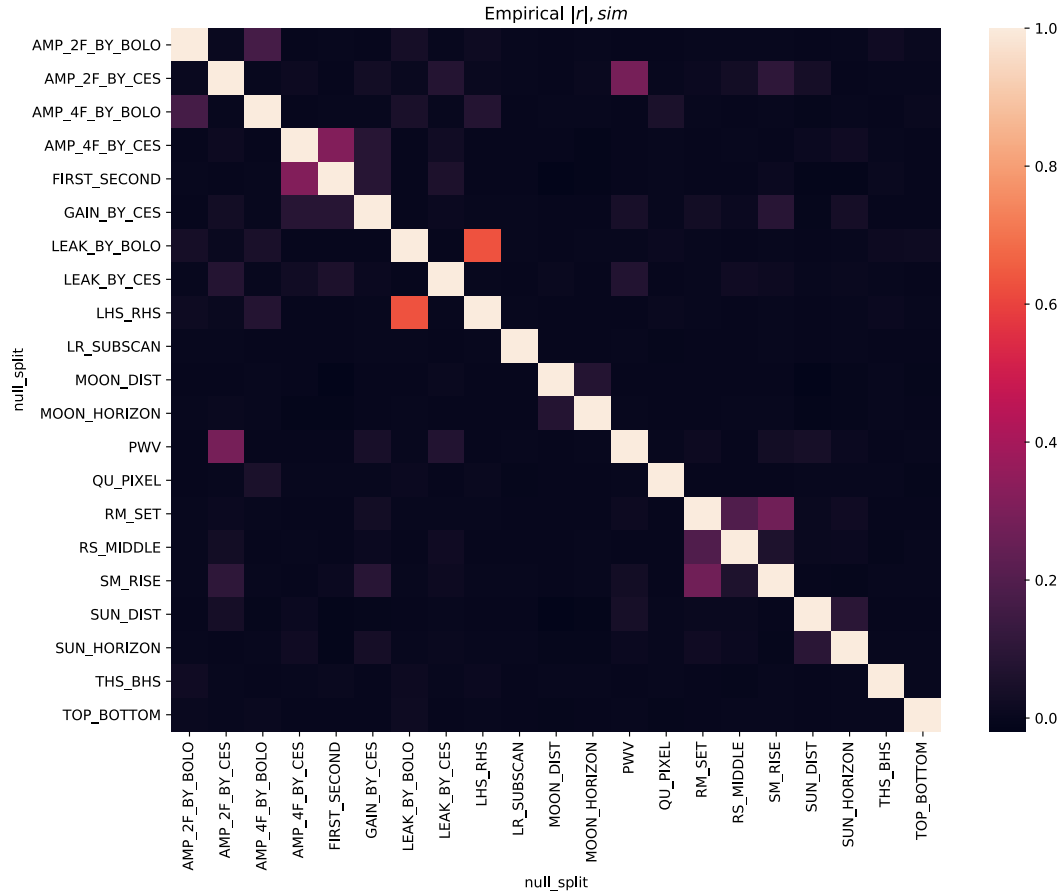}
\caption{Empirical correlation between null-splits estimated using v1.0.0 simulation}\label{fig:null-correlation}
\end{figure}

\begin{figure}
\centering
\includesvg[width=1\linewidth,height=\textheight,keepaspectratio]{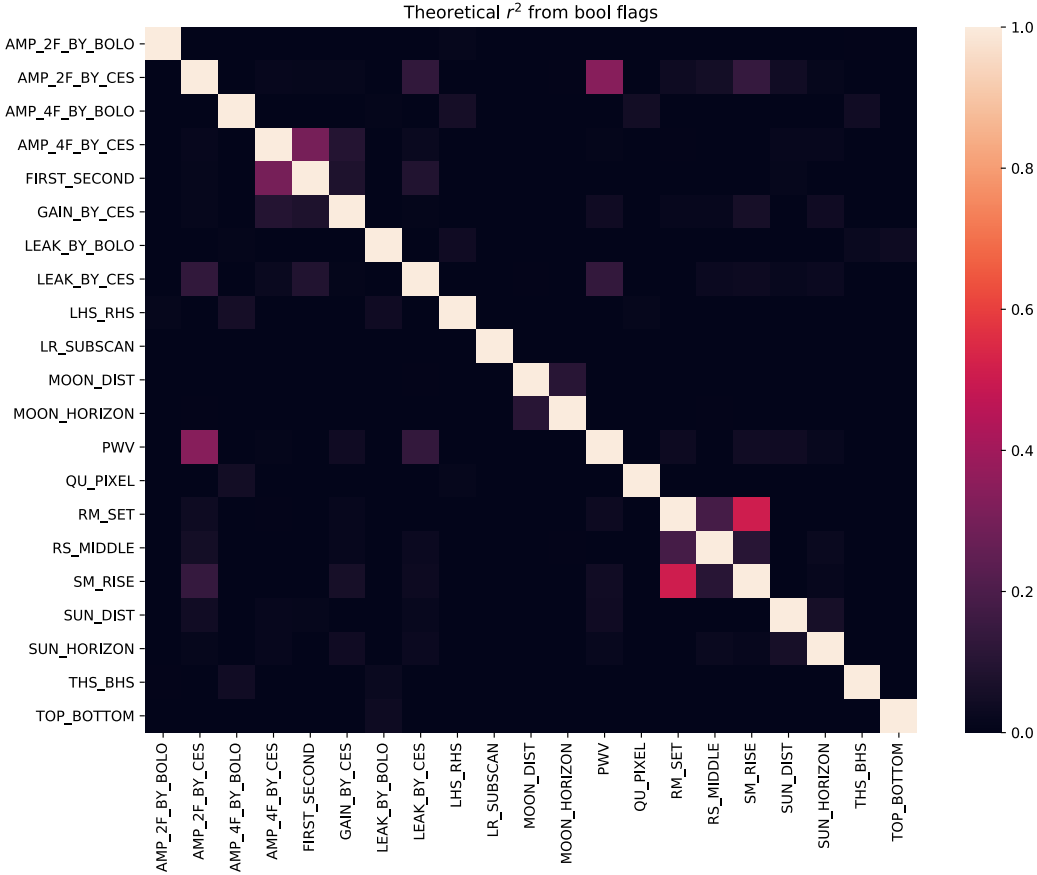}
\caption{Predicted correlation between null-splits estimated using v1.0.0 simulation}\label{fig:null-correlation-bool}
\end{figure}

\begin{figure}
\centering
\includesvg[width=1\linewidth,height=\textheight,keepaspectratio]{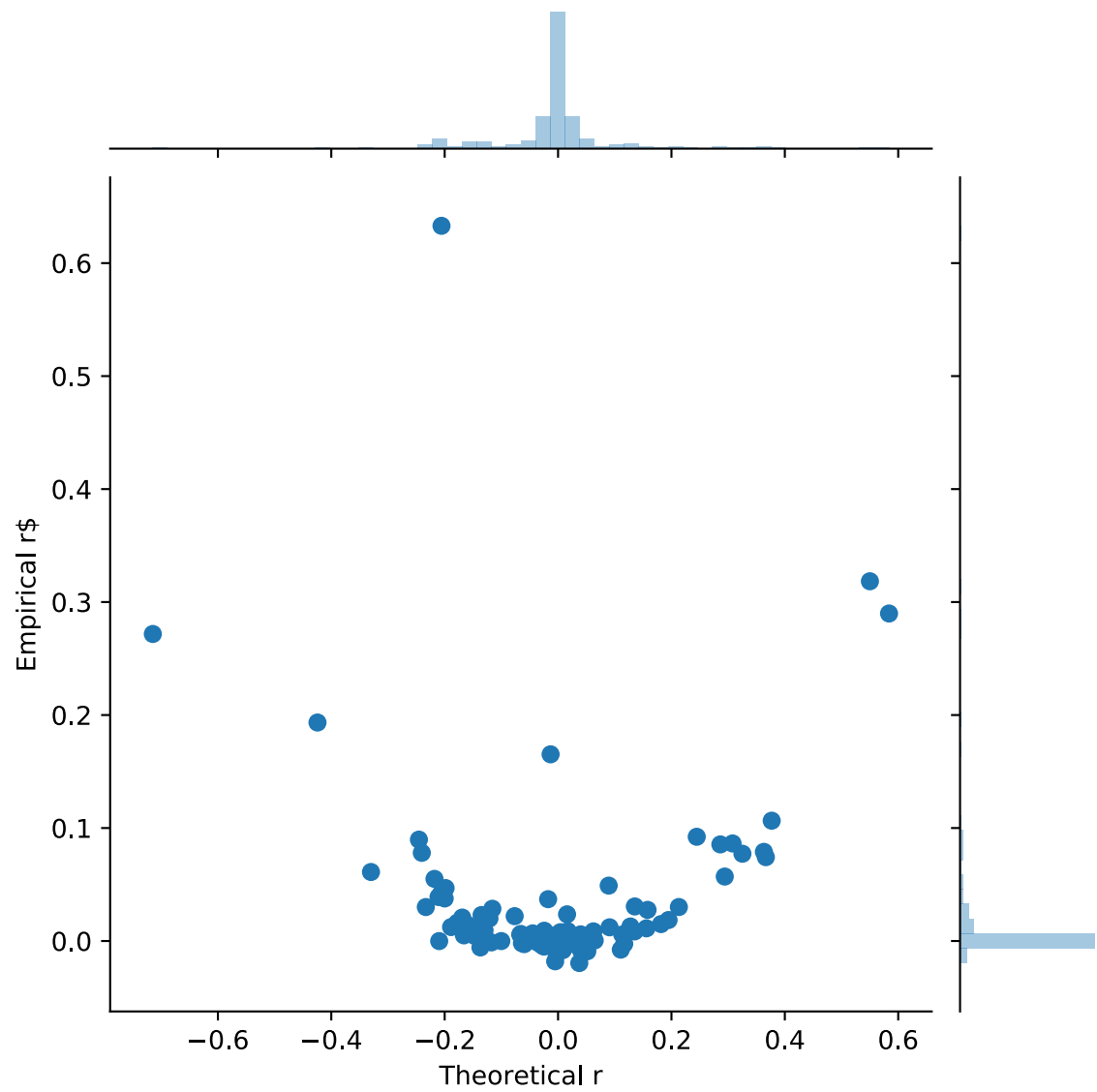}
\caption{Comparison between empirical correlation and predicted correlation}\label{fig:null-correlation-bool-vs-empirical}
\end{figure}

\Cref{fig:null-correlation} tells us the empirical correlations
between null-splits using their corresponding null-spectra.
\Cref{fig:null-correlation-bool} shows the square of the correlation between
the boolean flags of the null-split definition\footnote{Consider
  the boolean flags as a time-series of boolean values
  where each indicates if a sub-scan is chosen to be in that null-split or not.}.
Under the ideal condition that the TOD has identical noise properties,
we can predict that \(r_\text{empirical} \approx r_\text{boolean}^2\).
\Cref{fig:null-correlation-bool-vs-empirical} shows
that this is also empirically true.

As you can see,
since we are splitting the same dataset differently,
they are correlated.

\begin{figure}
\centering
\includesvg[width=1\linewidth,height=\textheight,keepaspectratio]{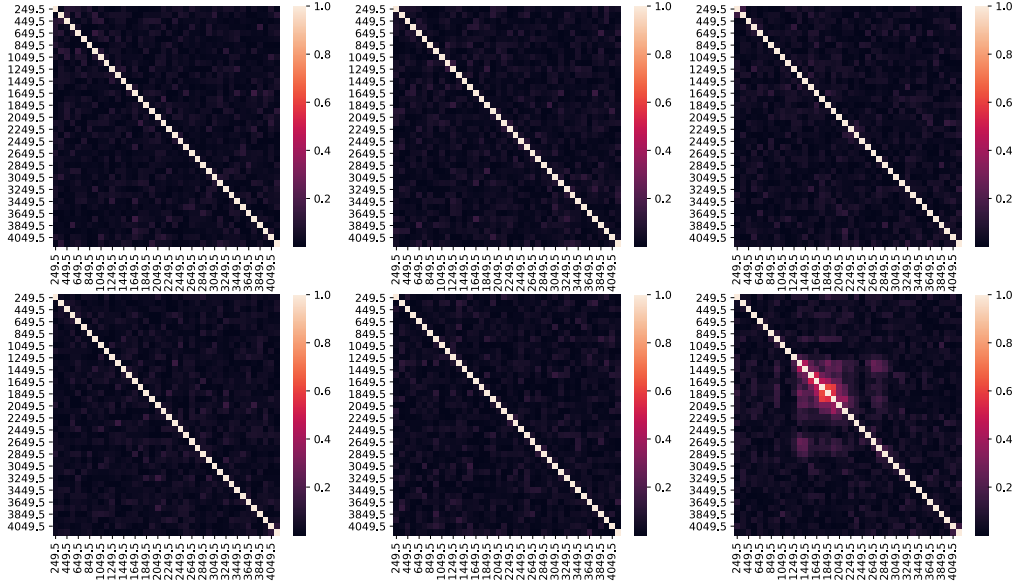}
\caption{Empirical correlation between \(b\)-bins estimated using v1.0.0 simulation}\label{fig:bin-correlation}
\end{figure}

Similarly, from \cref{fig:bin-correlation} we see that the bin-bin correlations are negligible in all but the \(TT\) power-spectrum. Note that the HWP \(2f\) frequency corresponds to \(\ell = 1800\).

And if 2 null-splits are highly correlated,
they are somewhat redundant with each other.
Since we are performing a kind of sum over them in null-statistics,
a natural question to ask is, given 2 of them have non-negligible correlations,
should we choose only one of them?
If we only choose one in the passing criteria,
how can we be sure that is the correct choice,
that is, the other one will equally pass the test?

\subsection{Null-test selection and partial correlation}\label{null-test-selection-and-partial-correlation}

First, independence implies no correlations,
but not vice versa.

As defined in definition~\ref{def-null-statistics},
our null-statistics are complicated functions of the
null random variable \(χ\).
It is a misconception among some physicists
that somehow the Pearson correlation coefficient is more physical,
and if we find it to be small between 2 null-splits,
their resultant \(5\) null-statistics will also be fairly independent.

Moreover, even if we find 2 null-splits to be highly
correlated, are they still that much correlated
after the correlation with other null-splits is taken into account?

A statistical way to quantify this is the
partial correlation matrix, which can be written as

\[ρ_{χ_i, χ_j | χ_{k \neq i, j}, \ldots},\]

which is defined as the correlations
between the residuals \(r_{χ_i | χ_{k \neq i, j}, \ldots}\)
and \(r_{χ_j | χ_{k \neq i, j}, \ldots}\).

\begin{figure}
\centering
\includesvg[width=1\linewidth,height=\textheight,keepaspectratio]{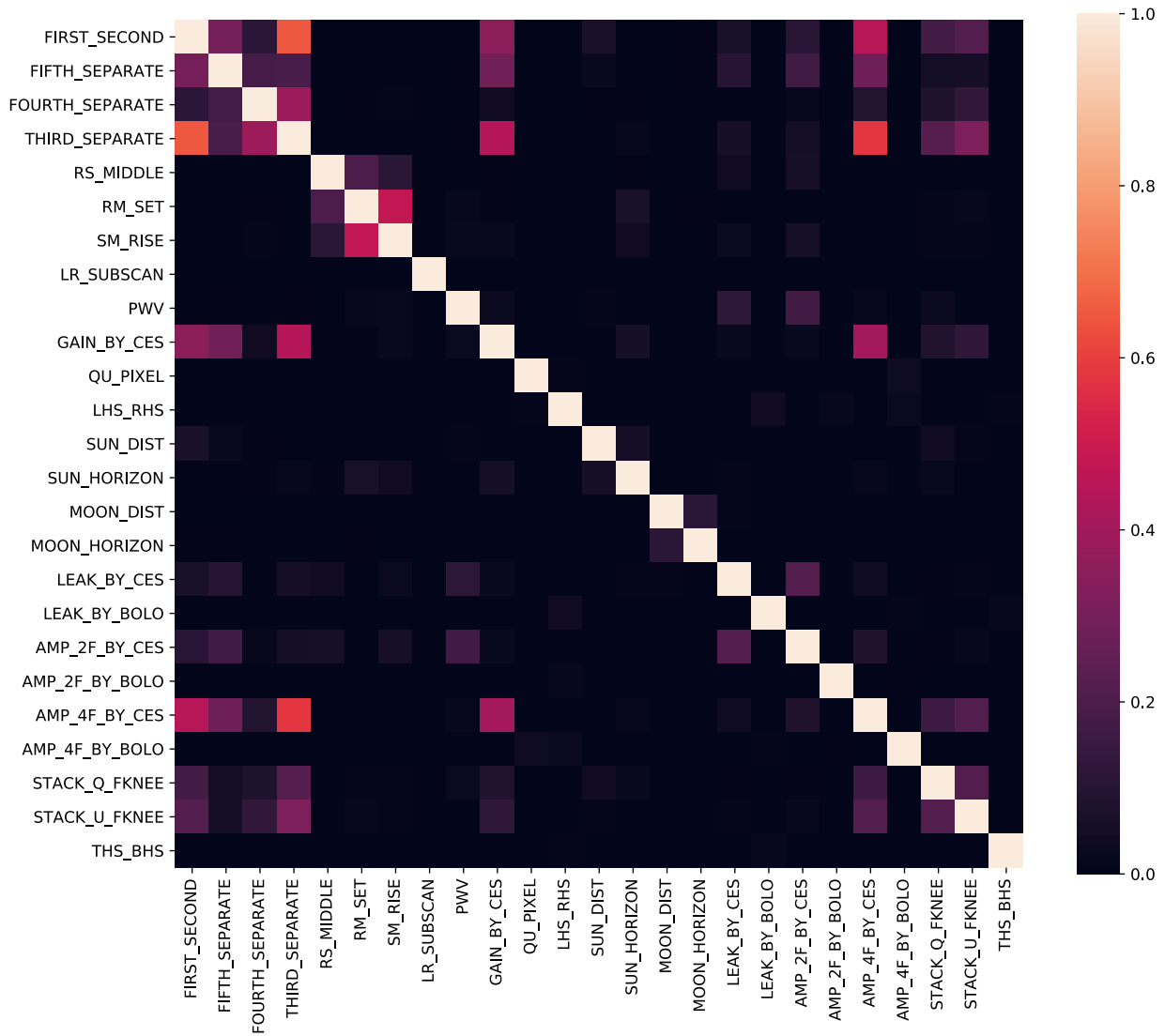}
\caption{Square of correlation between null-splits flags estimated using v0.2.0 simulation}\label{fig:null_corr_pearson}
\end{figure}

\begin{figure}
\centering
\includesvg[width=1\linewidth,height=\textheight,keepaspectratio]{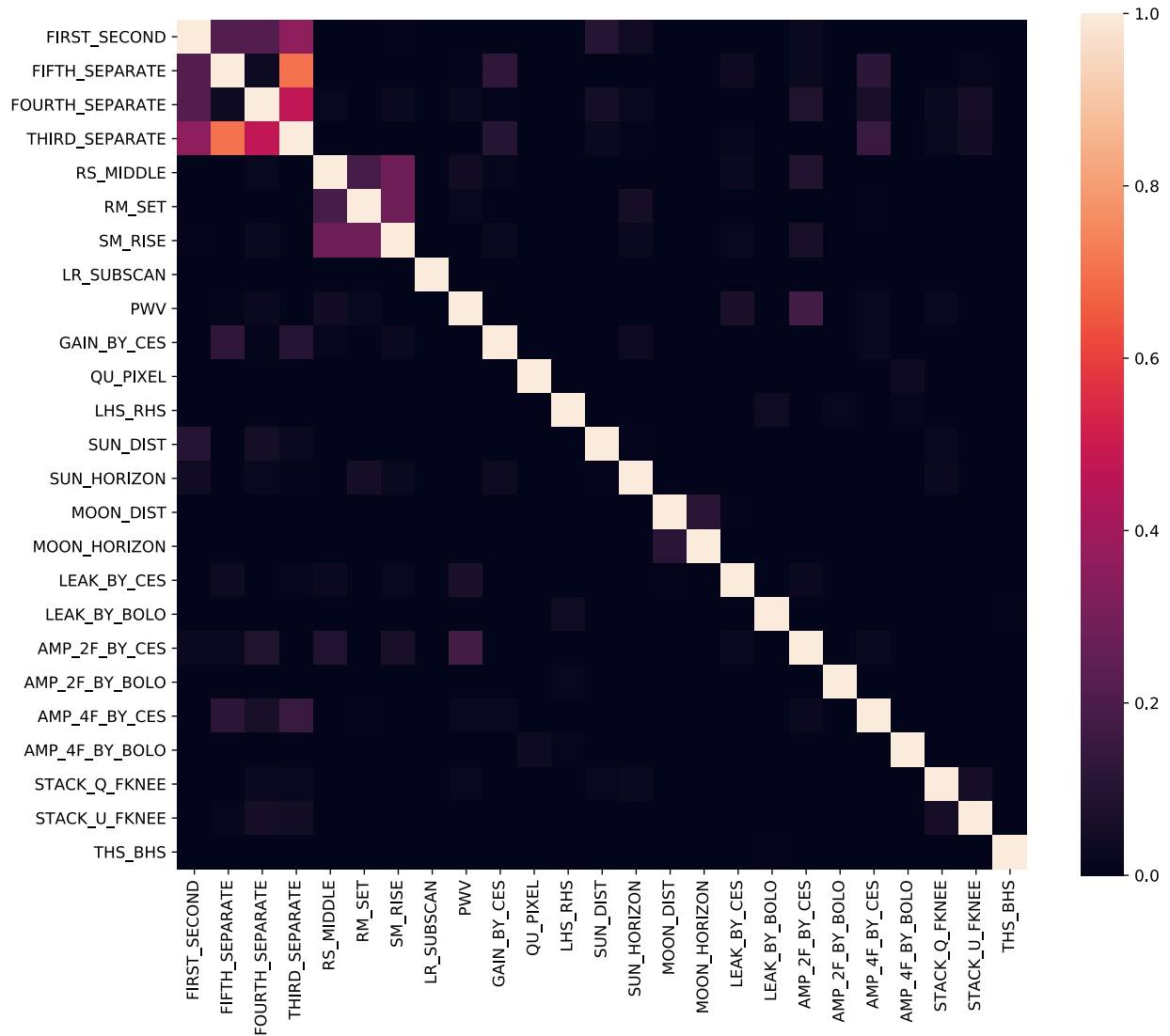}
\caption{Square of partial correlation between null-splits flags estimated using v0.2.0 simulation. Pseudo-inverse is used for this estimation because of numerical stability.}\label{fig:null_corr_pearson_partial}
\end{figure}

\cref{fig:null_corr_pearson}; \cref{fig:null_corr_pearson_partial}
show the difference between the correlation matrices.
The takeaway message here is that
2 null-splits can be apparently highly correlated,
but much of the correlations can be explained by the rest
of the null-splits. Hence, if one of them is excluded
in the final null-test suite, it actually helps less than expected in removing
dependencies.

In the end, the author suggests not to exclude
null-splits in passing criteria based on some apparent correlations
between them. Often, a null-split is defined by
some physical observables with clearly defined meaning
that should increase our confidence in the lack of
systematics if the null-test suite passes.
However, some can be excluded if there are strong reasons
behind them.
Here for example, we see that in earlier null-test v0.2.0 shown in \cref{fig:null_corr_pearson}, there were more null-splits
including \texttt{THIRD\_\allowbreak{}SEPARATE}, \texttt{FOURTH\_\allowbreak{}SEPARATE}, \texttt{FIFTH\_\allowbreak{}SEPARATE},
which are defined as the 3rd season of data vs.~the rest, and so on.
Since \texttt{FIRST\_\allowbreak{}SECOND} splits the data in the two halves
chronologically, these 4 null-splits are quite redundant with
each other. Since splitting the data in a half in data volume
maximizes the sensitivity of the null-spectra, only \texttt{FIRST\_\allowbreak{}SECOND}
is kept in the final analysis.

\subsection{Combined statistics---Family-wise Error Rate (FWER)}\label{combined-statisticsfamily-wise-error-rate-fwer}

Essentially,
we do not want to have false alarms of non-existent systematics.
It means that in terms of null-hypothesis testing,
we want to control the type I errors.
And since we have more than one such statistics,
we want to control the Family-Wise Error Rate (FWER).

There are many ways
to control the FWER to be of \(\alpha\)-level,
where we chose this to be \(\SI{5}{\percent}\) as our passing criteria.

Given 15 numbers, one way of controlling the Family-wise error rate (FWER) is the most basic Bonferroni correction---assign \(\frac{\alpha}{15}\) to each test. If none is below that level, it passes.
But this
can let us pass the null-test too easily.
Statistically, this is called less \emph{powerful}.

\subsubsection{\texorpdfstring{Combined statistics used in previous POLARBEAR \(B\)-mode publications}{Combined statistics used in previous POLARBEAR B-mode publications}}\label{combined-statistics-used-in-previous-polarbear-b-mode-publications}

While the previous POLARBEAR papers have not written down
how the statistics are combined,
here is its definition.

\begin{Definition}
Combined statistics used in previous POLARBEAR \(B\)-mode publications.

\[\mathcal{G}(s) \equiv \min_i (G_{Y_i}(Y_i(s))), s \in \Omega\]

\[\mathcal{F}(s) \equiv P(\mathcal{G} \leq \mathcal{G}(s))\]

\[\mathcal{F} \rightarrow U(0, 1)\]

\(\mathcal{F}\) is the combined statistics which is proved
to be distributed as \(U(0, 1)\),
consistent with what one would expect from an FWER.

Informally, it is defined as \(\mathcal{G}\)
which is the minimum of all \(p\)-values
and ask how often this minimum (as a random variable) is smaller than
the observed minimum.

Note that these functions are constructed over the sample space directly to avoid confusion. They take 1 single realization (the real observation) and map it to a number (indirectly through \(\chi, Y_i, G_i\).)
\end{Definition}

Computationally,
the distribution of \(\mathcal{G}\) is estimated empirically through Monte Carlo simulation and \(\mathcal{G}(s)\) from measured data.
Note that to obtain \(G\) empirically,
one would need to compare a value to other values obtained
using MC simulations---this value can be
the real data when we are computing the statistics,
or from an MC simulation when we are constructing the sample distribution
of \(G\).

Hence, in order to construct the empirical distribution of \(\mathcal{G}\),
the \(n_\text{sim} = 512\) simulations are needed to be used recursively,
say by comparing values from a simulation to the rest of \(511\) simulations, etc.

Note that then the empirical distribution of \(G\) is discretized,
from \(G = 0\) when the MC chosen is the maximum among all MC,
to \(G = \frac{511}{512}\) when the MC chosen is the minimum.
That is, it takes \(\frac{0, 1, \dots, 511}{512}\), exactly 512 different discrete values.

Then when we construct \(\mathcal{G}\),
which is the minimum among \(n_\text{family} = 15\) number of \(G\)-statistics,
can only be at most \(n_\text{sim} = 512\) different values.
And in practice the number of different values would be much less
by the mere fact that we are performing a minimum statistics here\footnote{Minimum statistics
  is a special case of ordered statistics,
  where the CDF can be computed analytically given the original CDF
  if they are independent.
  In this case the original CDF is from a uniform discrete distribution of \(\frac{0, 1, \dots, 511}{512}\). It can be shown that it is skewed towards the left
  and more skewed as \(n_\text{family}\) increases.}.

This is the subtle mathematical reason why this is inferior,
as we are losing resolution (in the sense of having fewer different discrete values)
when comparing the statistics obtained from real data \(\mathcal{G}(s)\)
to this empirical distribution.
In other words,
a small change in \(\mathcal{G}(s)\) can result in a large
change in \(\mathcal{F}(s)\) depending on the multiplicity in the empirical distribution.
This makes the estimated PTE unreliable,
or alternatively requires many more simulations
to counter this effect.

\subsubsection{Combined statistics using Holm--Bonferroni method}\label{sec:holm-bonferroni-method}

A better method is Holm--Bonferroni method. Mathematically, it is uniformly more powerful than
the Bonferroni correction\footnote{Loosely speaking, it means it does not pass as easily.},
and does not require any assumption on the statistical independence
between the statistics that is often imposed by other more powerful methods.

Note that this already solves our problem about the correlations between null-splits in \cref{sec:null-correlation}.
In short, we do not have to preselect null-splits based on arbitrary criteria;
we will choose all of them and let the mathematics take care of the rest.

Compared to the Bonferroni method,
instead of uniformly assigning \(\frac{\alpha}{15}\) to each test, it assigns \(\frac{\alpha}{15}\) to the smallest p-value, \(\frac{2 \alpha}{15}\) to the 2nd smallest, \(\frac{k \alpha}{15}\) to the k-th smallest, \(\alpha\) to the last (largest) p-value.

It can be proven, without any assumption on the correlation between the tests, that Holm--Bonferroni method guarantees \(\text{FWER} \leq \alpha\).

Also, a critical \(\alpha\) value, \(\alpha_c\), can be defined such that if the chosen \(\alpha \leq \alpha_c\), then it passes. This single number \(\alpha_c\) will be our combined statistics.

Moreover, Holm-Bonferroni method not only allows us to distill
many statistics into one \(\alpha_c\),
this procedure also tells us, given a critical \(\alpha\) value,
which of the \(n_\text{family}\) statistics are failing.

For example, with the \(Y_3\) statistics in definition~\ref{def-null-statistics},
before we take the final maximum,
it has the form \(\sum_b \chi^2(X, n, b)\)
depending on the null-split only.
By applying the Holm--Bonferroni method on
this \(n_n = 21\) number of statistics,
it can tell you exactly which null-splits are failing.

For example, since the beginning in null-test v0.1.0,
these null-splits are failing by inspection,

\begin{itemize}
\tightlist
\item
  \texttt{QU\_\allowbreak{}PIXEL}
\item
  \texttt{LEAK\_\allowbreak{}BY\_\allowbreak{}BOLO}
\item
  \texttt{AMP\_\allowbreak{}2F\_\allowbreak{}BY\_\allowbreak{}BOLO}
\item
  \texttt{AMP\_\allowbreak{}4F\_\allowbreak{}BY\_\allowbreak{}BOLO}
\end{itemize}

Running my statistics pipeline
with the Holm--Bonferroni method,
these are exactly our ``discoveries''.

This eliminates an error-prone procedure of manually looking into each null-spectra,
as sometimes we may be seeing things that statistically are just a fluke,
and it happens from time to time.

Finally, it should be mentioned that
the culture in POLARBEAR prefers simple statistics
that are simple to understand.
In response to that,
there is nothing simple in the previous section.
In fact it is very complicated and abstract to define that quantity
mathematically.
Computationally, we also showed with non-simple explanations
why subtly it is not well behaved.
Instead, Holm--Bonferroni method is easy to describe as above,
and only takes a few lines to prove that it controls the FWER\@.
It is first constructed in 1979,\footnote{\citeproc{ref-holm_simple_1979}{Holm, {``A Simple Sequentially Rejective Multiple Test Procedure''}}.} which is quite old.
Put it in other words,
we are not doing cutting edge research in statistics if we adopt
the Holm--Bonferroni method here,
we are simply catching up in putting a well-known good tool to use.
Cosmology is a science based on statistics\footnote{As we argued elsewhere
  we only have one observable universe
  that we cannot experiment on
  but to observe statistical properties from it.},
we should not be afraid to adopt better tools
offered by the vast field of statistics.

\section{Likelihood estimation}\label{sec:likelihood-estimation}

\subsection{Simultaneous calibration and fitting of the lensing amplitude}\label{simultaneous-calibration-and-fitting-of-the-lensing-amplitude}

Note that Einstein summation notation is implied in this section unless otherwise stated.

We can represent a rotated, scaled version of the true power-spectra by

\begin{align*}
C_{b, k l}^\text{scaled}    &= a_k a_l  C_{b, k l}^\text{CMB}, \text{no summation}  \\
C_{b, i j}^\text{observed}  &= R_i^k R_j^l C_{b, k l}^\text{scaled} \\
    &= \sum_{kl} R_i^k R_j^l a_k a_l  C_{b, k l}^\text{primordial}  \\
    &= T_{ij}^{kl} C_{b, k l}^\text{primordial} \\
T_{ij}^{kl} &\equiv R_i^k R_j^l a_k a_l, \text{no summation}    \\
\end{align*}

Here, \(i, j, k, l = T, E, B\), and

\[R_i^k (\theta) = \begin{pmatrix}
1   &0  &0  \\
0   &\cos 2 \theta  &\sin 2 \theta  \\
0   &-\sin 2 \theta &\cos 2 \theta  \\
\end{pmatrix}.\]

After expansion this agrees with Keating, Shimon, and Yadav,\footnote{\citeproc{ref-keating_self-calibration_2013}{{``SELF-CALIBRATION OF COSMIC MICROWAVE BACKGROUND POLARIZATION EXPERIMENTS''}}.} eq.~5 if we choose \(a_i = 1\). Written in this form it is trivial to perform the inverse transformation to transform the measured data instead.

Also, note that \(a_T = g\) is the absolute gain calibration, \(a_E = g \eta\) where \(\eta\) is the absolute polarization efficiency calibration, \(a_B = a_E\).

We are going to apply \(T^{-1}\) to the measured spectra below. That is,

\[\hat{C}_{b, ij}(a_T, a_E, \theta) \equiv (T^{-1})_{ij}^{kl} C_{b, kl}^\text{observed}\]

In the end, the \(C_{b, BB}\) from the theory will have an additional \(A_\text{lens}\) term to fit for lensing amplitude. That is,

\newcommand{\cth}[1]{C_{b, #1}^\text{th}}

\[C_{b, ij} = \begin{pmatrix}
C_{b, TT}^\text{th}    &C_{b, TE}^\text{th}   &0  \\
C_{b, TE}^\text{th}    &C_{b, EE}^\text{th}   &0  \\
0   &0  &A_\text{lens}C_{b, BB}^\text{th}  \\
\end{pmatrix}\]

\subsection{Likelihood procedure}\label{sec:likelihood-procedure}

\begin{enumerate}
\def\labelenumi{\arabic{enumi}.}
\item
  The effective d.o.f. \(ν\) used below is calculated according to \cref{sec:effDOF}.
\item
  Transform the measured (signal) spectra according to previous section
  to obtain \(\hat{C}_{b, ij}(a_T, a_E, \theta)\).
\item
  Noise spectra scaled with \(a_i\) as we are using sign-flip noise which is from the measured data.
\item
  \(BB\) leakage template is constructed using the \(TE\) noiseless simulation, that is, Pixel Window Function (PWF) should not be applied here and hence it has a slightly different Band-Pass Window Function (BPWF).
\end{enumerate}

\newcommand{\cL}{\mathcal{L}}

\newcommand{\bC}[1]{\mathbf{C}_{#1}}
\newcommand{\hbC}[1]{\mathbf{\hat{C}}_{#1}}
\newcommand{\hC}[1]{\hat{C}_{#1}}

\newcommand{\hCoverC}[1]{\hbC{b} \bC{b}^{-1}}

Following Hamimeche and Lewis,\footnote{\citeproc{ref-hamimeche_likelihood_2008}{{``Likelihood Analysis of CMB Temperature and Polarization Power Spectra''}}.} we define the likelihood \(\mathcal{L}\) as

\[-2 \ln \mathcal{L}( \mathbf{C}_{b} | \mathbf{\hat{C}}_{b}) =
\sum_b \nu_b \{
    \tr[\mathbf{\hat{C}}_{b} \mathbf{C}_{b}^{-1}] -
    \ln| \mathbf{\hat{C}}_{b} \mathbf{C}_{b}^{-1} | -
    n
\}
,\quad
n = 3\]

\newcommand{\matrixToComponent}[1]{\mathbf{#1}_b = #1_b^{XY, \text{ total}}}

\[\mathbf{C}_b = C_b^{XY, \text{ total}},\quad
\mathbf{\hat{C}}_b = \hat{C}_b^{XY, \text{ total}},\quad
X, Y = T, E, B\]

\newcommand{\BBonly}[1]{\hat{C}_b^{BB, \text{#1}} \delta^{XB} \delta^{YB}}

\newcommand{\componentForm}[2]{
    {#1}^{XY} =
        C_b^{XY, \text{ total, #2}} +
        \hat{N}_b^{XX} \delta^{XY}
}

\[
    {\hat{C}_{b}}^{XY} =
        C_b^{XY, \text{ total, measured}} +
        \hat{N}_b^{XX} \delta^{XY}
\]
\[\begin{split}
    {C_b}^{XY} =
        C_b^{XY, \text{ total, theory}} +
        \hat{N}_b^{XX} \delta^{XY}
 +
\hat{C}_b^{BB, \text{leakage-template}} \delta^{XB} \delta^{YB} + \\
\hat{C}_b^{BB, \text{foreground-reference}} \delta^{XB} \delta^{YB}\end{split}\]

\subsection{Notes on Absolute Calibrations}\label{notes-on-absolute-calibrations}

The procedures outlined above will be used to fit the final \(A_\text{lens}\) once. That is, the absolute angle and gain will not be used to rotate and scale the maps. The resulting science product, the \(BB\) spectrum, will be calculated using the MLE \(a_i, \theta\) on \((T^{-1})_{ij}^{kl}(a_i, \theta) C_{b,kl}^\text{measured}\).

This is a full-Bayesian approach to perform simultaneous self-calibration
and parameter estimation.

In previous POLARBEAR \(B\)-mode analyses,
the maps are rotated and scaled directly.
This means that gains and angle have to be predetermined, say by the MLE\@.
As this does not take the error of the gains and angle into account,
it requires manual error-propagation of them into the spectra.
Furthermore, the ratio between the \(EE\) subtracted in measured \(BB\), mainly by the rotation mechanism, is fixed. By having a full likelihood allowing all these to vary (gain, angle, \(A_\text{lens}\)), the likelihood of the \(A_\text{lens}\) is also sensitive to this unknown amount of \(EE\) rotated in the \(BB\) spectrum. This is better than just scaling the error-bar of the spectra, because \(EE\) is not simply a multiple of \(BB\).

This procedure can cause a misunderstanding that this likelihood is mixing the calibration factors into the cosmological factor \(A_\text{lens}\). It is not mixing them, because

\[\hat{C}_{b, ij}(a_T, a_E, \theta) \equiv (T^{-1})_{ij}^{kl} C_{b, kl}^\text{observed}\]

\[C_{b, ij} = \begin{pmatrix}
C_{b, TT}^\text{th}    &C_{b, TE}^\text{th}   &0  \\
C_{b, TE}^\text{th}    &C_{b, EE}^\text{th}   &0  \\
0   &0  &A_\text{lens}C_{b, BB}^\text{th}  \\
\end{pmatrix}\]

and

\(\ln \mathcal{L}( \mathbf{C}_{b} | \mathbf{\hat{C}}_{b})\) depends on \(\mathbf{\hat{C}}_{b} \mathbf{C}_{b}^{-1}\) only. Since the \(C_{b, ij}\) is block-diagonal, the inverse is easy to calculate in the \(BB\) corner. Expanding that term alone, one has

\[\mathbf{\hat{C}}_{b} \mathbf{C}_{b}^{-1} \sim \frac{(\dots)_{EE} C_{EE}^\text{measured} + (\dots)_{BB} C_{BB}^\text{measured}}{A_\text{lens} C_{BB}^\text{th}} + (\dots)_\text{others}\]

So it only couples the theory \(BB\) into the measured \(BB\) and \(EE\), through the gains and angle in \((\dots)_{BB \text{ or } EE}\), which is unavoidable regardless of approach. If we choose predetermined gains and angle, the \((\dots)_{BB \text{ or } EE}\) term is fixed. If we do not fix it, then it is equivalent to having \((\dots)_{BB \text{ or } EE}\) set by a prior distribution, where this prior distribution inevitably is still introduced by the \((\dots)_\text{others}\) terms.

So rather than mixing the calibration factors into the cosmological factor, this procedure automatically propagates the error-bar in angle and gains into the final estimation of \(A_\text{lens}\), and does this better than simply rescaling the error-bars (since \(EE\) is not proportional to \(BB\).)

The only caveat is about the pure B-mode estimator,
as performing rotation in the map can change this.
But given the angle is \(\sim \SI{0.4}{\degree}\), this should not be a problem.

\section{Analytical error-bar: Effective d.o.f.}\label{sec:effDOF}

From \cref{sec:Knox-formula}, we already see the term \(ν\) as the effective
degree of freedom (d.o.f.) in a \(b\)-bin across a range of \(\ell\).

To understand the Knox formula,
first note that

\[C_{\ell} \equiv \frac{1}{\left( 2 \ell + 1 \right)} \sum_{m \leq \ell} \left\langle \left| a_{\ell m} \right|^{2} \right\rangle ,\]

where we expect \(a_{\ell m}\) are independent Gaussian random variables
that share the same expectation value \(C_\ell\) for all \(m\) for each \(\ell\).
This is why we expect \(C_\ell\) to be a \(χ^2_ν\)
random variable.

Then define the following random variable such that

\begin{align}
\mathcal{C}_{\ell}	&	 \sim C_{\ell} \frac{\chi^{2}_{\nu_{\ell}}}{\nu_{\ell}}	\nonumber	\\
\mathbb{E} \mathopen{} \left [ \mathcal{C}_{\ell} \right ] \mathclose{}	&	 = C_{\ell}	\nonumber	\\
\operatorname{Var} \mathopen{} \left ( C_{\ell} \right ) \mathclose{}	&	 = \frac{2}{\nu} C_{\ell}^{2}	\label{eq:knox_var}	\\
\nu_{\ell}	&	 \equiv \left( 2 \ell + 1 \right) f_{\text{sky}}	\nonumber
\end{align}

Here, for simplicity, \(C\) represents the total power-spectrum
which includes both the signal and noise term.
This proves the Knox formula.

Then the key idea from Hivon et al.\footnote{\citeproc{ref-hivon_master_2002}{{``MASTER of the Cosmic Microwave Background Anisotropy Power Spectrum''}}.} mentioned in \cref{sec:Knox-formula},
is that in the case of cut-sky,
effectively we lose some amount of degree of freedom \(ν\),
scaled accordingly with respect to the fraction \(f_\text{sky}\).

\begin{figure}
\centering
\pandocbounded{\includegraphics[keepaspectratio,alt={The 600-700 bin of band pass window function of B-mode power-spectrum}]{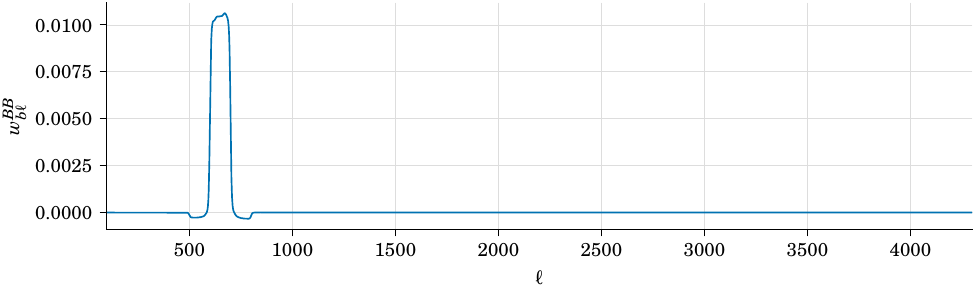}}
\caption{The 600-700 bin of band pass window function of \(B\)-mode power-spectrum}\label{fig:BPWF}
\end{figure}

Finally, we should notice that
the combined effect of the MASTER equation,
the binning and interpolation operators together
has the effect of,\footnote{\citeproc{ref-ade_measurement_2014-1}{The POLARBEAR Collaboration et al., {``A Measurement of the Cosmic Microwave Background \emph{B}-Mode Polarization Power Spectrum at Sub-Degree Scales with POLARBEAR''}}.}

\[\hat{C}_b = \sum_\ell w_{b \ell} C_\ell,\]

where \(w_{b\ell}\) is often called the Band-Pass Window Function (BPWF)
as shown in \cref{fig:BPWF}.

You can see some of the power is negative.
That is, rather than being distributed as a \(\chi^2_\nu\) distribution, it is more like

\begin{align*}
\mathcal{C}_{b , \text{eff.}}	&	 \sim a \left( \chi^{2}_{\nu^{+}_{b}} - \chi^{2}_{\nu^{-}_{b}} \right)	\\
\mathbb{E} \mathopen{} \left [ \mathcal{C}_{b , \text{eff.}} \right ] \mathclose{}	&	 = C_{b} \Rightarrow a \equiv \frac{C_{b}}{\nu^{+}_{b} - \nu^{-}_{b}}	\\
\operatorname{Var} \mathopen{} \left ( \mathcal{C}_{b , \text{eff.}} \right ) \mathclose{}	&	 = a^{2} \left( 2 \nu^{+}_{b} + 2 \nu^{-}_{b} \right) = 2 C_{b}^{2} \frac{\nu^{+}_{b} + \nu^{-}_{b}}{\left( \nu^{+}_{b} - \nu^{-}_{b} \right)^{2}}
\end{align*}

comparing this to \cref{eq:knox_var}, we can define an effective d.o.f.

\[\nu_b^\text{eff} = \frac{(\nu^+ - \nu^-)^2}{\nu^+ + \nu^-}.\]

The new formula improves the agreement between
MC error-bar and analytical error-bar, from \(\sim \SI{10}{\percent}\) to \(\sim \SI{3}{\percent}\) agreement for \(BB, EB, EE, TB, TE\) power-spectra, and from \(\sim \SI{20}{\percent}\) to \(\sim \SI{13}{\percent}\) agreement for \(TT\) power-spectrum\footnote{\(TT\) is not as i.i.d.\ as the others empirically, so the analytical model fails more here. See \cref{sec:null-correlation}.}.

This improvement is crucial,
as general agreement between the analytical error-bar
and the MC error-bar is a good cross-check,
and the likelihood used in \cref{sec:likelihood-procedure}
contains \(ν\) which implicitly uses the analytical error-bar
rather than the MC error-bar.

\subsection{Error propagation \& miscellaneous likelihood parameters}\label{error-propagation-miscellaneous-likelihood-parameters}

It is already mentioned that the likelihood is going to propagate
the uncertainties of calibration parameters.
Here we mention other sources of error.

As mentioned before,
the filter transfer functions are estimated
via 2 sets of 128 MC simulations,
one without the \(B\)-mode signal,
another without the \(E\)-mode signal.
This results in a relative error in \(F_\ell\)
peaked at \(\SI{1}{\percent}\)-level in the low-\(\ell\) end.
The final bin-width used is \(100\), making relative error of \(F_b\) to be at least an order of magnitude smaller.

Empirically, the agreement between the 2 sets of simulations above
and the theory with BPWF is peaked at \(10^{-6}\)-level in the low-\(\ell\) end.

Because of these,
and the fact that the statistical error
from noise is much larger,
errors from filter transfer function
are negligible and will be ignored here.

Error from beam will be fitted similarly as in previous POLARBEAR \(B\)-mode publications,
that is, an \(\ell\) dependent gain. The error-bar will be introduced in the likelihood equation automatically by having the following gain term,

\[a_i = a_{0, i} e^{-\frac{\ell (\ell +1) \sigma^2}{2}},\]

where \(σ\) is a free parameter in the likelihood.

\chapter{Running the Null-test suite}\label{running-the-null-test-suite}

\section{Null-spectra}\label{null-spectra}

\begin{figure}
\centering
\pandocbounded{\includegraphics[keepaspectratio,alt={Null test v0.6.2  null-spectra}]{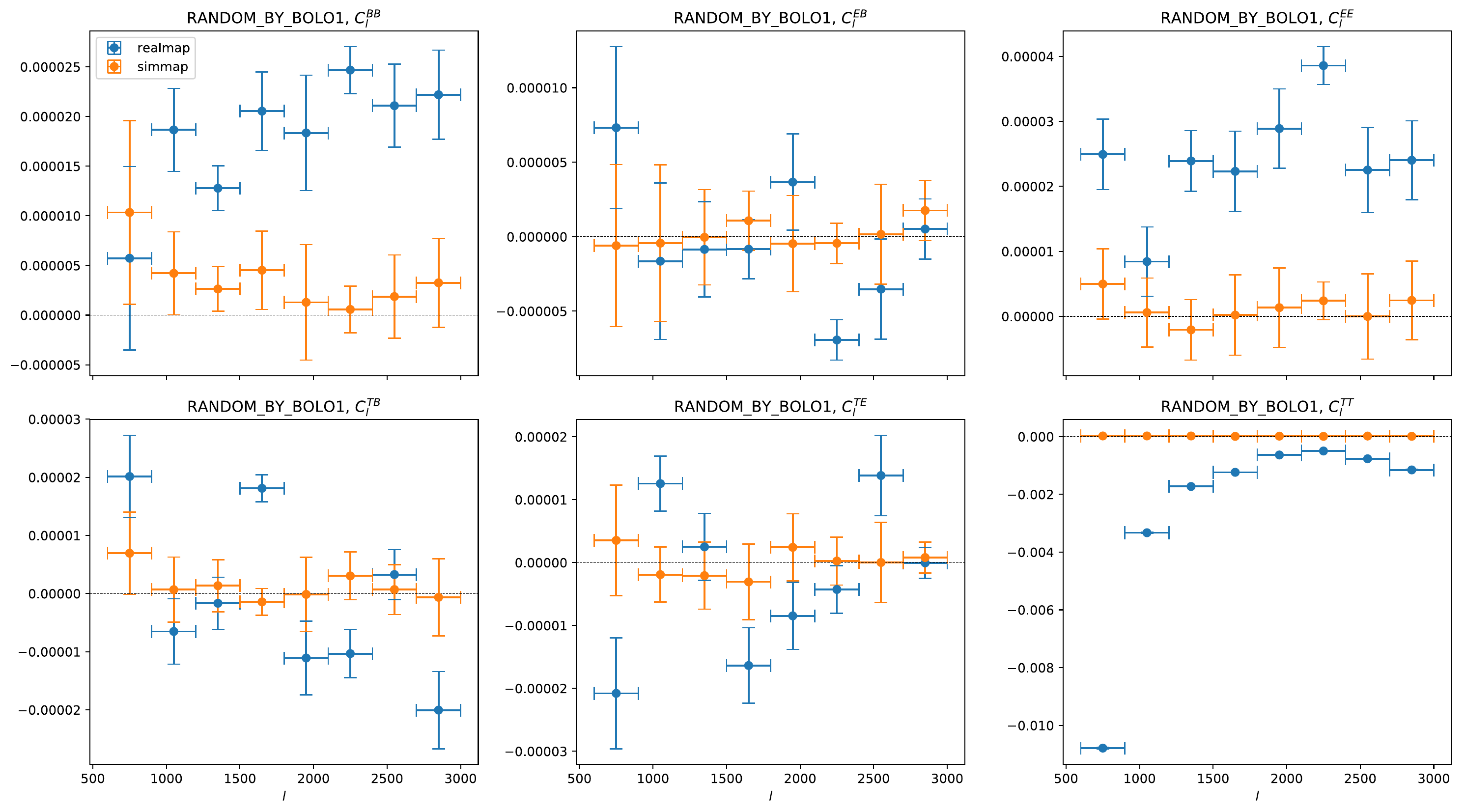}}
\caption{Null test v0.6.2 \texttt{RANDOM\_\allowbreak{}BY\_\allowbreak{}BOLO1} null-spectra}\label{fig:0_6_2-RANDOM_BY_BOLO1}
\end{figure}

Let us take a look at an example null-spectra in \cref{fig:0_6_2-RANDOM_BY_BOLO1}.

\begin{enumerate}
\def\labelenumi{\arabic{enumi}.}
\tightlist
\item
  The error-bars are estimated
  using \(n_\text{sim} = 8\) simulations
  that include fiducial signals
  and an uncorrelated white noise model with magnitude estimated from the TOD.
\item
  The values of the simulations
  are averaged from all simulations
  to show you that this null-estimator is unbiased
  as we expect it to be consistent with \(0\) statistically.
\item
  The values from the real data
  are the same null-estimator constructed using
  the observed TOD\@.
  Hence,
  this is where we might make ``discovery'' of unknown systematic effects.
\end{enumerate}

(Normally, the vertical axis is in \(D_\ell\) scale,
but here it is \(C_\ell B^2\ell\) for a reason that is going to be clear later.)

Just by looking at it,
we can see that something is wrong with the \(BB, EE\) spectra.
In fact,
something is very wrong,
as this null-split,
\texttt{RANDOM\_\allowbreak{}BY\_\allowbreak{}BOLO1},
is randomly splitting the data
into two halves according to which detectors it belongs to.
And we do not normally
expect a random-split to be failing.

Other null-test failures include

\begin{itemize}
\tightlist
\item
  \texttt{QU\_\allowbreak{}PIXEL}
\item
  \texttt{LEAK\_\allowbreak{}BY\_\allowbreak{}BOLO}
\item
  \texttt{AMP\_\allowbreak{}2F\_\allowbreak{}BY\_\allowbreak{}BOLO}
\item
  \texttt{AMP\_\allowbreak{}4F\_\allowbreak{}BY\_\allowbreak{}BOLO}
\end{itemize}

which are our discoveries in \cref{sec:holm-bonferroni-method}.

\section{Null-test failures}\label{sec:null-test-failures}

We named these failures as ``local-by-bolo'' splits,
as they are all null-split by detectors
that are separated locally on the focal plane.

For a long while, these null-test failures
were not understood,
and there was a long list of speculations
as to why they are failing.
And I checked each of these theories
and falsified them:

\begin{itemize}
\tightlist
\item
  I-to-P leakage?
\item
  Gain mismatch?
\item
  \(1/f\) noise aliasing into \(4f\) frequency band contaminating the polarization signal?
\item
  Requiring pairing of top and bottom detectors?
\end{itemize}

The actual problem and solution are discussed below.

The first hint
is from \cref{fig:0_6_2-RANDOM_BY_BOLO1}
when it was first plotted in
\(C_\ell B^2\ell\) scale.
Looking like white noise here is a huge hint.
So we now know that
this is a noise problem.

\begin{figure}
\centering
\pandocbounded{\includegraphics[keepaspectratio,alt={\textbackslash frac\{N\_\{l, \textbackslash text\{real\}\}\}\{N\_\{l, \textbackslash text\{sim.\}\}\} as a function of null-split}]{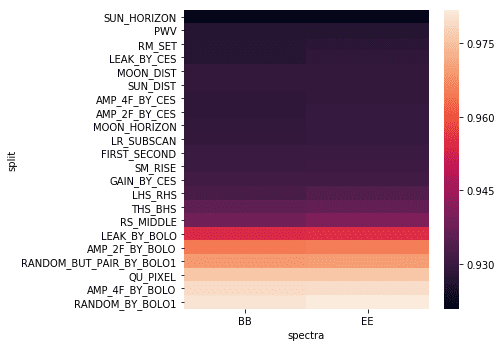}}
\caption{\(\frac{N_{l, \text{real}}}{N_{l, \text{sim.}}}\) as a function of null-split}\label{fig:real_over_sim_ratio_of_Nl}
\end{figure}

Further digging into the noise spectra,
when the ratio between the real noise spectra
and the simulated noise spectra
is plotted in \cref{fig:real_over_sim_ratio_of_Nl},
we have found the same set of null-splits
that stand out from the others\footnote{Do not worry about the overall
  normalization as it would be fixed later.
  Also,
  the two random null-splits are added
  when searching for hints on why this problem exists.}.

Digging further,
as pointed out by Goeckner-Wald\footnote{\citeproc{ref-goeckner-wald_expanding_2018}{{``Expanding Out the QUIET Cross Null Spectrum''}}.} in an internally circulated document,
there is a bug in AnalysisBackend
that added a term that should not exist
in the calculation of the null-spectra,
which normally would not be a problem
as the expected value of this term is zero
if the noise between the two splits is uncorrelated.
But the very bug also gave us a hint of the origin of the problem---correlated detector noise.
(Since different null-splits are splitting the same data
in different halves,
we know that the total noise level must be a constant
and therefore this detector noise correlation is anti-correlating
with some of the nearby detectors.)

\begin{figure}
\centering
\pandocbounded{\includegraphics[keepaspectratio,alt={\textbackslash left\textbar{} r\_\{ij\} \textbackslash right\textbar\^{}2, where r\_\{ij\} is the correlation between detectors TOD estimated using Q+iU with an example CES named . R.H.S. shows a heatmap with the detectors reordered to reveal the clustering. On the L.H.S., \textbackslash tilde\{R\_i\} \textbackslash equiv \textbackslash max\_j \textbackslash left\textbar{} r\_\{ij\} \textbackslash right\textbar\^{}2 is first computed, and the Kernel Density Estimation (KDE) of \textbackslash tilde\{R\_i\} is calculated and shown here.}]{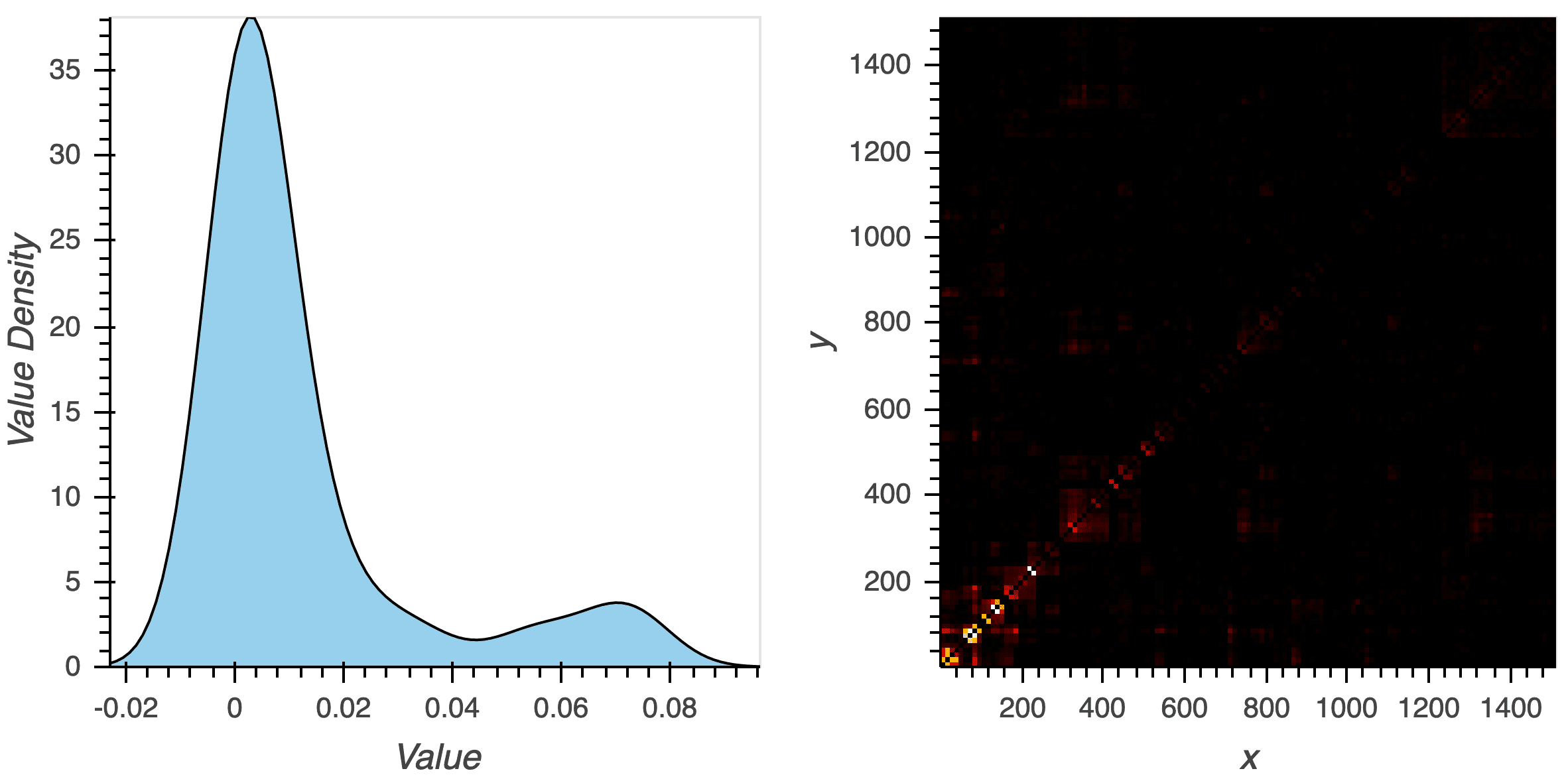}}
\caption{\(\left| r_{ij} \right|^2\), where \(r_{ij}\) is the correlation between detectors TOD estimated using \(Q+iU\) with an example CES named \texttt{20141211\_\allowbreak{}173206}.
R.H.S. shows a heatmap with
the detectors reordered to reveal the clustering.
On the L.H.S.,
\(\tilde{R_i} \equiv \max_j \left| r_{ij} \right|^2\) is first computed,
and the Kernel Density Estimation (KDE) of \(\tilde{R_i}\)
is calculated and shown here.}\label{fig:20141211_173206_squared_clustered}
\end{figure}

The detector-detector correlations are calculated
and a bad example CES
is shown in \cref{fig:20141211_173206_squared_clustered}.
These correlations
are used to calculate
a Kernel Density Estimation (K.D.E.) and there is a second peak.
This gives us a data selection criterion,
which is a classification problem,
using the feature on the K.D.E. as the classification rule.

Our rule is then to find this 2nd peak
per CES
and cut away the data from the valley between them.
This cuts away an additional amount of data volume at around \(\SI{5}{\percent}\).

\begin{figure}
\centering
\pandocbounded{\includegraphics[keepaspectratio,alt={Null test v0.10.0  null-spectra}]{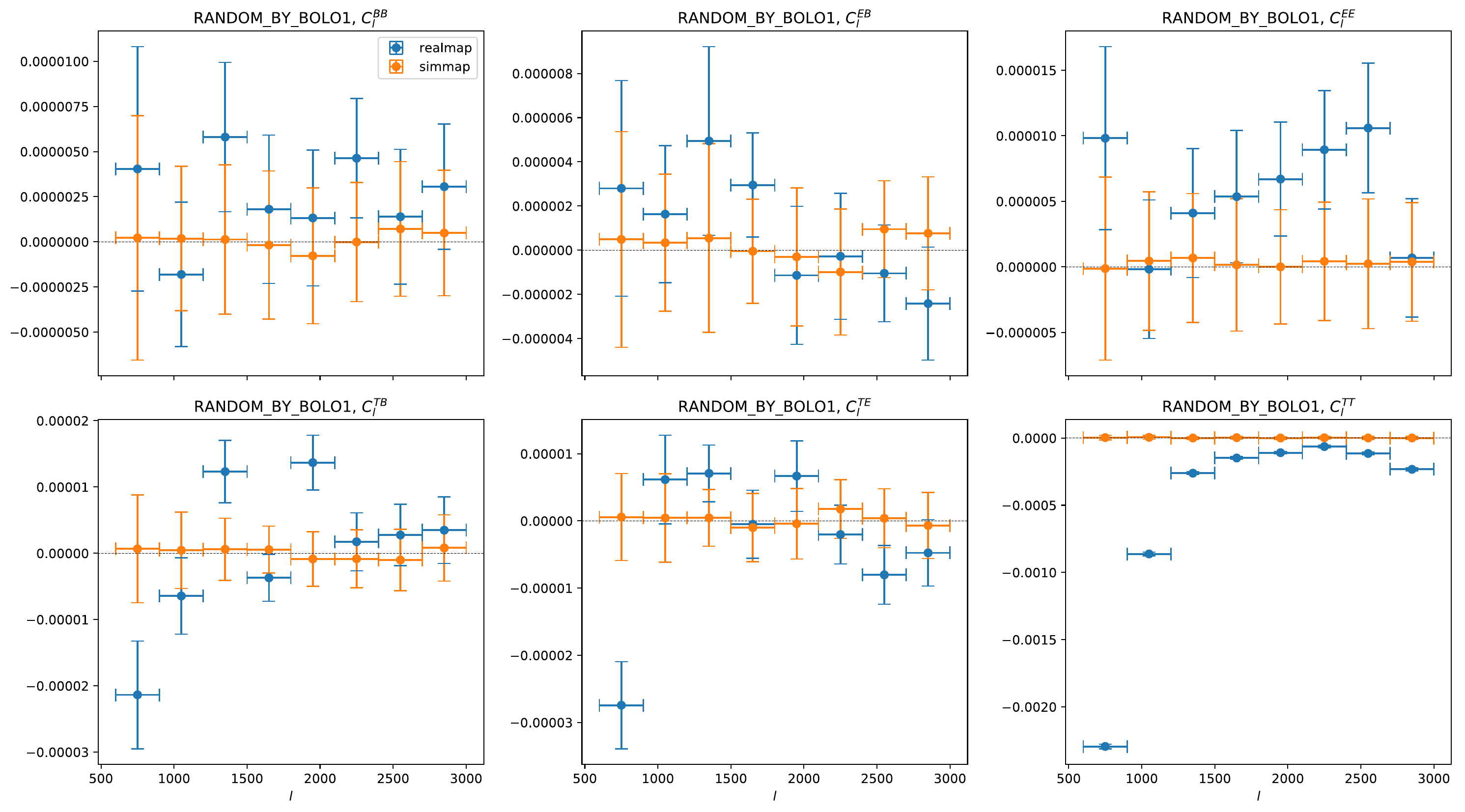}}
\caption{Null test v0.10.0 \texttt{RANDOM\_\allowbreak{}BY\_\allowbreak{}BOLO1} null-spectra}\label{fig:0_10_0-RANDOM_BY_BOLO1}
\end{figure}

With all these fixes in-place,
the null-spectra look like \cref{fig:0_10_0-RANDOM_BY_BOLO1}.

We can see it is a huge improvement over
\cref{fig:0_6_2-RANDOM_BY_BOLO1}
for all of our science products \(EE, EB, BB\) power-spectra\footnote{Note that
  the white noise model instead of the sign-flip noise
  is used, so the \(TT\) null-power-spectra are still not good. This is fixed in the final analysis. See \cref{sec:noise}.}.

\section{When can we unblind?}\label{sec:when-to-unblind}

As we are marching through the null-test iterations,
the situation is improving.

The actual passing criteria
are complicated,
which includes other criteria
such as Kolmogorov--Smirnov test
mentioned in \cref{sec:ks-test}.

But even focusing on the null-test suite alone,
when are we going to end this endless pursuit of hidden systematics
and be confident that we are good enough to unblind?
In other words,
what is the stopping criterion?

The first boring answer is
to increase the number of simulations
because they are not failing as dramatically
and hence we need more accurate statistics
to discern between fluke and discovery.
But how much is enough?

This is a statistical problem.
In short,
we want to quantify
how confident we are in passing the null-test
before running the final run with hundreds of simulations.

The answer is confidence interval.
But how do we estimate it?
In Chinone,\footnote{\citeproc{ref-chinone_polarbear_2014}{{``POLARBEAR Null-Test Framework''}}.} an erroneous argument is made
following Clopper and Pearson,\footnote{\citeproc{ref-clopper_use_1934}{{``THE USE OF CONFIDENCE OR FIDUCIAL LIMITS ILLUSTRATED IN THE CASE OF THE BINOMIAL''}}.} which uses an approximation
that fails at the two extremes
of the binomial, and is applied at one of the extremes \(p=0\).\footnote{See more from \citeproc{ref-agresti_approximate_1998}{Agresti and Coull, {``Approximate Is Better Than {`Exact'} for Interval Estimation of Binomial Proportions''}}; \citeproc{ref-thulin_cost_2014}{Thulin, {``The Cost of Using Exact Confidence Intervals for a Binomial Proportion''}}.}

But we do not even have to fix that formula,
instead, we can apply the bootstrap method
on the MC simulations from existing
preliminary null-test result using a smaller number of MC simulations.

For example, in null-test v0.13.4,
I have \(n_\text{sim} = 48\) simulations.
Using the bootstrap method,
I can resample the \(n_\text{sim}\) simulations
and estimate the confidence interval
\(P(\alpha_c \geq 0.05)\).

Applying this procedure,
the weakest statistics in v0.13.4
is from \(EB\),
where I have \(\alpha = \SI{6.2}{\percent}\)
and \(P(\alpha_c \geq 0.05) = \SI{81.4}{\percent}\).

Since we require all \(EE, EB, BB\) to be passing,
this is good enough and a sign that
we are good to go.

\chapter{The limitation of the null-test framework}\label{sec:null-test-limitation}

\section{Hypothesis testing, discovery, and default action}\label{hypothesis-testing-discovery-and-default-action}

As mentioned before,
the POLARBEAR null-test framework
is based on hypothesis testing in statistics
where our null-hypothesis \(H_0\) is the absence of systematics.
Hence, our discovery in this hypothesis testing is
the detection of systematics,
which contaminates our data.

But here we are using hypothesis testing
in a very different context
compared to how it is usually used in science.
Namely,
typically the null-hypothesis
is the lack of new discovery leading to new science.
To put it in other words,
as Oleszak\footnote{\citeproc{ref-oleszak_hypothesis_2021}{{``The Hypothesis Tester's Guide''}}.} put it as ``the null hypothesis reflects our default action''.
In the case of making new scientific discoveries,
when no discovery (null-hypothesis not rejected) is found,
the default action would be not to proceed any further including publications.\footnote{See \citeproc{ref-banerjee_hypothesis_2009}{Banerjee et al., {``Hypothesis Testing, Type I and Type II Errors''}} for more about the basics of good hypothesis testing.}

Here, as we made no discovery (null-test suite makes no discovery on systematic bias)
we proceed and go to unblind the data resulting in a publication.
We are taking a much more risky default action---unblind and publish.

On one hand,
certainly the null-test suite is not the only procedure to guard against biases.
But on the other hand,
biases can arise even after all these and passing any null-test criteria.

For example,
we found that
the absolute angle calibration
is too large after unblinding the data,
which is due to
not having deconvolved\cref{sec:TODmisc} the TOD
by the detector time constant,
which turns out to be an internal (lack of) communication error
in applying a single command line argument!

Despite the presence of this systematics,
the null-test suite passes anyway
because it represents a lag in the measured detector signal,
which has an effect that is degenerate from an additional
rotation (resulting in a large apparent absolute angle in calibration.)
This effect would not cause a null-test failure
(as rotating null still results in null.)

Besides being an example of a limitation in the null-test framework,
this perhaps reiterates the point in \cref{sec:reproducible-research}
that even reproducing a result within a collaboration
is crucial to avoid an error like this.
It is a systemic issue that cultivates a problem
as important as this to be overlooked.

\section{The action of unblinding}\label{the-action-of-unblinding}

Another limitation of the null-test framework
is implicit in its default action---if null-test is passed,
then we unblind the data.

But what is blindness?
It is a somewhat arbitrary entity with no clear definitions.
For example,
it is argued that looking at the signal maps is ok,
because the map itself is not the science product in our case.
And it is abstract enough that from looking at the map,
it is hard to draw conclusions.
Nonetheless, this makes defining blindness rigorously impossible.

Another more subtle case would be to look at the null-spectra.
Since they are expected to be zero,
there is no harm done by looking at them, right?
But the nature of systematics is that
we may not know every source of systematics beforehand,
and some, which can be as simple as a gain mismatch,
can result in the signal power-spectra leaking into the null-spectra.
One can be accidentally looking at a rescaled signal
when looking at null-spectra.

There is an even more philosophical example,
that when one runs the pipelines from AnalysisBackend,
by default it will print the numerical power-spectra on screen.
Those are usually scrolling so fast that a normal human's eyes
cannot catch up, but nonetheless this is an arbitrary blindness criterion
which relies on the human's incapability (in reading numbers fast enough).

Not being able to write down a formal definition of blindness has consequences.
In an early study of absolute gain calibration,
I accidentally used the real data to project the level of absolute gain expected.
In hindsight, it clearly violates the blindness criteria,
but it is easy to lump all these together and think that
``since the gain is not our science product, it would be fine.''

\section{After unblinding}\label{after-unblinding}

Another deficiency of the null-test framework is that it
implicitly assumes an unblinding procedure
comes from what can be changed after unblinding.

If one changes an analysis technique
for example, it can be a source of observer bias.
So ideally one should not change anything after unblinding.
But in our framework,
a practical problem is that we do not strongly specify
the analysis techniques involved after unblinding.

We already went to great lengths in \cref{sec:statistical-methods}
discussing the procedure involved,
but that is not a strong specification
in the sense that there are devils in the details
that can be changed.

To give one example,
recall that we use sign-flip noise in the final analysis.
It means that noise estimated from the real data
is injected into maps simulated using fiducial maps,
which are generated using the 2018 Planck cosmology.\footnote{\citeproc{ref-planck_collaboration_planck_2020-3}{Planck Collaboration, Aghanim, Akrami, Ashdown, et al., {``\emph{Planck} 2018 Results''}}.} Hence, the noise components and signal components have a different
Pixel Window Function (PWF)\@.
Admittedly, the implication or even existence
of this has not been well thought out before unblinding.
It also means after unblinding,
new code was written to take this effect into account.

A solution is offered by reproducible research\footnote{See \cref{sec:reproducible-research}.},
that we should have required the whole software stack and pipelines
to be reproducible before unblinding.
The unblinding stage should run exactly the same pipelines
over the newly generated datasets.

\section{Miscellaneous problems}\label{miscellaneous-problems}

Lastly, after the final analysis,
I found 2 programming bugs such that the counter term
of the pure \(B\)-mode estimator is incorrect.

This increases the amount of \(E\)-to-\(B\) leakage.
Phenomenologically,
\(\sim \SI{20}{\percent}\) of \(B\)-mode power comes from leakage
peaked at the lower-\(\ell\) end.
While the bug makes this larger,
it is much smaller than the total statistical error in this analysis.

One solution to problems like this
is to have maintainers for research software,
and to require code-review before merging,
which is a standard software engineering practice elsewhere.

\chapter{Our measurement}\label{sec:null-test-results}

In null-test v0.13.4\footnote{See \cref{sec:PB-null-changelog}.},
24 \(T, E\) noiseless simulations and
24 \(T, B\) noiseless simulations are used
to estimate the filter transfer functions.
48 \(T, E, B\) with sign-flip noise simulations
are used as MC simulations for the estimation
of statistical properties including the error-bar.
Null-test suite is passing preliminarily.
Following the reasoning outlined in \cref{sec:when-to-unblind},
it is scaled up to a set of \(128, 128, 512\) simulations respectively
in null-test v1.0.0.

The deconvolution problem mentioned in \cref{sec:null-test-limitation}
is fixed in v1.0.1.
There is another minor problem concerning how the ground-template
is constructed and is fixed in v1.0.2. This is the final analysis.
Below are its null-statistics and the measurements.

The \(\ell\)-range for \(BB, EB, EE\) is 600--3000, using a bin-width of 100.
The \(\ell\)-range for \(TT, TE, TB\) is 600--1600 instead
as the \(T\)-filter transfer function has numerical
stability issues above \(\ell \simeq 1600\).

\section{Passing Null Test suite}\label{passing-null-test-suite}

\(\alpha = \SI{5}{\percent}\) is required in the null-test suite passing requirements.

For Kolmogorov--Smirnov test,
when it is performed using null random variable \(χ(X, n, b)\) with \(X = BB, EB, EE\),
overall p-value is \(\SI{13.0}{\percent}\) with a confidence level of \(P(p > \alpha) = \SI{99.9}{\percent}\), estimated using \(1,000\) bootstrap.

\needspace{11\baselineskip}

\begin{longtable}[]{@{}llllll@{}}
\caption{PTE for the null-statistics defined in definition~\ref{def-null-statistics} estimated using MC simulations.}\label{tbl:p-values-final}\tabularnewline
\toprule\noalign{}
power-spectra & \(G_{Y_1}\) & \(G_{Y_2}\) & \(G_{Y_3}\) & \(G_{Y_4}\) & \(G_{Y_5}\) \\
\midrule\noalign{}
\endfirsthead
\toprule\noalign{}
power-spectra & \(G_{Y_1}\) & \(G_{Y_2}\) & \(G_{Y_3}\) & \(G_{Y_4}\) & \(G_{Y_5}\) \\
\midrule\noalign{}
\endhead
\bottomrule\noalign{}
\endlastfoot
BB & 72.07\% & 81.25\% & 75.39\% & 85.74\% & 14.45\% \\
EB & 80.27\% & 37.30\% & 63.28\% & 27.93\% & 6.05\% \\
EE & 37.30\% & 83.20\% & 80.66\% & 98.44\% & 65.82\% \\
TB & 8.01\% & 41.02\% & 99.22\% & 81.45\% & 80.66\% \\
TE & 11.33\% & 81.45\% & 31.45\% & 88.67\% & 48.83\% \\
TT & 29.10\% & 84.77\% & 40.63\% & 94.53\% & 91.60\% \\
\end{longtable}

\Cref{tbl:p-values-final}
shows the null-statistics as defined in definition~\ref{def-null-statistics}
estimated using MC simulations.
Using the 15 null-statistics with \(X = BB, EB, EE\),
overall \(p\)-value is at \(α_c = \SI{90.8}{\percent}\) with a confidence level of \(P(p > \alpha) = \SI{100}{\percent}\), estimated using \(10,000\) bootstrap.
This is the unblinding requirement.

For reference,
if all \(30\) null-statistics with \(X = BB, EB, EE, TE, TB, TT\) are to be included,
overall \(p\)-value is at \(α_c =  \SI{99.2}{\percent}\) with a confidence level of \(P(p > \alpha) = \SI{100}{\percent}\), estimated using \(10,000\) bootstrap.

\section{Likelihood \& posterior}\label{likelihood-posterior}

Method detailed in \cref{sec:likelihood-estimation} is followed to perform likelihood estimation.

\subsection{Foreground model}\label{foreground-model}

Prior of dust foreground is included following the same model
used in Sayre et al.\footnote{\citeproc{ref-sayre_measurements_2020}{{``Measurements of B-Mode Polarization of the Cosmic Microwave Background from 500 Square Degrees of SPTpol Data''}}.}

\[D_\ell = A_{d, \SI{150}{\giga\hertz}} \left(\frac{\ell}{80} \right)^{-0.58}.\]

A different Gaussian prior is used however,
as a slightly different numerical value is found
when it is derived from the source of the input in the model from Ade et al.\footnote{\citeproc{ref-ade_constraints_2018}{{``Constraints on Primordial Gravitational Waves Using \emph{Planck}, WMAP, and New BICEP2/\emph{Keck} Observations Through the 2015 Season''}}.}\footnote{Instead of
  \(A^\text{dust}_{\ell = 80; \SI{150}{\giga\hertz}} = \SI{0.0094 \pm 0.0021}{\micro\kelvin^2}\)
  from Sayre et al., \citeproc{ref-sayre_measurements_2020}{{``Measurements of B-Mode Polarization of the Cosmic Microwave Background from 500 Square Degrees of SPTpol Data''}}.},

\begin{align*}
\beta                                       &= \num{1.59 \pm 0.11}  \\
T                                           &= \SI{19.6}{\kelvin}   \\
A_{d, \SI{353}{\giga\hertz}}                &= \num{4.6 \pm 1.1}    \\
\Rightarrow A_{d, \SI{150}{\giga\hertz}}    &= \num{0.009554078346271361 \pm 0.002907862359357405}.
\end{align*}

Lastly, \(D_b \equiv \sum_\ell w_{b\ell}^{BB} D_\ell\).

\subsection{Calibration / nuisance parameters}\label{calibration-nuisance-parameters}

Using the likelihood procedure defined in \cref{sec:likelihood-estimation},
an affine-invariant ensemble sampler
for Markov chain Monte Carlo (MCMC)\footnote{\citeproc{ref-foreman-mackey_emcee_2013}{Foreman-Mackey et al., {``Emcee''}}.}
is used to estimate the posterior distribution
of the input parameters.

The posterior distribution of \(A^\text{dust}_{\ell = 80; \SI{150}{\giga\hertz}}\)
is similar to the prior.

Regarding calibration parameters,
the absolute gain of the intensity map is
\(a_T = \num{1.091 \pm 0.012}\),
and the absolute gain of the polarization maps is
\(a_E = \num{1.107 \pm 0.013}\).
The agreement of both means
the polarization efficiency calibration is unbiased.

For the absolute angle calibration,
\(\theta = \SI{-0.32 \pm 0.19}{\degree}\).
For the \(\sigma^2\) parameter in the beam model,
\(\sigma^2 = \SI{0.48 \pm 0.19}{\arcminute}^2\).

These are consistent with the result from the large-patch low-\(\ell\) pipeline.\footnote{\citeproc{ref-the_polarbear_collaboration_measurement_2020}{The POLARBEAR Collaboration et al., {``A Measurement of the Degree-Scale CMB \emph{B}-Mode Angular Power Spectrum with POLARBEAR''}}.}

\subsection{Cosmological parameter}\label{cosmological-parameter}

The cosmological parameter here is
\(A_\text{lens}\) which is the ratio between the measured
lensing \(B\)-mode and the theoretically expected amount of lensing.
Hence, its fiducial value is \(1\).
There are a couple of ways to interpret this result.

Starting with the MCMC chain, one can estimate that

\[P(A_\text{lens} > 0) = \SI{99.6}{\percent}.\]

Wilks' theorem can also be used\footnote{Since the assumption
  made in Wilks' theorem is not exactly true here,
  this is an approximation.} to convert the likelihood ratio
to a probability of \(p = \SI{99.2}{\percent}\).
The agreement between the two is reassuring,
and the result from MCMC is used in the end as it makes no assumption
on the behavior of the likelihood.

Lastly, \(A_\text{lens} = \num{3.90 \pm 1.51}\), where \(3.90\) is the MLE
and \(1.51\) is the standard deviation.
The two-sided \(\SI{90}{\percent}\) CI is \(1.46 < A_\text{lens} < 6.42\),
while the one-sided \(\SI{90}{\percent}\) CI upper bound is \(A_\text{lens} < 5.85\).

\part{The effect of detector crosstalk systematics on the CMB power-spectra in the example of the LiteBIRD experiment}\label{sec:LiteBIRD-research}

\chapter{The LiteBIRD satellite}\label{the-litebird-satellite}

LiteBIRD stands for Lite (Light) satellite for the studies of B-mode polarization and inflation from cosmic background Radiation Detection.

\section{Introduction}\label{introduction-1}

\begin{figure}
\centering
\pandocbounded{\includegraphics[keepaspectratio,alt={Scan strategy of LiteBIRD in a Lissajous orbit around L2.}]{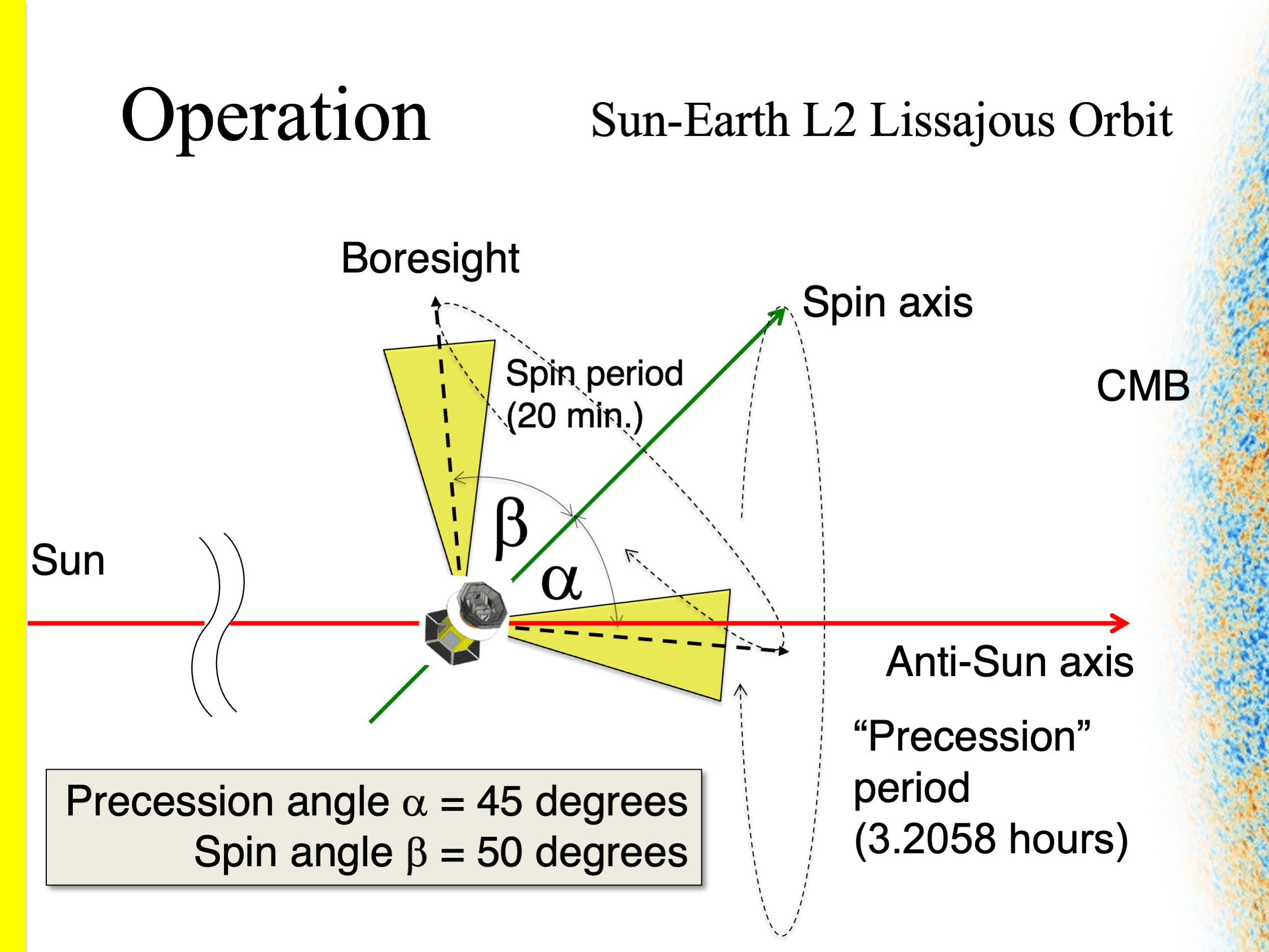}}
\caption[Scan strategy of LiteBIRD in a Lissajous orbit around L2.]{Scan strategy of LiteBIRD in a Lissajous orbit around L2.\footnotemark{}}\label{fig:LiteBIRD-scan}
\end{figure}
\footnotetext{\citeproc{ref-hazumi_litebird_2020}{Hazumi et al., {``LiteBIRD Satellite''}}.}

LiteBIRD is a satellite-based CMB experiment
expected to launch in the late 2020s.
It will be the successor of Planck and will become the 4th satellite
dedicated to CMB observation.
Like its predecessors since the 2nd generation WMAP,
it will orbit around L2 point
with a quasi-stable Lissajous orbit
as shown in \cref{fig:LiteBIRD-scan},
far away from the contamination by the Earth and the Moon.

\section{Science target}\label{science-target}

The primary science target of LiteBIRD
is the primordial \(B\)-mode
and hence it is a probe of inflationary physics.

\begin{figure}
\centering
\pandocbounded{\includegraphics[keepaspectratio,alt={Summary of present measurements of CMB power spectra and expected polarization sensitivity of LiteBIRD.}]{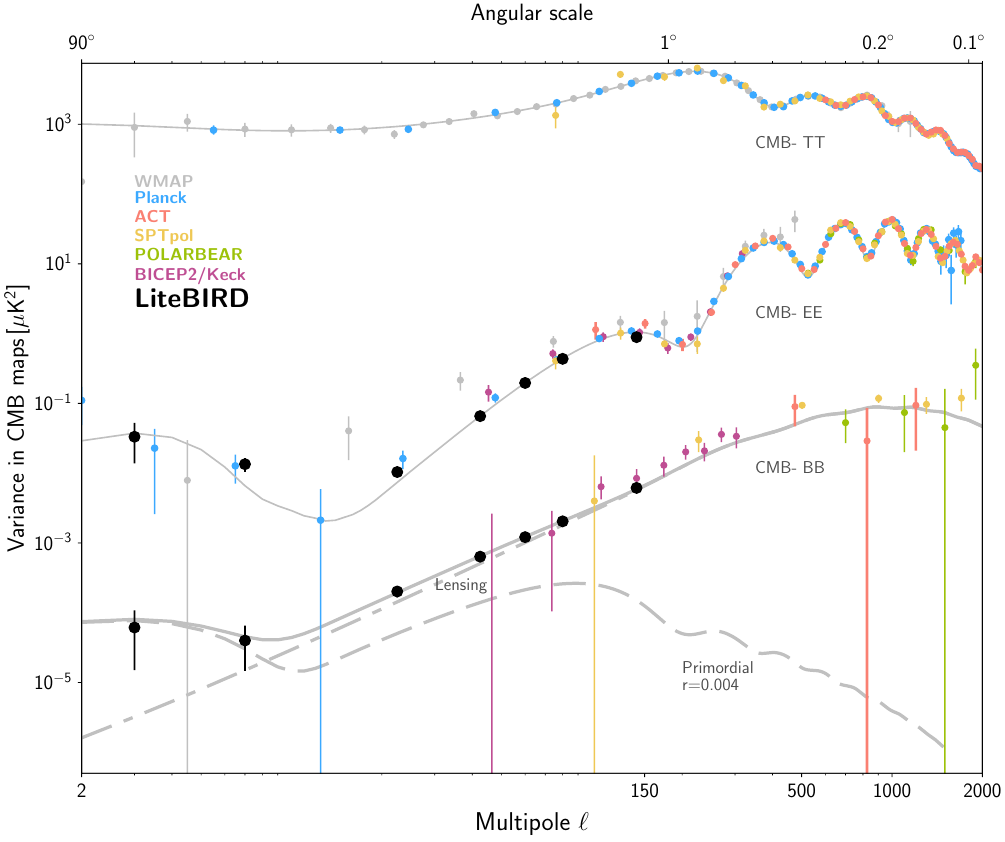}}
\caption[Summary of present measurements of CMB power spectra and expected polarization sensitivity of LiteBIRD.]{Summary of present measurements of CMB power spectra and expected polarization sensitivity of LiteBIRD.\footnotemark{}}\label{fig:LiteBIRD_sensitivity}
\end{figure}
\footnotetext{\citeproc{ref-hazumi_litebird_2020}{Hazumi et al., {``LiteBIRD Satellite''}}.}

\begin{figure}
\centering
\pandocbounded{\includegraphics[keepaspectratio,alt={Sensitivity contours of LiteBIRD and the Simons Observatory. Systematic uncertainties are not included here. The plot assumes that the actual value of r is 0.004.}]{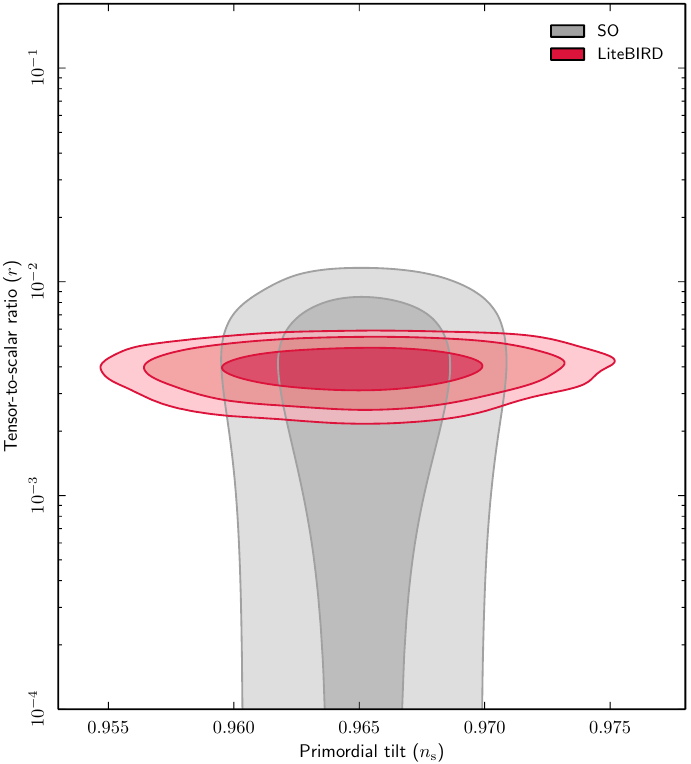}}
\caption[Sensitivity contours of LiteBIRD and the Simons Observatory. Systematic uncertainties are not included here. The plot assumes that the actual value of r is 0.004.]{Sensitivity contours of LiteBIRD and the Simons Observatory. Systematic uncertainties are not included here. The plot assumes that the actual value of r is 0.004.\footnotemark{}}\label{fig:LiteBIRD_r-n_s}
\end{figure}

In \cref{fig:LiteBIRD_sensitivity} \& \cref{fig:LiteBIRD_r-n_s},
which can be compared to
\cref{fig:2018_compilation_v5} \& \cref{fig:Planck2018_TestTestV4BK14LegacyV4_120mm},
they show the projected sensitivity of LiteBIRD
in terms of the B-mode spectrum,
parameterized by the tensor-to-scalar ratio \(r\).

\footnotetext{\citeproc{ref-hazumi_litebird_2020}{Hazumi et al., {``LiteBIRD Satellite''}}.}

The current constraint is \(r < 0.044\).
LiteBIRD aims to constrain
\(\delta r < \num{1e-3}\).
This can either make a discovery or rule out well-motivated inflationary models.\footnote{\citeproc{ref-hazumi_litebird_2020}{Hazumi et al., {``LiteBIRD Satellite''}}.}

This makes it one of the most exciting CMB experiments
in the coming decade.
And to date,
hundreds of scientists around the globe
are working on the project
to achieve this goal.

\section{Multiplexed readout system}\label{multiplexed-readout-system}

LiteBIRD is designed to be a 3-year mission,
with the current design holding 4508 detectors;
reading out data from all those detectors is challenging.
These TES bolometers
are daisy-chained together
using digital frequency-domain multiplexing (fMUX)
through superconducting quantum interference device (SQUID)
to achieve simultaneous readout.

For example,
in POLARBEAR,
8 bolometers are chained through SQUID\@.
In LiteBIRD,
the current baseline design is
to chain around 68 bolometers,
with possibly more in the order of 100s
as the design improves.

This is also a general trend in CMB experiments.
As seen in \cref{sec:next-stage},
the number of detectors is growing exponentially.
For reasons laid out in
Montgomery et al.\footnote{\citeproc{ref-montgomery_performance_2020}{{``Performance and Characterization of the SPT-3G Digital Frequency Multiplexed Readout System Using an Improved Noise and Crosstalk Model''}}.} one of the primary limitations in growing the number of detectors
is cryogenic cooling limitations,
where multiplexing is an essential part of the solution.
Hence, we are going to see increasingly larger
mux factor, translated into the size of the crosstalk matrix.
It makes the crosstalk systematics increasingly
more important to be understood and mitigated.

\section{Crosstalk between detectors}\label{crosstalk-between-detectors}

With so many detectors chained together,
crosstalk between them can contaminate our signal.

For example,
We would like to chain detectors
across different observing frequencies
for fault tolerance such that if a whole SQUID is down,
we are not knocking out a whole observing frequency or wafer.
The presence of crosstalk across different
frequency channels
will mix different amounts of foreground components
into our resultant maps.
We typically form
individual frequency maps
and from there we cross-correlate
to estimate the foreground systematics,
such as how it is analyzed in Planck.
This mixing will break our model
and systematics can bias our results.

To solve this problem,
it is a joint effort
between the hardware and software team.
The experimentalists can
reduce the crosstalk in the design,
and provide realistic simulations
on the amount of crosstalk expected
from a design.
The data scientists can
take the crosstalk matrix as an input
and simulate the amount of bias
injected into the signal.

This feeds back into the hardware design process
which informs us if it can achieve our science goal.

Also, in terms of software development,
we can devise mitigation strategies
in how we can reduce the amount of bias.

Moreover,
a given hardware design
has a huge parameter space to explore.
For example,
in our current design for a certain SQUID,
it is required for around 60 detectors
to be chained across 3 wafers with around 20 detectors each.
The exact ordering within a SQUID is totally free.
This is then an optimization problem for us
to simulate the amount of bias
subject to the same requirement
and optimize the configuration
to provide feedback to hardware design.

In \cref{sec:crosstalk-SQUID},
we will see how to generate the crosstalk matrix
from SQUID parameters.
In \cref{sec:analytic-crosstalk},
a map-domain crosstalk model is presented
together with a set of assumptions which
is fulfilled approximately in the case of LiteBIRD\@.
In \cref{sec:time-domain-crosstalk},
an implementation of crosstalk simulation
is presented in the framework of TOAST\@.
In \cref{sec:crosstalk-simulations},
crosstalk systematics simulations
in the example of LiteBIRD are performed,
and a mitigation strategy is presented.

\chapter{Modeling crosstalk from SQUID design}\label{sec:crosstalk-SQUID}

\section{Introduction}\label{introduction-2}

\begin{figure}
\centering
\pandocbounded{\includegraphics[keepaspectratio,alt={An example circuit diagram of the cryogenic network. This includes all relevant components used in the derivation of leakage current crosstalk and leakage power crosstalk.}]{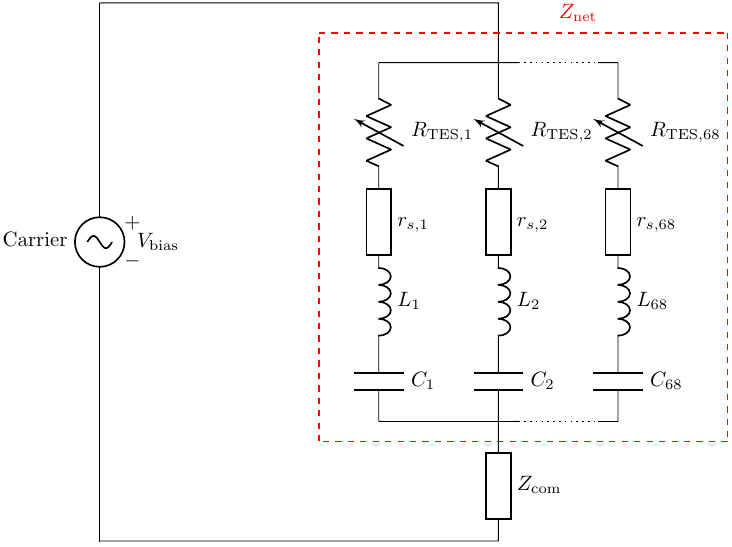}}
\caption[An example circuit diagram of the cryogenic network. This includes all relevant components used in the derivation of leakage current crosstalk and leakage power crosstalk.]{An example circuit diagram of the cryogenic network. This includes all relevant components used in the derivation of leakage current crosstalk and leakage power crosstalk.\footnotemark{}}\label{fig:SQUID-circuit}
\end{figure}

Montgomery et al.\footnote{\citeproc{ref-montgomery_performance_2020}{{``Performance and Characterization of the SPT-3G Digital Frequency Multiplexed Readout System Using an Improved Noise and Crosstalk Model''}}.} developed
a theoretical model describing how the SQUID design generates the crosstalk matrix,
and should be valid for
digital frequency-domain multiplexing (DfMUX)
with Digital Active Nulling (DAN),
and is validated using SPT-3G data.
See \cref{fig:SQUID-circuit} for the circuit model.

\footnotetext{\citeproc{ref-montgomery_performance_2020}{Montgomery et al., {``Performance and Characterization of the SPT-3G Digital Frequency Multiplexed Readout System Using an Improved Noise and Crosstalk Model''}}.}

\section{The crosstalk model and an exact solution}\label{the-crosstalk-model-and-an-exact-solution}

Montgomery et al.\footnote{\citeproc{ref-montgomery_performance_2020}{{``Performance and Characterization of the SPT-3G Digital Frequency Multiplexed Readout System Using an Improved Noise and Crosstalk Model''}}.} makes use of 3 approximations, 2 mathematical and 1 computational,

\begin{enumerate}
\def\labelenumi{\arabic{enumi}.}
\tightlist
\item
  Eq.~3 approximates the total (network) impedance by that at a leg, used in leakage power crosstalk;
\item
  Sec. 3.2 approximates the phase of the total signal by that of the detector signal; and
\item
  an implicit assumption in software that solves for resonance frequencies by finding extrema on a grid.
\end{enumerate}

Here, the same model is used without these 3 approximations.

This is released in the \texttt{util} module in open-source library \href{https://github.com/ickc/python-coscon/}{coscon}
which stands for Cosmological Convenient functions.
The \texttt{total\_\allowbreak{}crosstalk\_\allowbreak{}matrix} function implements
the model Montgomery2020 in Montgomery et al.\footnote{\citeproc{ref-montgomery_performance_2020}{{``Performance and Characterization of the SPT-3G Digital Frequency Multiplexed Readout System Using an Improved Noise and Crosstalk Model''}}.} whereas the \texttt{total\_\allowbreak{}crosstalk\_\allowbreak{}matrix\_\allowbreak{}exact} function implements
the same model without the 3 approximations, named Montgomery2020Exact here.

\section{Comparison}\label{comparison}

\begin{pandoccrossrefsubfigures}

\begin{subfigure}{0.5\linewidth}\centering\pandocbounded{\includegraphics[keepaspectratio,alt={Crosstalk matrix, Montgomery2020}]{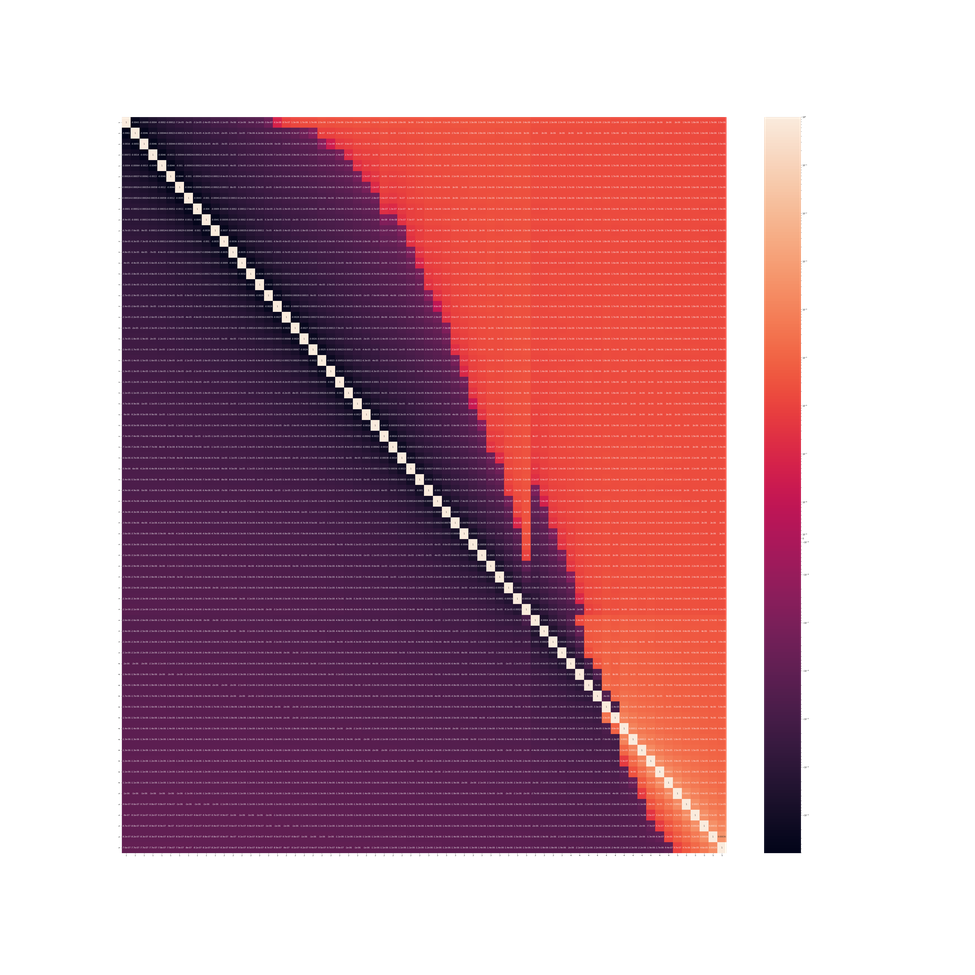}}\caption{Crosstalk matrix, Montgomery2020\label{fig:crosstalk_approx}}\end{subfigure}%
\begin{subfigure}{0.5\linewidth}\centering\pandocbounded{\includegraphics[keepaspectratio,alt={Crosstalk matrix, Montgomery2020Exact}]{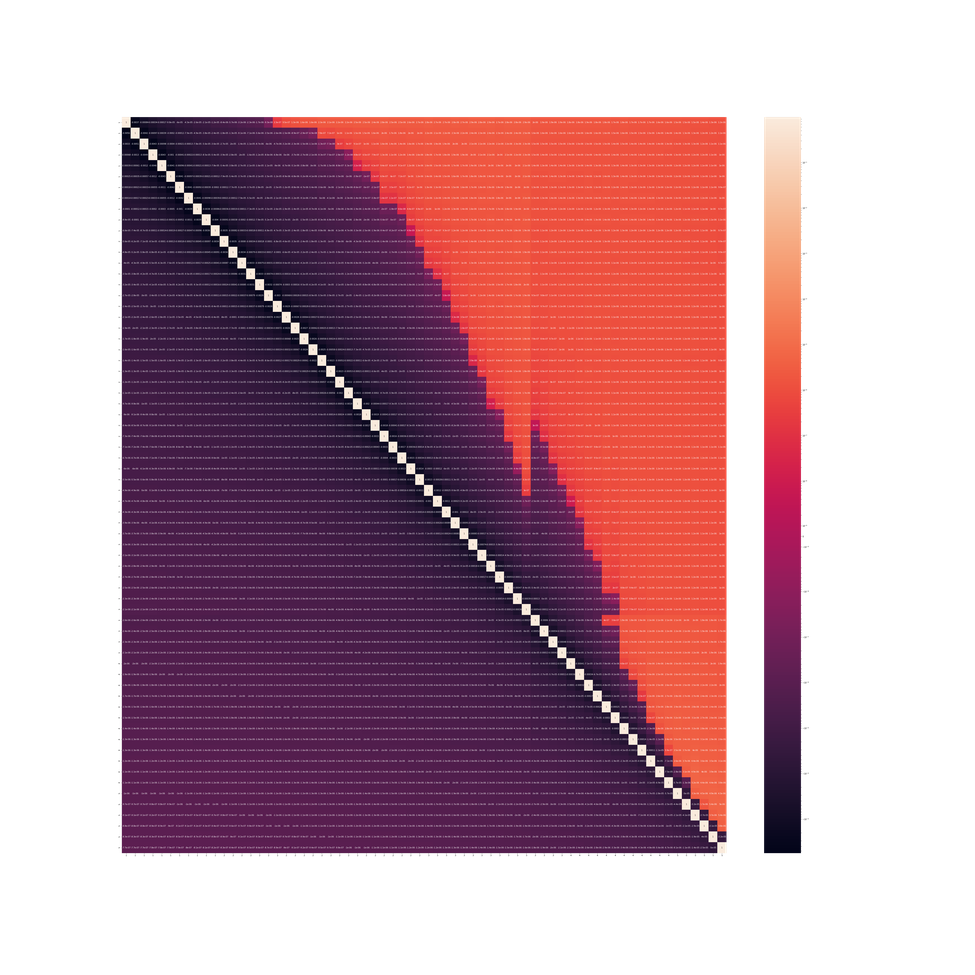}}\caption{Crosstalk matrix, Montgomery2020Exact\label{fig:crosstalk_exact}}\end{subfigure}%

\begin{subfigure}{0.5\linewidth}\centering\pandocbounded{\includegraphics[keepaspectratio,alt={Nearest neighbor crosstalk comparison}]{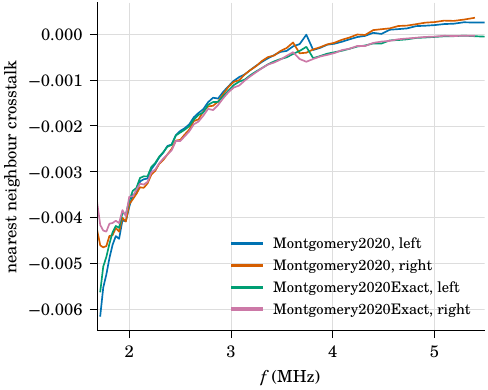}}\caption{Nearest neighbor crosstalk comparison\label{fig:crosstalk-nearest-neighbor-comparison}}\end{subfigure}%
\begin{subfigure}{0.5\linewidth}\centering\pandocbounded{\includegraphics[keepaspectratio,alt={Diagonal ``crosstalk'' comparison}]{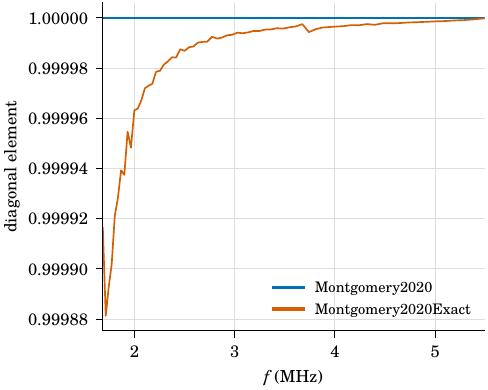}}\caption{Diagonal ``crosstalk'' comparison\label{fig:crosstalk-diagonal-comparison}}\end{subfigure}%

\caption{Comparison between Montgomery2020 and Montgomery2020Exact models.}\label{fig:crosstalk-matrix-comparison}

\end{pandoccrossrefsubfigures}

Using the same internal SPT-3G SQUID parameters as that in Montgomery et al.\footnote{\citeproc{ref-montgomery_performance_2020}{{``Performance and Characterization of the SPT-3G Digital Frequency Multiplexed Readout System Using an Improved Noise and Crosstalk Model''}}.} the crosstalk matrices using Montgomery2020 and Montgomery2020Exact models are compared in
\cref{fig:crosstalk-matrix-comparison}.

These are tiny corrections and the experimental variations are much bigger.\footnote{\citeproc{ref-montgomery_performance_2020}{Montgomery et al., {``Performance and Characterization of the SPT-3G Digital Frequency Multiplexed Readout System Using an Improved Noise and Crosstalk Model,''} fig. 7}.} The result from the Montgomery2020Exact model does have a few minor but interesting features:

\begin{enumerate}
\def\labelenumi{\arabic{enumi}.}
\tightlist
\item
  all nearest neighbors have negative crosstalk in this specific example.
\item
  the diagonal is no longer exactly 1, with correction peaked on the order of the crosstalk strength \(\sim \SI{0.01}{\percent}\) which is negligible.
  This correction originates from not using
  approximation (2).
  The change of phase contributed by crosstalk
  will change the magnitude slightly.
  In practice, the true sensitivity at calibration
  is measured with this contribution baked in\footnote{So the
    raw sensitivity should be what is measured
    divided by this factor.}.
\end{enumerate}

\chapter{A map domain model to predict crosstalk systematic effects on maps and power-spectra}\label{sec:analytic-crosstalk}

\chaptermark{A map domain model of crosstalk}

Crosstalk systematics are a time-domain problem.
It mixes the detector TOD according to\footnote{Matrices are specified with explicit indices throughout unless stated otherwise. Also, upper indices and lower indices are used for detectors and the discrete sampling points respectively.},

\[\tilde{d}^{i} = \sum_{j} X^{i j} d^{j} . \label{crosstalk_tod}\]

However, it is equivalent to a map-domain problem if the following holds:

\begin{Assumption}\label{asu-crosstalk-map-domain-model}
Assumptions of treating crosstalk systematics as a map-domain problem.

\begin{enumerate}
\def\labelenumi{\arabic{enumi}.}
\tightlist
\item
  In scanning strategy:

  \begin{enumerate}
  \def\labelenumii{\alph{enumii}.}
  \tightlist
  \item
    each detector has a uniform map coverage (that is, the weight-map per detector is constant over the sphere);
  \item
    (for fixed orientation) the focal plane has fixed orientation with respect to the sky locally;
  \item
    (or for isotropic orientation) the scanning strategy is isotropic in focal plane orientation\footnote{That is, (1a) assumes the scan is \(\mathbb{S}^{2}\) invariant, and (1c) in addition to (1a) assumes it is \(\operatorname{SO} \mathopen{} \left ( 3 \right ) \mathclose{}\) invariant.};
  \end{enumerate}
\item
  The mapmaking operation:

  \begin{enumerate}
  \def\labelenumii{\alph{enumii}.}
  \tightlist
  \item
    is linear in the observed TOD, and
  \item
    depends on the pointing matrix only;
  \item
    there is no noise such that the mapmaking operation can be considered as the inverse transform of the observation operation;
  \item
    where the resultant map is a weighted sum of individual detector maps (independence of detectors);
  \item
    each detector can resolve the \(I, Q, U\) components independently;
  \item
    has infinite resolution, that is, in the continuum limit;
  \end{enumerate}
\item
  In detector layout,

  \begin{enumerate}
  \def\labelenumii{\alph{enumii}.}
  \tightlist
  \item
    all detectors are fixed on the focal plane;
  \item
    the focal plane angular scale is \(\ll 1\).
  \end{enumerate}
\end{enumerate}

\end{Assumption}

\begin{proof}
With Assumption~\ref{asu-crosstalk-map-domain-model},
it can be proved that crosstalk systematics are map-domain systematics.

Observation and mapmaking are often written as discrete sums:

\begin{align*}
d_{t}	&	 = \sum_{p} A_{t p} m_{p} + n_{t}	\\
\hat{m}_{p}	&	 = \sum_{t} W_{p t} d_{t}
\end{align*}

Here, this is made abstract under the continuum limit (2f), and noise is ignored using assumption (2c):

\begin{align*}
\mathcal{O} \mathopen{} \left [ A \right ] \mathclose{} \mathopen{} \left \{ m \right \} \mathclose{} \mathopen{} \left ( t \right ) \mathclose{}	&	 \equiv d \mathopen{} \left ( t \right ) \mathclose{}	&		&	 = \oint A \mathopen{} \left ( t, \mathbf{r} \right ) \mathclose{} m \mathopen{} \left ( \mathbf{r} \right ) \mathclose{} \,\mathrm{d}^{2} \mathbf{r}	\\
\mathcal{M} \mathopen{} \left [ A \right ] \mathclose{} \mathopen{} \left \{ d \right \} \mathclose{} \mathopen{} \left ( \mathbf{r} \right ) \mathclose{}	&	 \equiv m \mathopen{} \left ( \mathbf{r} \right ) \mathclose{}	&		&	 = \int W \mathopen{} \left [ A \right ] \mathclose{} \mathopen{} \left ( \mathbf{r}, t \right ) \mathclose{} d \mathopen{} \left ( t \right ) \mathclose{} \,\mathrm{d} t
\end{align*}

where \(\mathcal{O} \mathopen{} \left [ A \right ] \mathclose{} , \mathcal{M} \mathopen{} \left [ A \right ] \mathclose{}\)\footnote{\(O[f]\{g\}(x)\) means an operator \(O\) is a higher-order function of \(f\)
  which transforms a function \(g\) to another function \(O\{g\}\) evaluated at \(x\).}
are linear transformations of \(m, d\) respectively (assumption 2a) that depend on \(A\)
(assumption 2b). They are inverse transformations of each other (assumption 2c).
Note that while
\(\mathcal{O} \mathopen{} \left [ A \right ] \mathclose{}\)
is linear in \(A\),
\(\mathcal{M} \mathopen{} \left [ A \right ] \mathclose{}\)
is not required to be linear in \(A\), and in practice they are not.

For the map \(m\)
, from assumption (2e),
we can consider either \(m = T\) as a real scalar field, or \(m = \mathcal{Q} = Q + \mathrm{i} U\)
as a complex spin-\(2\) field.
Explicitly, we can write \(A \mathopen{} \left ( t, \mathbf{r} \right ) \mathclose{} = \mathrm{e}^{- s \mathrm{i} \alpha \mathopen{} \left ( t \right ) \mathclose{}} \delta^{\left( 2 \right)} \mathopen{} \left ( \mathbf{r} - \mathbf{r} \mathopen{} \left ( t \right ) \mathclose{} \right ) \mathclose{}\)
where \(\alpha\)
is the orientation of the detector, \(s\) is the spin of the field (
\(s^{T} = 0 , s^{\mathcal{Q}} = 2\) ).
This can be used with assumption (3a) to rewrite the detector pointing with respect to the boresight pointing \(A^{i} \mathopen{} \left ( t, \mathbf{r} \right ) \mathclose{} = \mathrm{e}^{- s \mathrm{i} \Delta \alpha^{i}} A \mathopen{} \left ( t, R_{i}^{-1} \mathbf{r} \right ) \mathclose{}\), where \(R_{i} : \mathbf{r} \mathopen{} \left ( t \right ) \mathclose{} \mapsto \mathbf{r}^{i} \mathopen{} \left ( t \right ) \mathclose{} , \Delta \alpha^{i} \equiv \alpha^{i} \mathopen{} \left ( t \right ) \mathclose{} - \alpha \mathopen{} \left ( t \right ) \mathclose{}\)\footnote{Lower indices are used for the detector index in \(R\) and its variants for ease of typesetting.}
are independent of \(t\) by assumption (3a)\footnote{\(R\) can depend on \(\mathbf{r}\)
  by assumption (1b).}.
With this, note that we can rewrite an observation at \(A^i\) to an observation at \(A\):

\begin{align*}
\mathcal{O} \mathopen{} \left [ A^{i} \right ] \mathclose{} \mathopen{} \left \{ m \right \} \mathclose{} \mathopen{} \left ( t \right ) \mathclose{}	&	 = \oint A^{i} \mathopen{} \left ( t, \mathbf{r} \right ) \mathclose{} m \mathopen{} \left ( \mathbf{r} \right ) \mathclose{} \,\mathrm{d}^{2} \mathbf{r}	\\
	&	 = \mathrm{e}^{- s \mathrm{i} \Delta \alpha^{i}} \oint A \mathopen{} \left ( t, R_{i}^{-1} \mathbf{r} \right ) \mathclose{} m \mathopen{} \left ( \mathbf{r} \right ) \mathclose{} \,\mathrm{d}^{2} \mathbf{r}	\\
	&	 = \mathrm{e}^{- s \mathrm{i} \Delta \alpha^{i}} \oint A \mathopen{} \left ( t, \mathbf{r}^{\prime} \right ) \mathclose{} m \mathopen{} \left ( R_{i} \mathbf{r}^{\prime} \right ) \mathclose{} \,\mathrm{d}^{2} \mathbf{r}^{\prime}	\\
	&	 = \mathcal{O} \mathopen{} \left [ A \right ] \mathclose{} \mathopen{} \left \{ \mathcal{R}^{-1}_{i} \mathopen{} \left \{ m \right \} \mathclose{} \right \} \mathclose{} \mathopen{} \left ( t \right ) \mathclose{}
\end{align*}

where \(\mathcal{R} \mathopen{} \left \{ m \right \} \mathclose{} \mathopen{} \left ( \mathbf{r} \right ) \mathclose{} \equiv \mathrm{e}^{s \mathrm{i} \Delta \alpha^{}} m \mathopen{} \left ( R^{-1} \mathbf{r} \right ) \mathclose{}\)
is the active transform of the spin-\(s\) field \(m\)\footnote{This fixes the remaining d.o.f.\ of \(R\) in terms of \(\Delta \alpha\)
  .} (and hence \(\mathcal{R}^{-1} \mathopen{} \left \{ m \right \} \mathclose{} \mathopen{} \left ( \mathbf{r} \right ) \mathclose{} \equiv \mathrm{e}^{- s \mathrm{i} \Delta \alpha^{}} m \mathopen{} \left ( R \mathbf{r} \right ) \mathclose{}\)
is the passive transform.)

The resultant map per detector about the crosstalk in TOD would then be,

\begin{align}
\tilde{m}^{i}	&	 \equiv \mathcal{M} \mathopen{} \left [ A^{i} \right ] \mathclose{} \mathopen{} \left \{ \tilde{d}^{i} \right \} \mathclose{}	\nonumber	\\
\mathcal{O} \mathopen{} \left [ A^{i} \right ] \mathclose{} \mathopen{} \left \{ \tilde{m}^{i} \right \} \mathclose{}	&	 = \tilde{d}^{i} = \sum_{j} X^{i j} \mathcal{O} \mathopen{} \left [ A^{j} \right ] \mathclose{} \mathopen{} \left \{ m \right \} \mathclose{}	\nonumber	\\
\mathcal{O} \mathopen{} \left [ A \right ] \mathclose{} \mathopen{} \left \{ \mathcal{R}^{-1}_{i} \mathopen{} \left \{ \tilde{m}^{i} \right \} \mathclose{} \right \} \mathclose{}	&	 = \sum_{j} X^{i j} \mathcal{O} \mathopen{} \left [ A \right ] \mathclose{} \mathopen{} \left \{ \mathcal{R}^{-1}_{j} \mathopen{} \left \{ m \right \} \mathclose{} \right \} \mathclose{}	\nonumber	\\
\mathcal{R}^{-1}_{i} \mathopen{} \left \{ \tilde{m}^{i} \right \} \mathclose{}	&	 = \sum_{j} X^{i j} \mathcal{R}^{-1}_{j} \mathopen{} \left \{ m \right \} \mathclose{}	\nonumber	\\
\tilde{m}^{i}	&	 = \sum_{j} X^{i j} \mathcal{R}^{-1}_{i \rightarrow j} \mathopen{} \left \{ m \right \} \mathclose{} \text{ , or explicitly,}	\label{eq:crosstalk_map_pf}	\\
\mathrm{e}^{- s \mathrm{i} \Delta \alpha^{i}} \tilde{m}^{i} \mathopen{} \left ( R_{i} \mathbf{r} \right ) \mathclose{}	&	 = \sum_{j} X^{i j} \mathrm{e}^{- s \mathrm{i} \Delta \alpha^{j}} m \mathopen{} \left ( R_{j} \mathbf{r} \right ) \mathclose{}	\nonumber	\\
\tilde{m}^{i} \mathopen{} \left ( \mathbf{r} ' \right ) \mathclose{}	&	 = \sum_{j} X^{i j} \mathrm{e}^{- s \mathrm{i} \alpha^{i \rightarrow j}} m \mathopen{} \left ( R_{j} R_{i}^{-1} \mathbf{r} ' \right ) \mathclose{}	\nonumber	\\
\tilde{m}^{i} \mathopen{} \left ( \mathbf{r} \right ) \mathclose{}	&	 = \sum_{j} X^{i j} \mathrm{e}^{- s \mathrm{i} \alpha^{i \rightarrow j}} m \mathopen{} \left ( R_{i \rightarrow j} \mathbf{r} \right ) \mathclose{}	\nonumber
\end{align}

where \(\alpha^{i \rightarrow j} \equiv \Delta \alpha^{j} - \Delta \alpha^{i} , R_{i \rightarrow j} \equiv R_{j} R_{i}^{-1} , \mathcal{R}_{i \rightarrow j} \equiv \mathcal{R}_{j} \mathcal{R}_{i}^{-1}\). Finally, from assumption (2d) \& (1a),
the resultant map obtained in the mapmaking operation is,

\begin{align}
m^{\prime} \mathopen{} \left ( \mathbf{r} \right ) \mathclose{}	&	 = \sum_{i} w^{i} \tilde{m}^{i} \mathopen{} \left ( \mathbf{r} \right ) \mathclose{} , \quad \sum_{i} w^{i} = 1 \nonumber	\\
	&	 = \mathcal{X} m \mathopen{} \left ( \mathbf{r} \right ) \mathclose{} + \sum_{i , j \neq i} \mathcal{X}^{i j} \mathcal{R}_{i \rightarrow j}^{-1} \mathopen{} \left \{ m \right \} \mathclose{} \mathopen{} \left ( \mathbf{r} \right ) \mathclose{}	\label{crosstalkMapAbstract}	\\
	&	 = \mathcal{X} m \mathopen{} \left ( \mathbf{r} \right ) \mathclose{} + \sum_{i , j \neq i} \mathcal{X}^{i j} \mathrm{e}^{- s \mathrm{i} \alpha^{i \rightarrow j}} m \mathopen{} \left ( R_{i \rightarrow j} \mathbf{r} \right ) \mathclose{}	\label{crosstalkMap}
\end{align}

where \(\mathcal{X}^{i j} \equiv w^{i} X^{i j} , \mathcal{X} \equiv \operatorname{tr} \mathopen{} \left ( \mathcal{X}^{i j} \right ) \mathclose{}\). Note that in the limit that the crosstalk matrix is identity on the diagonal, \(\mathcal{X} = 1\)
by definition. That is, the first term above recovers the original map,
whereas the 2nd term is a first order correction as we normally have
\(X^{i j} \underset{i \neq j}{ \ll } 1\).
\end{proof}

This section is closed by observing when the assumptions are approximately true.
Note that the assumption that mapmaking is linear in \(d\) is true for most mapmaking methods,\footnote{Methods 1--8 in \citeproc{ref-tegmark_how_1997}{Tegmark, {``How to Make Maps from Cosmic Microwave Background Data Without Losing Information''}}, table 1.} while the assumption that it depends on \(A\) only is valid for mapmaking methods 1--3 of Tegmark,\footnote{\citeproc{ref-tegmark_how_1997}{{``How to Make Maps from Cosmic Microwave Background Data Without Losing Information''}}.} which include binned mapmaking (also known as naïve mapmaking)
and a maximal likelihood mapmaking.

\section{Flat sky, fixed orientation, isotropic map coverage}\label{sec:crosstalk-derivation-flat}

To apply \(\eqref{crosstalkMap}\) under flat-sky approximation with a fixed focal plane orientation, one substitutes \(R_{i} \rightarrow T_{i}\), the translation operator, and reinterprets \(\mathbf{r}\) as the 2-dimensional angle in radians. This implies \(R_{i \rightarrow j} \mathbf{r} \rightarrow T_{i \rightarrow j} \mathbf{r} = \mathbf{r} + \mathbf{r}^{i \rightarrow j}\),

\begin{align*}
m^{\prime} \mathopen{} \left ( \mathbf{r} \right ) \mathclose{}	&	 = \mathcal{X} m \mathopen{} \left ( \mathbf{r} \right ) \mathclose{} + \sum_{i , j \neq i} \mathcal{X}^{i j} \mathrm{e}^{- s \mathrm{i} \alpha^{i \rightarrow j}} m \mathopen{} \left ( \mathbf{r} + \mathbf{r}^{i \rightarrow j} \right ) \mathclose{}	\\
	&	 = \left( m \ast P \right) \mathopen{} \left ( \mathbf{r} \right ) \mathclose{} \equiv \int_{\mathbb{R}^{2}} m \mathopen{} \left ( \mathbf{r}^{\prime} \right ) \mathclose{} P \mathopen{} \left ( \mathbf{r} - \mathbf{r}^{\prime} \right ) \mathclose{} \,\mathrm{d}^{2} \mathbf{r} \text{, where}	\\
P \mathopen{} \left ( \mathbf{r} \right ) \mathclose{}	&	 = \mathcal{X} \delta^{\left( 2 \right)} \mathopen{} \left ( \mathbf{r} \right ) \mathclose{} + \sum_{i , j \neq i} \mathcal{X}^{i j} \mathrm{e}^{- s \mathrm{i} \alpha^{i \rightarrow j}} \delta^{\left( 2 \right)} \mathopen{} \left ( \mathbf{r} + \mathbf{r}^{i \rightarrow j} \right ) \mathclose{}
\end{align*}

is the point spread function\footnote{Note that the point spread function convolution here is analogous to beam convolution with \(P \mathopen{} \left ( \mathbf{r} \right ) \mathclose{} = B \mathopen{} \left ( - \mathbf{r} \right ) \mathclose{}\)
  .}. By convolution theorem\footnote{Physicists' convention is followed here. The same symbol
  to represent the function before or after Fourier Transform depends on the argument,
  and the Fourier Transform convention used is standard in the CMB literature,
  namely, defined with angular frequency in the non-unitary gauge
  where only the inverse transform has a factor of \(2\pi\).},
\(m^{\prime} \mathopen{} \left ( \mathbf{\ell} \right ) \mathclose{} = P \mathopen{} \left ( \mathbf{\ell} \right ) \mathclose{} m \mathopen{} \left ( \mathbf{\ell} \right ) \mathclose{}\), where

\begin{align*}
\mathcal{F} \mathopen{} \left \{ P \right \} \mathclose{} \mathopen{} \left ( \mathbf{\ell} \right ) \mathclose{}	&	 = \int_{\mathbb{R}^{2}} P \mathopen{} \left ( \mathbf{r} \right ) \mathclose{} \mathrm{e}^{- \mathrm{i} \mathbf{r} \cdot \mathbf{\ell}} \,\mathrm{d}^{2} \mathbf{r}	\\
{}_{{s}} P \mathopen{} \left ( \mathbf{\ell} \right ) \mathclose{}	&	 = \mathcal{X} + \sum_{i , j \neq i} \mathcal{X}^{i j} \mathrm{e}^{\mathrm{i} \left( \mathbf{r}^{i \rightarrow j} \cdot \mathbf{\ell} - s \alpha^{i \rightarrow j} \right)}
\end{align*}

Hence, this is how the fields are transformed under crosstalk,

\begin{align*}
T^{\prime} \mathopen{} \left ( \mathbf{\ell} \right ) \mathclose{}	&	 = {}_{{0}} P \mathopen{} \left ( \mathbf{\ell} \right ) \mathclose{} T \mathopen{} \left ( \mathbf{\ell} \right ) \mathclose{}	\\
\mathcal{Q}^{\prime} \mathopen{} \left ( \mathbf{\ell} \right ) \mathclose{}	&	 = {}_{{2}} P \mathopen{} \left ( \mathbf{\ell} \right ) \mathclose{} \mathcal{Q} \mathopen{} \left ( \mathbf{\ell} \right ) \mathclose{}
\end{align*}

With the following standard equations to propagate from \(\mathcal{Q}\) to \(E, B\), with \(\mathcal{E} \equiv E + \mathrm{i} B\),

\begin{align*}
\mathcal{E} \mathopen{} \left ( \mathbf{\ell} \right ) \mathclose{}	&	 = \mathcal{Q} \mathopen{} \left ( \mathbf{\ell} \right ) \mathclose{} \mathrm{e}^{-2 \mathrm{i} \phi_{\mathbf{\ell}}}	\\
E \mathopen{} \left ( \mathbf{\ell} \right ) \mathclose{}	&	 = \frac{1}{2} \left[ \mathcal{E} \mathopen{} \left ( \mathbf{\ell} \right ) \mathclose{} + \bar{\mathcal{E}} \mathopen{} \left ( - \mathbf{\ell} \right ) \mathclose{} \right]	\\
B \mathopen{} \left ( \mathbf{\ell} \right ) \mathclose{}	&	 = \frac{1}{2 \mathrm{i}} \left[ \mathcal{E} \mathopen{} \left ( \mathbf{\ell} \right ) \mathclose{} - \bar{\mathcal{E}} \mathopen{} \left ( - \mathbf{\ell} \right ) \mathclose{} \right]
\end{align*}

With some derivation and simplification, and defining

\[\mathcal{A} ^\mathsf{T} \mathopen{} \left ( \mathbf{\ell} \right ) \mathclose{} = \begin{bmatrix}
T \mathopen{} \left ( \mathbf{\ell} \right ) \mathclose{}	&	E \mathopen{} \left ( \mathbf{\ell} \right ) \mathclose{}	&	B \mathopen{} \left ( \mathbf{\ell} \right ) \mathclose{}
\end{bmatrix}\]

and expanding the sum inside \(P\)\footnote{\(\Re , \Im\) below are the real part and imaginary part operators.},

\begin{align}
\mathcal{A}^{\prime} \mathopen{} \left ( \mathbf{\ell} \right ) \mathclose{}	&	 = \begin{bmatrix}
{}_{{0}} P \mathopen{} \left ( \mathbf{\ell} \right ) \mathclose{}	&	0	&	0	\\
0	&	\Re \mathopen{} \left \{ {}_{{2}} P \mathopen{} \left ( \mathbf{\ell} \right ) \mathclose{} \right \} \mathclose{}	&	- \Im \mathopen{} \left \{ {}_{{2}} P \mathopen{} \left ( \mathbf{\ell} \right ) \mathclose{} \right \} \mathclose{}	\\
0	&	\Im \mathopen{} \left \{ {}_{{2}} P \mathopen{} \left ( \mathbf{\ell} \right ) \mathclose{} \right \} \mathclose{}	&	\Re \mathopen{} \left \{ {}_{{2}} P \mathopen{} \left ( \mathbf{\ell} \right ) \mathclose{} \right \} \mathclose{}
\end{bmatrix} \mathcal{A} \mathopen{} \left ( \mathbf{\ell} \right ) \mathclose{} \label{APA}	\\
	&	 = \left[ \mathcal{X} + \sum_{i , j \neq i} \mathcal{X}^{i j} \mathcal{I}^{i j} \mathopen{} \left ( \mathbf{\ell} \right ) \mathclose{} \right] \mathcal{A} \mathopen{} \left ( \mathbf{\ell} \right ) \mathclose{} \label{ACrosstalk}	\\
\mathcal{I} \mathopen{} \left ( \mathbf{r} \cdot \mathbf{\ell}, \alpha \right ) \mathclose{}	&	 \equiv \begin{bmatrix}
\mathrm{e}^{\mathrm{i} \mathbf{r} \cdot \mathbf{\ell}}	&	0	&	0	\\
0	&	\cos \mathopen{} \left ( \mathbf{r} \cdot \mathbf{\ell} - 2 \alpha \right ) \mathclose{}	&	- \sin \mathopen{} \left ( \mathbf{r} \cdot \mathbf{\ell} - 2 \alpha \right ) \mathclose{}	\\
0	&	\sin \mathopen{} \left ( \mathbf{r} \cdot \mathbf{\ell} - 2 \alpha \right ) \mathclose{}	&	\cos \mathopen{} \left ( \mathbf{r} \cdot \mathbf{\ell} - 2 \alpha \right ) \mathclose{}
\end{bmatrix} \nonumber	\\
\mathcal{I}^{i j} \mathopen{} \left ( \mathbf{\ell} \right ) \mathclose{}	&	 \equiv \mathcal{I} \mathopen{} \left ( \mathbf{r}^{i \rightarrow j} \cdot \mathbf{\ell}, \alpha^{i \rightarrow j} \right ) \mathclose{} \nonumber
\end{align}

Writing the usual power spectrum \(C_\ell\) in terms of \(\mathcal{A}\),

\begin{align*}
\mathbb{C} \mathopen{} \left ( \mathbf{\ell} \right ) \mathclose{}	&	 \equiv \mathcal{A} \mathopen{} \left ( \mathbf{\ell} \right ) \mathclose{} \mathcal{A} ^\dagger \mathopen{} \left ( \mathbf{\ell} \right ) \mathclose{}	\\
\mathbb{C}_{\ell}	&	 \equiv \left\langle \mathbb{C} \mathopen{} \left ( \mathbf{\ell} \right ) \mathclose{} \right\rangle_{\mathbf{\hat{\ell}}} \equiv \left\langle \mathcal{A} \mathopen{} \left ( \mathbf{\ell} \right ) \mathclose{} \mathcal{A} ^\dagger \mathopen{} \left ( \mathbf{\ell} \right ) \mathclose{} \right\rangle_{\mathbf{\hat{\ell}}}	\\
	&	 = \begin{bmatrix}
C^{TT}_{\ell}	&	C^{TE}_{\ell}	&	C^{TB}_{\ell}	\\
C^{ET}_{\ell}	&	C^{EE}_{\ell}	&	C^{EB}_{\ell}	\\
C^{BT}_{\ell}	&	C^{BE}_{\ell}	&	C^{BB}_{\ell}
\end{bmatrix}
\end{align*}

Expanding using \(\eqref{ACrosstalk}\), we have\footnote{h.c.\ stands for Hermitian Conjugate.},

\begin{align}
\mathbb{C}^{\prime}_{\ell}	&	 = \mathcal{X}^{2} \mathbb{C}_{\ell} + \mathcal{X} \sum_{i , j \neq i} \mathcal{X}^{i j} \left\langle \mathcal{I}^{i j} \mathopen{} \left ( \mathbf{\ell} \right ) \mathclose{} \mathbb{C} \mathopen{} \left ( \mathbf{\ell} \right ) \mathclose{} + \text{ h.c.} \right\rangle_{\mathbf{\hat{\ell}}} + \dots \nonumber	\\
	&	 = \mathcal{X}^{2} \mathbb{C}_{\ell} + \mathcal{X} \sum_{i , j \neq i} \mathcal{X}^{i j} \left( \mathcal{I}^{i j}_{\ell} \mathbb{C}_{\ell} + \text{ h.c.} \right) + \dots \label{CIC}	\\
\mathcal{I}^{i j}_{\ell}	&	 \equiv \left\langle \mathcal{I}^{i j} \mathopen{} \left ( \mathbf{\ell} \right ) \mathclose{} \right\rangle_{\mathbf{\hat{\ell}}} \nonumber	\\
	&	 = J_{0} \mathopen{} \left ( r^{i \rightarrow j} \ell \right ) \mathclose{} \begin{bmatrix}
1	&	0	&	0	\\
0	&	\cos 2 \alpha^{i \rightarrow j}	&	\sin 2 \alpha^{i \rightarrow j}	\\
0	&	- \sin 2 \alpha^{i \rightarrow j}	&	\cos 2 \alpha^{i \rightarrow j}
\end{bmatrix} \nonumber	\\
	&	 = J_{0} \mathopen{} \left ( r^{i \rightarrow j} \ell \right ) \mathclose{} R^{-1}_{T} \mathopen{} \left ( 2 \alpha^{i \rightarrow j} \right ) \mathclose{} \nonumber	\\
\mathbb{C}^{\prime}_{\ell}	&	 = \mathcal{X}^{2} \mathbb{C}_{\ell} + \mathcal{X} \sum_{i , j \neq i} \mathcal{X}^{i j} J_{0} \mathopen{} \left ( r^{i \rightarrow j} \ell \right ) \mathclose{} \left[ R^{-1}_{T} \mathopen{} \left ( 2 \alpha^{i \rightarrow j} \right ) \mathclose{} \mathbb{C}_{\ell} + \text{ h.c.} \right] + \dots \label{CXJRC}
\end{align}

where the 2nd line has the statistical average applied on the inner \(\mathbb{C} \mathopen{} \left ( \mathbf{\ell} \right ) \mathclose{}\),
\(J_{\alpha}\)
is the Bessel functions of the first kind,
, \(R_T\) is the rotational matrix over the \(T\)-axis,
and only up to first-order in \(\mathcal{X}^{i j}\)
is kept.
Note that it means that to first order,
each of the crosstalk \(i, j\) components is modulated
by a Bessel function and rotated in the \(EB\)-plane.
Unlike a global mis-calibration,
this rotation is not global as the individual contributions are different.

\section{Flat sky, isotropic orientation, isotropic map coverage}\label{sec:crosstalk-derivation-flat-isotropic}

In the previous section,
the focal plane is assumed to be fixed in orientation
with respect to the (flat-)sky when we replace \(R\) by \(T\).
However, in a more realistic situation,
it will have different orientations as it scans across the same point on the sky.
That effect can be considered analytically if the scanning strategy is assumed to be
isotropic in the focal plane orientation (assumption (1c)).
Due to this isotropic averaging effect, the effective PSF is now the
azimuthal average of the original PSF,

\begin{align*}
P^{\prime} \mathopen{} \left ( r \right ) \mathclose{}	&	 \equiv \left\langle P \mathopen{} \left ( \mathbf{r} \right ) \mathclose{} \right\rangle_{\mathbf{\hat{r}}} = \frac{1}{2 \pi} \int_{0}^{2 \pi} P \mathopen{} \left ( \mathbf{r} \right ) \mathclose{} \,\mathrm{d} \theta	\\
	&	 = \frac{1}{2 \pi} \int_{0}^{2 \pi} \mathcal{X} \frac{\delta \mathopen{} \left ( r \right ) \mathclose{}}{2 \pi r} \\  & + \sum_{i , j \neq i} \mathcal{X}^{i j} \mathrm{e}^{- s \mathrm{i} \alpha^{i \rightarrow j}} \frac{\delta \mathopen{} \left ( r + r^{i \rightarrow j} \right ) \mathclose{} \delta \mathopen{} \left ( \theta + \theta^{i \rightarrow j} \right ) \mathclose{}}{r} \,\mathrm{d} \theta	\\
	&	 = \frac{1}{2 \pi r} \left[ \mathcal{X} \delta \mathopen{} \left ( r \right ) \mathclose{} + \sum_{i , j \neq i} \mathcal{X}^{i j} \mathrm{e}^{- s \mathrm{i} \alpha^{i \rightarrow j}} \delta \mathopen{} \left ( r + r^{i \rightarrow j} \right ) \mathclose{} \right]
\end{align*}

\begin{align}
\mathcal{F} \mathopen{} \left \{ P^{\prime} \right \} \mathclose{} \mathopen{} \left ( \ell \right ) \mathclose{}	&	 = \int_{0}^{2 \pi} \int_{0}^{\infty} P^{\prime} \mathopen{} \left ( r \right ) \mathclose{} \mathrm{e}^{- \mathrm{i} r \ell \cos \theta} r \,\mathrm{d} r \,\mathrm{d} \theta \nonumber	\\
{}_{{s}} P^{\prime}_{\ell}	&	 = \mathcal{X} + \sum_{i , j \neq i} \mathcal{X}^{i j} \mathrm{e}^{- s \mathrm{i} \alpha^{i \rightarrow j}} J_{0} \mathopen{} \left ( \ell r^{i \rightarrow j} \right ) \mathclose{} \label{sPlAveraged}
\end{align}

Note that as \(P\) is azimuthally symmetric, it depends on \(\ell\) only.
Substituting \(\eqref{sPlAveraged}\) into \(\eqref{APA}\),

\begin{equation}
\mathcal{A}^{\prime} \mathopen{} \left ( \mathbf{\ell} \right ) \mathclose{} = \left[ \mathcal{X} + \sum_{i , j \neq i} \mathcal{X}^{i j} \mathcal{I}^{i j}_{\ell} \right] \mathcal{A} \mathopen{} \left ( \mathbf{\ell} \right ) \mathclose{} \label{ACrosstalkAveraged}
\end{equation}

Following the same procedure as in last section,
it is found that the resultant power-spectra are the same as those in \(\eqref{CXJRC}\)
up to first order.

Comparing the results of this section with the last,
it is found that the 2d-power-spectra in the previous section \(\eqref{ACrosstalk}\)
is not statistically isotropic, while here \(\eqref{ACrosstalkAveraged}\) is.
Furthermore, while the resultant 1d-power-spectra are the same up to 1st order,
the 2nd order term is different,
where in the previous section it is an average of products of \(\mathcal{I}^{i j} \mathopen{} \left ( \mathbf{\ell} \right ) \mathclose{}\), here it is a product of \(\mathcal{I}^{i j}_{\ell}\) which is already averaged.
This means the 2nd order term can be written explicitly,

\begin{multline*}
\mathbb{C}^{\prime}_{\ell} = \mathcal{X}^{2} \mathbb{C}_{\ell} + \mathcal{X} \sum_{i , j \neq i} \mathcal{X}^{i j} J_{0} \mathopen{} \left ( r^{i \rightarrow j} \ell \right ) \mathclose{} \left[ R^{-1}_{T} \mathopen{} \left ( 2 \alpha^{i \rightarrow j} \right ) \mathclose{} \mathbb{C}_{\ell} + \text{ h.c.} \right] \\  + \sum_{i , j \neq i , m , n \neq m} \mathcal{X}^{i j} \mathcal{X}^{m n} J_{0} \mathopen{} \left ( r^{i \rightarrow j} \ell \right ) \mathclose{} J_{0} \mathopen{} \left ( r^{m \rightarrow n} \ell \right ) \mathclose{} \times \\ R^{-1}_{T} \mathopen{} \left ( 2 \alpha^{i \rightarrow j} \right ) \mathclose{} \mathbb{C}_{\ell} R_{T} \mathopen{} \left ( 2 \alpha^{m \rightarrow n} \right ) \mathclose{}
\end{multline*}

\section{Curved sky, isotropic orientation, isotropic map coverage}\label{sec:crosstalk-derivation-curve-isotropic}

To generalize the above results to curved sky,
it is necessary to digress to some definitions and theorems.

First, CMB power-spectra in curved sky are expanded in
spin-weighted spherical harmonics (SWSH)\@.
Using a view that is popularized by,\footnote{\citeproc{ref-boyle_how_2016}{Boyle, {``How Should Spin-Weighted Spherical Functions Be Defined?''}}} define

\begin{align*}
{}_{s} \mathcal{Y}_{\ell m} \mathopen{} \left ( \phi, \theta, \psi \right ) \mathclose{}	&	 \equiv \mathrm{e}^{- \mathrm{i} s \psi} {}_{s} Y_{\ell m} \mathopen{} \left ( \theta, \phi \right ) \mathclose{}	\\
{}_{s} \mathcal{Y}_{\ell m} \mathopen{} \left ( R \right ) \mathclose{}	&	 \equiv \mathrm{e}^{- \mathrm{i} s \psi} {}_{s} Y_{\ell m} \mathopen{} \left ( R \mathbf{\hat{z}} \right ) \mathclose{} ,\quad R \equiv R \mathopen{} \left ( \phi, \theta, \psi \right ) \mathclose{}	\\
	&	 = \left( -1 \right)^{m} \sqrt{\frac{2 \ell + 1}{4 \pi}} D^{\left( \ell \right)}_{- m s} \mathopen{} \left ( R \right ) \mathclose{}
\end{align*}

\({}_{s} Y_{l m}\) is a spin-\(s\) function on the 2-sphere,
that is, it changes upon a local transformation.
\({}_{s} \mathcal{Y}_{l m}\) is defined instead as a scalar function such that \({}_{s} \mathcal{Y}_{l m}\): \(\operatorname{SO(3)} \rightarrow \mathbb{C}\)\footnote{\(\operatorname{Spin(3)}\) for half-integer spin.}.
\(R\) is a rotation parameterized by the Euler angles\footnote{Using convention commonly used in quantum mechanics,
  namely, the z-y-z, right-handed, active transformation.},
and \(D\) is the Wigner-\(D\) matrix in Wigner's original sign convention.

The key to deriving a curved-sky solution is to define
a delta function on this \(\operatorname{SO} \mathopen{} \left ( 3 \right ) \mathclose{}\) space.
Rewriting the delta function in \(\mathbb{R}^{2}\) as
\(\delta^{R^{\prime}} \mathopen{} \left ( \mathbf{\hat{n}} \right ) \mathclose{} \equiv \delta^{\left( 2 \right)} \mathopen{} \left ( R^{\prime} \mathbf{\hat{n}} - \mathbf{\hat{z}} \right ) \mathclose{}\) where \(R'\) is the pointing.
Define a \(2\)-dimensional spin-\(s\) delta function in \(\operatorname{SO} \mathopen{} \left ( 3 \right ) \mathclose{}\),

\[{}_{{s}} \delta^{R^{\prime}} \mathopen{} \left ( R \right ) \mathclose{} \equiv \delta^{\left( 2 \right)} \mathopen{} \left ( R^{\prime} R = R_{z} \mathopen{} \left ( \alpha \right ) \mathclose{} \right ) \mathclose{} \mathrm{e}^{- \mathrm{i} s \psi}\]

Note that \(α\) is an arbitrary angle, and hence
after an integral over \(\operatorname{SO} \mathopen{} \left ( 3 \right ) \mathclose{}\),
there remains a \(1\)-dimensional integral.

The convolution with this delta function will then be\footnote{See \cref{sec:generalized-convolution-theorem}.},

\begin{align*}
\left( {}_{{s}} f \ast {}_{{s}} \delta^{R^{\prime}} \right) \mathopen{} \left ( R \right ) \mathclose{}	&	 = \frac{1}{2 \pi} \int_{R^{\prime \prime} \in \operatorname{SO} \mathopen{} \left ( 3 \right ) \mathclose{}} {}_{{s}} f \mathopen{} \left ( R^{\prime \prime} \right ) \mathclose{} {}_{{s}} \delta^{R^{\prime}} \mathopen{} \left ( R^{\prime \prime -1} R \right ) \mathclose{} \,\mathrm{d} R^{\prime \prime}	\\
{}_{{s}} \delta^{R^{\prime}} \mathopen{} \left ( R^{\prime \prime -1} R \right ) \mathclose{}	&	 = \delta^{\left( 2 \right)} \mathopen{} \left ( R^{\prime} R^{\prime \prime -1} R = R_{z} \mathopen{} \left ( \alpha \right ) \mathclose{} \right ) \mathclose{} \mathrm{e}^{- \mathrm{i} s \psi^{\prime \prime \prime}}	\\
R^{\prime \prime \prime}	&	 \equiv R^{\prime \prime -1} R	\\
	&	 = R^{\prime -1} R_{z} \mathopen{} \left ( \alpha \right ) \mathclose{}	\\
R \mathopen{} \left ( \phi, \theta, \psi \right ) \mathclose{}	&	 = R_{z} \mathopen{} \left ( \phi \right ) \mathclose{} R_{y} \mathopen{} \left ( \theta \right ) \mathclose{} R_{z} \mathopen{} \left ( \psi \right ) \mathclose{} \Rightarrow \psi^{\prime \prime \prime} = - \phi^{\prime} + \alpha	\\
R^{\prime \prime}	&	 = R R_{z} \mathopen{} \left ( - \alpha \right ) \mathclose{} R^{\prime}	\\
\left( {}_{{s}} f \ast {}_{{s}} \delta^{R^{\prime}} \right) \mathopen{} \left ( R \right ) \mathclose{}	&	 = \frac{1}{2 \pi} \oint {}_{{s}} f \mathopen{} \left ( R R_{z} \mathopen{} \left ( - \alpha \right ) \mathclose{} R^{\prime} \right ) \mathclose{} \mathrm{e}^{\mathrm{i} s \left( \phi^{\prime} - \alpha \right)} \,\mathrm{d} \alpha	\\
	&	 = \frac{1}{2 \pi} \oint {}_{{s}} f \mathopen{} \left ( R R^{\prime} \mathopen{} \left ( \phi^{\prime} = \phi^{\prime \prime} \right ) \mathclose{} \right ) \mathclose{} \mathrm{e}^{\mathrm{i} s \phi^{\prime \prime}} \,\mathrm{d} \phi^{\prime \prime}
\end{align*}

Specializing at \(R = I\),

\begin{align*}
\left( {}_{{s}} f \ast {}_{{s}} \delta^{R^{\prime}} \right) \mathopen{} \left ( I \right ) \mathclose{}	&	 = \frac{1}{2 \pi} \oint {}_{{s}} f \mathopen{} \left ( R^{\prime} \mathopen{} \left ( \phi^{\prime} = \phi^{\prime \prime} \right ) \mathclose{} \right ) \mathclose{} \mathrm{e}^{\mathrm{i} s \phi^{\prime \prime}} \,\mathrm{d} \phi^{\prime \prime}	\\
	&	 = \frac{1}{2 \pi} \oint {}_{{s}} f \mathopen{} \left ( R_{z} \mathopen{} \left ( \phi^{\prime \prime} \right ) \mathclose{} R^{\prime} \mathopen{} \left ( \phi^{\prime} = 0 \right ) \mathclose{} \right ) \mathclose{} \mathrm{e}^{\mathrm{i} s \phi^{\prime \prime}} \,\mathrm{d} \phi^{\prime \prime}
\end{align*}

The above shows that it is performing a spin-weighted azimuthal average around \(R'\).
That is, with assumption (1c, 3b) and eq.~\eqref{eq:crosstalk_map_pf}, \eqref{crosstalkMap},
the convolved map is, using the \(\operatorname{SO} \mathopen{} \left ( 3 \right ) \mathclose{}\) notation,

\begin{align*}
{}_{{s}} m^{\prime} \mathopen{} \left ( R \right ) \mathclose{}	&	 = \left( m \ast {}_{{s}} P \right) \mathopen{} \left ( R \right ) \mathclose{} \text{, where}	\\
{}_{{s}} P \mathopen{} \left ( R \right ) \mathclose{}	&	 \equiv \mathcal{X} {}_{{s}} \delta^{I} \mathopen{} \left ( R \right ) \mathclose{} + \sum_{i , j \neq i} \mathcal{X}^{i j} {}_{{s}} \delta^{i \rightarrow j} \mathopen{} \left ( R \right ) \mathclose{}	\\
{}_{{s}} \delta^{i \rightarrow j}	&	 \equiv {}_{{s}} \delta^{R^{i \rightarrow j}}
\end{align*}

Assumption (1c) is implicit when the convolution is used
which has an azimuthal averaging effect as explained above,
corresponding to a scanning strategy with isotropic focal plane orientation.

Then, the harmonic expansion of this delta function can be derived,

\begin{align*}
{}_{{s}} \delta^{R^{\prime}}_{\ell m}	&	 = \oint {}_{{s}} \delta^{R^{\prime}} \mathopen{} \left ( R \right ) \mathclose{} {}_{{s}} \bar{\mathcal{Y}}_{\ell m} \mathopen{} \left ( R \right ) \mathclose{} \,\mathrm{d} \Omega	\\
	&	 = \oint \delta^{\left( 2 \right)} \mathopen{} \left ( R^{\prime} R = R_{z} \mathopen{} \left ( \alpha \right ) \mathclose{} \right ) \mathclose{} {}_{{s}} \bar{Y}_{\ell m} \mathopen{} \left ( \theta, \phi \right ) \mathclose{} \,\mathrm{d} \Omega	\\
	&	 = {}_{{s}} \bar{Y}_{\ell m} \mathopen{} \left ( R^{\prime -1} \hat{z} \right ) \mathclose{} = {}_{{s}} \bar{\mathcal{Y}}_{\ell m} \mathopen{} \left ( R^{\prime -1} \right ) \mathclose{} \vert_{\phi^{\prime} = 0}	\\
	&	 = \left( -1 \right)^{- m} \sqrt{\frac{2 \ell + 1}{4 \pi}} d^{\ell}_{s , - m} \mathopen{} \left ( \theta^{\prime} \right ) \mathclose{} \mathrm{e}^{- \mathrm{i} s \psi^{\prime}}
\end{align*}

And that of \(P\),

\[{}_{{s}} \left( {}_{{s}} f \ast {}_{{s}} P \right)_{\ell m} = {}_{{s}} f_{\ell m} \left( \mathcal{X} + \sum_{i , j \neq i} \mathcal{X}^{i j} d^{\ell}_{s s} \mathopen{} \left ( \theta^{i \rightarrow j} \right ) \mathclose{} \mathrm{e}^{- \mathrm{i} s \alpha^{i \rightarrow j}} \right)\]

where \(d\) is the Wigner \(d\) matrix.

Equipped with this, the effect of crosstalk on the spherical harmonics can be derived,

\begin{align*}
a^{\prime}_{\ell m}	&	 \equiv {}_{{0}} \left( T \ast {}_{{0}} P \right)_{\ell m} = a_{\ell m} \left( \mathcal{X} + \sum_{i , j \neq i} \mathcal{X}^{i j} \mathcal{P}_{\ell} \mathopen{} \left ( \cos \theta^{i \rightarrow j} \right ) \mathclose{} \right)	\\
{}_{{2}} a^{\prime}_{\ell m}	&	 \equiv {}_{{2}} \left( \mathcal{Q} \ast {}_{{2}} P \right)_{\ell m} = {}_{{2}} a_{\ell m} \left( \mathcal{X} + \sum_{i , j \neq i} \mathcal{X}^{i j} d^{\ell}_{22} \mathopen{} \left ( \theta^{i \rightarrow j} \right ) \mathclose{} \mathrm{e}^{-2 \mathrm{i} \alpha^{i \rightarrow j}} \right)	\\
{}_{{-2}} a^{\prime}_{\ell m}	&	 = {}_{{-2}} a_{\ell m} \left( \mathcal{X} + \sum_{i , j \neq i} \mathcal{X}^{i j} d^{\ell}_{22} \mathopen{} \left ( \theta^{i \rightarrow j} \right ) \mathclose{} \mathrm{e}^{2 \mathrm{i} \alpha^{i \rightarrow j}} \right)
\end{align*}

Defining \(\mathcal{A} ^\mathsf{T}_{\ell m} = \begin{bmatrix}
a^{T}_{\ell m}	&	a^{E}_{\ell m}	&	a^{B}_{\ell m}
\end{bmatrix}\), the effect of crosstalk on \(\mathcal{A}\) would be,

\begin{align*}
\mathcal{A}_{\ell m}^{\prime}	&	 = \left[ \mathcal{X} + \sum_{i , j \neq i} \mathcal{X}^{i j} \mathcal{I}^{i j}_{\ell} \right] \mathcal{A}_{\ell m}	\\
\mathcal{I}_{\ell} \mathopen{} \left ( \theta, \alpha \right ) \mathclose{}	&	 \equiv \begin{bmatrix}
\mathcal{P}_{\ell} \mathopen{} \left ( \cos \theta \right ) \mathclose{}	&	0	&	0	\\
0	&	d^{\ell}_{22} \mathopen{} \left ( \theta \right ) \mathclose{} \cos 2 \alpha	&	d^{\ell}_{22} \mathopen{} \left ( \theta \right ) \mathclose{} \sin 2 \alpha	\\
0	&	- d^{\ell}_{22} \mathopen{} \left ( \theta \right ) \mathclose{} \sin 2 \alpha	&	d^{\ell}_{22} \mathopen{} \left ( \theta \right ) \mathclose{} \cos 2 \alpha
\end{bmatrix}	\\
\mathcal{I}^{i j}_{\ell}	&	 \equiv \mathcal{I} \mathopen{} \left ( \theta^{i \rightarrow j}, \alpha^{i \rightarrow j} \right ) \mathclose{}
\end{align*}

Where \(\mathcal{P}\) is the Legendre polynomials\footnote{\(\mathcal{P}\) is used instead of the usual \(P\)
  as it is already used for the PSF above.}.
Defining the 2-point power-spectra as,

\begin{align*}
\mathbb{C}_{\ell}	&	 \equiv \left\langle \mathcal{A}_{\ell m} \mathcal{A} ^\dagger_{\ell m} \right\rangle_{m}	\\
	&	 \equiv \begin{bmatrix}
C^{TT}_{\ell}	&	C^{TE}_{\ell}	&	C^{TB}_{\ell}	\\
C^{ET}_{\ell}	&	C^{EE}_{\ell}	&	C^{EB}_{\ell}	\\
C^{BT}_{\ell}	&	C^{BE}_{\ell}	&	C^{BB}_{\ell}
\end{bmatrix}
\end{align*}

The resultant power-spectra with crosstalk is computed as,

\begin{equation}
\mathbb{C}^{\prime}_{\ell} = \mathcal{X}^{2} \mathbb{C}_{\ell} + \left[ \mathcal{X} \sum_{i , j \neq i} \mathcal{X}^{i j} \mathcal{I}^{i j}_{\ell} \mathbb{C}_{\ell} + \text{h.c.} \right] + \sum_{i , j \neq i , m , n \neq m} \mathcal{X}^{i j} \mathcal{X}^{m n} \mathcal{I}^{i j}_{\ell} \mathbb{C}_{\ell} \mathcal{I}^{\dagger m n}_{\ell} \label{eq:crosstalk_spectra_curve_sky}
\end{equation}

This can be compared with the derivation in \cref{sec:crosstalk-derivation-flat}
and the result in \cref{sec:crosstalk-derivation-flat-isotropic}.
Their relationship with each other is more apparent, given
\(\mathcal{P}_\ell(\cos θ) = d^\ell_{00}(θ)\) and\footnote{\citeproc{ref-technology_us_nist_2010}{Technology (U.S.), \emph{NIST Handbook of Mathematical Functions Hardback and CD-ROM}}, eq.~14.15.11.}

\[\mathcal{P}_{\ell} \mathopen{} \left ( \cos \theta \right ) \mathclose{} = \sqrt{\frac{\theta}{\sin \theta}} J_{0} \mathopen{} \left ( \left( \ell + \frac{1}{2} \right) \theta \right ) \mathclose{} \left( 1 + \operatorname{O} \mathopen{} \left ( \frac{1}{\ell} \right ) \mathclose{} \right) ,\]

where \(θ \ll 1\) from assumption (3b).

Lastly, note that
keeping \(\mathbb{C}'_\ell\) up to first order
is easier to understand the crosstalk effect on the resultant
power-spectra qualitatively,
while \(\mathcal{A}'_{\ell m}\) is more useful to evaluate
the exact\footnote{Exact as long as the assumptions are valid.
  But it is still an approximation
  if the assumptions are only approximately true.}
crosstalk effect.
Notice how the crosstalk effect
is independent of the signal and
is completely captured
by the hardware information only,
which includes the crosstalk matrix,
detector noise, focal plane configuration, etc.
This means that it can be computed very fast
without performing computationally expensive TOD simulations.
This prediction can be used to design the hardware
to minimize crosstalk effect with very fast iterations.

Below we will expand \cref{eq:crosstalk_spectra_curve_sky}
in a special case which will be useful later.

\subsection{2-detectors with equal weights}\label{detectors-with-equal-weights}

If we take a 2-detector case
with symmetric crosstalk matrix and equal weighting between them,

\[X = \begin{bmatrix}
1	&	2 x	\\
2 x	&	1
\end{bmatrix} \Rightarrow \mathcal{X}^{i j} = \begin{bmatrix}
\frac{1}{2}	&	x	\\
x	&	\frac{1}{2}
\end{bmatrix} \Rightarrow \mathcal{X} = 1 .\]

Assuming \(C^{TB}_\ell = C^{EB}_\ell = 0\),
expanding \cref{eq:crosstalk_spectra_curve_sky},

\begin{equation}\begin{aligned}
C_{\ell }^{\prime\text{TT}}     &=  C_{\ell }^{\text{TT}} \left(x P_{\ell }+1\right){}^2    \\
C_{\ell }^{\prime\text{EE}}     &=  d_{22}^2 x^2 \sin ^2(2 \alpha ) C_{\ell }^{\text{BB}}+C_{\ell }^{\text{EE}} \left(d_{22} x \cos (2 \alpha )+1\right){}^2    \\
C_{\ell }^{\prime\text{BB}}     &=  C_{\ell }^{\text{BB}} \left(d_{22} x \cos (2 \alpha )+1\right){}^2+d_{22}^2 x^2 \sin ^2(2 \alpha ) C_{\ell }^{\text{EE}}    \\
C_{\ell }^{\prime\text{TE}}     &=  C_{\ell }^{\text{TE}} \left(x P_{\ell }+1\right) \left(d_{22} x \cos (2 \alpha )+1\right)   \\
C_{\ell }^{\prime\text{TB}}     &=  -x d_{22} \sin (2 \alpha ) C_{\ell }^{\text{TE}} \left(x P_{\ell }+1\right)   \\
C_{\ell }^{\prime\text{EB}}     &=  d_{22} x \sin (2 \alpha ) \left(C_{\ell }^{\text{BB}}-C_{\ell }^{\text{EE}}\right) \left(d_{22} x \cos (2 \alpha )+1\right) \label{eq:2-detector-crosstalk-spectra-expanded}
\end{aligned}\end{equation}

with some of the functions suppressed, such as
\(P_\ell \equiv P_\ell(\cos θ)\), \(d_{22} \equiv d^\ell_{22}(θ)\),
where \(θ\) is the angular distance on sky between the 2 detectors,
and \(α\) is the angle between the polarization of the 2 detectors.
The expression is more simplified in the case of flat-sky approximation
where \(P_\ell(\cos θ) \sim d^\ell_{22}(θ) \sim J_0(\ell θ)\) as mentioned above.

There are a few features worth pointing out here.
\(α\) does not affect \(TT\), as expected.
For \(EE, BB\), there is a symmetry in the equations in swapping between them.
Lastly, the \(E \rightarrow B\) leakage
induced by crosstalk systematics
is 2nd order in \(x\),
which can still be significant
as \(C^{EE}_\ell \gg C^{BB}_\ell\).

\chapter{Simulating crosstalk systematics in the time-domain}\label{sec:time-domain-crosstalk}

\chaptermark{Simulating crosstalk in the time-domain}

In this section,
we will focus on how to simulate crosstalk systematics
from time domain simulations.

\section{Crosstalk as a computational problem}\label{crosstalk-as-a-computational-problem}

Fundamentally,
crosstalk simulation is a matrix multiplication problem.
So let us take a look at
in the computational details first.

The software of choice in our simulation
is \href{https://github.com/hpc4cmb/toast}{Time Ordered Astrophysics Scalable Tools (TOAST), a software framework simulating and processing TODs in CMB observations and performing data compression into maps}.
It is designed with High-Performance Computing (HPC) in mind
and is \href{https://web.archive.org/web/20220930122248/https://www.nersc.gov/news-publications/nersc-news/science-news/2017/a-toast-for-next-generation-cmb-experiments/}{endorsed by NERSC},
and for example, is used in the processing of the Planck satellite mission data.

\begin{figure}
\centering
\pandocbounded{\includegraphics[keepaspectratio,alt={TOAST data distribution}]{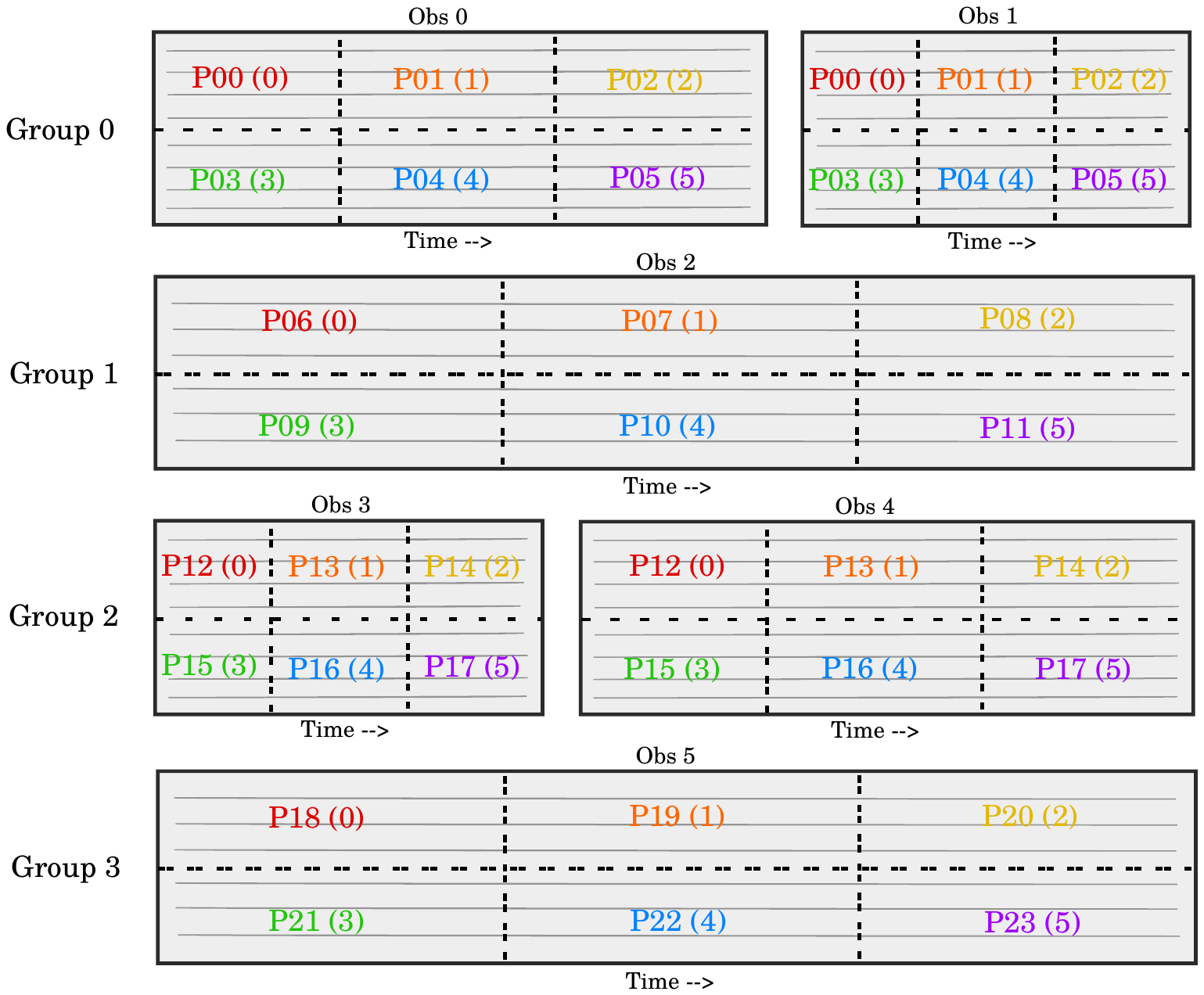}}
\caption[TOAST data distribution]{TOAST data distribution\footnotemark{}}\label{fig:toast_data_dist}
\end{figure}
\footnotetext{\citeproc{ref-kisner_toast_nodate}{Kisner, {``TOAST Workshop''}}.}

Needless to say,
TOAST is a massively parallel application.
For example,
\cref{fig:toast_data_dist} shows the data distribution
across MPI groups and ranks.
The main takeaway here is that
per group,
TODs from different detectors
are distributed across different ranks,
where each rank may hold data from a different total number of detectors.
Say if there are a total of 10 detectors
and 4 ranks,
some ranks may hold data from 2 or 3 detectors.

Our task is to apply a matrix multiplication
on all detectors under this data distribution pattern.
And from the hardware simulation using SQUID parameters in \cref{sec:crosstalk-SQUID},
the crosstalk matrix is dense\footnote{While the \(n\)-th diagonal in the crosstalk matrix is roughly \(n\)-th order,
  later we will see that the first few leading order effects
  can be approximately canceling.
  Hence, Without Loss of Generality (WLOG),
  it should be treated as dense
  and can only be treated as sparse
  when justified by simulations.}.
We should note that block-diagonal structure can be broken down into separate
dense matrix operations,
as we can loop over each
block (typically a wafer)
and perform the dense matrix multiplication
there.

Since we are having a massively parallel application,
and the fact that TOAST is already memory limited
for other operations such as the Madam maker,
the goal is to minimize the amount of MPI communications needed
and the memory overhead in computing the intermediate results.

\section{OpCrosstalk}\label{opcrosstalk}

Each detector owner,
the rank which holds the detector TOD locally,
eventually needs the TODs from all other detectors
and hence all other ranks within a group,
as per detector this is just a weighted sum of all detectors.
Because of this,
we have to loop over each detector, corresponding to a row in the crosstalk matrix.

Per row,
to minimize the MPI communication,
a weighted sum is first performed within a local rank
who owns a couple of detectors already,
and after that, a single MPI-Reduce is called
to perform the global sum, which costs around
\(O(\log n_\text{rank})\) communication time.

Since we loop over the number of detectors,
the total communication cost is
\(O(n_\text{det} \log n_\text{rank})\).

To minimize memory use,
note that as someone once said,
``the Python garbage collector is garbage.''
We use an internal facility from TOAST
to create new arrays.
While it has a Numpy array interface,
it is allocated in C++ directly
where we can manually free them
to avoid memory overhead.

With this,
\((2 n_\text{local dets} + 1) \times n_\text{samples}\) memory is needed,
comparing to \(n_\text{local dets} \times n_\text{samples}\) originally needed,
this requires double memory
as each rank needs to hold the intermediate results
during the whole matrix multiplication process.

However,
with an implementation detail in TOAST
that typically \(n_\text{obs}\) observations are performed
in a run, such as \(n_\text{obs} = 365\) for a year of satellite mission,
this doubling of memory requirements
only occur within each observation,
so the actual memory increase is negligible.

Currently,
technological limit places \(n_\text{det}\) per SQUID to be around \textasciitilde100.
And since \(n_\text{rank} \leq n_\text{det}\), the computational aspect
appears to be well under control.
An example run below with \(n_\text{det} \sim 60\)
takes a negligible amount of compute time compared to the total duration
with other expensive operations such as the Madam mapmaker.

OpCrosstalk is also general purpose
and is not tied to a particular CMB experiment,
and will benefit any future CMB experiments
that use TOAST such as SO and CMB-S4.
The pull request has been merged to TOAST and is available in \url{https://github.com/hpc4cmb/toast/pull/380}.

\chapter{Crosstalk systematics simulations with hardware based mitigation}\label{sec:crosstalk-simulations}

\chaptermark{Crosstalk and hardware based mitigation}

\section{Introduction}\label{introduction-3}

This section puts everything together and demonstrates the simulation of
the crosstalk systematics using the LiteBIRD Instrument Model (IMo) v1
as an example.

\subsection{Terminologies \& conventions of detectors}\label{terminologies-conventions-of-detectors}

\begin{figure}
\centering
\includesvg[width=1\linewidth,height=\textheight,keepaspectratio]{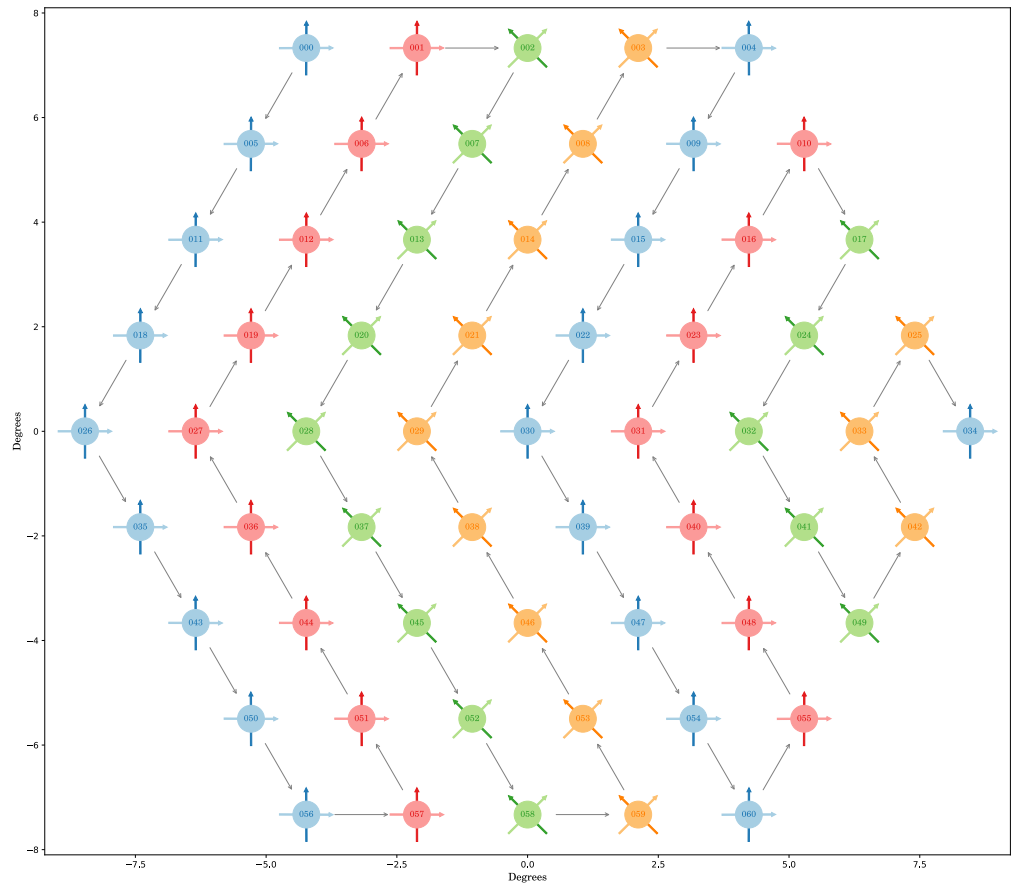}
\caption{An example focal plane
using the MFT central wafer at \(\SI{140}{\giga\hertz}\).}\label{fig:ArrowTopLeftFirst}
\end{figure}

\Cref{fig:ArrowTopLeftFirst} shows an example focal plane
using the MFT central wafer at \(\SI{140}{\giga\hertz}\).

\needspace{8\baselineskip}

\begin{longtable}[]{@{}lll@{}}
\caption{\(Q\)/\(U\)-pixel vs.~Top/Bottom-ness}\label{tbl:qu-tp}\tabularnewline
\toprule\noalign{}
& \(T\) & \(B\) \\
\midrule\noalign{}
\endfirsthead
\toprule\noalign{}
& \(T\) & \(B\) \\
\midrule\noalign{}
\endhead
\bottomrule\noalign{}
\endlastfoot
\(Q\) & \(\SI{0}{\degree}\) & \(\SI{90}{\degree}\) \\
\(U\) & \(\SI{45}{\degree}\) & \(\SI{135}{\degree}\) \\
color & light & dark \\
\end{longtable}

\needspace{7\baselineskip}

\begin{longtable}[]{@{}lll@{}}
\caption{\(Q\)/\(U\)-pixel \& handedness colors}\label{tbl:qu-handedness}\tabularnewline
\toprule\noalign{}
& \(A\) & \(B\) \\
\midrule\noalign{}
\endfirsthead
\toprule\noalign{}
& \(A\) & \(B\) \\
\midrule\noalign{}
\endhead
\bottomrule\noalign{}
\endlastfoot
\(Q\) & Blue & Red \\
\(U\) & Green & Orange \\
\end{longtable}

\(Q\) detectors are represented by arrows at \(\SI{0}{\degree}, \SI{90}{\degree}\),
and \(U\) detectors are represented by arrows at \(\SI{45}{\degree}, \SI{135}{\degree}\).
They are color-coded in brightness as summarized in \cref{tbl:qu-tp}.
Then there are 2 kinds of detector handedness, where one is
a mirror image of another. They are simply named \(A, B\).
The 4 combinations of \(Q, U\) and \(A, B\) are color-coded according to
\cref{tbl:qu-handedness}.
Hence, the colors blue, red, green, orange correspond to \texttt{QA}, \texttt{QB}, \texttt{UA}, \texttt{UB}.

Also note that \(n\)-th nearest neighbor can be in focal plane (angular distance) or in SQUID (frequency) depending on the context.

\section{2-detectors simulations}\label{detectors-simulations}

\subsection{Decomposition of crosstalk matrix}\label{decomposition-of-crosstalk-matrix}

First, a constant crosstalk matrix with only non-zero elements in the nearest neighbor (in SQUID) is considered.
In the limit of infinitely long SQUID, it can be decomposed in
a superposition of 2-detectors crosstalk matrix like this:

\[\begin{bmatrix}
1/2     &   x       &   0       &   \dots   \\
x       &   1/2     &   0       &   \dots   \\
0       &   0       &   0       &   \dots   \\
\vdots  &   \vdots  &   \vdots  &   \ddots  \\
\end{bmatrix}
+ \begin{bmatrix}
0       &   0       &   0       &   \dots   \\
0       &   1/2     &   x       &   \dots   \\
0       &   x       &   1/2     &   \dots   \\
\vdots  &   \vdots  &   \vdots  &   \ddots  \\
\end{bmatrix}
+ \dots
= \begin{bmatrix}
1/2     &   x       &   0       &   0       &   \dots   &   \dots   \\
x       &   1       &   x       &   0       &   \dots   &   \dots   \\
0       &   x       &   1       &   x       &   \dots   &   \dots   \\
0       &   0       &   x       &   1       &   \ddots  &   \dots   \\
\vdots  &   \vdots  &   \vdots  &   \ddots  &   \ddots  &   x       \\
\vdots  &   \vdots  &   \vdots  &   0       &   x       &   1/2     \\
\end{bmatrix}\]

The decomposition is exact in TOD\@.
It should be approximately true in map-domain and eventually power-spectra-domain as well
if \(x \ll 1\).

Moreover, in a realistic crosstalk matrix simulated from SQUID,
it is approximately true that the next nearest neighbors (in SQUID) are of lower orders,
whereas the nearest neighbor crosstalk is of the same order of magnitude. (See more in Montgomery\footnote{\citeproc{ref-montgomery_statistical_2019}{{``Statistical Scatter in LC Resonators and LiteBIRD Crosstalk Simulations''}}.} or below.)

This means that if the 2-detector case is understood, it can be generalized to a long SQUID approximately.

\subsection{2-detectors simulation results}\label{detectors-simulation-results}

Below are the systematic effects of a pair of detectors in \(\frac{C_\ell^\text{crosstalk}}{C_\ell}\),

\begin{figure}
\centering
\includesvg[width=2.08333in,height=\textheight,keepaspectratio]{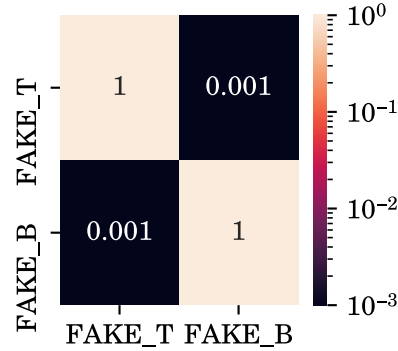}
\caption{2-detectors crosstalk matrix}
\end{figure}

\begin{itemize}
\tightlist
\item
  Using IMO v1.0 parameters,
\item
  Modified from the central pixel of MFT at \(\SI{140}{\giga\hertz}\),
\item
  Scanning from a \texttt{synfast} input map using Planck 2018 best-fit ΛCDM cosmology with 183 days of observations,
\item
  At an angular distance of \(\SI{2.114}{\degree}\) (corresponding to the mean lattice angular distance of the MFT hexagonal wafer),
\item
  varying the polarization angles of the 1st and 2nd detectors.
\end{itemize}

Below, 1st number is 1st detector polarization angle, last number is 2nd detector polarization angle.

\begin{pandoccrossrefsubfigures}

\begin{subfigure}{1.0\linewidth}\centering\pandocbounded{\includegraphics[keepaspectratio,alt={BB}]{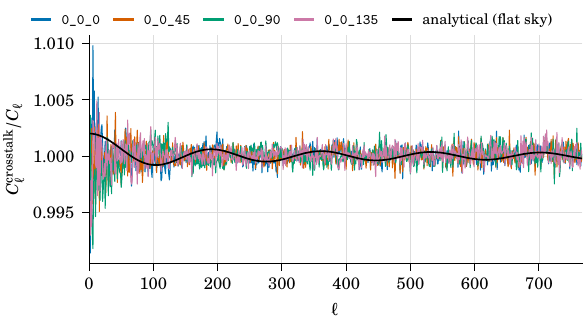}}\caption{\(BB\)}\end{subfigure}%

\begin{subfigure}{1.0\linewidth}\centering\pandocbounded{\includegraphics[keepaspectratio,alt={EE}]{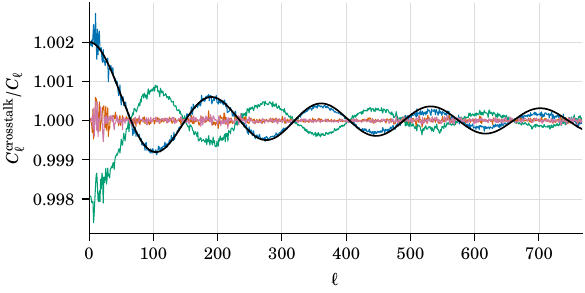}}\caption{\(EE\)}\end{subfigure}%

\begin{subfigure}{1.0\linewidth}\centering\pandocbounded{\includegraphics[keepaspectratio,alt={TT}]{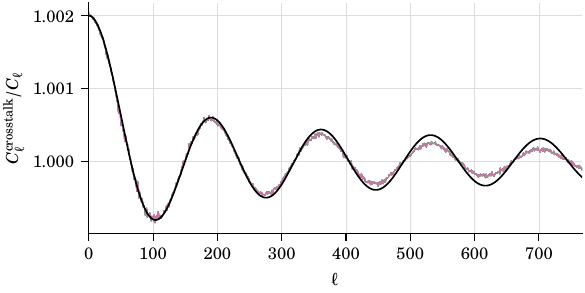}}\caption{\(TT\)}\end{subfigure}%

\caption{2-detectors crosstalk systematics expressed as \(\frac{C_\ell^\text{crosstalk}}{C_\ell}\).
The legend in (a) applies to all panels.}\label{fig:2-detectors-crosstalk}

\end{pandoccrossrefsubfigures}

\cref{fig:2-detectors-crosstalk} shows
the result of the simulations,
presented as a transfer function of the power-spectra.
These can be compared to the analytic formula in \cref{sec:crosstalk-derivation-curve-isotropic} shown in the plot.
There is a high level of agreement,
with the exception of damping towards high-\(\ell\),
which is an effect similar to the PWF as
the analytical model assumes a delta function,
while in mapmaking, finite size pixels are used.

\subsection{Lesson learnt---rule of thumb of wafer ordering}\label{lesson-learntrule-of-thumb-of-wafer-ordering}

Combining the empirical simulation results
and the analytical behavior expected from~\eqref{eq:2-detector-crosstalk-spectra-expanded},
the crosstalk systematics have minimal effects
on the resultant \(EE, EB, BB\) power-spectra
in these special cases,

\begin{enumerate}
\def\labelenumi{\arabic{enumi}.}
\tightlist
\item
  \(\theta = 0\), or
\item
  for \(\alpha\) to be either

  \begin{enumerate}
  \def\labelenumii{\arabic{enumii}.}
  \tightlist
  \item
    \(\alpha = \SI{45}{\degree}\) or \(\SI{135}{\degree} \rightarrow \frac{C_\ell^\text{crosstalk}}{C_\ell} = 0\)
  \item
    \(\frac{C_\ell^\text{crosstalk} (\alpha = \SI{0}{\degree})}{C_\ell} = -\frac{C_\ell^\text{crosstalk} (\alpha = \SI{90}{\degree})}{C_\ell}\)
  \end{enumerate}
\end{enumerate}

Now, we can apply what we learnt in wafer \& SQUID \emph{design} using the following empirical rules.
\emph{Design} here means that given a wafer, there is a considerable degree of freedom in how to wire the detectors together (ordering in SQUID) and how to \emph{walk} the wafer.
Here is an empirical algorithm in doing so.

\begin{Corollary}\label{cor-wafer-walking-rule}
Rule-of-thumb to design the wafer to minimize crosstalk systematics.

\begin{enumerate}
\def\labelenumi{\arabic{enumi}.}
\item
  Walk within the same pixel first, as this has no bias (\(\theta = 0\)).
\item
  Then walk to the nearest neighbors on the focal plane (note that in a hexagonal wafer this always injects systematics at the same angular scale \(\theta\), but a rectangular one has 2 angular scales). This rule comes from limitations in how they can be wired physically.

  Calculate \(\alpha\), the angle turned when walking from one pixel to another, following the following preferences:

  \begin{enumerate}
  \def\labelenumii{\alph{enumii})}
  \item
    Prefer \(\SI{45}{\degree}\) or \(\SI{135}{\degree}\), as~\eqref{eq:2-detector-crosstalk-spectra-expanded} shows this has no bias up to 1st order\footnote{Note, however, that this maximizes the \(E \rightarrow B\) leakage at 2nd order in \(x\). In practice, which rule is more advantageous depends on the separation of scales comparing higher-order crosstalk effects to the \(E\) to \(B\) hierarchy.}, otherwise,
  \item
    if a pair has a rotation of \(\SI{0}{\degree}\) (\(\cos 0 = 1\)), balance it by another pair that rotates by \(\SI{90}{\degree}\) (\(\cos \SI{90}{\degree} = -1\)). Essentially, for the whole SQUID, \(\sum_i \cos (2 \alpha_i) = 0\) is desired as a rule of thumb\footnote{Assuming 1st order crosstalk is roughly constant.}. If there are multiple angular scales in walking the focal plane (for example, in LFT), balance this for each angular scale individually.
  \end{enumerate}
\end{enumerate}

\end{Corollary}

\section{\texorpdfstring{Full wafer simulations---Realistic wafer simulation with the example of the central wafer of MFT at \(\SI{140}{\giga\hertz}\) with 2 SQUIDs each having 61 detectors}{Full wafer simulations---Realistic wafer simulation with the example of the central wafer of MFT at \textbackslash SI\{140\}\{\textbackslash giga\textbackslash hertz\} with 2 SQUIDs each having 61 detectors}}\label{full-wafer-simulationsrealistic-wafer-simulation-with-the-example-of-the-central-wafer-of-mft-at-si140gigahertz-with-2-squids-each-having-61-detectors}

\subsection{Wafer mockups}\label{wafer-mockups}

\begin{pandoccrossrefsubfigures}

\begin{subfigure}{0.5\linewidth}\centering\pandocbounded{\includesvg[keepaspectratio]{media-dissertation/crosstalk-order-angle/ArrowTopLeftFirst-order-angle.svg}}\caption{ArrowTopLeftFirst\label{ArrowTopLeftFirst}}\end{subfigure}%
\begin{subfigure}{0.5\linewidth}\centering\pandocbounded{\includesvg[keepaspectratio]{LiteBIRD-presentations/media/crosstalk-order-angle/ArrowTopLeftFirst-wire_TB.svg}}\caption{ArrowTopLeftFirst-wire\_TB\label{ArrowTopLeftFirst-wire_TB}}\end{subfigure}%

\begin{subfigure}{0.5\linewidth}\centering\pandocbounded{\includesvg[keepaspectratio]{LiteBIRD-presentations/media/crosstalk-order-angle/ArrowTopLeftFirst_TB-alternate-anti-alignment-wire_TB.svg}}\caption{ArrowTopLeftFirst\_TB-alternate-anti-alignment-wire\_TB\label{ArrowTopLeftFirst_TB-alternate-anti-alignment-wire_TB}}\end{subfigure}%
\begin{subfigure}{0.5\linewidth}\centering\pandocbounded{\includesvg[keepaspectratio]{LiteBIRD-presentations/media/crosstalk-order-angle/ArrowTopLeftFirst_TB-anti-alignment-wire_TB.svg}}\caption{ArrowTopLeftFirst\_TB-anti-alignment-wire\_TB\label{ArrowTopLeftFirst_TB-anti-alignment-wire_TB}}\end{subfigure}%

\caption{\texttt{ArrowTopLeftFirst}: a wafer design to walk the wafer.
The top-left plot shows the ordering in pixel order only.
The other 3 plots show the \(T\)/\(B\)-wiring explicitly,
each with a different \(T\)/\(B\) ordering.}\label{fig:ArrowTopLeftFirstAll}

\end{pandoccrossrefsubfigures}

\begin{pandoccrossrefsubfigures}

\begin{subfigure}{0.5\linewidth}\centering\pandocbounded{\includesvg[keepaspectratio]{LiteBIRD-presentations/media/crosstalk-order-angle/AlternateQU.svg}}\caption{AlternateQU\label{AlternateQU}}\end{subfigure}%
\begin{subfigure}{0.5\linewidth}\centering\pandocbounded{\includesvg[keepaspectratio]{LiteBIRD-presentations/media/crosstalk-order-angle/AlternateQU-wire_TB.svg}}\caption{AlternateQU-wire\_TB\label{AlternateQU-wire_TB}}\end{subfigure}%

\begin{subfigure}{0.5\linewidth}\centering\pandocbounded{\includesvg[keepaspectratio]{LiteBIRD-presentations/media/crosstalk-order-angle/AlternateQU_TB-alternate-anti-alignment-wire_TB.svg}}\caption{AlternateQU\_TB-alternate-anti-alignment-wire\_TB\label{AlternateQU_TB-alternate-anti-alignment-wire_TB}}\end{subfigure}%
\begin{subfigure}{0.5\linewidth}\centering\pandocbounded{\includesvg[keepaspectratio]{LiteBIRD-presentations/media/crosstalk-order-angle/AlternateQU_TB-anti-alignment-wire_TB.svg}}\caption{AlternateQU\_TB-anti-alignment-wire\_TB\label{AlternateQU_TB-anti-alignment-wire_TB}}\end{subfigure}%

\caption{\texttt{AlternateQU}: a wafer design to walk the wafer.
The top-left plot shows the ordering in pixel order only.
The other 3 plots show the \(T\)/\(B\)-wiring explicitly,
each with a different \(T\)/\(B\) ordering.}\label{fig:AlternateQUAll}

\end{pandoccrossrefsubfigures}

\begin{pandoccrossrefsubfigures}

\begin{subfigure}{0.5\linewidth}\centering\pandocbounded{\includesvg[keepaspectratio]{LiteBIRD-presentations/media/crosstalk-order-angle/SpiralCounterclockwise_QUAB.svg}}\caption{SpiralCounterclockwise\_QUAB\label{SpiralCounterclockwise_QUAB}}\end{subfigure}%
\begin{subfigure}{0.5\linewidth}\centering\pandocbounded{\includesvg[keepaspectratio]{LiteBIRD-presentations/media/crosstalk-order-angle/SpiralCounterclockwise_QUAB-wire_TB.svg}}\caption{SpiralCounterclockwise\_QUAB-wire\_TB\label{SpiralCounterclockwise_QUAB-wire_TB}}\end{subfigure}%

\begin{subfigure}{0.5\linewidth}\centering\pandocbounded{\includesvg[keepaspectratio]{LiteBIRD-presentations/media/crosstalk-order-angle/SpiralCounterclockwise_QUAB_TB-alternate-anti-alignment-wire_TB.svg}}\caption{SpiralCounterclockwise\_QUAB\_TB-alternate-anti-alignment-wire\_TB\label{SpiralCounterclockwise_QUAB_TB-alternate-anti-alignment-wire_TB}}\end{subfigure}%
\begin{subfigure}{0.5\linewidth}\centering\pandocbounded{\includesvg[keepaspectratio]{LiteBIRD-presentations/media/crosstalk-order-angle/SpiralCounterclockwise_QUAB_TB-anti-alignment-wire_TB.svg}}\caption{SpiralCounterclockwise\_QUAB\_TB-anti-alignment-wire\_TB\label{SpiralCounterclockwise_QUAB_TB-anti-alignment-wire_TB}}\end{subfigure}%

\caption{\texttt{SpiralCounterclockwise\_\allowbreak{}QUAB}: a wafer design to walk the wafer.
The top-left plot shows the ordering in pixel order only.
The other 3 plots show the \(T\)/\(B\)-wiring explicitly,
each with a different \(T\)/\(B\) ordering.}\label{fig:SpiralCounterclockwise_QUABAll}

\end{pandoccrossrefsubfigures}

In \cref{fig:ArrowTopLeftFirstAll},
\cref{fig:AlternateQUAll},
\cref{fig:SpiralCounterclockwise_QUABAll},
we look at 3 different wafer mockups \texttt{ArrowTopLeftFirst}, \texttt{AlternateQU}, \& \texttt{SpiralCounterclockwise\_\allowbreak{}QUAB}.

In the first 2 cases, \texttt{ArrowTopLeftFirst}, \texttt{AlternateQU}, the IMO v1 MFT central wafer at \(\SI{140}{\giga\hertz}\) is taken as is.

In \texttt{ArrowTopLeftFirst}, it walks the focal plane by following the same \texttt{QA} line first, etc., that is, most of the walking across pixels happens at \(\SI{0}{\degree}\) or \(\SI{90}{\degree}\), this follows Corollary~\ref{cor-wafer-walking-rule} 2b.

In \texttt{AlternateQU}, when crossing between pixels, it tries to go from a \(Q\)-pixel to a \(U\)-pixel as much as possible.
This follows Corollary~\ref{cor-wafer-walking-rule} 2a.
Note that since the wafer has more Q-pixels than U-pixels,
the last few detectors cannot fulfill this.

In the last case \texttt{SpiralCounterclockwise\_\allowbreak{}QUAB},
the wafer is redesigned in a way that given a SQUID-ordering on the focal-plane,
the ordering is alternated between \(Q\) and \(U\)-pixels.
This follows Corollary~\ref{cor-wafer-walking-rule} 2a exactly by construction.
Note that this has a considerable degree of freedom
and the only requirement is that the numbers of \(Q\) and \(U\)-pixels
are balanced.

``TB-alignment'' represents when walking to the next detector, how to wire between the Top \& Bottom detectors.

\begin{description}
\tightlist
\item[normal]
It always walks from a Bottom detector to a Top detector in a new location.

\(0T \rightarrow 0B \rightarrow 1T \rightarrow 1B \rightarrow 2T \rightarrow 2B \rightarrow 3T \rightarrow 3B \ldots\)

This results in \(\sum_i \cos (2 \alpha_i)\) maximally negative.
\item[TB-anti-alignment]
It always walks from a Bottom detector to a Bottom detector in a new location, similarly for Top to Top.

\(0T \rightarrow 0B \rightarrow 1B \rightarrow 1T \rightarrow 2T \rightarrow 2B \rightarrow 3B \rightarrow 3T \ldots\)

This results in \(\sum_i \cos (2 \alpha_i)\) maximally positive.
\item[TB-alternate-anti-alignment]
It mixes the above 2 strategies alternatively.

\(0T \rightarrow 0B \rightarrow 1B \rightarrow 1T \rightarrow 2B \rightarrow 2T \rightarrow 3T \rightarrow 3B \ldots\)

This results in \(\sum_i \cos (2 \alpha_i) \approx 0\) by alternating between the 2.
\end{description}

These serve as examples. Other orderings following similar schemes are explored, which seem to indicate the general characteristics in the final power-spectra do not depend on the exact details of how they are ordered, as long as Corollary~\ref{cor-wafer-walking-rule} is followed.

\subsection{Full wafer simulation results}\label{full-wafer-simulation-results}

Similar to the 2-detectors case, but now with the wafer defined above, with 122 detectors and 4 days of observations (with a sky coverage around \(60–\SI{70}{\percent}\).)

For the crosstalk matrix, the following SQUID parameters are used, this is similar to the SPT-3G design,\footnote{\citeproc{ref-montgomery_performance_2020}{Montgomery et al., {``Performance and Characterization of the SPT-3G Digital Frequency Multiplexed Readout System Using an Improved Noise and Crosstalk Model''}}.} slightly modified for the LiteBIRD case provided by Joshua Montgomery, with \(L_\text{com}\) chosen with the expectation of using \(\unit{\micro\kelvin}\)-SQUID.

\begin{align*}
R_\text{TES}    &= \SI{0.6}{\ohm}           \\
r_s             &= \SI{0.3}{\ohm}           \\
L               &= \SI{60}{\micro\henry}    \\
L_\text{com}    &= \SI{9}{\nano\henry}      \\
f_\text{min}    &= \SI{1.69}{\mega\hertz}   \\
f_\text{max}    &= \SI{5.49}{\mega\hertz}
\end{align*}

with 68 detectors with SQUID frequencies spacing linearly between its min/max.

\begin{figure}
\centering
\includesvg[width=1\linewidth,height=\textheight,keepaspectratio]{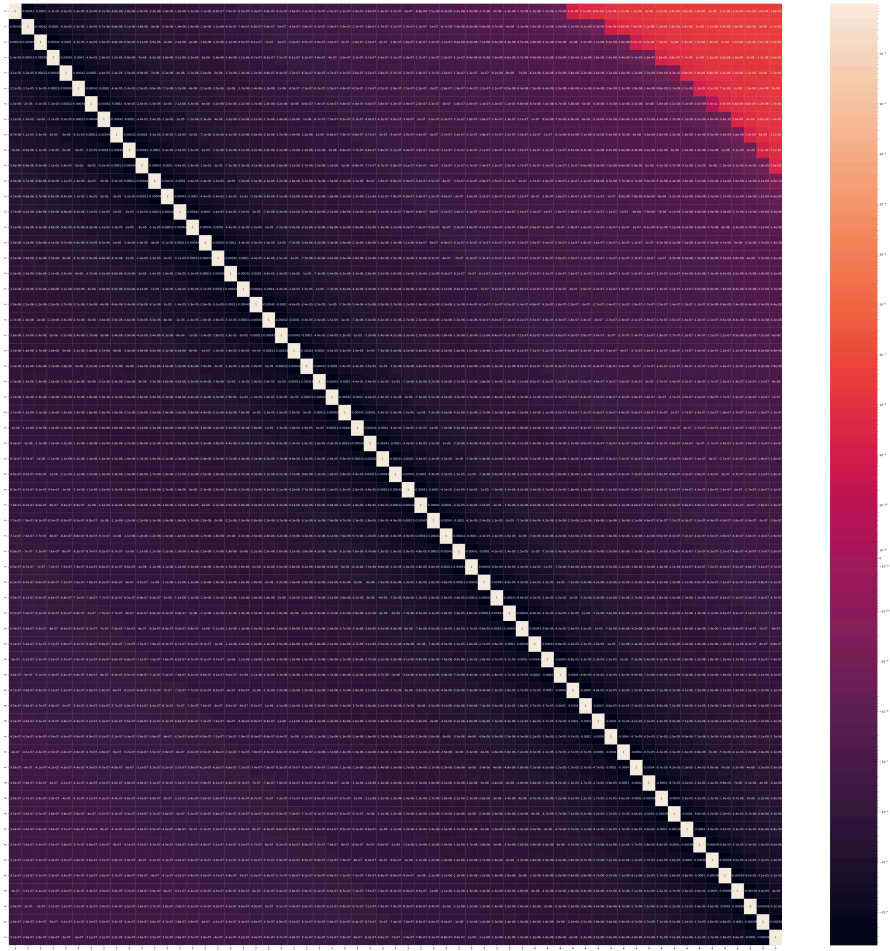}
\caption{Analytical crosstalk matrix simulated from SQUID parameters}\label{fig:SQUID-crosstalk}
\end{figure}

Then the highest 7 frequencies are removed to have a crosstalk matrix of 61 detectors, as shown in \cref{fig:SQUID-crosstalk}.

And 2 such SQUIDs are put in the focal plane in the previous sections. That is, the first 61 detectors belong to the 1st SQUID, the last 61 to the 2nd SQUID.

\begin{figure}
\centering
\pandocbounded{\includegraphics[keepaspectratio,alt={Full wafer crosstalk systematics expressed as \textbackslash frac\{C\_\textbackslash ell\^{}\textbackslash text\{crosstalk\}\}\{C\_\textbackslash ell\}, a row per power-spectrum and a column per wafer mockup; the colours are the T/B-wirings.}]{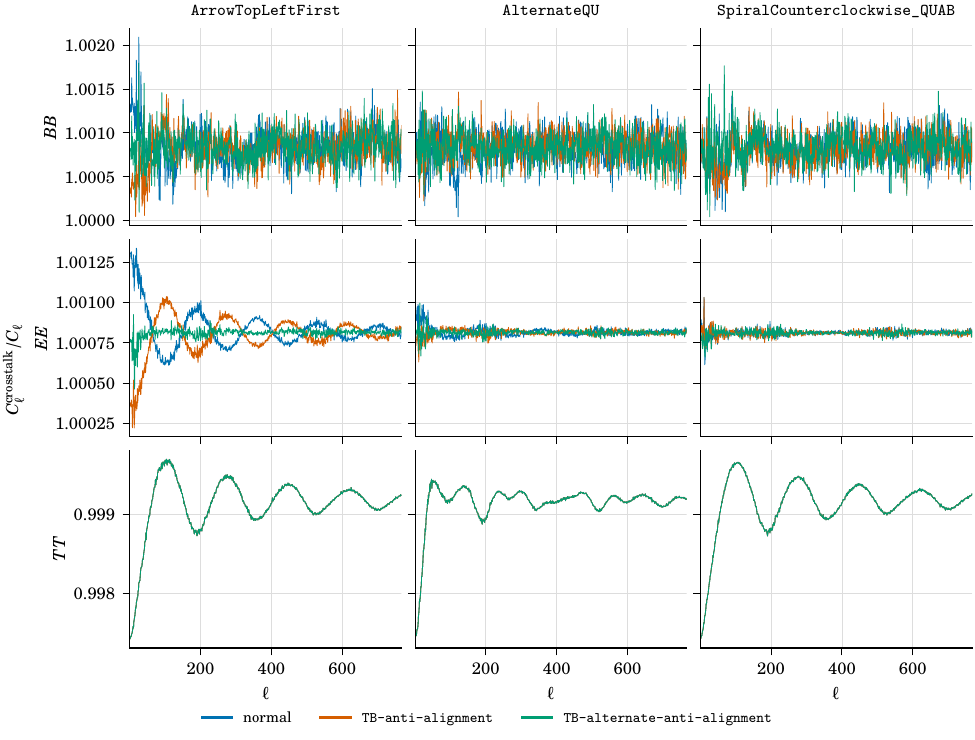}}
\caption{Full wafer crosstalk systematics expressed as \(\frac{C_\ell^\text{crosstalk}}{C_\ell}\), a row per power-spectrum and a column per wafer mockup; the colours are the \(T\)/\(B\)-wirings.}\label{fig:full-wafer-simulation}
\end{figure}

From \cref{fig:full-wafer-simulation},
one can observe,

\begin{itemize}
\item
  in polarizations,

  \begin{itemize}
  \item
    the \(B\)-mode power-spectra are noisier compared to the \(E\)-mode, but they follow a similar trend.
  \item
    For walking order, \texttt{SpiralCounterclockwise\_\allowbreak{}QUAB} \textgreater{} \texttt{AlternateQU} \textgreater{} \texttt{ArrowTopLeftFirst}. It is basically determined by the amount of \(Q\)-\(U\) pixel crossing when walking the wafer.
  \item
    for \(TB\)-alignment, \texttt{TB-\allowbreak{}anti-\allowbreak{}alignment} is as bad as normal-ordering, and \texttt{TB-\allowbreak{}alternate-\allowbreak{}anti-\allowbreak{}alignment}-ordering is the best. As \(\sum_i \cos (2 \alpha_i) \approx 0\).
  \end{itemize}
\item
  in \(TT\),

  \begin{itemize}
  \tightlist
  \item
    \texttt{AlternateQU} is better. As systematics are injected at different angular scales, so the Legendre polynomial term partially cancelled between different contributions. In fact, if we walk the wafer randomly, this effect would be maximized.
  \end{itemize}
\end{itemize}

\section{Conclusions}\label{conclusions}

The \(QU\)-ordering optimization effectively removes crosstalk systematic bias in the resultant \(E\)-mode and \(B\)-mode power-spectra.

One reason that the \texttt{SpiralCounterclockwise\_\allowbreak{}QUAB} family is so effective is that we can consider the ordering in a SQUID (below we use \texttt{SpiralCounterclockwise\_\allowbreak{}QUAB\_\allowbreak{}TB-\allowbreak{}alternate-\allowbreak{}anti-\allowbreak{}alignment-\allowbreak{}wire\_\allowbreak{}TB} as an explicit example),

\[1QAT \rightarrow 1QAB \rightarrow 2UAB \rightarrow 2UAT \rightarrow 3QBT \rightarrow 3QBB \rightarrow 4UBB \rightarrow 4UBT \dots\]

Note that not only the nearest neighbor (in SQUID) is a nearest neighbor (in focal plane) crossing between \(Q\) \& \(U\) pixels (such as \(1QAB\)-\(2UAB\)), but it is true for the next nearest neighbor too (such as \(1QAT\)--\(2UAB\)). So effectively this has mitigated the leading 2 orders of crosstalk in the matrices.

It is also superior in terms of physically wiring the wafer. For example, \texttt{ArrowTopLeftFirst\_\allowbreak{}TB-\allowbreak{}alternate-\allowbreak{}anti-\allowbreak{}alignment-\allowbreak{}wire\_\allowbreak{}TB} also has small bias due to \(\sum_i \cos (2 \alpha_i) \approx 0\), but it is considerably harder to wire because it requires a lot of crossing. Not to mention the other case \texttt{AlternateQU\_\allowbreak{}TB-\allowbreak{}alternate-\allowbreak{}anti-\allowbreak{}alignment-\allowbreak{}wire\_\allowbreak{}TB} which needs to go back and forth.

Instead, \texttt{SpiralCounterclockwise\_\allowbreak{}QUAB} designs the wafer starting from how one wants to wire/walk them on the focal plane first, once that is done, \(Q\)/\(U\)-ness of the detectors is assigned accordingly. As the \(T\)/\(B\)-ordering does not matter in this design, one is free to choose the \(TB\)-ordering due to physical limitation (minimize crossing.)

We should emphasize that this mitigation technique depends on the presence of CRHWP,
as its design is based on Corollary~\ref{cor-wafer-walking-rule},
which is derived from \cref{eq:crosstalk_spectra_curve_sky},
a consequence of an analytic model with a set of assumptions listed in Assumption~\ref{asu-crosstalk-map-domain-model}.
How well it generalizes to other CMB experiments
remains to be explored.
It is likely that any CMB experiments with CRHWP
with a fairly large sky coverage and isotropic scanning strategy (which are typical design requirements) such as the SAT in SO
can take advantage of this, because they fulfill Assumption~\ref{asu-crosstalk-map-domain-model}
approximately.

If this mitigation technique is applied,
the crosstalk requirements may, for example, be relaxed (which may allow a larger multiplexing factor and in turn lower the payload \& cost) by using this mitigation method,
but the \(BB\) systematics have to be studied more in detail at large crosstalk limit.

Note that a popular technique to mitigate
crosstalk systematics is to perform matrix inversion
to remove the crosstalk in the TOD\@.
This requires an accurate measurement of the crosstalk matrix,
and the systematic uncertainties will be dominated by
how accurate the crosstalk matrix can be measured.
The proposed technique here can be used together
with the matrix inversion technique,
such that any residual crosstalk in the TOD
can be mitigated by this wafer ordering technique.

There remains more crosstalk systematics to be explored,
such as time-dependent crosstalk, and cross-frequency crosstalk.

Lastly, another possible exploration of the analytical structure seen here
is to perform likelihood analysis using the power-spectra
with the \(P_\ell(\cos θ)\), \(d^\ell_{22}(θ)\) as templates
with \(θ\) at some of the characteristic angular scales of the focal plane on the sky.
This gives an advantage to remove residual crosstalk systematics,
but reduces the statistical power of the experiment.

\part{Conclusion}\label{sec:conclusion}

\chapter{Summary}\label{summary}

In \cref{sec:beginning-universe},
a brief overview of the standard model of Big Bang cosmology
is given.
Unresolved problems such as fine-tuning problems
can be resolved by introducing cosmic inflation.
Inflation is generally believed to be true
but has not been experimentally or observationally verified.
Due to the extreme high energy involved,
observation of the CMB radiation offers
one of the best chances to falsify inflationary models.

In \cref{sec:CMB-radiation},
the physics of the CMB is briefly presented,
showing the anisotropies in the intensity and polarization of the CMB
hold a wealth of information to be extracted
to better characterize and understand the universe.
In particular,
the primordial \(B\)-mode
holds the key to falsify inflationary models.

In \cref{sec:CMB-observations},
an overview of CMB observations is given.
It ranges from
the intrinsic properties of the CMB that affect experimental designs,
to the mathematics of CMB data analysis,
to the computational aspects including simulations and systematics mitigations,
to the statistical aspects which enable us to extract information of the universe from the data.
It concludes with the current measurements of the CMB and what to expect next.

In \cref{sec:POLARBEAR},
the POLARBEAR experiment is introduced.
Some of the key details on existing data analysis in POLARBEAR
are given, such as
the null-test framework,
mapmaking method including CRHWP demodulation,
and power-spectra estimation method.

Building on top of that,
\cref{sec:polarbear-research} shows original research
that builds upon the observed data and existing data analysis framework.
In particular,
a measurement of the high-\(\ell\) \(B\)-mode in the range of 600--3000
using 3rd--5th seasons of POLARBEAR data is presented.
It started with the expected sensitivity that should have been known before taking the observation.
It also pointed out the major challenges
experienced in this research
including computational efficiency,
reproducible research,
and null-test failures.
Among other details,
the solutions to these major obstacles are presented.
It also presented improved statistical techniques,
including
the definition of null-statistics and its combined statistics,
a full-Bayesian likelihood procedure,
and an improved formula to calculate the analytical statistical uncertainty.
The limitation of the null-test framework is documented
where some are general and some are specific to POLARBEAR\@.
In the end,
the lensing amplitude is detected with \(P(A_\text{lens} > 0) = \SI{99.6}{\percent}\).

In \cref{sec:LiteBIRD-research}, the crosstalk systematic is studied in detail.
An analytical model on the map-domain is developed to predict the effect of crosstalk on the power-spectra
given a set of conditions.
A TOAST operator is developed to enable time-domain simulations of crosstalk systematics
in the framework of the software Time Ordered Astrophysics Scalable Tools (TOAST)\@.
The two predictions are compared and show a high degree of agreement in the example of
the LiteBIRD experiment.
Exploiting the analytical behavior in the model,
a hardware based mitigation technique is developed to minimize its effect on the 2pt-power-spectra.
This mitigation technique is general as long as the set of conditions is fulfilled,
and a simple rule of thumb is presented that is easy to use for experimentalists.
Given a new focal plane design, the effect can be predicted without time-consuming time-domain simulations
for fast prototyping.
One limitation of this model and mitigation technique is the requirement of the HWP.

\chapter{Discussions and future prospects}\label{discussions-and-future-prospects}

\section{Use of HWP in CMB observations}\label{use-of-hwp-in-cmb-observations}

Incidentally,
the HWP is central to both parts of my research in \cref{sec:polarbear-research,sec:LiteBIRD-research}.
The HWP enables the POLARBEAR experiment to measure a wide \(\ell\)-range
using the same instrument,
and some core analysis techniques are built around the presence of the CRHWP such as demodulation.
The HWP in \cref{sec:LiteBIRD-research} plays a central role
that the analytical model is only valid with its presence,
which is a prerequisite for the proposed hardware-based mitigation strategy to work.

The HWP should continue to play an important role in CMB experiments in the near future.
Important upcoming CMB experiments including the LiteBIRD experiment,
the Small Aperture Telescopes (SATs) of the Simons Observatory,\footnote{\citeproc{ref-kiuchi_simons_2020}{Kiuchi et al., {``The Simons Observatory Small Aperture Telescope Overview''}}.} etc.\ will include the HWP.

However,
it should be stressed that while the HWP
eliminates a class of systematics,
it introduces some of its own.
And having moving parts such as the HWP
on mission critical experiments such as the LiteBIRD
is a significant risk that warrants careful considerations.

\section{Control of systematics and null-test framework in CMB experiments}\label{control-of-systematics-and-null-test-framework-in-cmb-experiments}

For any scientific research worth publishing,
a high degree of assurance of the lack of biases is an essential ingredient.
Among these biases are systematic effects,
observer bias, and confirmation bias.

To control for systematic effects,
a systematic effect has to be identified,
and such effects have to be measured and/or simulated to ensure they are subdominant to the signal.
Mitigation strategies can also be devised to suppress it.

The null-test framework provides a way to discover systematic effects.
As long as there is a physical variable that can be used to split the data into two-halves,
a null-test can be performed.

While my research has shown examples of how it can be achieved
in current generations of CMB experiments,
it will be an ongoing challenge for future generations of CMB experiments.
It follows simply from the fact that CMB datasets are growing exponentially,
and our target signal is fainter than ever as we march towards the uncharted territory.
With fainter signals,
any previously subdominant systematics can become dominant.
This means that the number of systematic effects
we need to consider also grows in size.
And for well-known systematics,
better analysis and mitigation techniques may be required.

This exploding size of datasets also
makes the number of possible ways to cut the data into two-halves
grow exponentially.
The naïve mapmaker is very suitable for running null-tests,
as TOD processing can potentially be shared for all null-tests, but
memory and disk requirements to coadd and process the null-split maps
will grow accordingly, on top of the already growing dataset size.

This may mean a more carefully selected set of null-splits is needed.
And this diminishes the usefulness of the null-test framework to discover
unknown systematics.
This goes back to the previous point that identifying, measuring and/or simulating
systematics are going to be more important.

\section{CMB data analysis as a statistical problem}\label{cmb-data-analysis-as-a-statistical-problem}

As a random example,
people studying network topology used to be able to just look at the graph
and discover patterns about it.
As networks grow bigger, visualizing the networks becomes increasingly difficult, if not impossible,
and abstract, formal statistical analysis was developed to address this problem.

Similarly,
for datasets as small as the POLARBEAR observation,
there are many things that can still be done with the intuition of a researcher.
To name a few,
we can look at each of the maps by brute force and see if there are any abnormal patterns.
Since there are 38 map bundles, 21 null-splits,
there are \(1535\) null-split maps\footnote{See \cref{sec:low-lying-fruits}.}.
If it is converted into a video at 24 fps,
it will take slightly more than 1 minute to watch all of them\footnote{Do not try this at home if you have photosensitive epilepsy!
  Even for normal people, they may experience dizziness, nausea, and vomiting.}.
But what if we want to look at each individual CES\@?
There are 3391 of them, which becomes around \(3391 \times 21 \times 2 = 142,422\).
This equates to more than an hour of video, making it virtually impossible to complete.

Another example would be null-power-spectra and many of its associated visualization
such as chi-square plots.
Again some researchers can eyeball them and use intuition to discover problems.
But the number of null-power-spectra increases with the number of null-splits chosen,
possibly making the task intractable by humans.

Some solutions to problems like this have been proposed in my research,
for example by using the Holm--Bonferroni method to decide objectively.
But there are more areas that can benefit from proper statistical procedure
to automate some analyses which rely on human intuition, such as data selection.

One potential avenue to tackle these problems (such as data selection and visualizing maps)
that is in very high demand right now is statistical machine learning.
They can be well-suited to these sorts of problems as they are good at pattern recognition,
and they are only making discoveries on potential biases that can be further investigated
using more stringent methods.
Application of machine learning in science and in CMB experiments, in particular,
is already actively explored by many researchers.

\section{CMB data analysis as a computation problem}\label{cmb-data-analysis-as-a-computation-problem}

As have been shown briefly in \cref{sec:next-stage} with \cref{fig:exponential-cmb-data-growth},
and also in the above context,
CMB data analysis will continue to be challenging computationally.
Even for existing algorithms,
they need to be updated constantly to keep up with state-of-the-art HPC systems.
This includes adaptation to continually changing paradigms such as different CPU architectures and/or topologies,
adoption of accelerators such as GPUs;
challenges in scaling to increasingly more nodes and/or heterogeneous computing;
and energy-constrained computing.
On top of this,
more systematic simulations and more sophisticated analysis techniques are required.

Moreover,
as the complexity of this software increases,
wide-scale adoption of good software engineering skills is needed in the CMB community.
This includes proper maintenance of the code base,
proper collaboration skills in programming to scale beyond a few contributors,
and proper structuring to nurture growth in complexity while maintaining correctness.
From the limited observation of the author,
only a handful of researchers do this right.
Of all the skills needed in
CMB data analysis---physics,
mathematics, statistics, and software engineering---the last
is the most neglected and underrated.

Education may be part of the answer.
It is safe to say no physicist can enter graduate school
without a proper training in mathematics and some basic training in statistics.
This may not be true in programming, and in general computer science and software engineering.
Often these are self-taught, yours truly included.
With no quality control and no requirement,
it is easy for it to be swept under the rug.

It can also be an attitude problem.
No physicist will gloss over the importance of mathematics,
perhaps because a mathematical model can only be right or wrong.
But it is easy for them to gloss over the importance of statistics and software engineering.
Often they have a mindset to avoid complexity
and only opt for things simple to understand.
One solution may be to let specialized people handle specialized areas.
For example, in the industry,
it is not uncommon for data scientists to prototype models
and successful models will be productionized by professional programmers.

\section{Reproducible research}\label{reproducible-research}

The problem of reproducible research is briefly discussed in \cref{sec:reproducible-research}.
The author makes an effort to move closer in this direction,
but it is far from completely reproducible.

I expect this to continue to be an important area to explore
and a challenge to overcome.

Paradigm shifts happen all the time in science,
and while it is too early to tell,
we may be experiencing a paradigm shift in reproducible research.
Here, the paradigm is not how we understand the world
but how we carry out science.

Today,
we find it unacceptable to call something science without publishing
its methodology and results.
If indeed, we are going through such a paradigm shift,
then in the future we may find it equally unacceptable
to call something science without being reproducible
(in the strict sense of what computationally reproducible research means.)

But as the CMB datasets are growing exponentially,
we are facing a challenge in making these available.
And even if an individual can access all the data and software,
few can have the computational resource to reproduce it.
Maybe because of this,
the paradigm shift of reproducible research would be restricted
in science with smaller datasets such as biology.
Or maybe a smaller subset of reproducibility is defined to make it more manageable.
Or maybe, similar to how mathematicians accept formal proofs,
we develop an ``in-principle'' reproducible research method
where at least one independent party is required to reproduce the result.
Suffice it to say that reproducible research in CMB will look very different
from the reproducible research practiced elsewhere nowadays.

\section{Crosstalk systematics}\label{crosstalk-systematics}

The crosstalk simulation method presented here is general,
while computationally it may have room for improvements
to handle future CMB experiments where the crosstalk matrix
can be much larger.

The analytical model comes with a set of assumptions,
where the hardware-based mitigation method relies on its validity.
The biggest limitation would be the requirement of the HWP
so that each individual detector can resolve the \(Q, U\) components by itself.

There are other assumptions that are only approximately fulfilled.
For example, isotropy in both scanning strategy and attack angle is unrealistic,
while a realistic one may approximate it well.
It will be interesting to see how well this model holds in other CMB experiments.
A natural next step would be to compare time-domain simulation to the prediction of this model
using the specification of CMB experiments with HWP and a very different scanning strategy
such as the SO SATs.

There are also other aspects of crosstalk systematics not covered in this research,
for example, time dependent crosstalk,
and crosstalk between detectors observing at different frequencies.
These will be explored in the future.

\backmatter

\part{References}\label{sec:references}

These figures are circulated internally in different presentations and the actual sources are unknown to the author:
\cref{fig:atmosphericalWindows},
\cref{fig:PWV},
\cref{fig:003-003},
\cref{fig:image20},
\cref{fig:sample_ASD-6462}.

\bigskip

\protect\phantomsection\label{refs-references}

\part{Bibliography}\label{bibliography}

\protect\phantomsection\label{refs-bibliography}

\part{Appendix}\label{appendix}

\protect\phantomsection\label{c0b8cc9b}
\chapter{The Mathematics of CMB flat sky approximation}\label{sec:CMBFlatSkyMath}

\section{Flat-sky approximation}\label{flat-sky-approximation}

The following are derived from,\footnote{\citeproc{ref-zaldarriaga_all-sky_1997}{Zaldarriaga and Seljak, {``All-Sky Analysis of Polarization in the Microwave Background,''} sec. V}.} including its sign conventions.

\protect\phantomsection\label{853aca0a}

\protect\phantomsection\label{cc4ace29}
Also, we follow the physicists' convention that the same symbol with a
different argument actually indicates a different function after a
transform. For example,
\(f \mathopen{} \left ( x \right ) \mathclose{} , f \mathopen{} \left ( k \right ) \mathclose{}\)
.

\protect\phantomsection\label{7488ce2a}
\subsection{Definitions}\label{definitions}

\protect\phantomsection\label{aed9dd36}
\begin{align*}
\mathcal{Q} \mathopen{} \left ( \symbf{\theta} \right ) \mathclose{}    &    \equiv Q \mathopen{} \left ( \symbf{\theta} \right ) \mathclose{} + \mathrm{i} U \mathopen{} \left ( \symbf{\theta} \right ) \mathclose{}  \\
\mathcal{E} \mathopen{} \left ( \mathbf{\ell} \right ) \mathclose{} &    \equiv E \mathopen{} \left ( \mathbf{\ell} \right ) \mathclose{} + \mathrm{i} B \mathopen{} \left ( \mathbf{\ell} \right ) \mathclose{}
\end{align*}

\protect\phantomsection\label{30ada3a6}
Note that \(Q , U\) are real fields, \(\mathcal{Q} , \mathcal{E} , E , B\)
are complex fields.

\protect\phantomsection\label{4885b319}
\subsection{From curved sky to flat sky}\label{from-curved-sky-to-flat-sky}

We replace

\protect\phantomsection\label{81ec8955}
\[T \mathopen{} \left ( \mathbf{\hat{n}} \right ) \mathclose{} = \sum_{\ell , m} a_{\ell m} Y_{\ell m} \mathopen{} \left ( \mathbf{\hat{n}} \right ) \mathclose{} \rightarrow \int T \mathopen{} \left ( \mathbf{\ell} \right ) \mathclose{} \mathrm{e}^{\mathrm{i} \mathbf{\ell} \cdot \symbf{\theta}} \,\mathrm{d}^{2} \mathbf{\ell}\]

\protect\phantomsection\label{3b9105d9}
that is, using the angular frequency, non-unitary convention of Fourier
transform,

\protect\phantomsection\label{04c4381d}

\protect\phantomsection\label{785d5c0d}

\protect\phantomsection\label{43f77356}
\[\begin{cases}
T \mathopen{} \left ( \mathbf{\ell} \right ) \mathclose{} &\equiv \int T \mathopen{} \left ( \symbf{\theta} \right ) \mathclose{} \mathrm{e}^{- \mathrm{i} \mathbf{\ell} \cdot \symbf{\theta}} \,\mathrm{d}^{2} \symbf{\theta}  \\
\mathcal{Q} \mathopen{} \left ( \mathbf{\ell} \right ) \mathclose{} &\equiv \int \mathcal{Q} \mathopen{} \left ( \symbf{\theta} \right ) \mathclose{} \mathrm{e}^{- \mathrm{i} \mathbf{\ell} \cdot \symbf{\theta}} \,\mathrm{d}^{2} \symbf{\theta}
\end{cases}\]

\protect\phantomsection\label{04d5c60e}
\begin{equation}
\begin{cases}
T \mathopen{} \left ( \symbf{\theta} \right ) \mathclose{} &= \frac{1}{\left( 2 \pi \right)^{2}} \int T \mathopen{} \left ( \mathbf{\ell} \right ) \mathclose{} \mathrm{e}^{\mathrm{i} \mathbf{\ell} \cdot \symbf{\theta}} \,\mathrm{d}^{2} \mathbf{\ell}   \\
\mathcal{Q} \mathopen{} \left ( \symbf{\theta} \right ) \mathclose{} &= \frac{1}{\left( 2 \pi \right)^{2}} \int \mathcal{Q} \mathopen{} \left ( \mathbf{\ell} \right ) \mathclose{} \mathrm{e}^{\mathrm{i} \mathbf{\ell} \cdot \symbf{\theta}} \,\mathrm{d}^{2} \mathbf{\ell}   \label{TQDef}
\end{cases}
\end{equation}

\protect\phantomsection\label{6a826cbf}
Note that
\(\mathcal{Q} \mathopen{} \left ( \mathbf{\ell} \right ) \mathclose{}\) is
defined above, without giving a Physical meaning.

\protect\phantomsection\label{b4f108b2}
\subsection{\texorpdfstring{\(\mathcal{Q}\) as spin-\(2\) field}{\textbackslash mathcal\{Q\} as spin-2 field}}\label{mathcalq-as-spin-2-field}

\protect\phantomsection\label{c0bcf9c5}
As
\(\mathcal{Q} \mathopen{} \left ( \symbf{\theta} \right ) \mathclose{}\)
is a spin-\(2\) field (vice versa for
\(\bar{\mathcal{Q}} \mathopen{} \left ( \symbf{\theta} \right ) \mathclose{}\)
),

\protect\phantomsection\label{2001874b}

\protect\phantomsection\label{5e44bcf0}
\begin{align}
\mathcal{Q} \mathopen{} \left ( \symbf{\theta} \right ) \mathclose{} &= \frac{1}{\left( 2 \pi \right)^{2}} \int \left[ E \mathopen{} \left ( \mathbf{\ell} \right ) \mathclose{} + \mathrm{i} B \mathopen{} \left ( \mathbf{\ell} \right ) \mathclose{} \right] \frac{-1}{\ell^{2}} \eth^{2} \mathrm{e}^{\mathrm{i} \mathbf{\ell} \cdot \symbf{\theta}} \,\mathrm{d}^{2} \mathbf{\ell}  \\
\bar{\mathcal{Q}} \mathopen{} \left ( \symbf{\theta} \right ) \mathclose{} &= \frac{1}{\left( 2 \pi \right)^{2}} \int \left[ E \mathopen{} \left ( \mathbf{\ell} \right ) \mathclose{} - \mathrm{i} B \mathopen{} \left ( \mathbf{\ell} \right ) \mathclose{} \right] \frac{-1}{\ell^{2}} \bar{\eth}^{2} \mathrm{e}^{\mathrm{i} \mathbf{\ell} \cdot \symbf{\theta}} \,\mathrm{d}^{2} \mathbf{\ell}
\end{align}

\protect\phantomsection\label{4aae34e8}
In the small scale limit,

\protect\phantomsection\label{2b7b1091}
\begin{align*}
\frac{-1}{\ell^{2}} \eth^{2} \mathrm{e}^{\mathrm{i} \mathbf{\ell} \cdot \symbf{\theta}} &    \approx \mathrm{e}^{-2 \mathrm{i} \left( \phi - \phi_{\ell} \right)} \mathrm{e}^{\mathrm{i} \mathbf{\ell} \cdot \symbf{\theta}}    \\
\frac{-1}{\ell^{2}} \bar{\eth}^{2} \mathrm{e}^{\mathrm{i} \mathbf{\ell} \cdot \symbf{\theta}}   &    \approx \mathrm{e}^{2 \mathrm{i} \left( \phi - \phi_{\ell} \right)} \mathrm{e}^{\mathrm{i} \mathbf{\ell} \cdot \symbf{\theta}}
\end{align*}

\protect\phantomsection\label{a85ac98f}
That is,

\protect\phantomsection\label{d2f41ce3}
\begin{align}
\mathcal{Q} \mathopen{} \left ( \symbf{\theta} \right ) \mathclose{} &\approx \frac{1}{\left( 2 \pi \right)^{2}} \int \left[ E \mathopen{} \left ( \mathbf{\ell} \right ) \mathclose{} + \mathrm{i} B \mathopen{} \left ( \mathbf{\ell} \right ) \mathclose{} \right] \mathrm{e}^{-2 \mathrm{i} \left( \phi - \phi_{\ell} \right)} \mathrm{e}^{\mathrm{i} \mathbf{\ell} \cdot \symbf{\theta}} \,\mathrm{d}^{2} \mathbf{\ell}    \label{EBDef}   \\
\bar{\mathcal{Q}} \mathopen{} \left ( \symbf{\theta} \right ) \mathclose{} &\approx \frac{1}{\left( 2 \pi \right)^{2}} \int \left[ E \mathopen{} \left ( \mathbf{\ell} \right ) \mathclose{} - \mathrm{i} B \mathopen{} \left ( \mathbf{\ell} \right ) \mathclose{} \right] \mathrm{e}^{2 \mathrm{i} \left( \phi - \phi_{\ell} \right)} \mathrm{e}^{\mathrm{i} \mathbf{\ell} \cdot \symbf{\theta}} \,\mathrm{d}^{2} \mathbf{\ell}   \label{EBDefConj}
\end{align}

\protect\phantomsection\label{12731940}
Comparing~\eqref{TQDef}, \eqref{EBDef}, we have

\protect\phantomsection\label{3ac2069c}
\[\mathcal{Q} \mathopen{} \left ( \mathbf{\ell} \right ) \mathclose{} \approx \mathcal{E} \mathopen{} \left ( \mathbf{\ell} \right ) \mathclose{} \mathrm{e}^{-2 \mathrm{i} \left( \phi - \phi_{\ell} \right)}\]

\protect\phantomsection\label{0c1d0b46}
\subsection{Further approximation}\label{further-approximation}

\protect\phantomsection\label{3f190305}
Where \(\phi_{\ell}\) is the complex phase of \(\ell\) (when we interpret
\(\ell \in \mathbb{R}^{2} \cong \mathbb{C}\).)

\protect\phantomsection\label{9742c0e0}
We can further set \(\phi = 0\) by rotating from
\(\left( \symbf{\hat{\theta}} , \symbf{\hat{\phi}} \right)\) to
\(\left( \mathbf{\hat{x}} , \mathbf{\hat{y}} \right)\) on the
\(\mathbf{\hat{z}}\)-tangent plane. Hence,

\protect\phantomsection\label{010b4fd5}
\begin{equation}
\begin{cases}
\mathcal{Q} \mathopen{} \left ( \mathbf{\ell} \right ) \mathclose{} &    \approx \mathcal{E} \mathopen{} \left ( \ell \right ) \mathclose{} \mathrm{e}^{2 \mathrm{i} \phi_{\ell}}   \\
\mathcal{E} \mathopen{} \left ( \mathbf{\ell} \right ) \mathclose{} &    \approx \mathcal{Q} \mathopen{} \left ( \ell \right ) \mathclose{} \mathrm{e}^{-2 \mathrm{i} \phi_{\ell}}
\end{cases} \label{eq:QErelation}
\end{equation}

\protect\phantomsection\label{64363471}
\subsection{\texorpdfstring{Decomposing \(\mathcal{E}\) in terms of \(E , B\)}{Decomposing \textbackslash mathcal\{E\} in terms of E , B}}\label{decomposing-mathcale-in-terms-of-e-b}

\protect\phantomsection\label{95504bdf}
Starting from~\eqref{EBDefConj}, take the c.c., and
\(\mathbf{\ell} \mapsto - \mathbf{\ell}\), we have

\protect\phantomsection\label{6a5a1b22}
\begin{equation}
\mathcal{Q} \mathopen{} \left ( \symbf{\theta} \right ) \mathclose{} \approx \frac{1}{\left( 2 \pi \right)^{2}} \int \left[ \bar{E} \mathopen{} \left ( - \mathbf{\ell} \right ) \mathclose{} + \mathrm{i} \bar{B} \mathopen{} \left ( - \mathbf{\ell} \right ) \mathclose{} \right] \mathrm{e}^{-2 \mathrm{i} \left( \phi - \phi_{\ell} \right)} \mathrm{e}^{\mathrm{i} \mathbf{\ell} \cdot \symbf{\theta}} \,\mathrm{d}^{2} \mathbf{\ell} \label{EBDefAlt}
\end{equation}

\protect\phantomsection\label{98895289}
Comparing~\eqref{EBDefAlt}, \eqref{EBDef}, we have

\protect\phantomsection\label{662e1a6e}
\[\begin{cases}
\mathcal{E} \mathopen{} \left ( \mathbf{\ell} \right ) \mathclose{} &    \equiv E \mathopen{} \left ( \mathbf{\ell} \right ) \mathclose{} + \mathrm{i} B \mathopen{} \left ( \mathbf{\ell} \right ) \mathclose{} \approx \bar{E} \mathopen{} \left ( - \mathbf{\ell} \right ) \mathclose{} + \mathrm{i} \bar{B} \mathopen{} \left ( - \mathbf{\ell} \right ) \mathclose{}   \\
\mathcal{E} \mathopen{} \left ( - \mathbf{\ell} \right ) \mathclose{}   &    \approx E \mathopen{} \left ( \mathbf{\ell} \right ) \mathclose{} - \mathrm{i} B \mathopen{} \left ( \mathbf{\ell} \right ) \mathclose{}
\end{cases}\]

\protect\phantomsection\label{b9a5df75}
That is,

\protect\phantomsection\label{efeacadd}
\begin{equation}
\begin{cases}
E \mathopen{} \left ( \mathbf{\ell} \right ) \mathclose{}   &    \approx \frac{1}{2} \left[ \mathcal{E} \mathopen{} \left ( \mathbf{\ell} \right ) \mathclose{} + \bar{\mathcal{E}} \mathopen{} \left ( - \mathbf{\ell} \right ) \mathclose{} \right]   \\
B \mathopen{} \left ( \mathbf{\ell} \right ) \mathclose{}   &    \approx \frac{1}{2 \mathrm{i}} \left[ \mathcal{E} \mathopen{} \left ( \mathbf{\ell} \right ) \mathclose{} - \bar{\mathcal{E}} \mathopen{} \left ( - \mathbf{\ell} \right ) \mathclose{} \right]
\end{cases} \label{eq:realEB}
\end{equation}

\protect\phantomsection\label{5f7b9c9d}
Note that \(E, B\) have the following property accordingly,

\protect\phantomsection\label{e78f4506}
\begin{align*}
\bar{E} \mathopen{} \left ( - \mathbf{\ell} \right ) \mathclose{} &= E \mathopen{} \left ( \mathbf{\ell} \right ) \mathclose{}  \\
\bar{B} \mathopen{} \left ( - \mathbf{\ell} \right ) \mathclose{} &= B \mathopen{} \left ( \mathbf{\ell} \right ) \mathclose{}
\end{align*}

\protect\phantomsection\label{958923ac}
That is, the field corresponding to
\(E \mathopen{} \left ( \mathbf{\ell} \right ) \mathclose{} , B \mathopen{} \left ( \mathbf{\ell} \right ) \mathclose{}\)
, namely,
\(E \mathopen{} \left ( \symbf{\theta} \right ) \mathclose{} , B \mathopen{} \left ( \symbf{\theta} \right ) \mathclose{}\)
, are real, as they fulfill the reality condition.

\protect\phantomsection\label{cdcc7c34}
Also, using \(\eqref{eq:realEB}\), we can derive the relationships
directly between the real-fields\footnote{using the reality condition on real fields. See proof in the
  companion file \texttt{flat-\allowbreak{}sky-\allowbreak{}approximation.nb}}:

\protect\phantomsection\label{79235616}
\begin{align*}
\mathbb{E} \mathopen{} \left ( \mathbf{\ell} \right ) \mathclose{}  &    \equiv \begin{bmatrix}
E \mathopen{} \left ( \mathbf{\ell} \right ) \mathclose{}   \\
B \mathopen{} \left ( \mathbf{\ell} \right ) \mathclose{}
\end{bmatrix}   \\
\mathbb{Q} \mathopen{} \left ( \mathbf{\ell} \right ) \mathclose{}  &    \equiv \begin{bmatrix}
Q \mathopen{} \left ( \mathbf{\ell} \right ) \mathclose{}   \\
U \mathopen{} \left ( \mathbf{\ell} \right ) \mathclose{}
\end{bmatrix}   \\
R \mathopen{} \left ( \phi \right ) \mathclose{}    &    \equiv \begin{bmatrix}
\cos \phi   &   - \sin \phi \\
\sin \phi   &   \cos \phi
\end{bmatrix}   \\
\mathbb{E} \mathopen{} \left ( \mathbf{\ell} \right ) \mathclose{}  &    \approx R \mathopen{} \left ( -2 \phi_{\ell} \right ) \mathclose{} \mathbb{Q} \mathopen{} \left ( \mathbf{\ell} \right ) \mathclose{}  \\
\mathbb{Q} \mathopen{} \left ( \mathbf{\ell} \right ) \mathclose{}  &    \approx R \mathopen{} \left ( 2 \phi_{\ell} \right ) \mathclose{} \mathbb{E} \mathopen{} \left ( \mathbf{\ell} \right ) \mathclose{}
\end{align*}

\protect\phantomsection\label{a001115c}
which can be compared to \(\eqref{eq:QErelation}\).

\protect\phantomsection\label{d1fe59a8}
\subsection{Conclusion}\label{conclusion-1}

That is, we can go between

\protect\phantomsection\label{493297a2}
\[\left( Q \mathopen{} \left ( \symbf{\theta} \right ) \mathclose{} , U \mathopen{} \left ( \symbf{\theta} \right ) \mathclose{} \right) \leftrightarrow \mathcal{Q} \mathopen{} \left ( \symbf{\theta} \right ) \mathclose{} \leftrightarrow \mathcal{Q} \mathopen{} \left ( \mathbf{\ell} \right ) \mathclose{} \leftrightarrow \mathcal{E} \mathopen{} \left ( \mathbf{\ell} \right ) \mathclose{} \leftrightarrow \left( E \mathopen{} \left ( \mathbf{\ell} \right ) \mathclose{} , B \mathopen{} \left ( \mathbf{\ell} \right ) \mathclose{} \right)\]

\chapter{Summary of convolution and cross-correlation}\label{summary-of-convolution-and-cross-correlation}

\section{\texorpdfstring{\(\mathbb{R}^n\)}{\textbackslash mathbb\{R\}\^{}n}}\label{mathbbrn}

See \href{https://en.wikipedia.org/wiki/Convolution\#Domain_of_definition}{Convolution}:

\[{\displaystyle (f*g)(x)=\int _{\mathbb {R} ^{d}}f(y)g(x-y)\,dy=\int _{\mathbb {R} ^{d}}f(x-y)g(y)\,dy,}\]

\subsection{\texorpdfstring{Convolution theorem on \(\mathbb{R}^n\) convolution}{Convolution theorem on \textbackslash mathbb\{R\}\^{}n convolution}}\label{convolution-theorem-on-mathbbrn-convolution}

\[{\displaystyle {\mathcal {F}}\{f*g\}=k\cdot {\mathcal {F}}\{f\}\cdot {\mathcal {F}}\{g\}},\]

where \(k\) depends on the gauge. See \href{https://en.wikipedia.org/wiki/Fourier_transform\#Functional_relationships,_one-dimensional}{Fourier transform}. \(k=1\) for the 2 commonly used conventions, unitary with ordinary frequency \& non-unitary with angular frequency. See also \cref{sec:fourier}.

\section{Group}\label{group}

See \href{https://en.wikipedia.org/wiki/Convolution\#Convolutions_on_groups}{Convolution}.

\[{\displaystyle (f*g)(x)=\int _{G}f(y)g\left(y^{-1}x\right)\,d\lambda (y).}\]

where \(G\) is a locally compact Hausdorff topological group and \(λ\) is a (left-) Haar measure.

It has the property of

\[{\displaystyle L_{h}(f*g)=(L_{h}f)*g.}\]

where \(L\) is the left translation of the group.

If \(G\) is commutative, then

\[{\displaystyle \tau _{x}(f*g)=(\tau _{x}f)*g=f*(\tau _{x}g)}\]

Specializing to the translation group in \(\mathbb{R}^n\), we recover the above special case of \(\mathbb{R}^n\). Here, the translation group is defined as

\[{\displaystyle (\tau _{x}f)(y)=f(y-x).}\]

\subsection{Properties}\label{properties}

A few useful properties are

\[{\displaystyle f*\delta =f}\]

\[\tau_x f = f * \tau_x \delta\]

with some conditions, where \(\delta\) is the delta distribution. See \href{https://en.wikipedia.org/wiki/Convolution\#Translational_equivariance}{Convolution}.

\chapter{References to Fourier series and transform}\label{sec:fourier}

\section{Conventions}\label{conventions}

\href{https://en.wikipedia.org/wiki/Fourier_transform\#Other_conventions}{Fourier transform - Wikipedia} has a nice table summarizing (non-)unitary and ordinary/angular frequency definition.

The unitary, ordinary frequency definition is the simplest, and most commonly used in Mathematics.

In Quantum mechanics, angular frequency with the non-unitary definition is used. The reason for angular frequency is probably for the canonical momentum relationship.

In \href{https://www.johndcook.com/blog/fourier-theorems/}{Fourier transform definition conventions and formulas}, John Cook unifies all different conventions in a parameterized definition, which can have 8 combinations. Mathematica has a different parametrization in \href{https://reference.wolfram.com/language/ref/FourierTransform.html}{FourierTransform---Wolfram Language Documentation}. He pointed out the relationship between the 2 parameterizations in \href{https://www.johndcook.com/blog/2021/03/17/fourier-transforms-in-mathematica/}{Fourier transforms and conventions in Mathematica}.

\section{Definitions \& formula}\label{definitions-formula}

Copied from Wikipedia for completeness:

ordinary frequency ξ (Hz), unitary:

\[{\displaystyle {\begin{aligned}{\hat {f}}_{1}(\xi )\ &{\stackrel {\mathrm {def} }{=}}\ \int _{\mathbb {R} ^{n}}f(x)e^{-2\pi ix\cdot \xi }\,dx=(2\pi )^{\frac {n}{2}}{\hat {f}}_{2}(2\pi \xi )={\hat {f}}_{3}(2\pi \xi )\\f(x)&=\int _{\mathbb {R} ^{n}}{\hat {f}}_{1}(\xi )e^{2\pi ix\cdot \xi }\,d\xi \end{aligned}}}\]

angular frequency ω (rad/s), unitary:

\[{\displaystyle {\begin{aligned}{\hat {f}}_{2}(\omega )\ &{\stackrel {\mathrm {def} }{=}}\ {\frac {1}{(2\pi )^{\frac {n}{2}}}}\int _{\mathbb {R} ^{n}}f(x)e^{-i\omega \cdot x}\,dx={\frac {1}{(2\pi )^{\frac {n}{2}}}}{\hat {f}}_{1}\!\left({\frac {\omega }{2\pi }}\right)={\frac {1}{(2\pi )^{\frac {n}{2}}}}{\hat {f}}_{3}(\omega )\\f(x)&={\frac {1}{(2\pi )^{\frac {n}{2}}}}\int _{\mathbb {R} ^{n}}{\hat {f}}_{2}(\omega )e^{i\omega \cdot x}\,d\omega \end{aligned}}}\]

angular frequency ω (rad/s), non-unitary:

\[{\displaystyle {\begin{aligned}{\hat {f}}_{3}(\omega )\ &{\stackrel {\mathrm {def} }{=}}\ \int _{\mathbb {R} ^{n}}f(x)e^{-i\omega \cdot x}\,dx={\hat {f}}_{1}\left({\frac {\omega }{2\pi }}\right)=(2\pi )^{\frac {n}{2}}{\hat {f}}_{2}(\omega )\\f(x)&={\frac {1}{(2\pi )^{n}}}\int _{\mathbb {R} ^{n}}{\hat {f}}_{3}(\omega )e^{i\omega \cdot x}\,d\omega \end{aligned}}}\]

\protect\phantomsection\label{e6780e68}
\chapter{The Mathematics of telescope beam}\label{the-mathematics-of-telescope-beam}

\section{Intro}\label{intro}

Beam convolution and definitions are confusing.

First of all, how should the beam be defined? Is it the Point Spread
Function (PSF)?

2 different pictures of possible definitions of beam can be related by

Curved sky:

\protect\phantomsection\label{585c8561}
\[B \mathopen{} \left ( \theta, \phi \right ) \mathclose{} = P \mathopen{} \left ( \theta, \pi + \phi \right ) \mathclose{} \Rightarrow B \mathopen{} \left ( \mathbf{\hat{n}} \right ) \mathclose{} = P \mathopen{} \left ( R_{z} \mathopen{} \left ( \frac{\pi}{2} \right ) \mathclose{} \mathbf{\hat{n}} \right ) \mathclose{}\]

\protect\phantomsection\label{a90cc75e}
Flat sky:

\protect\phantomsection\label{3a723a7e}
\[B \mathopen{} \left ( \mathbf{r} \right ) \mathclose{} = P \mathopen{} \left ( - \mathbf{r} \right ) \mathclose{}\]

\protect\phantomsection\label{e9cc1643}
In the latter case, the relationship in Fourier domain is simple:

\protect\phantomsection\label{afb26d97}
\[\mathcal{F} \mathopen{} \left \{ \pi f \right \} \mathclose{} \mathopen{} \left ( \mathbf{k} \right ) \mathclose{} = \mathcal{F} \mathopen{} \left \{ f \right \} \mathclose{} \mathopen{} \left ( - \mathbf{k} \right ) \mathclose{}\]

\protect\phantomsection\label{7a3f7d5b}
where

\protect\phantomsection\label{1e0f8ba6}
\[\left( π f \right) \mathopen{} \left ( x \right ) \mathclose{} \equiv f \mathopen{} \left ( - x \right ) \mathclose{}\]

\protect\phantomsection\label{f8038e22}
In the case of symmetric beam (which means that the beam depends only on
the ``distance'' to the center, not the angular direction), this subtlety
does not exist, which may explain why the definition is not clear in
the literature.

In integral form, in flat sky,

\protect\phantomsection\label{d7ee3189}
\begin{align*}
m_{\text{obs}} \mathopen{} \left ( \mathbf{x} \right ) \mathclose{} &    = \int_{\mathbb{R}^{2}} P \mathopen{} \left ( \mathbf{x} - \mathbf{y} \right ) \mathclose{} m \mathopen{} \left ( \mathbf{y} \right ) \mathclose{} \,\mathrm{d}^{2} \mathbf{y} ,   &   \mathbf{x} , \mathbf{y} \in \mathbb{R}^{2}  \\
    &    = \int_{\mathbb{R}^{2}} B \mathopen{} \left ( \mathbf{y} - \mathbf{x} \right ) \mathclose{} m \mathopen{} \left ( \mathbf{y} \right ) \mathclose{} \,\mathrm{d}^{2} \mathbf{y}
\end{align*}

\protect\phantomsection\label{6554db51}

\protect\phantomsection\label{4700e16c}
\chapter{The Mathematics of spin-weighted spherical harmonics and related concepts}\label{the-mathematics-of-spin-weighted-spherical-harmonics-and-related-concepts}

\chaptermark{Spin-weighted spherical harmonics}

\protect\phantomsection\label{187158db}
\section{Legendre polynomials}\label{legendre-polynomials}

\subsection{Condon-Shortley phase Convention}\label{condon-shortley-phase-convention}

The following convention agrees with \href{https://en.wikipedia.org/wiki/Associated_Legendre_polynomials}{Associated Legendre polynomials -
Wikipedia},
and is also the choice in Gorski et al.\footnote{\citeproc{ref-gorski_healpix_1999}{{``The HEALPix Primer''}}.} revision 3.80.

Also compare to Technology (U.S.)\footnote{\citeproc{ref-technology_us_nist_2010}{\emph{NIST Handbook of Mathematical Functions Hardback and CD-ROM}}.} 14.7.8, 14.7.10, 14.7.13
which shares the same sign convention here.

Mathematica also shares the same convention in \href{https://reference.wolfram.com/language/ref/LegendreP.html}{LegendreP---Wolfram
Language
Documentation}.

\protect\phantomsection\label{2e9e7efc}
Note that \(P^{m}_{\ell}\) below is sometimes denoted as \(P_{\ell m}\),
for example in CMB.

\protect\phantomsection\label{c58cc86e}
\subsection{Definition}\label{definition}

\protect\phantomsection\label{b40d226d}
The general Legendre equation (which reduces to the Legendre equation
when \(m = 0\)),

\protect\phantomsection\label{b5991ef9}
\[(1-x^{2}){\frac {d^{2}}{dx^{2}}}P_{\ell }^{m}(x)-2x{\frac {d}{dx}}P_{\ell }^{m}(x)+\left[\ell (\ell +1)-{\frac {m^{2}}{1-x^{2}}}\right]P_{\ell }^{m}(x)=0\]

\protect\phantomsection\label{27f9bc77}
\[x \in \left[ -1 , 1 \right] , \ell , m \in \mathbb{N} \cup \left\{ 0 \right\} , - \ell \leq m \leq \ell\]

\protect\phantomsection\label{03341be7}
Note that the equation is invariant under \(\ell \rightarrow -\ell -1\),
hence negative values are ignored WLOG.

\protect\phantomsection\label{4e7bcc20}
It has a solution when \(m = 0\) (Rodrigues' formula),

\protect\phantomsection\label{ea0bcd52}
\[P_{\ell }(x)={\frac {1}{2^{\ell }\,\ell !}}\ {\frac {d^{\ell }}{dx^{\ell }}}\left[(x^{2}-1)^{\ell }\right]\]

\protect\phantomsection\label{6edac5c0}
And for general \(m\), the solution is the associated Legendre
polynomials, with the Condon-Shortley phase,

\protect\phantomsection\label{bef26d55}
\[P_{\ell }^{m}(x)=(-1)^{m}(1-x^{2})^{m/2}{\frac {d^{m}}{dx^{m}}}\left(P_{\ell }(x)\right) , m \geq 0\]

\protect\phantomsection\label{56114dc3}
Note that if \(m\) is odd, the associated Legendre ``polynomial'' is not a
polynomial despite the name.

\protect\phantomsection\label{8307b725}
\[P_{\ell }^{m}(x)={\frac {(-1)^{m}}{2^{\ell }\ell !}}(1-x^{2})^{m/2}\ {\frac {d^{\ell +m}}{dx^{\ell +m}}}(x^{2}-1)^{\ell } , - \ell \leq m \leq \ell\]

\protect\phantomsection\label{b46aba34}
Note that the differential equation is invariant under
\(m \rightarrow -m\), hence the 2 sets of solutions are proportional to
each other:

\protect\phantomsection\label{e294afa9}
\[P_{\ell }^{-m}(x)=(-1)^{m}{\frac {(\ell -m)!}{(\ell +m)!}}P_{\ell }^{m}(x)\]

\protect\phantomsection\label{83cbb646}
\subsection{Closed form}\label{closed-form}

\protect\phantomsection\label{c34a21e9}
\[P_{l}^{m}(x)=(-1)^{m}\cdot 2^{l}\cdot (1-x^{2})^{m/2}\cdot \sum _{k=m}^{l}{\frac {k!}{(k-m)!}}\cdot x^{k-m}\cdot {\binom {l}{k}}{\binom {\frac {l+k-1}{2}}{l}}\]

\protect\phantomsection\label{f1ae540a}
Here, we can see that

\protect\phantomsection\label{baed802c}
\[\lim_{x \rightarrow 1^{-}} \operatorname{sgn} \mathopen{} \left ( P^{m}_{\ell} \mathopen{} \left ( x \right ) \mathclose{} \right ) \mathclose{} \rightarrow \left( -1 \right)^{m}\]

\protect\phantomsection\label{a848600d}
\section{Spherical harmonics}\label{sec:sphericalHarmonics}

\subsection{Condon-Shortley phase Convention}\label{condon-shortley-phase-convention-1}

The following convention is from the ``quantum mechanics'' convention in
\href{https://en.wikipedia.org/wiki/Spherical_harmonics\#Orthogonality_and_normalization}{Spherical harmonics -
Wikipedia}.
Note that the formula looks different, but is identical as we absorb the
Condon-Shortley phase into the Associated Legendre polynomials.

It also shares the same convention in Gorski et al.\footnote{\citeproc{ref-gorski_healpix_1999}{{``The HEALPix Primer''}}.} revision
3.80 after simplifications, albeit it does not define the Associated
Legendre polynomials when \(m < 0\).

Mathematica also shares the same convention in
\href{https://reference.wolfram.com/language/ref/SphericalHarmonicY.html}{SphericalHarmonicY---Wolfram Language
Documentation}.

\protect\phantomsection\label{c8ac11aa}
Note that \(Y^{m}_{\ell}\) below is sometimes denoted as \(Y_{\ell m}\),
for example in CMB.

\protect\phantomsection\label{935308c4}
\subsection{Definition}\label{definition-1}

\protect\phantomsection\label{3dcec499}
\[Y_{\ell }^{m}(\theta ,\varphi )={\sqrt {{\frac {(2\ell +1)}{4\pi }}{\frac {(\ell -m)!}{(\ell +m)!}}}}\,P_{\ell }^{m}(\cos {\theta })\,e^{im\varphi }\]

\protect\phantomsection\label{3b82ae8d}
Here, we can see that

\protect\phantomsection\label{06a949d3}
\[\lim_{\theta \rightarrow 0^{+}} \operatorname{sgn} \mathopen{} \left ( Y^{m}_{\ell} \mathopen{} \left ( \theta, 0 \right ) \mathclose{} \right ) \mathclose{} \rightarrow \left( -1 \right)^{m}\]

\protect\phantomsection\label{1909a98e}
\section{Spin-weighted spherical harmonics (SWSH)}\label{spin-weighted-spherical-harmonics-swsh}

See \href{https://en.wikipedia.org/wiki/Spin-weighted_spherical_harmonics}{Spin-weighted spherical harmonics -
Wikipedia}
for details.

\subsection{Spin-weighted functions}\label{spin-weighted-functions}

A function is called spin-weighted \(s\) if it transforms under rotation
on the North pole by angle \(\psi\),

\protect\phantomsection\label{8e69ba7c}
\[\eta_s \rightarrow e^{is\psi }\eta_s\]

\protect\phantomsection\label{2ace15e1}
\subsubsection{\texorpdfstring{Operator \(\eth\)}{Operator \textbackslash eth}}\label{operator-eth}

There exists an operator that can raise the spin-weight of a function.

Explicitly,

\protect\phantomsection\label{f00749b2}
\begin{align*}
\eth \eta   &   =-\left(\sin {\theta }\right)^{s}\left\{{\frac {\partial }{\partial \theta }}+{\frac {i}{\sin {\theta }}}{\frac {\partial }{\partial \phi }}\right\}\left[\left(\sin {\theta }\right)^{-s}\eta \right]  \\
{\bar {\eth }}\eta  &   =-\left(\sin {\theta }\right)^{-s}\left\{{\frac {\partial }{\partial \theta }}-{\frac {i}{\sin {\theta }}}{\frac {\partial }{\partial \phi }}\right\}\left[\left(\sin {\theta }\right)^{s}\eta \right]
\end{align*}

\protect\phantomsection\label{ca22fd1e}
Note that

\protect\phantomsection\label{84446ef1}
\[\bar{\eth}^n g_s = \overline{\eth^n \overline{g_s}}\]

\protect\phantomsection\label{9123361b}
as \(\overline{g_s}\) is of spin-weight \(-s\).

pf.\ by Mathematical Induction.

\protect\phantomsection\label{bcefc33e}
\subsection{Definition}\label{definition-2}

\protect\phantomsection\label{a602b053-3d61-48a2-819a-4863c2741a01}
\[{}_{s}Y_{lm} = {
    \begin{cases}
        {\sqrt {\frac {(l-s)!}{(l+s)!}}}\ \eth ^{s}Y_{lm},    & 0\leq s\leq l;    \\
        {\sqrt {\frac {(l+s)!}{(l-s)!}}}\ \left(-1\right)^{s}{\bar {\eth }}^{-s}Y_{lm},    & -l\leq s\leq 0;    \\
        0, & l<|s|.
    \end{cases}
}\]

\protect\phantomsection\label{da908f3a}
\subsection{Closed form}\label{closed-form-1}

Spin-weighted Spherical Harmonics with Condon-Shortley phase. See
\href{https://en.wikipedia.org/wiki/Spin-weighted_spherical_harmonics\#Calculating}{Spin-weighted spherical harmonics -
Wikipedia}.

\protect\phantomsection\label{dbc8ceb9}
\begin{align*}
{}_{s}Y_{lm}(\theta ,\phi )     &=\left(-1\right)^{m}{\sqrt {\frac {(l+m)!(l-m)!(2l+1)}{4\pi (l+s)!(l-s)!}}}\sin ^{2l}\left({\frac {\theta }{2}}\right) \\
    &\times \sum _{r=0}^{l-s}{l-s \choose r}{l+s \choose r+s-m}\left(-1\right)^{l-r-s}e^{im\phi }\cot ^{2r+s-m}\left({\frac {\theta }{2}}\right).
\end{align*}

\protect\phantomsection\label{ad9102e7}
\subsection{Note on Condon-Shortley phase Convention}\label{note-on-condon-shortley-phase-convention}

Note that there exist different sign conventions. Healpix does not
explicitly declare the sign convention used in their
documentation/Primer, but

\begin{enumerate}
\def\labelenumi{\arabic{enumi}.}
\tightlist
\item
  since its definition of Spherical Harmonics agrees with above,
\item
  we consistently include the Condon-Shortley phase here in both the
  associated Legendre polynomials and spherical harmonics, and
\item
  Spin-weighted spherical harmonics defined through eth of spherical
  harmonics does not seem to have different sign convention
  (for example, compare Zaldarriaga and Seljak,\footnote{\citeproc{ref-zaldarriaga_all-sky_1997}{{``All-Sky Analysis of Polarization in the Microwave Background''}}.} no.~A2 to the eth defined
  above.)
\item
  Zaldarriaga and Seljak,\footnote{\citeproc{ref-zaldarriaga_all-sky_1997}{{``All-Sky Analysis of Polarization in the Microwave Background''}}.} no.~A6 agrees with the closed form
  formula above.
\end{enumerate}

We should be confident that the above definition agrees with
Healpix and the CMB literature.

\section{Wigner matrices}\label{wigner-matrices}

\subsection{Conventions}\label{conventions-1}

Note that different authors use different conventions, sometimes without
any notes or explanations.

The main source of confusion probably comes from active vs.~passive
transformation. Below we use the active transformation commonly found
in Quantum Mechanics. See citations below.

\subsection{Definitions}\label{definitions-1}

With the convention of z-y-z convention, right-handed frame, right-hand
screw rule, active interpretation

\protect\phantomsection\label{62880260}
\[D_{m'm}^{j}(\alpha ,\beta ,\gamma )\equiv \langle jm'|{\mathcal {R}}(\alpha ,\beta ,\gamma )|jm\rangle =e^{-im'\alpha }d_{m'm}^{j}(\beta )e^{-im\gamma }\]

\protect\phantomsection\label{859e108d}
See proofs in Sakurai and Napolitano,\footnote{\citeproc{ref-sakurai_modern_2017}{\emph{Modern Quantum Mechanics}}.} p.~175, 177, 196, 198.

Note that this has a different sign convention from Mathematica.\footnote{\href{https://reference.wolfram.com/language/ref/WignerD.html}{WignerD---Wolfram Language
  Documentation}}

The Wigner d-matrix has closed form formula below, with the sign
convention given by Wigner,\footnote{\citeproc{ref-wigner_group_1959}{\emph{Group Theory and Its Application to the Quantum Mechanics of Atomic Spectra}}.} which is also used in
Sakurai and Napolitano,\footnote{\citeproc{ref-sakurai_modern_2017}{\emph{Modern Quantum Mechanics}}.} p.~238. See \href{https://en.wikipedia.org/wiki/Wigner_D-matrix\#Wigner_(small)_d-matrix}{Wigner D-matrix -
Wikipedia}.

\protect\phantomsection\label{f95a6019}
\begin{align*}
d_{m'm}^{j}(\beta )     &= [(j+m')!(j-m')!(j+m)!(j-m)!]^{\frac {1}{2}}    \\
    &\times \sum _{s=s_{\mathrm {min} }}^{s_{\mathrm {max} }}\left[{\frac {(-1)^{m'-m+s}\left(\cos {\frac {\beta }{2}}\right)^{2j+m-m'-2s}\left(\sin {\frac {\beta }{2}}\right)^{m'-m+2s}}{(j+m-s)!s!(m'-m+s)!(j-m'-s)!}}\right]
\end{align*}

\protect\phantomsection\label{35ff41e3}
where

\protect\phantomsection\label{6c0bc8b8}
\begin{align*}
s_{\mathrm {min} }  &   =\mathrm {max} (0,m-m') \\
s_{\mathrm {max} }  &   =\mathrm {min} (j+m,j-m')
\end{align*}

\protect\phantomsection\label{59cd51e6}
\subsection{Note on Convention}\label{note-on-convention}

Note that Mathematica uses a different convention

\begin{enumerate}
\def\labelenumi{\arabic{enumi}.}
\item
  in terms of the sign

\begin{Shaded}
\begin{Highlighting}[]
\NormalTok{WignerDPhysics}\OperatorTok{[\{}\AttributeTok{j\_}\OperatorTok{,} \AttributeTok{m1\_}\OperatorTok{,} \AttributeTok{m2\_}\OperatorTok{\},}\NormalTok{ θ\_}\OperatorTok{]} \ExtensionTok{=}\NormalTok{ (}\SpecialCharTok{{-}}\DecValTok{1}\NormalTok{)}\SpecialCharTok{\^{}}\NormalTok{(m1 }\SpecialCharTok{{-}}\NormalTok{ m2) }\FunctionTok{WignerD}\OperatorTok{[\{}\FunctionTok{j}\OperatorTok{,}\NormalTok{ m1}\OperatorTok{,}\NormalTok{ m2}\OperatorTok{\},}\NormalTok{ θ}\OperatorTok{]}
\end{Highlighting}
\end{Shaded}
\item
  and also in label convention: the 3 argument form \((ψ, θ, φ)\) are
  the 3 Euler angles, and is commonly defined as \((α, β, γ)\), or
  \((φ, θ, ψ)\) instead.
\end{enumerate}

\subsection{Relationship with SWSH}\label{relationship-with-swsh}

In Mathematica,

\begin{Shaded}
\begin{Highlighting}[]
\NormalTok{SpinWeightedSphericalHarmonicY}\OperatorTok{[}\AttributeTok{s\_}\OperatorTok{,} \AttributeTok{l\_}\OperatorTok{,} \AttributeTok{m\_}\OperatorTok{,}\NormalTok{ θ\_}\OperatorTok{,}\NormalTok{ φ\_}\OperatorTok{]} \ExtensionTok{=} \SpecialCharTok{\textbackslash{}}
\NormalTok{    (}\SpecialCharTok{{-}}\DecValTok{1}\NormalTok{)}\SpecialCharTok{\^{}}\FunctionTok{m} \FunctionTok{Sqrt}\OperatorTok{[}\NormalTok{(}\DecValTok{2} \FunctionTok{l} \SpecialCharTok{+} \DecValTok{1}\NormalTok{) }\SpecialCharTok{/}\NormalTok{ (}\DecValTok{4}\NormalTok{ π)}\OperatorTok{]}\NormalTok{ WignerDPhysics}\OperatorTok{[\{}\FunctionTok{l}\OperatorTok{,}\SpecialCharTok{{-}}\FunctionTok{m}\OperatorTok{,}\FunctionTok{s}\OperatorTok{\},}\NormalTok{ θ}\OperatorTok{]} \FunctionTok{E}\SpecialCharTok{\^{}}\NormalTok{(}\FunctionTok{I} \FunctionTok{m}\NormalTok{ φ)}

\CommentTok{(* with Mathematica\textquotesingle{}s sign convention: *)}
\NormalTok{SpinWeightedSphericalHarmonicY}\OperatorTok{[}\AttributeTok{s\_}\OperatorTok{,} \AttributeTok{l\_}\OperatorTok{,} \AttributeTok{m\_}\OperatorTok{,}\NormalTok{ θ\_}\OperatorTok{,}\NormalTok{ φ\_}\OperatorTok{]} \ExtensionTok{=} \SpecialCharTok{\textbackslash{}}
\NormalTok{    (}\SpecialCharTok{{-}}\DecValTok{1}\NormalTok{)}\SpecialCharTok{\^{}}\NormalTok{(}\SpecialCharTok{{-}}\FunctionTok{s}\NormalTok{) }\FunctionTok{Sqrt}\OperatorTok{[}\NormalTok{(}\DecValTok{2} \FunctionTok{l} \SpecialCharTok{+} \DecValTok{1}\NormalTok{) }\SpecialCharTok{/}\NormalTok{ (}\DecValTok{4}\NormalTok{ π)}\OperatorTok{]} \FunctionTok{WignerD}\OperatorTok{[\{}\FunctionTok{l}\OperatorTok{,}\SpecialCharTok{{-}}\FunctionTok{m}\OperatorTok{,}\FunctionTok{s}\OperatorTok{\},}\NormalTok{ θ}\OperatorTok{]} \FunctionTok{E}\SpecialCharTok{\^{}}\NormalTok{(}\FunctionTok{I} \FunctionTok{m}\NormalTok{ φ)}
\end{Highlighting}
\end{Shaded}

\protect\phantomsection\label{46a9ee01}
Mathematically,

\begin{align*}
{}_{s} Y_{\ell m} \mathopen{} \left ( \theta, \phi \right ) \mathclose{}    &    = \left( -1 \right)^{m} \sqrt{\frac{2 \ell + 1}{4 \pi}} d^{\ell}_{- m s} \mathopen{} \left ( \theta \right ) \mathclose{} \mathrm{e}^{\mathrm{i} m \phi}   \\
    &    = \left( -1 \right)^{m} \sqrt{\frac{2 \ell + 1}{4 \pi}} D^{\ell}_{- m s} \mathopen{} \left ( \phi, \theta, 0 \right ) \mathclose{}
\end{align*}

\chapter{\texorpdfstring{Generalized convolution theorem on \(\operatorname{SO} \mathopen{} \left ( 3 \right ) \mathclose{}\)}{Generalized convolution theorem on \textbackslash operatorname\{SO\} \textbackslash mathopen\{\} \textbackslash left ( 3 \textbackslash right ) \textbackslash mathclose\{\}}}\label{sec:generalized-convolution-theorem}

\begin{Definition}\label{eq:SO3Convolution}
\[\left( {}_{{s}} f \ast {}_{{s^{\prime}}} g \right) \mathopen{} \left ( R \right ) \mathclose{} \equiv \frac{1}{2 \pi} \int_{R^{\prime} \in \operatorname{SO} \mathopen{} \left ( 3 \right ) \mathclose{}} {}_{{s}} f \mathopen{} \left ( R^{\prime} \right ) \mathclose{} {}_{{s^{\prime}}} g \mathopen{} \left ( R^{\prime -1} R \right ) \mathclose{} \,\mathrm{d} R^{\prime}\]
\end{Definition}

\begin{Theorem}
\[{}_{{s^{\prime}}} \left( {}_{{s}} f \ast {}_{{s^{\prime}}} g \right)_{\ell m} = \left( -1 \right)^{- s} \sqrt{\frac{4 \pi}{2 \ell + 1}} {}_{{s}} f_{\ell m} {}_{{s^{\prime}}} g_{\ell , - s}\]
\end{Theorem}

\begin{proof}

\begin{align*}
\left( {}_{{s}} f \ast {}_{{s^{\prime}}} g \right) \mathopen{} \left ( R \right ) \mathclose{}	&	 \equiv \frac{1}{2 \pi} \int_{R^{\prime} \in \operatorname{SO} \mathopen{} \left ( 3 \right ) \mathclose{}} {}_{{s}} f \mathopen{} \left ( R^{\prime} \right ) \mathclose{} {}_{{s^{\prime}}} g \mathopen{} \left ( R^{\prime -1} R \right ) \mathclose{} \,\mathrm{d} R^{\prime}	\\
	&	 = \sum_{\ell m \ell^{\prime} m^{\prime}} \frac{1}{2 \pi} \int_{R^{\prime} \in \operatorname{SO} \mathopen{} \left ( 3 \right ) \mathclose{}} {}_{{s}} f_{\ell m} {}_{s} Y_{l m} \mathopen{} \left ( R^{\prime} \right ) \mathclose{} {}_{{s^{\prime}}} g_{\ell^{\prime} m^{\prime}} {}_{s^{\prime}} Y_{\ell^{\prime} m^{\prime}} \mathopen{} \left ( R^{\prime -1} R \right ) \mathclose{} \,\mathrm{d} R^{\prime}	\\
	&	 = \sum_{\ell m \ell^{\prime} m^{\prime}} \frac{1}{2 \pi} \int_{R^{\prime} \in \operatorname{SO} \mathopen{} \left ( 3 \right ) \mathclose{}} {}_{{s}} f_{\ell m} {}_{{s^{\prime}}} g_{\ell^{\prime} m^{\prime}}  \\ & \times \left( -1 \right)^{m} \sqrt{\frac{2 \ell + 1}{4 \pi}} D^{\left( \ell \right)}_{- m s} \mathopen{} \left ( R^{\prime} \right ) \mathclose{} \sum_{m^{\prime \prime}} \left[ D^{\left( \ell^{\prime} \right)}_{m^{\prime} m^{\prime \prime}} \mathopen{} \left ( R^{\prime -1} \right ) \mathclose{} \right]^{ \ast } {}_{s^{\prime}} Y_{\ell^{\prime} m^{\prime \prime}} \mathopen{} \left ( R \right ) \mathclose{} \,\mathrm{d} R^{\prime}	\\
\left[ D^{\left( \ell^{\prime} \right)}_{m^{\prime} m^{\prime \prime}} \mathopen{} \left ( R^{\prime -1} \right ) \mathclose{} \right]^{ \ast }	&	 = D^{\left( \ell^{\prime} \right)}_{m^{\prime \prime} m^{\prime}} \mathopen{} \left ( R^{\prime} \right ) \mathclose{} = \left( -1 \right)^{m^{\prime} - m^{\prime \prime}} \left[ D^{\left( \ell^{\prime} \right)}_{- m^{\prime \prime} , - m^{\prime}} \mathopen{} \left ( R^{\prime} \right ) \mathclose{} \right]^{ \ast }	\\
\left( {}_{{s}} f \ast {}_{{s^{\prime}}} g \right) \mathopen{} \left ( R \right ) \mathclose{}	&	 = \sum_{\ell m \ell^{\prime} m^{\prime} m^{\prime \prime}} \frac{1}{2 \pi} {}_{{s}} f_{\ell m} {}_{{s^{\prime}}} g_{\ell^{\prime} m^{\prime}} \left( -1 \right)^{m + m^{\prime} - m^{\prime \prime}} \sqrt{\frac{2 \ell + 1}{4 \pi}} {}_{s^{\prime}} Y_{\ell^{\prime} m^{\prime \prime}} \mathopen{} \left ( R \right ) \mathclose{}  \\ & \times \int_{R^{\prime} \in \operatorname{SO} \mathopen{} \left ( 3 \right ) \mathclose{}} D^{\left( \ell \right)}_{- m s} \mathopen{} \left ( R^{\prime} \right ) \mathclose{} \left[ D^{\left( \ell^{\prime} \right)}_{- m^{\prime \prime} , - m^{\prime}} \mathopen{} \left ( R^{\prime} \right ) \mathclose{} \right]^{ \ast } \,\mathrm{d} R^{\prime}	\\
	&	 = \sum_{\ell m \ell^{\prime} m^{\prime} m^{\prime \prime}} \frac{1}{2 \pi} {}_{{s}} f_{\ell m} {}_{{s^{\prime}}} g_{\ell^{\prime} m^{\prime}} \left( -1 \right)^{m + m^{\prime} - m^{\prime \prime}} \sqrt{\frac{2 \ell + 1}{4 \pi}} {}_{s^{\prime}} Y_{\ell^{\prime} m^{\prime \prime}} \mathopen{} \left ( R \right ) \mathclose{}  \\ & \times \frac{8 \pi^{2}}{2 \ell + 1} \delta_{\ell , \ell^{\prime}} \delta_{- m , - m^{\prime \prime}} \delta_{s , - m^{\prime}}	\\
	&	 = \left( -1 \right)^{- s} \sqrt{\frac{4 \pi}{2 \ell + 1}} \sum_{\ell , m} {}_{{s}} f_{\ell m} \, {}_{s^{\prime}} g_{\ell, - s} \, {}_{s^{\prime}} Y_{\ell m} \mathopen{} \left ( R \right ) \mathclose{}
\end{align*}

\end{proof}

\begin{Definition}\label{eq:S2Convolution}
\[\left( {}_{{s}} f \odot {}_{{s}} g \right) \mathopen{} \left ( R \right ) \mathclose{} \equiv \oint {}_{{s}} f \mathopen{} \left ( R \mathbf{\hat{n}} \right ) \mathclose{} {}_{{s}} g^{ \ast } \mathopen{} \left ( \mathbf{\hat{n}} \right ) \mathclose{} \,\mathrm{d} \mathbf{\hat{n}}\]
\end{Definition}

\begin{Theorem}
\[{}_{{s}} f \odot {}_{{s}} g \mathopen{} \left ( \phi, \theta, \psi \right ) \mathclose{} = \sum_{\ell m m^{\prime}} {}_{{s}} f_{\ell m} \left( -1 \right)^{- m^{\prime}} \sqrt{\frac{4 \pi}{2 \ell + 1}} {}_{- m^{\prime}} Y_{\ell m} \mathopen{} \left ( \theta, \phi \right ) \mathclose{} \mathrm{e}^{\mathrm{i} m^{\prime} \psi} {}_{{s}} g^{ \ast }_{\ell m^{\prime}}\]
\end{Theorem}

\begin{proof}

\begin{align*}
\left( {}_{{s}} f \odot {}_{{s}} g \right) \mathopen{} \left ( R \right ) \mathclose{}	&	 \equiv \oint {}_{{s}} f \mathopen{} \left ( R \mathbf{\hat{n}} \right ) \mathclose{} {}_{{s}} g^{ \ast } \mathopen{} \left ( \mathbf{\hat{n}} \right ) \mathclose{} \,\mathrm{d} \mathbf{\hat{n}}	\\
	&	 = \sum_{\ell m} \int {}_{{s}} f_{\ell m} {}_{s} Y_{\ell m} \mathopen{} \left ( R \mathbf{\hat{n}} \right ) \mathclose{} {}_{{s}} g^{ \ast } \mathopen{} \left ( \mathbf{\hat{n}} \right ) \mathclose{} \,\mathrm{d} \mathbf{\hat{n}}	\\
	&	 = \sum_{\ell m m^{\prime}} \int {}_{{s}} f_{\ell m} \left[ D^{\left( \ell \right)}_{m m^{\prime}} \mathopen{} \left ( R \right ) \mathclose{} \right]^{ \ast } {}_{s} Y_{\ell m^{\prime}} \mathopen{} \left ( \mathbf{\hat{n}} \right ) \mathclose{} {}_{{s}} g^{ \ast } \mathopen{} \left ( \mathbf{\hat{n}} \right ) \mathclose{} \,\mathrm{d} \mathbf{\hat{n}}	\\
	&	 = \sum_{\ell \ell^{\prime} m m^{\prime} m^{\prime \prime}} \int {}_{{s}} f_{\ell m} \left[ D^{\left( \ell \right)}_{m m^{\prime}} \mathopen{} \left ( R \right ) \mathclose{} \right]^{ \ast } {}_{s} Y_{\ell m^{\prime}} \mathopen{} \left ( \mathbf{\hat{n}} \right ) \mathclose{} {}_{{s}} g^{ \ast }_{\ell^{\prime} m^{\prime \prime}} {}_{s} Y_{\ell^{\prime} m^{\prime \prime}}^{ \ast } \mathopen{} \left ( \mathbf{\hat{n}} \right ) \mathclose{} \,\mathrm{d} \mathbf{\hat{n}}	\\
	&	 = \sum_{\ell \ell^{\prime} m m^{\prime} m^{\prime \prime}} {}_{{s}} f_{\ell m} \left[ D^{\left( \ell \right)}_{m m^{\prime}} \mathopen{} \left ( R \right ) \mathclose{} \right]^{ \ast } {}_{{s}} g^{ \ast }_{\ell^{\prime} m^{\prime \prime}} \delta_{\ell \ell^{\prime}} \delta_{m^{\prime} m^{\prime \prime}}	\\
	&	 = \sum_{\ell m m^{\prime}} {}_{{s}} f_{\ell m} \left[ D^{\left( \ell \right)}_{m m^{\prime}} \mathopen{} \left ( R \right ) \mathclose{} \right]^{ \ast } {}_{{s}} g^{ \ast }_{\ell m^{\prime}}	\\
{}_{{s}} f \odot {}_{{s}} g \mathopen{} \left ( \phi, \theta, \psi \right ) \mathclose{}	&	 = \sum_{\ell m m^{\prime}} {}_{{s}} f_{\ell m} \left( -1 \right)^{- m^{\prime}} \sqrt{\frac{4 \pi}{2 \ell + 1}} {}_{s} Y_{\ell m} \mathopen{} \left ( \theta, \phi \right ) \mathclose{} \mathrm{e}^{\mathrm{i} m^{\prime} \psi} {}_{{s}} g^{ \ast }_{\ell m^{\prime}}
\end{align*}

\end{proof}

\chapter{\texorpdfstring{Final configuration of the null-test v1.0.2 used in POLARBEAR large-patch high-\(\ell\) analysis}{Final configuration of the null-test v1.0.2 used in POLARBEAR large-patch high-\textbackslash ell analysis}}\label{sec:POLARBEAR-configuration}

\chaptermark{Final configuration of the null-test v1.0.2}

The configuration used in the final analysis in \cref{sec:null-test-results} is presented below.

\section{Configuration for real maps}\label{configuration-for-real-maps}

\begin{Shaded}
\begin{Highlighting}[]
\CommentTok{\# "variables" to be reuse later}
\FunctionTok{common}\KeywordTok{:}

\AttributeTok{  }\FunctionTok{mapsplits}\KeywordTok{:}\AttributeTok{ }\OtherTok{\&mapsplits}
\AttributeTok{    }\FunctionTok{mapsplits}\KeywordTok{:}
\AttributeTok{      }\KeywordTok{{-} }\AttributeTok{FIRST\_SECOND}
\AttributeTok{      }\CommentTok{\# {-} FIFTH\_SEPARATE}
\AttributeTok{      }\CommentTok{\# {-} FOURTH\_SEPARATE}
\AttributeTok{      }\CommentTok{\# {-} THIRD\_SEPARATE}
\AttributeTok{      }\KeywordTok{{-} }\AttributeTok{RS\_MIDDLE}
\AttributeTok{      }\KeywordTok{{-} }\AttributeTok{RM\_SET}
\AttributeTok{      }\KeywordTok{{-} }\AttributeTok{SM\_RISE}
\AttributeTok{      }\KeywordTok{{-} }\AttributeTok{LR\_SUBSCAN}
\AttributeTok{      }\KeywordTok{{-} }\AttributeTok{PWV}
\AttributeTok{      }\KeywordTok{{-} }\AttributeTok{GAIN\_BY\_CES}
\AttributeTok{      }\KeywordTok{{-} }\AttributeTok{QU\_PIXEL}
\AttributeTok{      }\KeywordTok{{-} }\AttributeTok{LHS\_RHS}
\AttributeTok{      }\KeywordTok{{-} }\AttributeTok{SUN\_DIST}
\AttributeTok{      }\KeywordTok{{-} }\AttributeTok{SUN\_HORIZON}
\AttributeTok{      }\KeywordTok{{-} }\AttributeTok{MOON\_DIST}
\AttributeTok{      }\KeywordTok{{-} }\AttributeTok{MOON\_HORIZON}
\AttributeTok{      }\KeywordTok{{-} }\AttributeTok{LEAK\_BY\_CES}
\AttributeTok{      }\KeywordTok{{-} }\AttributeTok{LEAK\_BY\_BOLO}
\AttributeTok{      }\KeywordTok{{-} }\AttributeTok{AMP\_2F\_BY\_CES}
\AttributeTok{      }\KeywordTok{{-} }\AttributeTok{AMP\_2F\_BY\_BOLO}
\AttributeTok{      }\KeywordTok{{-} }\AttributeTok{AMP\_4F\_BY\_CES}
\AttributeTok{      }\KeywordTok{{-} }\AttributeTok{AMP\_4F\_BY\_BOLO}
\AttributeTok{      }\CommentTok{\# {-} RANDOM\_BY\_BOLO1}
\AttributeTok{      }\CommentTok{\# {-} RANDOM\_BY\_BOLO2}
\AttributeTok{      }\CommentTok{\# {-} RANDOM\_BY\_BOLO3}
\AttributeTok{      }\CommentTok{\# {-} RANDOM\_BY\_BOLO4}
\AttributeTok{      }\CommentTok{\# {-} RANDOM\_BY\_BOLO5}
\AttributeTok{      }\CommentTok{\# {-} RANDOM\_BY\_BOLO6}
\AttributeTok{      }\CommentTok{\# {-} RANDOM\_BY\_BOLO7}
\AttributeTok{      }\CommentTok{\# {-} RANDOM\_BY\_BOLO8}
\AttributeTok{      }\CommentTok{\# {-} RANDOM\_BY\_CES1}
\AttributeTok{      }\CommentTok{\# {-} RANDOM\_BY\_CES2}
\AttributeTok{      }\CommentTok{\# {-} RANDOM\_BY\_CES3}
\AttributeTok{      }\CommentTok{\# {-} RANDOM\_BY\_CES4}
\AttributeTok{      }\CommentTok{\# {-} RANDOM\_BY\_CES5}
\AttributeTok{      }\CommentTok{\# {-} RANDOM\_BY\_CES6}
\AttributeTok{      }\CommentTok{\# {-} RANDOM\_BY\_CES7}
\AttributeTok{      }\CommentTok{\# {-} RANDOM\_BY\_CES8}
\AttributeTok{      }\KeywordTok{{-} }\AttributeTok{THS\_BHS}
\AttributeTok{      }\KeywordTok{{-} }\AttributeTok{TOP\_BOTTOM}
\AttributeTok{      }\CommentTok{\# {-} RANDOM\_BUT\_PAIR\_BY\_BOLO1}


\CommentTok{\# mapmaking sim iter}

\FunctionTok{mapbolo\_sim}\KeywordTok{:}\AttributeTok{ }\OtherTok{\&boloargs}
\AttributeTok{  }\FunctionTok{gain\_folder}\KeywordTok{:}\AttributeTok{ $SCRATCH/data/largepatch4/gain\_181224\_renorm}
\AttributeTok{  }\FunctionTok{psd\_folder}\KeywordTok{:}\AttributeTok{ $SCRATCH/data/largepatch4/psd\_iota\_poleff\_181224\_renorm}
\AttributeTok{  }\FunctionTok{filterfolder}\KeywordTok{:}\AttributeTok{ $SCRATCH/data/largepatch4/filterdb\_iota\_final}
\AttributeTok{  }\FunctionTok{alt\_flag\_folder}\KeywordTok{:}\AttributeTok{ $SCRATCH/data/largepatch4/iota\_dataselection\_final\_highell\_corr}
\AttributeTok{  }\CommentTok{\# projection args}
\AttributeTok{  }\FunctionTok{width}\KeywordTok{:}\AttributeTok{ }\DecValTok{45}
\AttributeTok{  }\FunctionTok{fixed\_offset}\KeywordTok{:}\AttributeTok{ BICEP}
\AttributeTok{  }\FunctionTok{pixelsize}\KeywordTok{:}\AttributeTok{ }\DecValTok{2}
\AttributeTok{  }\FunctionTok{hwppos}\KeywordTok{:}\AttributeTok{ }\DecValTok{45}
\AttributeTok{  }\FunctionTok{projection}\KeywordTok{:}\AttributeTok{ OLEA}
\AttributeTok{  }\CommentTok{\# filter args}
\AttributeTok{  }\FunctionTok{detrend}\KeywordTok{:}\AttributeTok{ }\DecValTok{2}
\AttributeTok{  }\KeywordTok{? }\FunctionTok{groundlr}
\AttributeTok{  }\FunctionTok{groundtemplate}\KeywordTok{:}\AttributeTok{ }\FloatTok{0.24}
\AttributeTok{  }\FunctionTok{poly}\KeywordTok{:}\AttributeTok{ }\DecValTok{9}
\AttributeTok{  }\FunctionTok{polpoly}\KeywordTok{:}\AttributeTok{ }\DecValTok{9}
\AttributeTok{  }\KeywordTok{? }\FunctionTok{commonmode}
\AttributeTok{  }\KeywordTok{? }\FunctionTok{notch\_lines}
\AttributeTok{  }\FunctionTok{notch\_line\_width}\KeywordTok{:}\AttributeTok{ }\FloatTok{0.01}
\AttributeTok{  }\FunctionTok{mask\_radius}\KeywordTok{:}\AttributeTok{ }\FloatTok{3.5}
\AttributeTok{  }\CommentTok{\# null only}
\AttributeTok{  }\FunctionTok{null\_folder}\KeywordTok{:}\AttributeTok{ $SCRATCH/data/largepatch4/null\_test\_flags\_iota\_highell\_corr\_pair}
\AttributeTok{  }\FunctionTok{\textless{}\textless{}}\KeywordTok{:}\AttributeTok{ }\OtherTok{*mapsplits}

\FunctionTok{mapbolo\_sim\_iter}\KeywordTok{:}\AttributeTok{ }\OtherTok{\&boloinput}
\AttributeTok{  }\FunctionTok{f}\KeywordTok{:}\AttributeTok{ $SCRATCH/data/largepatch4/hdf5\_ces\_dmds}

\CommentTok{\# coadd}

\FunctionTok{coadd}\KeywordTok{:}
\AttributeTok{  }\FunctionTok{basedir}\KeywordTok{:}\AttributeTok{ $SCRATCH/largepatch\_null/high\_1\_0\_2/maps}
\AttributeTok{  }\CommentTok{\# outsubdir: coadd}
\AttributeTok{  }\FunctionTok{\textless{}\textless{}}\KeywordTok{:}\AttributeTok{ }\OtherTok{*mapsplits}

\CommentTok{\# for coadd\_mpi\_iter.py}
\FunctionTok{coadd\_sim}\KeywordTok{:}
\AttributeTok{  }\CommentTok{\# outsubdir: coadd}
\AttributeTok{  }\FunctionTok{\textless{}\textless{}}\KeywordTok{:}\AttributeTok{ }\OtherTok{*mapsplits}

\CommentTok{\# pseudo}

\FunctionTok{pseudo}\KeywordTok{:}
\AttributeTok{  }\CommentTok{\# weight\_name: realmap}
\AttributeTok{  }\FunctionTok{lmax}\KeywordTok{:}\AttributeTok{ }\DecValTok{4901}
\AttributeTok{  }\FunctionTok{pixel\_size}\KeywordTok{:}\AttributeTok{ }\DecValTok{2}

\AttributeTok{  }\FunctionTok{selectionpath}\KeywordTok{:}\AttributeTok{ $SCRATCH/largepatch\_null/high\_1\_0\_2/weight\_nauto.hdf5}
\AttributeTok{  }\FunctionTok{maskpath}\KeywordTok{:}\AttributeTok{ $SCRATCH/largepatch\_null/high\_1\_0\_2/masks.hdf5}
\AttributeTok{  }\FunctionTok{weightdir}\KeywordTok{:}\AttributeTok{ $SCRATCH/largepatch\_null/high\_1\_0\_2/coadd/null\_realmaps}

\FunctionTok{pseudo\_iter}\KeywordTok{:}
\AttributeTok{  }\FunctionTok{\textless{}\textless{}}\KeywordTok{:}\AttributeTok{ }\OtherTok{*mapsplits}

\CommentTok{\# null map to full map}

\FunctionTok{null2full}\KeywordTok{:}
\AttributeTok{  }\FunctionTok{basedir}\KeywordTok{:}\AttributeTok{ $SCRATCH/largepatch\_null/high\_1\_0\_2/coadd}
\AttributeTok{  }\FunctionTok{outbasedir}\KeywordTok{:}\AttributeTok{ $SCRATCH/largepatch\_null/high\_1\_0\_2/coadd\_full}
\CommentTok{\# cat args{-}common.yml before this}

\CommentTok{\# "variables" to be reuse later}
\FunctionTok{template}\KeywordTok{:}

\AttributeTok{  }\FunctionTok{iter}\KeywordTok{:}\AttributeTok{ }\OtherTok{\&iter}
\AttributeTok{    }\FunctionTok{mapcase}\KeywordTok{:}
\AttributeTok{      }\KeywordTok{{-} }\AttributeTok{null\_realmaps}
\AttributeTok{    }\FunctionTok{nreal}\KeywordTok{:}
\AttributeTok{      }\KeywordTok{{-} }\CharTok{null}
\AttributeTok{    }\FunctionTok{save\_weight}\KeywordTok{:}
\AttributeTok{      }\KeywordTok{{-} }\CharTok{True}

\CommentTok{\# mapmaking{-}real}

\FunctionTok{mapbolo\_whwp}\KeywordTok{:}
\AttributeTok{  }\FunctionTok{\textless{}\textless{}}\KeywordTok{:}\AttributeTok{ }\OtherTok{*boloargs}
\AttributeTok{  }\KeywordTok{? }\FunctionTok{removeleakage\_pca}
\AttributeTok{  }\FunctionTok{leak\_fmax}\KeywordTok{:}\AttributeTok{ }\FloatTok{0.4}
\AttributeTok{  }\FunctionTok{correct\_poleff}\KeywordTok{:}\AttributeTok{ $SCRATCH/data/largepatch4/analysis\_data\_iota/cal/lp/combined\_poleff\_20181221.pkl}
\AttributeTok{  }\KeywordTok{? }\FunctionTok{deconvolve}
\FunctionTok{mapbolo\_whwp\_iter}\KeywordTok{:}
\AttributeTok{  }\FunctionTok{\textless{}\textless{}}\KeywordTok{:}\AttributeTok{ }\OtherTok{*boloinput}
\AttributeTok{  }\FunctionTok{tssim}\KeywordTok{:}
\AttributeTok{    }\KeywordTok{{-} }\CharTok{null}
\AttributeTok{  }\FunctionTok{whitenoise}\KeywordTok{:}
\AttributeTok{    }\KeywordTok{{-} }\CharTok{False}
\AttributeTok{  }\FunctionTok{\textless{}\textless{}}\KeywordTok{:}\AttributeTok{ }\OtherTok{*iter}

\CommentTok{\# coadd{-}real}

\FunctionTok{coadd\_iter}\KeywordTok{:}\AttributeTok{ }\OtherTok{\&coadd\_iter}
\AttributeTok{  }\FunctionTok{name}\KeywordTok{:}
\AttributeTok{    }\KeywordTok{{-} }\AttributeTok{realmap}
\AttributeTok{  }\FunctionTok{\textless{}\textless{}}\KeywordTok{:}\AttributeTok{ }\OtherTok{*iter}

\CommentTok{\# coadd{-}real{-}final}

\FunctionTok{coadd\_final}\KeywordTok{:}
\AttributeTok{  }\FunctionTok{basedir}\KeywordTok{:}\AttributeTok{ $SCRATCH/largepatch\_null/high\_1\_0\_2/coadd}
\AttributeTok{  }\CommentTok{\# outsubdir: coadd}
\AttributeTok{  }\FunctionTok{\textless{}\textless{}}\KeywordTok{:}\AttributeTok{ }\OtherTok{*mapsplits}
\FunctionTok{coadd\_final\_iter}\KeywordTok{:}
\AttributeTok{  }\FunctionTok{\textless{}\textless{}}\KeywordTok{:}\AttributeTok{ }\OtherTok{*coadd\_iter}

\CommentTok{\# coadd{-}sign flip}

\FunctionTok{coadd\_sign}\KeywordTok{:}
\AttributeTok{  }\FunctionTok{basedir}\KeywordTok{:}\AttributeTok{ $SCRATCH/largepatch\_null/high\_1\_0\_2/maps/null\_realmaps}
\AttributeTok{  }\FunctionTok{signpath}\KeywordTok{:}\AttributeTok{ $SCRATCH/data/largepatch\_high/obs\_paths\_by\_dir\_0\_13\_sign\_flip\_512.hdf5}
\AttributeTok{  }\FunctionTok{name}\KeywordTok{:}\AttributeTok{ realmap}
\AttributeTok{  }\FunctionTok{nreal}\KeywordTok{:}\AttributeTok{ }\DecValTok{512}
\AttributeTok{  }\CommentTok{\# outsubdir: coadd}
\AttributeTok{  }\FunctionTok{\textless{}\textless{}}\KeywordTok{:}\AttributeTok{ }\OtherTok{*mapsplits}

\CommentTok{\# null map to full map}

\FunctionTok{null2full\_iter}\KeywordTok{:}
\AttributeTok{  }\FunctionTok{mapcase}\KeywordTok{:}
\AttributeTok{    }\KeywordTok{{-} }\AttributeTok{null\_realmaps}
\AttributeTok{  }\FunctionTok{name}\KeywordTok{:}
\AttributeTok{    }\KeywordTok{{-} }\AttributeTok{realmap}
\AttributeTok{  }\FunctionTok{nreal}\KeywordTok{:}
\AttributeTok{    }\KeywordTok{{-} }\CharTok{null}
\AttributeTok{  }\FunctionTok{final}\KeywordTok{:}
\AttributeTok{    }\KeywordTok{{-} }\CharTok{True}
\AttributeTok{  }\FunctionTok{rawmaps}\KeywordTok{:}
\AttributeTok{    }\KeywordTok{{-} {-} }\AttributeTok{d0}
\AttributeTok{      }\KeywordTok{{-} }\AttributeTok{d4r}
\AttributeTok{      }\KeywordTok{{-} }\AttributeTok{d4i}
\AttributeTok{      }\KeywordTok{{-} }\AttributeTok{w0}
\AttributeTok{      }\KeywordTok{{-} }\AttributeTok{w4}
\end{Highlighting}
\end{Shaded}

\section{Configuration for simulated maps}\label{configuration-for-simulated-maps}

\begin{Shaded}
\begin{Highlighting}[]
\CommentTok{\# "variables" to be reuse later}
\FunctionTok{common}\KeywordTok{:}

\AttributeTok{  }\FunctionTok{mapsplits}\KeywordTok{:}\AttributeTok{ }\OtherTok{\&mapsplits}
\AttributeTok{    }\FunctionTok{mapsplits}\KeywordTok{:}
\AttributeTok{      }\KeywordTok{{-} }\AttributeTok{FIRST\_SECOND}
\AttributeTok{      }\CommentTok{\# {-} FIFTH\_SEPARATE}
\AttributeTok{      }\CommentTok{\# {-} FOURTH\_SEPARATE}
\AttributeTok{      }\CommentTok{\# {-} THIRD\_SEPARATE}
\AttributeTok{      }\KeywordTok{{-} }\AttributeTok{RS\_MIDDLE}
\AttributeTok{      }\KeywordTok{{-} }\AttributeTok{RM\_SET}
\AttributeTok{      }\KeywordTok{{-} }\AttributeTok{SM\_RISE}
\AttributeTok{      }\KeywordTok{{-} }\AttributeTok{LR\_SUBSCAN}
\AttributeTok{      }\KeywordTok{{-} }\AttributeTok{PWV}
\AttributeTok{      }\KeywordTok{{-} }\AttributeTok{GAIN\_BY\_CES}
\AttributeTok{      }\KeywordTok{{-} }\AttributeTok{QU\_PIXEL}
\AttributeTok{      }\KeywordTok{{-} }\AttributeTok{LHS\_RHS}
\AttributeTok{      }\KeywordTok{{-} }\AttributeTok{SUN\_DIST}
\AttributeTok{      }\KeywordTok{{-} }\AttributeTok{SUN\_HORIZON}
\AttributeTok{      }\KeywordTok{{-} }\AttributeTok{MOON\_DIST}
\AttributeTok{      }\KeywordTok{{-} }\AttributeTok{MOON\_HORIZON}
\AttributeTok{      }\KeywordTok{{-} }\AttributeTok{LEAK\_BY\_CES}
\AttributeTok{      }\KeywordTok{{-} }\AttributeTok{LEAK\_BY\_BOLO}
\AttributeTok{      }\KeywordTok{{-} }\AttributeTok{AMP\_2F\_BY\_CES}
\AttributeTok{      }\KeywordTok{{-} }\AttributeTok{AMP\_2F\_BY\_BOLO}
\AttributeTok{      }\KeywordTok{{-} }\AttributeTok{AMP\_4F\_BY\_CES}
\AttributeTok{      }\KeywordTok{{-} }\AttributeTok{AMP\_4F\_BY\_BOLO}
\AttributeTok{      }\CommentTok{\# {-} RANDOM\_BY\_BOLO1}
\AttributeTok{      }\CommentTok{\# {-} RANDOM\_BY\_BOLO2}
\AttributeTok{      }\CommentTok{\# {-} RANDOM\_BY\_BOLO3}
\AttributeTok{      }\CommentTok{\# {-} RANDOM\_BY\_BOLO4}
\AttributeTok{      }\CommentTok{\# {-} RANDOM\_BY\_BOLO5}
\AttributeTok{      }\CommentTok{\# {-} RANDOM\_BY\_BOLO6}
\AttributeTok{      }\CommentTok{\# {-} RANDOM\_BY\_BOLO7}
\AttributeTok{      }\CommentTok{\# {-} RANDOM\_BY\_BOLO8}
\AttributeTok{      }\CommentTok{\# {-} RANDOM\_BY\_CES1}
\AttributeTok{      }\CommentTok{\# {-} RANDOM\_BY\_CES2}
\AttributeTok{      }\CommentTok{\# {-} RANDOM\_BY\_CES3}
\AttributeTok{      }\CommentTok{\# {-} RANDOM\_BY\_CES4}
\AttributeTok{      }\CommentTok{\# {-} RANDOM\_BY\_CES5}
\AttributeTok{      }\CommentTok{\# {-} RANDOM\_BY\_CES6}
\AttributeTok{      }\CommentTok{\# {-} RANDOM\_BY\_CES7}
\AttributeTok{      }\CommentTok{\# {-} RANDOM\_BY\_CES8}
\AttributeTok{      }\KeywordTok{{-} }\AttributeTok{THS\_BHS}
\AttributeTok{      }\KeywordTok{{-} }\AttributeTok{TOP\_BOTTOM}
\AttributeTok{      }\CommentTok{\# {-} RANDOM\_BUT\_PAIR\_BY\_BOLO1}


\CommentTok{\# mapmaking sim iter}

\FunctionTok{mapbolo\_sim}\KeywordTok{:}\AttributeTok{ }\OtherTok{\&boloargs}
\AttributeTok{  }\FunctionTok{gain\_folder}\KeywordTok{:}\AttributeTok{ $SCRATCH/data/largepatch4/gain\_181224\_renorm}
\AttributeTok{  }\FunctionTok{psd\_folder}\KeywordTok{:}\AttributeTok{ $SCRATCH/data/largepatch4/psd\_iota\_poleff\_181224\_renorm}
\AttributeTok{  }\FunctionTok{filterfolder}\KeywordTok{:}\AttributeTok{ $SCRATCH/data/largepatch4/filterdb\_iota\_final}
\AttributeTok{  }\FunctionTok{alt\_flag\_folder}\KeywordTok{:}\AttributeTok{ $SCRATCH/data/largepatch4/iota\_dataselection\_final\_highell\_corr}
\AttributeTok{  }\CommentTok{\# projection args}
\AttributeTok{  }\FunctionTok{width}\KeywordTok{:}\AttributeTok{ }\DecValTok{45}
\AttributeTok{  }\FunctionTok{fixed\_offset}\KeywordTok{:}\AttributeTok{ BICEP}
\AttributeTok{  }\FunctionTok{pixelsize}\KeywordTok{:}\AttributeTok{ }\DecValTok{2}
\AttributeTok{  }\FunctionTok{hwppos}\KeywordTok{:}\AttributeTok{ }\DecValTok{45}
\AttributeTok{  }\FunctionTok{projection}\KeywordTok{:}\AttributeTok{ OLEA}
\AttributeTok{  }\CommentTok{\# filter args}
\AttributeTok{  }\FunctionTok{detrend}\KeywordTok{:}\AttributeTok{ }\DecValTok{2}
\AttributeTok{  }\KeywordTok{? }\FunctionTok{groundlr}
\AttributeTok{  }\FunctionTok{groundtemplate}\KeywordTok{:}\AttributeTok{ }\FloatTok{0.24}
\AttributeTok{  }\FunctionTok{poly}\KeywordTok{:}\AttributeTok{ }\DecValTok{9}
\AttributeTok{  }\FunctionTok{polpoly}\KeywordTok{:}\AttributeTok{ }\DecValTok{9}
\AttributeTok{  }\KeywordTok{? }\FunctionTok{commonmode}
\AttributeTok{  }\KeywordTok{? }\FunctionTok{notch\_lines}
\AttributeTok{  }\FunctionTok{notch\_line\_width}\KeywordTok{:}\AttributeTok{ }\FloatTok{0.01}
\AttributeTok{  }\FunctionTok{mask\_radius}\KeywordTok{:}\AttributeTok{ }\FloatTok{3.5}
\AttributeTok{  }\CommentTok{\# null only}
\AttributeTok{  }\FunctionTok{null\_folder}\KeywordTok{:}\AttributeTok{ $SCRATCH/data/largepatch4/null\_test\_flags\_iota\_highell\_corr\_pair}
\AttributeTok{  }\FunctionTok{\textless{}\textless{}}\KeywordTok{:}\AttributeTok{ }\OtherTok{*mapsplits}

\FunctionTok{mapbolo\_sim\_iter}\KeywordTok{:}\AttributeTok{ }\OtherTok{\&boloinput}
\AttributeTok{  }\FunctionTok{f}\KeywordTok{:}\AttributeTok{ $SCRATCH/data/largepatch4/hdf5\_ces\_dmds}

\CommentTok{\# coadd}

\FunctionTok{coadd}\KeywordTok{:}
\AttributeTok{  }\FunctionTok{basedir}\KeywordTok{:}\AttributeTok{ $SCRATCH/largepatch\_null/high\_1/maps}
\AttributeTok{  }\CommentTok{\# outsubdir: coadd}
\AttributeTok{  }\FunctionTok{\textless{}\textless{}}\KeywordTok{:}\AttributeTok{ }\OtherTok{*mapsplits}

\CommentTok{\# for coadd\_mpi\_iter.py}
\FunctionTok{coadd\_sim}\KeywordTok{:}
\AttributeTok{  }\CommentTok{\# outsubdir: coadd}
\AttributeTok{  }\FunctionTok{\textless{}\textless{}}\KeywordTok{:}\AttributeTok{ }\OtherTok{*mapsplits}

\CommentTok{\# pseudo}

\FunctionTok{pseudo}\KeywordTok{:}
\AttributeTok{  }\CommentTok{\# weight\_name: realmap}
\AttributeTok{  }\FunctionTok{lmax}\KeywordTok{:}\AttributeTok{ }\DecValTok{4901}
\AttributeTok{  }\FunctionTok{pixel\_size}\KeywordTok{:}\AttributeTok{ }\DecValTok{2}

\AttributeTok{  }\FunctionTok{selectionpath}\KeywordTok{:}\AttributeTok{ $SCRATCH/largepatch\_null/high\_1/weight\_nauto.hdf5}
\AttributeTok{  }\FunctionTok{maskpath}\KeywordTok{:}\AttributeTok{ $SCRATCH/largepatch\_null/high\_1/masks.hdf5}
\AttributeTok{  }\FunctionTok{weightdir}\KeywordTok{:}\AttributeTok{ $SCRATCH/largepatch\_null/high\_1/coadd/null\_realmaps}

\FunctionTok{pseudo\_iter}\KeywordTok{:}
\AttributeTok{  }\FunctionTok{\textless{}\textless{}}\KeywordTok{:}\AttributeTok{ }\OtherTok{*mapsplits}

\CommentTok{\# null map to full map}

\FunctionTok{null2full}\KeywordTok{:}
\AttributeTok{  }\FunctionTok{basedir}\KeywordTok{:}\AttributeTok{ $SCRATCH/largepatch\_null/high\_1/coadd}
\AttributeTok{  }\FunctionTok{outbasedir}\KeywordTok{:}\AttributeTok{ $SCRATCH/largepatch\_null/high\_1/coadd\_full}
\CommentTok{\# cat args{-}common.yml before this}

\CommentTok{\# "variables" to be reuse later}
\FunctionTok{template}\KeywordTok{:}

\AttributeTok{  }\FunctionTok{iter}\KeywordTok{:}\AttributeTok{ }\OtherTok{\&iter}
\AttributeTok{    }\FunctionTok{mapcase}\KeywordTok{:}
\AttributeTok{      }\KeywordTok{{-} }\AttributeTok{null\_simmaps\_TTBB}
\AttributeTok{      }\KeywordTok{{-} }\AttributeTok{null\_simmaps\_TTEE}
\AttributeTok{      }\KeywordTok{{-} }\AttributeTok{null\_simmaps\_TTEEBB}
\AttributeTok{    }\FunctionTok{nreal}\KeywordTok{:}
\AttributeTok{      }\KeywordTok{{-} }\DecValTok{128}
\AttributeTok{      }\KeywordTok{{-} }\DecValTok{128}
\AttributeTok{      }\KeywordTok{{-} }\DecValTok{512}
\AttributeTok{    }\FunctionTok{save\_weight}\KeywordTok{:}
\AttributeTok{      }\KeywordTok{{-} }\CharTok{False}
\AttributeTok{      }\KeywordTok{{-} }\CharTok{False}
\AttributeTok{      }\KeywordTok{{-} }\CharTok{False}

\CommentTok{\# mapmaking{-}sim}

\FunctionTok{mapbolo\_whwp}\KeywordTok{:}
\AttributeTok{  }\FunctionTok{\textless{}\textless{}}\KeywordTok{:}\AttributeTok{ }\OtherTok{*boloargs}
\FunctionTok{mapbolo\_whwp\_iter}\KeywordTok{:}
\AttributeTok{  }\FunctionTok{\textless{}\textless{}}\KeywordTok{:}\AttributeTok{ }\OtherTok{*boloinput}
\AttributeTok{  }\FunctionTok{tssim}\KeywordTok{:}
\AttributeTok{    }\KeywordTok{{-} }\AttributeTok{$SCRATCH/data/largepatch\_high/simmaps\_base\_2018\_plikHM\_TTTEEE\_lowl\_lowE\_lensing\_cl\_lensd/maps\_ttbb}
\AttributeTok{    }\KeywordTok{{-} }\AttributeTok{$SCRATCH/data/largepatch\_high/simmaps\_base\_2018\_plikHM\_TTTEEE\_lowl\_lowE\_lensing\_cl\_lensd/maps\_ttee}
\AttributeTok{    }\KeywordTok{{-} }\AttributeTok{$SCRATCH/data/largepatch\_high/simmaps\_base\_2018\_plikHM\_TTTEEE\_lowl\_lowE\_lensing\_cl\_lensd/maps\_tteebb}
\AttributeTok{  }\FunctionTok{whitenoise}\KeywordTok{:}
\AttributeTok{    }\KeywordTok{{-} }\CharTok{False}
\AttributeTok{    }\KeywordTok{{-} }\CharTok{False}
\AttributeTok{    }\KeywordTok{{-} }\CharTok{False}
\AttributeTok{  }\FunctionTok{\textless{}\textless{}}\KeywordTok{:}\AttributeTok{ }\OtherTok{*iter}

\CommentTok{\# coadd{-}sim}

\FunctionTok{coadd\_iter}\KeywordTok{:}
\AttributeTok{  }\FunctionTok{name}\KeywordTok{:}
\AttributeTok{    }\KeywordTok{{-} }\AttributeTok{sim}
\AttributeTok{    }\KeywordTok{{-} }\AttributeTok{sim}
\AttributeTok{    }\KeywordTok{{-} }\AttributeTok{sim}
\AttributeTok{  }\FunctionTok{\textless{}\textless{}}\KeywordTok{:}\AttributeTok{ }\OtherTok{*iter}

\FunctionTok{null2full\_iter}\KeywordTok{:}
\AttributeTok{  }\FunctionTok{mapcase}\KeywordTok{:}
\AttributeTok{    }\KeywordTok{{-} }\AttributeTok{null\_simmaps\_TTBB}
\AttributeTok{    }\KeywordTok{{-} }\AttributeTok{null\_simmaps\_TTEE}
\AttributeTok{    }\KeywordTok{{-} }\AttributeTok{null\_simmaps\_TTEEBB\_noise}
\AttributeTok{  }\FunctionTok{name}\KeywordTok{:}
\AttributeTok{    }\KeywordTok{{-} }\AttributeTok{sim}
\AttributeTok{    }\KeywordTok{{-} }\AttributeTok{sim}
\AttributeTok{    }\KeywordTok{{-} }\AttributeTok{sim}
\AttributeTok{  }\FunctionTok{nreal}\KeywordTok{:}
\AttributeTok{    }\KeywordTok{{-} }\DecValTok{128}
\AttributeTok{    }\KeywordTok{{-} }\DecValTok{128}
\AttributeTok{    }\KeywordTok{{-} }\DecValTok{512}
\AttributeTok{  }\FunctionTok{final}\KeywordTok{:}
\AttributeTok{    }\KeywordTok{{-} }\CharTok{False}
\AttributeTok{    }\KeywordTok{{-} }\CharTok{False}
\AttributeTok{    }\KeywordTok{{-} }\CharTok{False}
\AttributeTok{  }\FunctionTok{rawmaps}\KeywordTok{:}
\AttributeTok{    }\KeywordTok{{-} {-} }\AttributeTok{d0}
\AttributeTok{      }\KeywordTok{{-} }\AttributeTok{d4r}
\AttributeTok{      }\KeywordTok{{-} }\AttributeTok{d4i}
\AttributeTok{    }\KeywordTok{{-} {-} }\AttributeTok{d0}
\AttributeTok{      }\KeywordTok{{-} }\AttributeTok{d4r}
\AttributeTok{      }\KeywordTok{{-} }\AttributeTok{d4i}
\AttributeTok{    }\KeywordTok{{-} {-} }\AttributeTok{d0}
\AttributeTok{      }\KeywordTok{{-} }\AttributeTok{d4r}
\AttributeTok{      }\KeywordTok{{-} }\AttributeTok{d4i}
\end{Highlighting}
\end{Shaded}

\chapter{Changelog of null-tests performed in POLARBEAR large-patch analysis}\label{sec:PB-null-changelog}

\chaptermark{Changelog of the POLARBEAR null-tests}

Below is the changelog of all null-test runs performed in the POLARBEAR analysis in \cref{sec:polarbear-research}.
They are reproduced here for reference.

Note:

\begin{itemize}
\tightlist
\item
  Semantic versioning is used: \texttt{major.minor.bugfix}. Major versions all share the same map-making, for example v0.6 through v0.6.6. Minor versions differ in post-mapmaking, for example how they are grouped to coadd, how the spectra are generated, etc.
\item
  Since v0.13.2.1: revert to plotting \(C_l\) instead.
\item
  Since v0.6.2, \(C_l B_l^2\) is plotted instead.
\item
  Since v0.2.1, the spectra are plotted \emph{without} the dof factor of \(\frac{l (l+1)}{2 \pi}\). That is, the previous spectra are the so-called \(D_l\).
\item
  If you see \texttt{a\ +\ bj}, interpret \texttt{a} as the value and \texttt{b} as the error-bar. It is just a shortcut to represent elements of \(\mathbb{R}^2\) as elements of \(\mathbb{C}\).
\item
  Changelog:
\item
  v1.0.2: The only bug fix is the patch that fix ground template mentioned below:

  \begin{itemize}
  \tightlist
  \item
    Dominic found that tweaking how AnalysisBackend construct the ground template can remove the stripes. I applied a patch at AnalysisBackend commit 07b6610. Merge to cleanup?
  \item
    What the patch does:

    \begin{itemize}
    \tightlist
    \item
      before the patch, a point-source masked TOD is used to construct the ground template
    \item
      with the patch, TOD without point-source mask is used to construct the ground template
    \end{itemize}
  \end{itemize}
\item
  v1.1.1: (superseded by fix in v1.0.2) Only real-mapmaking for debugging stripes in I-maps: change group template from 0.5 to 1.
\item
  v1.1: (superseded by fix in v1.0.2) Only real-mapmaking for debugging stripes in I-maps: change group template from 0.24 to 0.5.
\item
  v1.0.1: add \texttt{-\/-deconvolve} to real mapmaking, everything else identical. That is, sign-flip-noise from real map is not regenerated.
\item
  v1.0.0: identical to v1.0.0-beta except no.\ of TTEEBB simulation increased from 358 to 512.
\item
  v1.0.0-beta:

  \begin{itemize}
  \tightlist
  \item
    updated theory spectra input using Planck 2018 values and CLASS.
  \item
    point source mask:

    \begin{itemize}
    \tightlist
    \item
      boolean mask in mapmaking changed to 3.5 arcmin (from default 5 arcmin)
    \item
      map to pseudo-spectra mask changed to Gaussian (from \(C^\infty\) taper with inner and outer radius) for mode-coupling niceness, with FWHM at 10.5 arcmin.
    \end{itemize}
  \item
    removed null tests that involve random splits, reducing from 24 to 21 null tests.
  \item
    128 TTEE, TTBB simulations; 512 TTEEBB simulations with 358 only included in this report.
  \end{itemize}
\item
  v0.13.4:

  \begin{itemize}
  \tightlist
  \item
    zero-padding on maps changed from 1.3x to 2x
  \item
    ``naïve'' Wiener filter on temperature weights removed
  \end{itemize}
\item
  v0.13.3: use sign-flip noise in TT/EE/BB/noise simulations
\item
  v0.13.2.1: null-spectra plotting changing from \(C_l B_l^2\) to \(C_l\)
\item
  v0.13.2: Fixed a bug that made v0.13.1 wrong. That is, this one has the correct output from \texttt{-\/-correct\_\allowbreak{}poleff}
\item
  v0.13.1: \texttt{-\/-correct\_\allowbreak{}poleff} is applied to real map to correct for the polarization efficiency which effectively rescale the TOD.
\item
  v0.13:

  \begin{itemize}
  \tightlist
  \item
    add point sources: \texttt{fixed\_\allowbreak{}sources\_\allowbreak{}pccsinlp} from \texttt{pointing/\allowbreak{}sources.py}
  \item
    OLEA projection
  \item
    completely new pipeline for post-mapmaking

    \begin{itemize}
    \tightlist
    \item
      no intermediate binning, only 1 single binning step is used in the final step
    \item
      mode-coupling:

      \begin{itemize}
      \tightlist
      \item
        bug fixed
      \item
        with and without point-source mask are compared
      \end{itemize}
    \item
      pseudo-spectra:

      \begin{itemize}
      \tightlist
      \item
        fully implement \texttt{no\_\allowbreak{}diag} variant for every case (also the cross cases has been unified to avoid bugs)
      \item
        should be numerically more stable (due to how the old one binning things)
      \end{itemize}
    \item
      new filter transfer function definition: \(\widetilde{C}_l = \sum_{\ell'} F_\ell M_{\ell \ell'} B_{\ell'}^2 C_{\ell'}\)
    \item
      there's only 1 single mask (per temperature/polarization, that is, the final mask) used across all observations (``10''-day cycles). Previously, each observation has their own mask, where as the mode-coupling matrices are calculated once using the final mask, which is what's done in the low-\(\ell\) pipeline. But since the low-\(\ell\) pipeline calculated the mode-coupling by simulation rather than analytical like the high-\(\ell\) pipeline, this shouldn't be a problem for low-\(\ell\).
    \item
      post mapmaking was more expensive than mapmaking, and it is now much cheaper. Debug QoS can almost accommodate the whole post-mapmaking calculation.
    \end{itemize}
  \item
    binning is done slightly differently in null-spectra
  \end{itemize}
\item
  v0.12:

  \begin{itemize}
  \tightlist
  \item
    use 1 arcmin.
  \item
    pseudo-spectra pipeline requires more memory than Cori I has available per node. So only maps are produced but no spectra.
  \end{itemize}
\item
  v0.11:

  \begin{itemize}
  \tightlist
  \item
    largepatch4 datasets
  \item
    merged \texttt{AnalysisBackend/\allowbreak{}cleanup/\allowbreak{}4254c87} to the high-l post-making pipelines
  \item
    applied the high-l data selection on top of it: No.~of sub-scans: before: 50902628; after: 47995975. \textasciitilde94\% efficiency (requiring pairing would result in only 40405768 sub-scans. \textasciitilde84\% efficiency.)
  \end{itemize}
\item
  v0.10: Same as v0.9, except:

  \begin{itemize}
  \tightlist
  \item
    regenerate null-flags \emph{after} data-selection to ensure its correctness (on splitting the available data in a half)
  \item
    no data-selection that requires top-bottom pairing.
  \item
    24 simulations for TTEE/TTBB; 48 simulations for TTEEBB\@. This is primarily for better p-values, and also as a scaling test.
  \item
    No \texttt{THS\_\allowbreak{}BHS} null-test because of some unknown bug.
  \end{itemize}
\item
  v0.9.1: same as v0.9, except not using \texttt{nodiag} variant mentioned in v0.6.5.
\item
  v0.9: similar to v0.8, but apply an intelligent data-selection on correlations, and also add the pairing data-selection from v0.6.
\item
  v0.8.1: same as v0.8, except not using \texttt{nodiag} variant mentioned in v0.6.5.
\item
  v0.8: apply data selection to cut half amount of data by relations between detectors. Caveats:

  \begin{itemize}
  \tightlist
  \item
    this aggressive data-cut is for showing if it does move the null-spectra or not. Different data-selection schemes will be tested to preserve more data.
  \item
    null flags aren't regenerated (that is, this additional data selection is done after the original data-selection and null-flag generations)
  \item
    \texttt{LEAK\_\allowbreak{}BY\_\allowbreak{}BOLO}, \texttt{LHS\_\allowbreak{}RHS}, \texttt{THS\_\allowbreak{}BHS} cannot be generated (probably just a data volume problem between null-splits)
  \item
    the last l-bin can be very large (again perhaps just data volume problem)
  \item
    \texttt{QU\_\allowbreak{}PIXEL} with fine bin is very bad (perhaps it's the same problem with the mismatch between fine-bin and large-bin result that we can't believe the fine-bin spectra)
  \end{itemize}
\item
  v0.7.1: same as v0.7, except not using \texttt{nodiag} variant mentioned in v0.6.5.
\item
  v0.7: turn off common-mode filter, otherwise same as v0.6.5.
\item
  v0.6.6: same as v0.6.4, except not using \texttt{nodiag} variant mentioned in v0.6.5.
\item
  v0.6.5:

  \begin{itemize}
  \tightlist
  \item
    back to v0.6.2 config, except
  \item
    in real-maps, replace \texttt{form\_\allowbreak{}2d\_\allowbreak{}cross\_\allowbreak{}spectra} with \texttt{nodiag} variant. If the noise between 2 null-splits are correlated, this would improve it. See Neil's ``Expanding out the QUIET cross null spectrum''.
  \end{itemize}
\item
  v0.6.4: use roughly quarter volume of data to generate null-spectra, to test if some of the null-spectra is really noise-like (\texttt{THS\_\allowbreak{}BHS} is missing due to an error)
\item
  v0.6.3: Coadd by season (instead of by 10-day cycle), that is, from 37 observations down to 3. Note that the vertical axis is still \(C_l B_l^2\).
\item
  v0.6.2: plot \(C_l B_l^2\) instead.
\item
  v0.6.1: add noise to TTEEBB sim.
\item
  v0.6:

  \begin{itemize}
  \tightlist
  \item
    back to v0.3's configuration, except
  \item
    with most null-tests in v0.2 plus new ones
  \item
    new gain from Anh
  \item
    add data-selection that requires t/b-pairing
  \item
    add a null-test \texttt{RANDOM\_\allowbreak{}BUT\_\allowbreak{}PAIR\_\allowbreak{}BY\_\allowbreak{}BOLO1} that is random but both t/b-pair always come together.
  \item
    \texttt{-\/-width} changed to 45 (between 45 to 70.3 there's only tiny bit of signals that's not worth saving, considering the extra computation requirements that would impose.)
  \item
    changed definition of \texttt{get\_\allowbreak{}cross\_\allowbreak{}weight} in my pseudo-spectra pipeline from \(\frac{1}{w} = \frac{1}{2} \left( \frac{1}{w_1} + \frac{1}{w_2} \right)\) to \(\frac{1}{w} = \frac{1}{w_1} + \frac{1}{w_2}\)
  \item
    (accidentally) no white-noise in the TTEEBB simulations
  \item
    backward incompatible HDF5 maps: new-ish mode-coupling and pseudo-spectra pipelines with a different HDF5 maps structure (for example maps are saved in 2d array.)
  \end{itemize}
\item
  v0.5:

  \begin{itemize}
  \tightlist
  \item
    applies heavy data-selection on T, in the hopes that it suppress the I to Q/U leakage in high-\(l\).
  \item
    add \texttt{TOP\_\allowbreak{}BOTTOM} split.
  \item
    (\texttt{-\/-mask-\allowbreak{}cross} hasn't been applied due to a misunderstanding of what the code does. But the diff.\ is at \textasciitilde4\% level in the apodization masks, also known as \texttt{tpweight}/\texttt{tpmask}, which should not make too much difference in the perceived severe failing of the null-tests.)
  \end{itemize}
\item
  v0.4: identical with v0.3 except for \texttt{-\/-notch\_\allowbreak{}line\_\allowbreak{}width:\ 0.04}, to see if it helps the null spectra at all.
\item
  v0.3:

  \begin{itemize}
  \tightlist
  \item
    config:

    \begin{itemize}
    \tightlist
    \item
      null-tests covers only the 4 worst failing null-test from v0.2: \texttt{QU\_\allowbreak{}PIXEL}, \texttt{LEAK\_\allowbreak{}BY\_\allowbreak{}BOLO}, \texttt{AMP\_\allowbreak{}2F\_\allowbreak{}BY\_\allowbreak{}BOLO}, \texttt{AMP\_\allowbreak{}4F\_\allowbreak{}BY\_\allowbreak{}BOLO}

      \begin{itemize}
      \tightlist
      \item
        add \texttt{RANDOM\_\allowbreak{}BY\_\allowbreak{}BOLO1}
      \end{itemize}
    \item
      \texttt{-\/-groundlr} to separately subtract the ground template on L/R scan
    \item
      \texttt{-\/-notch\_\allowbreak{}lines\ -\/-notch\_\allowbreak{}line\_\allowbreak{}width:\ 0.01} instead of \texttt{-\/-nofft}: perform FFT and then applies notch filtering
    \end{itemize}
  \item
    data:

    \begin{itemize}
    \tightlist
    \item
      Satoru's new HDF5
    \item
      Anh's new patch center (should be affecting the output maps only but not the spectra)
    \item
      Anh's new gain
    \item
      Neil's notch filter
    \item
      Kolen's new coadd (then doesn't save the I/Q/U and have a bunch of computational improvements)
    \end{itemize}
  \end{itemize}
\item
  v0.2.3.: ``gain matched'' between the two splits for each spectra
\item
  v0.2.2:

  \begin{itemize}
  \tightlist
  \item
    large bin at 300, fine bin at 30 (just above correlation length)
  \item
    add EE \(\chi\)-statistics.
  \end{itemize}
\end{itemize}

\chapter{Data released internally in POLARBEAR}\label{sec:POLARBEAR-data-release}

\section{Data Release---Pseudo-spectra}\label{data-releasepseudo-spectra}

\subsection{Introduction}\label{introduction-4}

The data version uniquely specified the beam spectra, pseudo spectra, mode-coupling matrices, and theory spectra. The content of these files has never been changed since the simulation runs at NERSC\@. (While in earlier internal versioning, it is in a different container.)

3 versions are released, v1.0.0, v1.0.1, v1.0.2, together with the masks:

\begin{verbatim}
high_1.hdf5
high_1_0_1.hdf5
high_1_0_2.hdf5
masks.hdf5
\end{verbatim}

It is released at NERSC in \texttt{/\allowbreak{}global/\allowbreak{}cfs/\allowbreak{}cdirs/\allowbreak{}polar/\allowbreak{}www/\allowbreak{}khcheung/\allowbreak{}largepatch\_\allowbreak{}new/\allowbreak{}}.

\texttt{high\_\allowbreak{}1\_\allowbreak{}0\_\allowbreak{}1.hdf5} and \texttt{high\_\allowbreak{}1\_\allowbreak{}0\_\allowbreak{}2.hdf5} uses HDF5 external symlinks to point to data in \texttt{high\_\allowbreak{}1.hdf5}. In practice it means both \texttt{high\_\allowbreak{}1.hdf5} and \texttt{high\_\allowbreak{}1\_\allowbreak{}0\_\allowbreak{}2.hdf5} have to be in the same directory in order to read v1.0.2 data.

\subsection{Configurations}\label{configurations}

\begin{itemize}
\tightlist
\item
  The mapmaking is done using the AnalysisBackend \texttt{khcheung} branch
\item
  The post-mapmaking pipeline (masks, pseudo-spectra, mode-coupling) is done using the software TAIL below
\item
  The configuration of the software is summarized in \cref{sec:POLARBEAR-configuration} to reproduce v1.0.0. \Cref{sec:PB-null-changelog} has recipe to reproduce v1.0.1 and v1.0.2.
\end{itemize}

\subsection{Specifications---General}\label{specificationsgeneral}

All \texttt{l} start at 0, and no binning. That is, if a shape is 4901 then it spans the l-range from 0-4900.

\subsection{Specifications---Masks}\label{specificationsmasks}

\texttt{p} is the polarization mask. \texttt{t} is the temperature mask. Both in map-domain.

\begin{Shaded}
\begin{Highlighting}[]
\ExtensionTok{$}\OperatorTok{\textgreater{}}\NormalTok{ h5ls }\ExtensionTok{{-}r}\NormalTok{ masks.hdf5 }
\ExtensionTok{/}\NormalTok{                        Group}
\ExtensionTok{/p}\NormalTok{                       Dataset \{2880, 2880\}}
\ExtensionTok{/t}\NormalTok{                       Dataset \{2880, 2880\}}
\end{Highlighting}
\end{Shaded}

\subsection{Specifications---Beam and theory spectra}\label{specificationsbeam-and-theory-spectra}

The data is put in a simple HDF5 container and is programming language agnostic.

This session should be self-explanatory, the datasets within each group are all one-dimensional with the same size. \texttt{l} gives the l-value and for example \texttt{TT} gives the \(TT\) spectrum at those \texttt{l}-values.

\begin{Shaded}
\begin{Highlighting}[]
\ExtensionTok{$}\OperatorTok{\textgreater{}}\NormalTok{ h5ls }\ExtensionTok{{-}r}\NormalTok{ high\_1.hdf5}
\ExtensionTok{...}
\ExtensionTok{/beam\_spectra}\NormalTok{            Group}
\ExtensionTok{/beam\_spectra/10.1}\NormalTok{       Dataset \{293\}}
\ExtensionTok{/beam\_spectra/10.2}\NormalTok{       Dataset \{293\}}
\ExtensionTok{/beam\_spectra/10.3}\NormalTok{       Dataset \{293\}}
\ExtensionTok{/beam\_spectra/10.4}\NormalTok{       Dataset \{293\}}
\ExtensionTok{/beam\_spectra/10.5}\NormalTok{       Dataset \{293\}}
\ExtensionTok{/beam\_spectra/8.2.0}\NormalTok{      Dataset \{293\}}
\ExtensionTok{/beam\_spectra/9.4}\NormalTok{        Dataset \{293\}}
\ExtensionTok{/beam\_spectra/all}\NormalTok{        Dataset \{293\}}
\ExtensionTok{/beam\_spectra/l}\NormalTok{          Dataset \{293\}}
\ExtensionTok{...}
\ExtensionTok{/theory\_spectra}\NormalTok{          Group}
\ExtensionTok{/theory\_spectra/BB}\NormalTok{       Dataset \{4901\}}
\ExtensionTok{/theory\_spectra/EE}\NormalTok{       Dataset \{4901\}}
\ExtensionTok{/theory\_spectra/TE}\NormalTok{       Dataset \{4901\}}
\ExtensionTok{/theory\_spectra/TT}\NormalTok{       Dataset \{4901\}}
\ExtensionTok{/theory\_spectra/dE}\NormalTok{       Dataset \{4901\}}
\ExtensionTok{/theory\_spectra/dT}\NormalTok{       Dataset \{4901\}}
\ExtensionTok{/theory\_spectra/dd}\NormalTok{       Dataset \{4901\}}
\ExtensionTok{/theory\_spectra/dens[1]{-}dens[1]}\NormalTok{ Dataset \{4901\}}
\ExtensionTok{/theory\_spectra/dens[1]{-}lens[1]}\NormalTok{ Dataset \{4901\}}
\ExtensionTok{/theory\_spectra/l}\NormalTok{        Dataset \{4901\}}
\ExtensionTok{/theory\_spectra/lens[1]{-}lens[1]}\NormalTok{ Dataset \{4901\}}
\ExtensionTok{/theory\_spectra/phi{-}dens[1]}\NormalTok{ Dataset \{4901\}}
\ExtensionTok{...}
\end{Highlighting}
\end{Shaded}

\subsection{Specifications---Mode-coupling matrices}\label{specificationsmode-coupling-matrices}

The names should be self-explanatory. The difference between \texttt{EBEB} and \texttt{EBEB\_\allowbreak{}pseudo} follows \texttt{AnalysisBackend} (although generated in TAIL.)

\begin{Shaded}
\begin{Highlighting}[]
\ExtensionTok{$}\OperatorTok{\textgreater{}}\NormalTok{ h5ls }\ExtensionTok{{-}r}\NormalTok{ high\_1.hdf5}
\ExtensionTok{...}
\ExtensionTok{/mode\_coupling}\NormalTok{           Group}
\ExtensionTok{/mode\_coupling/BBBB}\NormalTok{      Dataset \{4901, 4901\}}
\ExtensionTok{/mode\_coupling/EBEB}\NormalTok{      Dataset \{4901, 4901\}}
\ExtensionTok{/mode\_coupling/EBEB\_pseudo}\NormalTok{ Dataset \{4901, 4901\}}
\ExtensionTok{/mode\_coupling/EEEE}\NormalTok{      Dataset \{4901, 4901\}}
\ExtensionTok{/mode\_coupling/TBTB}\NormalTok{      Dataset \{4901, 4901\}}
\ExtensionTok{/mode\_coupling/TETE}\NormalTok{      Dataset \{4901, 4901\}}
\ExtensionTok{/mode\_coupling/TTTT}\NormalTok{      Dataset \{4901, 4901\}}
\ExtensionTok{...}
\end{Highlighting}
\end{Shaded}

\subsection{Specifications---pseudo-spectra}\label{specificationspseudo-spectra}

\begin{itemize}
\tightlist
\item
  full/null indicates it is from the full map data or null-split data.
\item
  \texttt{realmaps} is from real map data, \texttt{simmaps} is from simulated map data.
\item
  \texttt{TTBB}/\texttt{TTEE}/\texttt{TTEEBB} represent the power included in the input map.
\item
  \texttt{noise}: generated from random sign-flip-noise from \texttt{realmaps}
\item
  Each of these datasets are of 6 dimensions. In this order:

  \begin{itemize}
  \tightlist
  \item
    spectra: the 6 different spectra: \(TT\), \(TE\), \ldots{}
  \item
    null-split: the 21 null-split: \texttt{AMP\_\allowbreak{}2F\_\allowbreak{}BY\_\allowbreak{}BOLO}, \ldots{}
  \item
    sub-split: 0/1/2. 0 is the 1st-half, 1 is the 2nd-half, 2 is the cross-spectrum between the 1st- and 2nd-half.
  \item
    l: self-explanatory
  \item
    sub-spectra: Cl, Nl. Cl is the signal spectra (using the cross-spectrum estimator), Nl is the noise spectra.
  \item
    n: n-th realization
  \end{itemize}
\end{itemize}

See more on \texttt{meta} below.

\begin{Shaded}
\begin{Highlighting}[]
\ExtensionTok{$}\OperatorTok{\textgreater{}}\NormalTok{ h5ls }\ExtensionTok{{-}r}\NormalTok{ high\_1.hdf5}
\ExtensionTok{...}
\ExtensionTok{/full}\NormalTok{                    Group}
\ExtensionTok{/full/pseudo\_spectra}\NormalTok{     Group}
\ExtensionTok{/full/pseudo\_spectra/realmaps}\NormalTok{ Dataset \{6, 1, 1, 4901, 2, 1\}}
\ExtensionTok{/full/pseudo\_spectra/simmaps\_TTBB}\NormalTok{ Dataset \{6, 1, 1, 4901, 2, 128\}}
\ExtensionTok{/full/pseudo\_spectra/simmaps\_TTEE}\NormalTok{ Dataset \{6, 1, 1, 4901, 2, 128\}}
\ExtensionTok{/full/pseudo\_spectra/simmaps\_TTEEBB\_noise}\NormalTok{ Dataset \{6, 1, 1, 4901, 2, 512\}}
\ExtensionTok{...}
\ExtensionTok{/null}\NormalTok{                    Group}
\ExtensionTok{/null/pseudo\_spectra}\NormalTok{     Group}
\ExtensionTok{/null/pseudo\_spectra/realmaps}\NormalTok{ Dataset \{6, 21, 3, 4901, 2, 1\}}
\ExtensionTok{/null/pseudo\_spectra/simmaps\_TTBB}\NormalTok{ Dataset \{6, 21, 3, 4901, 2, 128\}}
\ExtensionTok{/null/pseudo\_spectra/simmaps\_TTEE}\NormalTok{ Dataset \{6, 21, 3, 4901, 2, 128\}}
\ExtensionTok{/null/pseudo\_spectra/simmaps\_TTEEBB\_noise}\NormalTok{ Dataset \{6, 21, 3, 4901, 2, 512\}}
\ExtensionTok{...}
\end{Highlighting}
\end{Shaded}

\subsection{Specifications---meta}\label{specificationsmeta}

\begin{itemize}
\tightlist
\item
  \texttt{null\_\allowbreak{}split}: the names of the 21 null-split
\item
  \texttt{spectra}: the names of the 6 spectra
\item
  \texttt{sub\_\allowbreak{}spectra}: the names as Cl, Nl
\end{itemize}

\begin{Shaded}
\begin{Highlighting}[]
\ExtensionTok{$}\OperatorTok{\textgreater{}}\NormalTok{ h5ls }\ExtensionTok{{-}r}\NormalTok{ high\_1.hdf5}
\ExtensionTok{...}
\ExtensionTok{/meta}\NormalTok{                    Group}
\ExtensionTok{/meta/null\_split}\NormalTok{         Dataset \{21\}}
\ExtensionTok{/meta/spectra}\NormalTok{            Dataset \{6\}}
\ExtensionTok{/meta/sub\_spectra}\NormalTok{        Dataset \{2\}}
\ExtensionTok{...}
\end{Highlighting}
\end{Shaded}

\section{Data Release---Spectra---final freeze}\label{data-releasespectrafinal-freeze}

\subsection{Final freeze configuration}\label{final-freeze-configuration}

This data is released as \texttt{high\_\allowbreak{}1\_\allowbreak{}0\_\allowbreak{}2\_\allowbreak{}spectra\_\allowbreak{}final.hdf5}. It contains both bin-width of 50 and 100.

Using TAIL,

\begin{Shaded}
\begin{Highlighting}[]
\ExtensionTok{solve\_spectra}\NormalTok{ high\_1\_0\_2.hdf5 }\AttributeTok{{-}o}\NormalTok{ high\_1\_0\_2\_spectra\_final.hdf5 }\AttributeTok{{-}{-}scaling} \AttributeTok{{-}{-}sim{-}noise{-}matching} \AttributeTok{{-}b}\NormalTok{ 50 100 }\AttributeTok{{-}{-}return{-}w}
\ExtensionTok{iterative\_map}\NormalTok{ high\_1\_0\_2.hdf5 masks.hdf5 }\AttributeTok{{-}o}\NormalTok{ high\_1\_0\_2\_spectra\_final.hdf5 }\AttributeTok{{-}{-}scaling} \AttributeTok{{-}{-}sim{-}noise{-}matching} \AttributeTok{{-}b}\NormalTok{ 100 }\AttributeTok{{-}{-}l{-}T{-}cutoff}\NormalTok{ 1600 }\AttributeTok{{-}n}\NormalTok{ 100}
\ExtensionTok{iterative\_map}\NormalTok{ high\_1\_0\_2.hdf5 masks.hdf5 }\AttributeTok{{-}o}\NormalTok{ high\_1\_0\_2\_spectra\_final.hdf5 }\AttributeTok{{-}{-}scaling} \AttributeTok{{-}{-}sim{-}noise{-}matching} \AttributeTok{{-}b}\NormalTok{  50 }\AttributeTok{{-}{-}l{-}T{-}cutoff}\NormalTok{ 1550 }\AttributeTok{{-}n}\NormalTok{ 100}
\end{Highlighting}
\end{Shaded}

Whenever MC error-bar is calculated, leave \texttt{Spectra} at default attributes: \texttt{ev\_\allowbreak{}est=\textquotesingle{}signal\textquotesingle{},\ ddof=None}.

Note: difference between final freeze and stage 2 unblinding spectra

\begin{itemize}
\tightlist
\item
  all options mentioned in {[}Changes and configurable options{]} is changed here comparing to those in {[}Reproducing null-spectra and full-spectra at unblinding{]}.
\item
  although between now and unblinding, sometimes bin-width of 50 is used in presentations, here bin-width of 100 is used to be consistent with the unblinding configuration. For the reason of optimizing bin-width, see the section ``Choosing ℓ-min, ℓ-max, bin-width'' of the internal POLARBEAR document \emph{Unblinding Proposal---Stage 1}
\end{itemize}

\subsection{Specifications---Filter Transfer Functions}\label{specificationsfilter-transfer-functions}

\begin{itemize}
\tightlist
\item
  full/null indicates it is from the full map data or null-split data.
\item
  \texttt{/\allowbreak{}full/\allowbreak{}filter\_\allowbreak{}transfer/\allowbreak{}l} and \texttt{/\allowbreak{}null/\allowbreak{}filter\_\allowbreak{}transfer/\allowbreak{}l} specify the filter l-range. Here the 2 elements are 50, 4300, meaning the l-range is from 50--4300 (no binning, that is, dl=1)
\item
  \texttt{/\allowbreak{}null/\allowbreak{}filter\_\allowbreak{}transfer/\allowbreak{}F} and \texttt{/\allowbreak{}full/\allowbreak{}filter\_\allowbreak{}transfer/\allowbreak{}F} are 4d-array, same as first 4 dimensions from pseudo-spectra. See more in \hyperref[specificationspseudo-spectra]{Specifications---pseudo-spectra}.
\item
  The first dimension (spectra) order is specified in \texttt{/\allowbreak{}meta/\allowbreak{}filter\_\allowbreak{}transfer}. For example, \texttt{{[}...,\ BB,\ EE,\ TT{]}} specifies the 3 auto-spectra cases.
\end{itemize}

\begin{Shaded}
\begin{Highlighting}[]
\ExtensionTok{$}\OperatorTok{\textgreater{}}\NormalTok{ h5ls }\ExtensionTok{{-}r}\NormalTok{ high\_1\_0\_2\_spectra\_final.hdf5 }
\ExtensionTok{/}\NormalTok{                        Group}
\ExtensionTok{/full}\NormalTok{                    Group}
\ExtensionTok{/full/filter\_transfer}\NormalTok{    Group}
\ExtensionTok{/full/filter\_transfer/F}\NormalTok{  Dataset \{3, 1, 1, 4250\}}
\ExtensionTok{/full/filter\_transfer/l}\NormalTok{  Dataset \{2\}}
\ExtensionTok{...}
\ExtensionTok{/meta}\NormalTok{                    Group}
\ExtensionTok{/meta/filter\_transfer}\NormalTok{    Dataset \{5\}}
\ExtensionTok{...}
\ExtensionTok{/null}\NormalTok{                    Group}
\ExtensionTok{/null/filter\_transfer}\NormalTok{    Group}
\ExtensionTok{/null/filter\_transfer/F}\NormalTok{  Dataset \{3, 21, 3, 4250\}}
\ExtensionTok{/null/filter\_transfer/l}\NormalTok{  Dataset \{2\}}
\ExtensionTok{...}
\end{Highlighting}
\end{Shaded}

\subsection{Specifications---spectra}\label{specificationsspectra}

\begin{itemize}
\tightlist
\item
  similar to pseudo-spectra, see more in \hyperref[specificationspseudo-spectra]{Specifications---pseudo-spectra}. Same 6 dimensions except that l becomes b (bin). \texttt{meta} is also the same.
\item
  there's one more sub-level, \texttt{lmin\_\allowbreak{}lmax\_\allowbreak{}binwidth} such as \texttt{50\_\allowbreak{}4300\_\allowbreak{}50}, indicating the l-min, l-max and bin-width used when solving for spectra.
\item
  \texttt{/\allowbreak{}*/\allowbreak{}spectra/\allowbreak{}*/\allowbreak{}thetas\_\allowbreak{}map} are the maximum a posteriori thetas found from the likelihood code that transform the spectra (including absolute gain and angle.)
\item
  \texttt{/\allowbreak{}*/\allowbreak{}spectra/\allowbreak{}*/\allowbreak{}w\_\allowbreak{}bl}: band pass window function. 5 dimension: spectra, null-split, sub-split, b (bin), l.
\end{itemize}

\begin{Shaded}
\begin{Highlighting}[]
\ExtensionTok{$}\OperatorTok{\textgreater{}}\NormalTok{ h5ls }\ExtensionTok{{-}r}\NormalTok{ high\_1\_0\_2\_spectra\_final.hdf5 }
\ExtensionTok{...}
\ExtensionTok{/full/spectra}\NormalTok{            Group}
\ExtensionTok{/full/spectra/100\_4300\_100}\NormalTok{ Group}
\ExtensionTok{/full/spectra/100\_4300\_100/map\_cases}\NormalTok{ Group}
\ExtensionTok{/full/spectra/100\_4300\_100/map\_cases/realmaps}\NormalTok{ Dataset \{6, 1, 1, 42, 2, 1\}}
\ExtensionTok{/full/spectra/100\_4300\_100/map\_cases/simmaps\_TTBB}\NormalTok{ Dataset \{6, 1, 1, 42, 2, 128\}}
\ExtensionTok{/full/spectra/100\_4300\_100/map\_cases/simmaps\_TTEE}\NormalTok{ Dataset \{6, 1, 1, 42, 2, 128\}}
\ExtensionTok{/full/spectra/100\_4300\_100/map\_cases/simmaps\_TTEEBB\_noise}\NormalTok{ Dataset \{6, 1, 1, 42, 2, 512\}}
\ExtensionTok{/full/spectra/100\_4300\_100/thetas\_map}\NormalTok{ Dataset \{6\}}
\ExtensionTok{/full/spectra/100\_4300\_100/w\_bl}\NormalTok{ Dataset \{6, 1, 1, 42, 4200\}}
\ExtensionTok{/full/spectra/50\_4300\_50}\NormalTok{ Group}
\ExtensionTok{/full/spectra/50\_4300\_50/map\_cases}\NormalTok{ Group}
\ExtensionTok{/full/spectra/50\_4300\_50/map\_cases/realmaps}\NormalTok{ Dataset \{6, 1, 1, 85, 2, 1\}}
\ExtensionTok{/full/spectra/50\_4300\_50/map\_cases/simmaps\_TTBB}\NormalTok{ Dataset \{6, 1, 1, 85, 2, 128\}}
\ExtensionTok{/full/spectra/50\_4300\_50/map\_cases/simmaps\_TTEE}\NormalTok{ Dataset \{6, 1, 1, 85, 2, 128\}}
\ExtensionTok{/full/spectra/50\_4300\_50/map\_cases/simmaps\_TTEEBB\_noise}\NormalTok{ Dataset \{6, 1, 1, 85, 2, 512\}}
\ExtensionTok{/full/spectra/50\_4300\_50/thetas\_map}\NormalTok{ Dataset \{6\}}
\ExtensionTok{/full/spectra/50\_4300\_50/w\_bl}\NormalTok{ Dataset \{6, 1, 1, 85, 4250\}}
\ExtensionTok{/meta}\NormalTok{                    Group}
\ExtensionTok{...}
\ExtensionTok{/meta/null\_split}\NormalTok{         Dataset \{21\}}
\ExtensionTok{/meta/spectra}\NormalTok{            Dataset \{6\}}
\ExtensionTok{/meta/sub\_spectra}\NormalTok{        Dataset \{2\}}
\ExtensionTok{...}
\ExtensionTok{/null/spectra}\NormalTok{            Group}
\ExtensionTok{/null/spectra/100\_4300\_100}\NormalTok{ Group}
\ExtensionTok{/null/spectra/100\_4300\_100/map\_cases}\NormalTok{ Group}
\ExtensionTok{/null/spectra/100\_4300\_100/map\_cases/realmaps}\NormalTok{ Dataset \{6, 21, 3, 42, 2, 1\}}
\ExtensionTok{/null/spectra/100\_4300\_100/map\_cases/simmaps\_TTBB}\NormalTok{ Dataset \{6, 21, 3, 42, 2, 128\}}
\ExtensionTok{/null/spectra/100\_4300\_100/map\_cases/simmaps\_TTEE}\NormalTok{ Dataset \{6, 21, 3, 42, 2, 128\}}
\ExtensionTok{/null/spectra/100\_4300\_100/map\_cases/simmaps\_TTEEBB\_noise}\NormalTok{ Dataset \{6, 21, 3, 42, 2, 512\}}
\ExtensionTok{/null/spectra/100\_4300\_100/w\_bl}\NormalTok{ Dataset \{6, 21, 3, 42, 4200\}}
\ExtensionTok{/null/spectra/50\_4300\_50}\NormalTok{ Group}
\ExtensionTok{/null/spectra/50\_4300\_50/map\_cases}\NormalTok{ Group}
\ExtensionTok{/null/spectra/50\_4300\_50/map\_cases/realmaps}\NormalTok{ Dataset \{6, 21, 3, 85, 2, 1\}}
\ExtensionTok{/null/spectra/50\_4300\_50/map\_cases/simmaps\_TTBB}\NormalTok{ Dataset \{6, 21, 3, 85, 2, 128\}}
\ExtensionTok{/null/spectra/50\_4300\_50/map\_cases/simmaps\_TTEE}\NormalTok{ Dataset \{6, 21, 3, 85, 2, 128\}}
\ExtensionTok{/null/spectra/50\_4300\_50/map\_cases/simmaps\_TTEEBB\_noise}\NormalTok{ Dataset \{6, 21, 3, 85, 2, 512\}}
\ExtensionTok{/null/spectra/50\_4300\_50/w\_bl}\NormalTok{ Dataset \{6, 21, 3, 85, 4250\}}
\end{Highlighting}
\end{Shaded}

\section{Data Release---Real maps}\label{data-releasereal-maps}

It is released at NERSC in \texttt{/\allowbreak{}global/\allowbreak{}cfs/\allowbreak{}cdirs/\allowbreak{}polar/\allowbreak{}www/\allowbreak{}khcheung/\allowbreak{}largepatch\_\allowbreak{}null/\allowbreak{}high\_\allowbreak{}1\_\allowbreak{}0\_\allowbreak{}2/\allowbreak{}}.

It contains all the maps created from the real data (but none of the simulated data.) For example, to find the final coadded maps, it is located in \texttt{/\allowbreak{}global/\allowbreak{}cfs/\allowbreak{}cdirs/\allowbreak{}polar/\allowbreak{}www/\allowbreak{}khcheung/\allowbreak{}largepatch\_\allowbreak{}null/\allowbreak{}high\_\allowbreak{}1\_\allowbreak{}0\_\allowbreak{}2/\allowbreak{}coadd\_\allowbreak{}full/\allowbreak{}null\_\allowbreak{}realmaps/\allowbreak{}coadd/\allowbreak{}realmap.hdf5} at NERSC.

\subsection{Specifications}\label{specifications}

\begin{Shaded}
\begin{Highlighting}[]
\ExtensionTok{$}\OperatorTok{\textgreater{}}\NormalTok{ h5ls }\ExtensionTok{{-}r}\NormalTok{ /global/cfs/cdirs/polar/www/khcheung/largepatch\_null/high\_1\_0\_2/coadd\_full/null\_realmaps/coadd/realmap.hdf5}
\ExtensionTok{/}\NormalTok{                        Group}
\ExtensionTok{/d0}\NormalTok{                      Dataset \{1351, 1351\}}
\ExtensionTok{/d4i}\NormalTok{                     Dataset \{1351, 1351\}}
\ExtensionTok{/d4r}\NormalTok{                     Dataset \{1351, 1351\}}
\ExtensionTok{/w0}\NormalTok{                      Dataset \{1351, 1351\}}
\ExtensionTok{/w4}\NormalTok{                      Dataset \{1351, 1351\}}
\end{Highlighting}
\end{Shaded}

They are all 2D-array, with \texttt{w} indicating the total weights, and \texttt{d} indicating the coadded signal. \texttt{0} is the 0-f component, \texttt{4} is the 4-f component, \texttt{r} is the real part, \texttt{i} is the imaginary part.

For example,

\[\begin{aligned}
I   &= \frac{d_0}{w_0}  \\
Q   &= \frac{d_4^r}{w_4}  \\
U   &= \frac{d_4^i}{w_4}  \\
\end{aligned}\]

These maps are projected using the OLEA projection, as requested from the high-\(\ell\) \(EE\) pipeline. An example implementation of the projector is in \url{https://github.com/ickc/TAIL/blob/3b4e49d6d5a4/tail/pointsource.py\#L10-L22}.

Also, for null-maps, there may be additional HDF5 group level with the name of respective null-split, which would be self-explanatory.

\backmatter
\end{document}